\pdfoutput=1
\documentclass[12pt,a4paper,floatfix]{article}
\usepackage{lineno}    
\usepackage{graphicx}  
\usepackage{xspace}    
\usepackage{color}
\usepackage{colortbl}
\usepackage[fleqn]{amsmath} 
\usepackage{ifthen}         
\usepackage{dcolumn}
\usepackage{tabularx}
\usepackage{amssymb}
\usepackage{amsfonts}
\usepackage{upgreek}     
\usepackage[bookmarks=false]{hyperref}
\usepackage[all]{hypcap} 
\newcolumntype{d}[1]{D{.}{.}{#1}}
\newboolean{articletitles}
\setboolean{articletitles}{true} 
\newboolean{uprightparticles}
\setboolean{uprightparticles}{false} 

\def\ux85 {UX85\xspace}

\def\babar  {BaBar\xspace}

\ifthenelse{\boolean{uprightparticles}}%
{

 \def\PDelta      {\ensuremath{\Delta}\xspace}                 
 \def\PXi      {\ensuremath{\Xi}\xspace}                 
 \def\PLambda      {\ensuremath{\Lambda}\xspace}                 
 \def\PSigma      {\ensuremath{\Sigma}\xspace}                 
 \def\POmega      {\ensuremath{\Omega}\xspace}                 
 \def\PUpsilon      {\ensuremath{\Upsilon}\xspace}                 
 
 \def\PB      {\ensuremath{\mathrm{B}}\xspace}                 
                  
 \def\PD      {\ensuremath{\mathrm{D}}\xspace}

 \def\PK      {\ensuremath{\mathrm{K}}\xspace}

 \def\Pi      {\ensuremath{\mathrm{i}}\xspace}

}
{

 \mathchardef\PDelta="7101
 \mathchardef\PXi="7104
 \mathchardef\PLambda="7103
 \mathchardef\PSigma="7106
 \mathchardef\POmega="710A
 \mathchardef\PUpsilon="7107
                  
 \def\PB      {\ensuremath{B}\xspace}                 
                  
 \def\PD      {\ensuremath{D}\xspace}

 \def\PK      {\ensuremath{K}\xspace}

 \def\Pi      {\ensuremath{i}\xspace}

}

\def\kaon  {\ensuremath{\PK}\xspace}
  \def\Kbar  {\kern 0.2em\overline{\kern -0.2em \PK}{}\xspace}

\def\Kz    {\ensuremath{\kaon^0}\xspace}
\def\Kzb   {\ensuremath{\Kbar^0}\xspace}
\def\KzKzb {\ensuremath{\Kz \kern -0.16em \Kzb}\xspace}
\def\Kp    {\ensuremath{\kaon^+}\xspace}
\def\Km    {\ensuremath{\kaon^-}\xspace}

\def\KpKm  {\ensuremath{\Kp \kern -0.16em \Km}\xspace}

  \def\Dbar    {\kern 0.2em\overline{\kern -0.2em \PD}{}\xspace}
\def\D       {\ensuremath{\PD}\xspace}

\def\Dz      {\ensuremath{\D^0}\xspace}
\def\Dzb     {\ensuremath{\Dbar^0}\xspace}
\def\DzDzb   {\ensuremath{\Dz {\kern -0.16em \Dzb}}\xspace}
\def\Dp      {\ensuremath{\D^+}\xspace}
\def\Dm      {\ensuremath{\D^-}\xspace}

\def\DpDm    {\ensuremath{\Dp {\kern -0.16em \Dm}}\xspace}

\def\B       {\ensuremath{\PB}\xspace}
  \def\Bbar    {\kern 0.18em\overline{\kern -0.18em \PB}{}\xspace}

  \def\Y#1S{\ensuremath{\PUpsilon{(#1S)}}\xspace}

\def\Lbar {\ensuremath{\kern 0.1em\overline{\kern -0.1em\Lambda\kern -0.05em}\kern 0.05em{}}\xspace}

\newcommand{\decay}[2]{\ensuremath{#1\!\to #2}\xspace}         

\def\to                 {\ensuremath{\rightarrow}\xspace}

\def\CP                {\ensuremath{C\!P}\xspace}

\def\AT#1     {\ensuremath{A_{\mathrm{T}}^{#1}}\xspace}           

\def\C#1      {\ensuremath{\mathcal{C}_{#1}}\xspace}                       
\def\Cp#1     {\ensuremath{\mathcal{C}_{#1}^{'}}\xspace}                    
\def\Ceff#1   {\ensuremath{\mathcal{C}_{#1}^{\mathrm{(eff)}}}\xspace}        
\def\Cpeff#1  {\ensuremath{\mathcal{C}_{#1}^{'\mathrm{(eff)}}}\xspace}       
\def\Ope#1    {\ensuremath{\mathcal{O}_{#1}}\xspace}                       
\def\Opep#1   {\ensuremath{\mathcal{O}_{#1}^{'}}\xspace}                    

\newcommand{\bra}[1]{\ensuremath{\langle #1|}}             
\newcommand{\ket}[1]{\ensuremath{|#1\rangle}}              

\newcommand{\tev}{\ensuremath{\mathrm{\,Te\kern -0.1em V}}\xspace}
\newcommand{\gev}{\ensuremath{\mathrm{\,Ge\kern -0.1em V}}\xspace}
\newcommand{\mev}{\ensuremath{\mathrm{\,Me\kern -0.1em V}}\xspace}
\newcommand{\kev}{\ensuremath{\mathrm{\,ke\kern -0.1em V}}\xspace}
\newcommand{\ev}{\ensuremath{\mathrm{\,e\kern -0.1em V}}\xspace}
\newcommand{\gevc}{\ensuremath{{\mathrm{\,Ge\kern -0.1em V\!/}c}}\xspace}
\newcommand{\mevc}{\ensuremath{{\mathrm{\,Me\kern -0.1em V\!/}c}}\xspace}
\newcommand{\gevcc}{\ensuremath{{\mathrm{\,Ge\kern -0.1em V\!/}c^2}}\xspace}
\newcommand{\gevgevcccc}{\ensuremath{{\mathrm{\,Ge\kern -0.1em V^2\!/}c^4}}\xspace}
\newcommand{\mevcc}{\ensuremath{{\mathrm{\,Me\kern -0.1em V\!/}c^2}}\xspace}

\newcommand{\chisq}{\ensuremath{\chi^2}\xspace}

\def\gsim{{~\raise.15em\hbox{$>$}\kern-.85em
          \lower.35em\hbox{$\sim$}~}\xspace}
\def\lsim{{~\raise.15em\hbox{$<$}\kern-.85em
          \lower.35em\hbox{$\sim$}~}\xspace}

\def\tell1  {TELL1\xspace}
\def\ukl1   {UKL1\xspace}

\ifthenelse{\boolean{uprightparticles}}%
{
 
}
{
 
}

\renewcommand\Re{\operatorname{Re}}

\newcommand{\imi}{\mathsf{i}}
\def\SIG{\ensuremath{\sigma}\xspace}
\def\A {\ensuremath{{\cal A}}\xspace}   
\def\O {\ensuremath{{\cal O}}\xspace}   
\def\CP {\ensuremath{{\cal CP}}\xspace}
\def\Pull {\ensuremath{{\rm Pull}}\xspace}
\def\Re {\ensuremath{\mathcal{R}e}\xspace}
\def\Im {\ensuremath{\mathcal{I}m}\xspace}
\def\chisq {\ensuremath{\chi^2}\xspace}
\def\and {\xspace\ensuremath{\cup}\xspace}
\def\ewp {\ensuremath{\textsc{ewp}}}
\usepackage{xparse}
\NewDocumentCommand\su{m}{\ensuremath{{SU}\mkern-2mu(#1)}}
\NewDocumentCommand\val{mmgg}{\ensuremath{ 
    \IfNoValueTF{#4}{}{(}
    \IfNoValueTF{#3}{#1 \pm #2}{ #1^{\scriptscriptstyle #2}_{\scriptscriptstyle #3}}
    \IfNoValueTF{#4}{}{)\textrm{#4}}
  }
}
\NewDocumentCommand\err{mg}{\ensuremath{ 
    \IfNoValueTF{#2}{\pm #1}{ {}^{\scriptscriptstyle #1}_{\scriptscriptstyle #2}}
  }
}

\NewDocumentCommand\pull{m}{\tiny (#1$\sigma$)}\NewDocumentCommand\gain{m}{\tiny (#1\%)}
\NewDocumentCommand\U{gg}{\ensuremath{{\cal U}\IfNoValueTF{#1}{}{_{\textrm{\tiny #1}}}\IfNoValueTF{#2}{}{^{\textrm{\tiny #2}}}}\xspace}
\NewDocumentCommand\I{gg}{\ensuremath{{\cal I}\IfNoValueTF{#1}{}{_{\textrm{\tiny #1}}}\IfNoValueTF{#2}{}{^{\textrm{\tiny #2}}}}\xspace}
\NewDocumentCommand\Obs{mgg}{\ensuremath{{#1}\IfNoValueTF{#2}{}{^{\textrm{\tiny #2}}}\IfNoValueTF{#3}{}{_{\tiny #3}}}\xspace}
\newcommand{\ofrac}[2]{\ensuremath{
  \raisebox{+.4ex}{\footnotesize $#1$}\negthinspace\slash\negthinspace
   \raisebox{-.6ex}{\footnotesize $#2$}}}
\def\B {\ensuremath{{\cal B}}\xspace}   
\def\C {\ensuremath{{\cal C}}\xspace}   
\def\S {\ensuremath{{\cal S}}\xspace}   

\begin{document}
\renewcommand{\thefootnote}{\fnsymbol{footnote}}
\setcounter{footnote}{1}

\begin{titlepage}

\vspace*{-1.5cm}

\hspace*{-0.5cm}
\begin{tabular*}{\linewidth}{lc@{\extracolsep{\fill}}r}
 & & LPT-Orsay-17-07 \\  
 & & \today \\ 
 & & \\
\hline
\end{tabular*}

\vspace*{.5cm}

{\bf\boldmath\huge
\begin{center} 
Isospin analysis of charmless $B$-meson decays
 \end{center}
}

\vspace*{.5cm}

\begin{center}
  J. Charles${}^a$,
  O. Deschamps${}^b$~\footnote{
    Contact author: Olivier Deschamps, 
   \href{mailto: Olivier.Deschamps@cern.ch}{Olivier.Deschamps@cern.ch} 
    } , S. Descotes-Genon${}^c$, V. Niess${}^b$\\[0.25cm]
   {\small [for the CKMfitter group]}\\[0.5cm]
   {
     \small\it 
     ${}^a$ CNRS, Aix Marseille Univ, Universit\'e de Toulon, CPT\\13288 Marseille Cedex 9, France\\[0.25cm]
     ${}^b$ Laboratoire de Physique de Clermont, CNRS/Universit\'e Clermont Auvergne, UMR 6533,\\Campus des C\'ezeaux, 24 Av. des Landais, F-63177 Aubi\`ere Cedex, France\\[0.25cm]
     ${}^c$ Laboratoire de Physique Th\'eorique (UMR 8627), CNRS, Univ. Paris-Sud,\\Universit\'e Paris-Saclay, 91405 Orsay Cedex, France
   }

\end{center}

\vspace{\fill}

\begin{abstract}
We discuss the determination of the CKM angle $\alpha$ using the non-leptonic two-body decays
$B\to \pi\pi$, $B\to \rho\rho$ and $B\to \rho\pi$ using the latest data available. We illustrate the methods used in each case and extract the corresponding value of $\alpha$. Combining all these elements, we obtain the determination $\alpha_{\rm dir}=(\val{86.2}{+4.4}{-4.0} \and \val{178.4}{+3.9}{-5.1})^\circ$. We assess the uncertainties associated to the breakdown of the isospin hypothesis and the choice of statistical framework in detail. We also determine the hadronic amplitudes (tree and penguin) describing the QCD dynamics involved in these decays, briefly comparing our results with theoretical expectations. For each observable of interest in the $B\to \pi\pi$, $B\to \rho\rho$ and $B\to \rho\pi$ systems, we perform  an indirect determination based on the constraints from all the other observables available and we discuss the compatibility between indirect and direct determinations. Finally, we review the impact of future improved measurements on the determination of $\alpha$.
\end{abstract}

\vspace*{1.5cm}

\begin{center}
Published in Eur.Phys.J. C77 (2017) no.8, 574 
\end{center}
\vspace{\fill}

\end{titlepage}

\pagestyle{empty}  


\newpage
\setcounter{page}{2}
\mbox{~}

\cleardoublepage

\renewcommand{\thefootnote}{\arabic{footnote}}
\setcounter{footnote}{0}


\pagestyle{plain} 
\setcounter{page}{1}
\pagenumbering{arabic}
\section{Introduction \label{sec:introduction} }

Over the last few decades, our understanding of $CP$ violation has made great progress, with many new constraints from \babar, Belle and LHCb experiments among others \cite{Bevan:2014iga,Bediaga:2012py}. These constraints were shown to be in remarkable agreement with each other and to support the Kobayashi--Maskawa mechanism of $CP$ violation at work within the Standard Model (SM) with three generations \cite{Cabibbo:1963yz,Kobayashi:1973fv}. This has led to an accurate determination of the Cabibbo--Kobayashi--Maskawa matrix (CKM) encoding the pattern of $CP$ violation as well as the strength of the weak transitions among quarks of different generations \cite{Hocker:2001xe,Charles:2004jd,Charles:2011va,Charles:2015gya,Koppenburg:2017mad,thePapIII}. These constraints prove also essential in assessing the viability of New Physics models with well-motivated flavour structures \cite{Deschamps:2009rh,Lenz:2010gu,Lenz:2012az,Charles:2013aka}.

As the CKM matrix is related to quark-flavour transitions, most of these constraints are significantly affected by hadronic uncertainties due to QCD binding quarks into the observed hadrons. However, some of these constraints have the very interesting feature of being almost free from such uncertainties. 
This is in particular the case for the constraints on the CKM angle $\alpha$ that are derived from the isospin analysis of the charmless decay modes $B\to\pi\pi$, $B\to\rho\rho$ and $B\to\rho\pi$. 
Indeed, assuming the isospin symmetry and neglecting the electroweak penguin contributions, the amplitudes of the \su{2}-conjugated modes are related. The measured branching fractions and asymmetries in the $\B^{\pm,0}\to(\pi\pi)^{\pm,0}$ and $B^{\pm,0}\to(\rho\rho)^{\pm,0}$ modes and the bilinear form factors in the Dalitz analysis of the $B^{0}\to(\rho\pi)^0$ decays provide enough observables to simultaneously extract the weak phase $\beta+\gamma=\pi-\alpha$ together with the hadronic tree and penguin contributions to each mode \cite{Gronau:1990ka,Lipkin:1991st,Harrison:1998yr}. Therefore, these modes probe two different corners of the SM: on one side, they yield information on $\alpha$ that is a powerful constraint on the Kobayashi--Maskawa mechanism (and the CKM matrix) involved in weak interactions, and on the other side, they provide a glimpse on the strong interaction and especially the hadronic dynamics of charmless two-body $B$-decays.

In the following, we will provide a thorough analysis of these decays, based on the data accumulated up to the conferences of Winter 2017. Combining the experimental data for the three decay modes above, we obtain the world-average value at 68\% Confidence Level (CL): 
\begin{equation} \label{eq:alphadir}
\alpha_{\rm dir} = (\val{86.2}{+4.4}{-4.0} \and \val{178.4}{+3.9}{-5.1})^\circ\,.
\end{equation}
The solution near $90^\circ$ is in  good agreement with the indirect determination obtained by the global fit of the flavour data performed by the CKMfitter group in Summer 2016 \cite{CKMfitterSummer16}:
\begin{equation}
\alpha_{\rm ind}  = \val{92.5}{+1.5}{-1.1}{$^\circ$}\,.\label{eq:alphaInd}
\end{equation}
 Considered separately, the $B\to\pi\pi$ and $B\to\rho\rho$ decays yield direct determinations of $\alpha$ in very good agreement with the indirect determination Eq.~(\ref{eq:alphaInd}), whereas the $B\to \rho\pi$ decay exhibits a 3$\sigma$ discrepancy.
This discrepancy affects only marginally the combination Eq.~(\ref{eq:alphadir}), which is dominated by the results from $\rho\rho$ decays, and to a lesser extent $\pi\pi$ decays, whereas $\rho\pi$ modes play only a limited role. At this level of accuracy, it proves interesting to assess uncertainties neglected up to now, namely the sources of violation of the assumptions underlying these determinations ($\Delta I=\ofrac{3}{2}$ electroweak penguins, $\pi^0-\eta-\eta'$ mixing, $\rho$ width) and the role played by the statistical framework used to extract the confidence intervals.  These effects may shift the central value of $\alpha_{\rm dir}$ by around $2^\circ$, while keeping the uncertainty around $4^\circ$ to $5^\circ$, thus remaining within the statistical uncertainty quoted in Eq.~(\ref{eq:alphadir}).

Besides the CKM angle $\alpha$, the isospin analysis of the charmless decays data provides a determination of the hadronic tree and penguin parameters for each mode. The penguin-to-tree and colour-suppression ratios in the 
\decay{B}{\pi\pi} and \decay{B}{\rho\rho} decay modes can be determined precisely: they show overall good agreement with theoretical expectations for $\rho\rho$ modes, whereas the ratio between colour-allowed and -suppressed tree contributions for $\pi\pi$ modes do not agree well with theoretical expectations (both for the modulus and the phase). We also perform indirect predictions of the experimental observables, using all the other available measurements to predict the value of a given observable, and we can compare 
our predictions to the existing measurements: a very good compatibility is observed for the $\pi\pi$ and $\rho\rho$ channels, whereas discrepancies occur in some of the observables describing the Dalitz plot for $B\to\rho\pi$ decays.
Among many other quantities, the yet-to-be-measured mixing-induced $CP$ asymmetry in the \decay{B^0}{\pi^0\pi^0} decay is predicted at 68\% CL:
\begin{equation}
\S^{00}_{\pi\pi}  = \val{0.65}{0.13}. 
\end{equation}
using as an input the indirect value of $\alpha$, see Eq.~(\ref{eq:alphaInd}).
Finally, we study how the determination of $\alpha$ would be affected if the accuracy of specific subsets of observables is improved through new measurements. We find that an improved accuracy for the time-dependent asymmetries in $B^0\to \rho^0\rho^0$ and the measurement of $\S^{00}_{\pi\pi}$ would reduce the uncertainty on $\alpha$ in a noticeable way.

The rest of this article goes as follows.
In Sec.~\ref{sec:isospin}, we discuss the basics of isospin analysis for charmless $B$-meson decays. In Sec.~\ref{sec:alpha}, we provide details on the extraction of the $\alpha$ angle, focusing on the $B\to \pi\pi$, $B\to \rho\rho$ and $B\to \rho\pi$ modes in turn, before combining these extractions in a world average. In Sec.~\ref{sec:syst}, we consider the uncertainties attached to the extraction of $\alpha$. First, we discuss
the \su{2} isospin framework underlying these analyses, considering three sources of corrections: the presence of $\Delta I=\ofrac{3}{2}$ electroweak penguins (isospin breaking due to different charges for the $u$ and $d$ quarks), $\pi^0-\eta-\eta'$ mixing (isospin breaking due to different masses for the $u$ and $d$ quarks), $\rho$ width (additional amplitude to include in the isospin relations for $B\to\rho\rho$). In addition, we discuss the statistical issues related to our frequentist framework. In Sec.~\ref{sec:nuisance}, we extract the hadronic  (tree and penguin) amplitudes associated with each decay, comparing them briefly with theoretical expectations. We use our framework to perform indirect predictions of observables in each channel and discuss the compatibility with the available measurements in Sec.~\ref{sec:observables}. We perform a prospective study to determine the impact of reducing the experimental uncertainties on specific observables in Sec.~\ref{sec:prospective}, before drawing our conclusions. Dedicated appendices gather additional numerical results for observables in three modes, separate analyses using either the \babar or Belle inputs only, and  a brief discussion of the quasi-two-body analysis  of the charmless $B^0\to a_1^\pm\pi^\mp$ that may provide some further information on $\alpha$.

\section{Isospin decomposition of charmless two-body $B$ decays\label{sec:isospin} }

\begin{figure}[t]
\begin{center}
  \includegraphics[width=26pc]{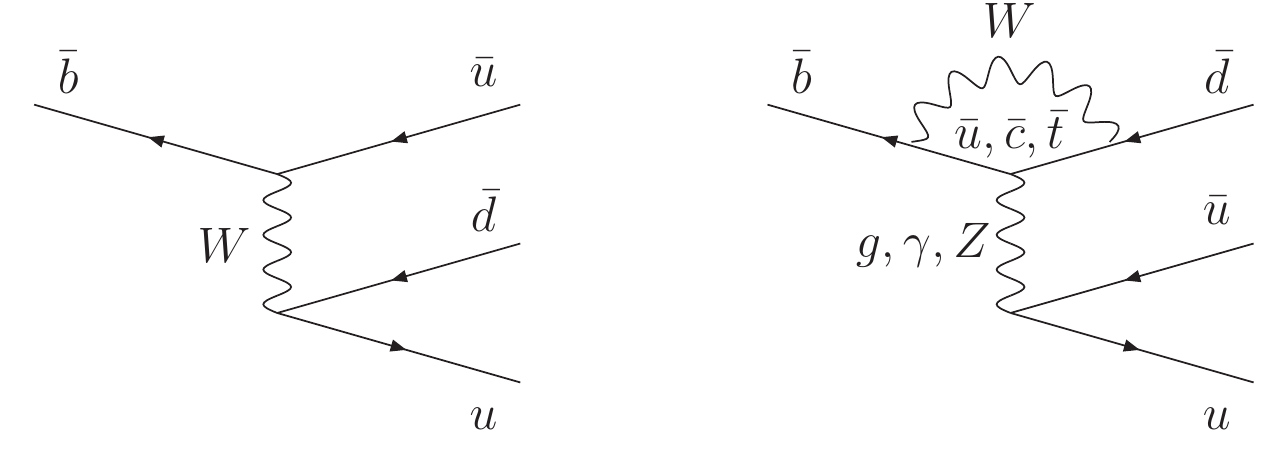}
\caption{\it\small Tree (left) and penguin (right) topology diagrams for the transition $\bar{b}\to u\bar u \bar{d}$.\label{fig:diagrams}}
\end{center}       
\end{figure}

The decay of  neutral $B_d$ and charged $B_u$ mesons into a pair of light unflavoured isovector mesons $B\to h_1^ih_2^j$ ($h$=$\pi,\rho$ and $i,j=-,0,+$) is described by the weak transition  $\bar b\to \bar u u\bar d$, followed by the hadronisation of the $(\bar u u\bar d,q)$ system, where $q=d(u)$ is the spectator quark in the neutral (charged) component of the mesons isodoublet. The weak process receives dominant contributions from both the tree-level $\bar b\to \bar u(u\bar d)$ charged transition  and the flavour-changing neutral current penguin transition, $\bar b\to \bar d (u\bar u)$, whose topologies are shown in Fig.~\ref{fig:diagrams}. 

\subsection{Penguin pollution\label{sec:amplitudes}}

We can write the charmless $B_q \to (\bar u u\bar d,q)\to h_1^ih_2^j$  transition amplitude, accounting for the CKM factors for the tree and penguin diagrams with three different up-type quark flavours $(u,c,t)$ occurring in the $W$ loop:
\begin{equation}\label{eq:CKMdecampl}
\mathcal{A}^{ij}=\bra{h_1^ih_2^j}{\cal H}_{\rm eff}\ket{B_d}=V_{ud}V^*_{ub}(\mathcal{T}^{ij}_u+\mathcal{P}^{ij}_u) + V_{cd}V^*_{cb}\mathcal{P}^{ij}_c + V_{td}V^*_{tb}\mathcal{P}^{ij}_t
\end{equation}
where ${\cal H}_{\rm eff}$ is the effective Hamiltonian describing the transition. The amplitudes ${\cal T}^{ij}_u$ and ${\cal P}^{ij}_{u,c,t}$ represents the tree-level and $(u,c,t)$-loop mediated topologies, respectively.

The unitarity of the CKM matrix can be used to eliminate one of the three terms in Eq.~(\ref{eq:CKMdecampl}),  resulting in three conventions $\mathcal{U}$, $\mathcal{C}$ and $\mathcal{T}$ defined as:
\begin{eqnarray}
{\cal U}-{\rm convention}: {\cal A}^{ij}&=& V_{cd}V^*_{cb} ({\cal P}^{ij}_c-{\cal T}^{ij}_u-{\cal P}^{ij}_u) + V_{td}V^*_{tb}({\cal P}^{ij}_t-{\cal T}^{ij}_u-{\cal P}^{ij}_u)\,,\nonumber\\
{\cal C}-{\rm convention}: {\cal A}^{ij}&=& V_{ud}V^*_{ub} ({\cal T}^{ij}_u+{\cal P}^{ij}_u-{\cal P}^{ij}_c) + V_{td}V^*_{tb}({\cal P}^{ij}_t-{\cal P}^{ij}_c)\,,\\
{\cal T}-{\rm convention}: {\cal A}^{ij}&=& V_{ud}V^*_{ub} ({\cal T}^{ij}_u+{\cal P}^{ij}_u-{\cal P}^{ij}_t) + V_{cd}V^*_{cb}({\cal P}^{ij}_c-{\cal P}^{ij}_t)\,.\nonumber
\end{eqnarray}
The $\mathcal{C}$-convention is adopted in the following, so that the amplitudes can be rewritten as:
\begin{equation}
\mathcal{A}^{ij}=V_{ud}V^*_{ub}\mathcal{\tilde T}^{ij} +  V_{td}V^*_{tb}\mathcal{\tilde P}^{ij}
\end{equation}
where the terms $\mathcal{\tilde T}^{ij}=\mathcal{T}^{ij}_u+\mathcal{P}^{ij}_u-\mathcal{P}^{ij}_c$ and $\mathcal{\tilde P}^{ij}=\mathcal{P}^{ij}_t-\mathcal{P}^{ij}_c$ associated with specific CKM factors will be referred to as 'tree' and 'penguin' amplitudes, respectively, following a conventional abuse of language (these amplitudes are not physical, in the sense that both mix contributions of distinct topologies and are not renormalisation invariant).
The choice of convention is arbitrary and it has no physical implication on the determination of the weak phase affecting the transitions. However, the particular choice sets the dynamical content of the hadronic tree and penguin amplitudes: for instance, it fixes the definition of the penguin-to-tree ratio and the other hadronic quantities discussed in Sec.~\ref{sec:nuisance}.

Pulling out the weak phases $\gamma=\arg\left(-\frac{V_{ud}V^*_{ub}}{V_{cd}V^*_{cb}}\right)$ and $\beta=\arg\left(-\frac{V_{cd}V^*_{cb}}{V_{td}V^*_{tb}}\right)$, the amplitude can be rewritten as:
\begin{equation}
\mathcal{A}^{ij} = -e^{\imi \gamma} T^{ij}  + e^{-\imi \beta}  P^{ij}\,,
\end{equation}
where the magnitude of the CKM products $R_u=|V_{ud}V^*_{ub}|$ and $R_t=|V_{td}V^*_{tb}|$ 
is included in the redefined amplitudes $T^{ij}=R_u{\cal\tilde T}^{ij}$ and $P^{ij}=-R_t{\cal\tilde P}^{ij}$.

Similarly, the decay amplitudes of the $CP$-conjugate isodoublet ($\bar B^0$, $B^-$) can be expressed as
\begin{equation}
\frac{p}{q}\mathcal{\bar A}^{ij} =-e^{-\imi \gamma}  T^{ij}  +  e^{\imi \beta}  P^{ij}\,,
\end{equation}
where the factor $p/q\sim e^{2\imi \beta}$ is included to take into account the $B^0-\bar B^0$ mixing phase of neutral $B$-meson that arises naturally in physical observables. 
 For consistency between all the \su{2}-related decay modes considered in the isospin analysis, the same phase convention has been applied to define the amplitudes for the charged $B$ meson. The $CP$ invariance of the strong interaction means that the same hadronic amplitudes $T^{ij}$ and $P^{ij}$ are involved in the $CP$-conjugate processes, whereas a complex conjugation is applied to the weak phases.

Rotating all amplitudes by the weak phase $\beta$ through the redefinition $A^{ij}=e^{\imi \beta}\mathcal{A}^{ij}$ and ${\bar A}^{ij}=e^{\imi \beta}\mathcal{\bar A}^{ij}$, the parameter $\alpha=\pi-\beta-\gamma$  appears as half of the phase difference between  the tree contributions to the $CP$-conjugate amplitudes:
\begin{equation}
e^{2\imi \alpha} = \frac{e^{\imi\alpha}T^{ij}}{e^{-\imi\alpha}T^{ij}}=\frac{\bar A^{ij}-P^{ij}}{A^{ij} - P^{ij}}.
\end{equation}
In the absence of penguin contributions, $\alpha$ can be related to the relative phase of $CP$-conjugate amplitudes describing the $B^0$ and $\bar{B}^0$ mesons decaying into the same final state $h_1^ih_2^j$. In particular, the time-dependent analysis of the $B^0/{\bar B^0}\to h_1^+h_2^-$ decay yields the $CP$ asymmetry:
\begin{equation}
a_{CP}(t)=\frac{\Gamma(\bar{B}^0(t)\to h_1^+h_2^-)-\Gamma(B^0(t)\to h_1^+h_2^-)}{\Gamma(\bar{B}^0(t)\to h_1^+h_2^-)+\Gamma(B^0(t)\to h_1^+h_2^-)}
  ={\cal S}^{+-} \sin(\Delta m_d\ t) - {\cal C}^{+-} \cos(\Delta m_d\ t)\,,\label{eq:acp}
\end{equation}
where $\Delta m_d$ is the $B^0-\bar{B}^0$ oscillation frequency, and $t$ is either the decay time of the meson, or (in the case of $B$-factories) the time difference between the $CP$- and tag-side decays. The coefficients can be expressed as
\begin{equation}
\lambda=\frac{\bar{A}^{+-}}{A^{+-}}\,,
\qquad {\cal S}^{+-}=\frac{2 {\rm Im}\lambda}{1+|\lambda|^2}\,,
\qquad {\cal C}^{+-}=\frac{1-|\lambda|^2}{1+|\lambda|^2}\,.
\end{equation}
Therefore, in the absence of the penguin contributions, the measurement of $\S^{+-}$ would yield $\sin(2\alpha)$.  The a priori non-negligible penguin contributions to the $B$ charmless decays modify this picture: if we introduce the effective angle corresponding to the phase of $\lambda$, we have \cite{Charles:1998qx}
\begin{equation}\label{eq:alphaeff}
\lambda=|\lambda|e^{2\imi\alpha_{\rm eff}}\,,
\qquad {\cal S}^{+-}=\sqrt{1-({\cal C}^{+-})^2} \sin(2\alpha_{\rm eff})\,.
\end{equation}
The non-negligible penguin contributions prevent us from identifying $\alpha$ and $\alpha_{\rm eff}$ obtained from the sole consideration of $B^0(t)\to h_1^+h_2^-$. However, since isospin is conserved during the hadronisation process, we can write useful relations among the hadronic amplitudes in $\su{2}$-related modes. These relationships will help us to determine the amount of penguin pollution from the data and thus to extract $\alpha$ in spite of hadronic effects \cite{Gronau:1990ka,Lipkin:1991st}.

\subsection{General isospin decomposition and application to the  $\rho \pi$ final-state}

One can factorise the decay amplitudes in two parts. First, the weak decay $\bar{b}\to u\bar{u} \bar{d}$ (common to all $B_q\to h_1^ih_2^j$ decay processes) corresponds to
a shift of isospin $\Delta I$. The hadronisation into two light mesons
can then be described as $\bra{h_1^ih_2^j}{\cal H}_s\ket{u\bar u\bar d,q}$ where ${\cal H}_s$ represents the isospin-conserving strong interaction Hamiltonian. The Wigner-Eckart theorem can be used to express
these amplitudes in tersm of reduced matrix elements, $A_{\Delta I,I_f}$, identified by the shift $\Delta I$ and  the final-state isospin $I_f$  ($I_f=0,1,2$)\footnote{Considering only the valence quarks, the isospin shift $\Delta I$ can only take the values $\ofrac{3}{2}$ and $\ofrac{1}{2}$ in the $\bar b\to (\bar u u\bar d)$ transition.  Tree and strong penguin topologies correspond to $\Delta I=\ofrac{1}{2}$, whereas electroweak penguins contain both $\Delta I=\ofrac{1}{2}$ and $\Delta I=\ofrac{3}{2}$ contributions.
For completeness, the possible $\Delta I=\ofrac{5}{2}$ contribution due to long-distance rescattering effects \cite{Gardner:2001gc,Gronau:2005pq}  is also reported in Tab.~\ref{tab:GeneralIsospin}. This contribution, suppressed by a factor $\alpha_{em}\sim 1/127$, will be neglected until Sec.~\ref{sec:breaking}.}. 
Tab.~\ref{tab:GeneralIsospin} yields the  decomposition of the $A^{ij}$ amplitudes in the general case of two distinguishable isovector mesons  ($h_1\ne h_2$).

\begin{table}[t]
\begin{center}
\begin{tabular}{|l||c|c|c|c|c|}
\hline
$A^{ij} = \bra{h_1^ih_2^j}{\cal H}_s\ket{u\bar u\bar d, q}$& $A_{\frac{5}{2},2}$ &$A_{\frac{3}{2},2}$ & $A_{\frac{3}{2},1}$ & $A_{\frac{1}{2},1}$ & $A_{\frac{1}{2},0}$\\ 
\hline
$A^{+0} = \bra{h_1^+h_2^0}{\cal H}_s\ket{u\bar u\bar d,u}$ &$-\sqrt{1/6}$  &$+\sqrt{3/8}$  & $-\sqrt{1/8}$  & $+\sqrt{\ofrac{1}{2}}$& 0\\
$A^{0+} = \bra{h_1^0h_2^+}{\cal H}_s\ket{u\bar u\bar d,u}$ &$-\sqrt{1/6}$  &$+\sqrt{3/8}$  & $+\sqrt{1/8}$  & $-\sqrt{\ofrac{1}{2}}$& 0\\
\hline
$A^{+-} = \bra{h_1^+h_2^-}{\cal H}_s\ket{u\bar u\bar d,d}$ &$+\sqrt{1/12}$ &$+\sqrt{1/12}$ & $+1/2$   & $+1/2$       &$ +\sqrt{1/6}$ \\
$A^{-+} = \bra{h_1^-h_2^+}{\cal H}_s\ket{u\bar u\bar d,d}$ &$+\sqrt{1/12}$ &$+\sqrt{1/12}$ & $-1/2$   & $-1/2$       &$ +\sqrt{1/6}$ \\
$A^{00} = \bra{h_1^0h_2^0}{\cal H}_s\ket{u\bar u\bar d,d}$ &$+\sqrt{1/3}$  &$+\sqrt{1/3}$  &       0  &     0        &$ -\sqrt{1/6}$ \\
\hline
\end{tabular}
\caption{\it\small General decomposition of the amplitudes $A^{ij}=\bra{h_1^ih_2^j}{\cal H}_s\ket{u\bar u\bar d,q}$ ($q=u,d$ ; $i,j=-,0,+$)  in terms of the reduced matrix elements $A_{\Delta I,I_f}$ for a pair of distinguishable isovector mesons $h_1^i$ and $h_2^j$.\label{tab:GeneralIsospin}}
\end{center}
\end{table}

This general decomposition applies for instance to the $B\to\rho\pi$ system. Neglecting the $\Delta I =\ofrac{5}{2}$ transition, the four remaining isospin amplitudes constrain  the five decays amplitudes to follow the pentagonal relation:
\begin{equation}
A^{+-}+A^{-+}+2A^{00}=\sqrt{2}(A^{+0}+A^{0+}).\label{eqn:pentagonal}
\end{equation}
The same identity applies to the $CP$-conjugate amplitudes.
Moreover, the sum of the decay amplitudes of the charged modes $(A^{+0}+A^{0+})$ is a  pure $A_{\frac{3}{2},2}$ isospin amplitude. 

A usual simplifying assumption consists in neglecting the $\Delta I=\ofrac{3}{2}$ contribution from electroweak penguins, as they are expected to be small: in this limit, all penguins (gluonic and electroweak) are mediated only by the $\Delta I=\ofrac{1}{2}$ transition $\bar b\to\bar d(u\bar u)_{I=0}$. The isospin relation for the amplitudes Eq.~(\ref{eqn:pentagonal}) may then be projected onto the penguin amplitudes, leading to the triangular relations:
\begin{equation}\label{eq:Prhopi1}
P^{+-} + P^{-+} + 2P^{00} = \sqrt{2}(P^{+0} + P^{0+}) =0 \,,
\end{equation}
Both combinations are identical and vanish under our assumptions, as they involve only $\Delta I=\ofrac{3}{2}$ amplitudes.
Under this assumption, some combinations of the decay amplitudes are free of penguin contributions. The parameter $\alpha$ can thus be identified as half of the phase difference between the sum of the $CP$-conjugate amplitudes of the charged modes,
or equivalently as a function of the amplitudes of the neutral modes only:
\begin{equation}
e^{2\imi \alpha} = \frac{\bar A^{+0}+\bar A^{0+}}{A^{+0} + A^{0+}}=\frac{\bar A^{+-}+\bar A^{-+}+2\bar A^{00}}{A^{+-}+A^{-+}+2A^{00}}.\label{eqn:alpha}
\end{equation}
Similarly, the combination of amplitudes  $(A^{+0}-A^{0+})-\sqrt{2}(A^{+-}-A^{-+})$ receives a pure $A_{\frac{3}{2},1}$  contribution and is thus free from penguin contributions, so that:
\begin{equation}\label{eq:Prhopi2}
P^{+0}-P^{0+}=\sqrt{2}(P^{+-}-P^{-+})\,,
\end{equation}
and the phase $\alpha$ can also be obtained from the pure tree combination:
\begin{equation}
e^{2\imi \alpha} = \frac{ (\bar A^{+0}-\bar A^{0+})-\sqrt{2}(\bar A^{+-}-\bar A^{-+}) }{( A^{+0}- A^{0+})-\sqrt{2}( A^{+-}- A^{-+})}\,.\label{eqn:alpha3}
\end{equation}
Combining Eqs.~(\ref{eq:Prhopi1}) and (\ref{eq:Prhopi2}), one can see that there are actually only two independent penguin amplitudes involved:
\begin{equation}\label{eq:Prhopi3}
P^{+0}= - P^{0+}=(P^{+-}-P^{-+})/\sqrt{2}, \qquad
P^{00}= -(P^{+-}+P^{-+})/2,
\end{equation}
illustrating the power of the \su{2} isospin approach to reduce the number of independent hadronic quantities to be determined from the data in order to extract a constraint on $\alpha$.

\subsection{Application to the $\pi\pi$ and $\rho\rho$ cases}

\begin{table}[t]
\begin{center}
\begin{tabular}{|l||c|c|c|c|c|}
\hline
$A^{ij} = \bra{h^ih^j}{\cal H}_s\ket{u\bar u\bar d,q}$& $A_{\frac{5}{2},2}$ &$A_{\frac{3}{2},2}$ &  $A_{\frac{1}{2},0}$\\ 
\hline
$A^{+0} = \bra{\pi^+\pi^0}{\cal H}_s\ket{u\bar u\bar d,u}$  &$-1$         &$\ofrac{3}{2}$         & $0$ \\
\hline
$A^{+-} = \bra{\pi^+\pi^-}{\cal H}_s\ket{u\bar u\bar d,d}$ &$\sqrt{1/2}$ &$+\sqrt{1/2}$ &$ +1$ \\
$A^{00} = \bra{\pi^0\pi^0}{\cal H}_s\ket{u\bar u\bar d,d}$ &$+1$          &$+1$          &  $-\sqrt{1/2}$ \\
\hline
\end{tabular}
\caption{\it\small Decomposition of the amplitude $\bra{h^ih^j}{\cal H}_s\ket{u\bar u\bar d,q}$ ($q=u,d$ ; $i,j=-,0,+$) in terms of the isospin amplitudes $A_{\Delta I,I_f}$ for indistinguishable mesons in the final state ($h_1=h_2=h$). A global factor $\sqrt{3}$ is applied to all the coefficients with respect to the general coefficients given in Tab.~\ref{tab:GeneralIsospin}. \label{tab:SpecificIsospin}}
\end{center}
\end{table}

In the cases where the two isovector mesons in the final state are indistinguishable from the point of view of isospin symmetry ($h_1=h_2=h$), as for the $B\to\pi\pi$ and $B\to\rho\rho$ systems, only the even amplitudes $I_f=0,2$ are allowed due to the Bose--Einstein statistics.
Defining the symmetrised amplitudes:
\begin{equation}
A^{+-}_{h_1=h_2}=\frac{A^{+-}+A^{-+}}{\sqrt{2}}\,,\qquad
A^{+0}_{h_1=h_2}=\frac{A^{+0}+A^{0+}}{\sqrt{2}}\,,
\end{equation}
the  isospin decomposition gets simplified as reported in Tab.~\ref{tab:SpecificIsospin}, leading 
to the triangular identity:
\begin{equation}
A^{+0}=\frac{A^{+-}}{\sqrt{2}}+A^{00},\label{eqn:triangular}
\end{equation}
with a similar identity for the $CP$-conjugate amplitudes \footnote{Unless otherwise stated, the subscript $h_1=h_2$ is dropped and  $A^{ij}$  implicitly refers to the symmetrised amplitude in the case of a decay into two mesons of same type.}. In the case of the $\rho\rho$ channel, one should consider  a different set of independent amplitudes for each of the three possible polarisations (which is identical for the two $\rho$ mesons).

\begin{figure}[t]
\begin{center}
  \includegraphics[width=30pc]{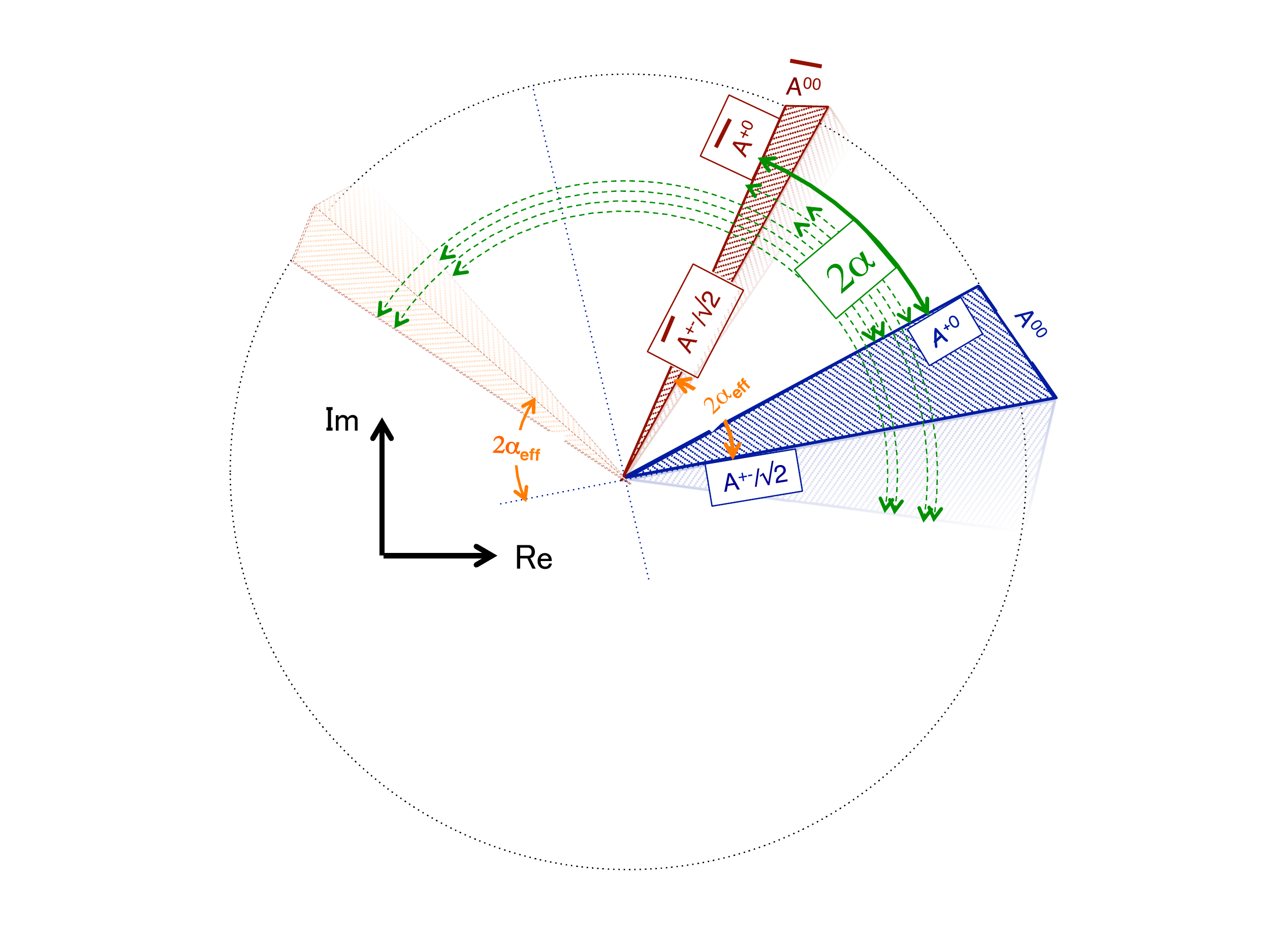}
\caption{\it\small Geometrical representation of the isospin triangular relation $A^{+0}={A^{+-}}/{\sqrt{2}}+A^{00}$ and its $CP$-conjugate equivalent  in the complex plane of $B\to hh$ amplitudes (red and blue shaded areas). The angle between the $CP$-conjugate charged amplitudes $A^{+-}$ and $\bar A^{+-}$ corresponds to twice the weak phase $\alpha_{\rm eff}$ (yellow solid arrows), whereas the angle between the $CP$-conjugate charged amplitudes $A^{+0}$ and $\bar A^{+0}$ corresponds to twice the weak phase $\alpha$ (green solid arrows). The other triangles with a lighter shade represents the mirror solutions allowed by the discrete ambiguities in the observables, with the corresponding values for $\alpha$ represented with green dashed arrows.\label{fig:triangles}}
\end{center}       
\end{figure}

Under the assumption of negligible contributions from $\Delta I=\ofrac{3}{2}$ electroweak penguins,
the penguin relations Eq.~(\ref{eq:Prhopi3}) reduce down to
$P^{+0}=0$ and $\ofrac{P^{+-}}{\sqrt{2}}=-P^{00}$. The total amplitude of the charged modes $A^{+0}=e^{-\imi \alpha}T^{+0}$ is free from penguin contributions, and the angle $\alpha$ can be derived from:
\begin{equation}
e^{-2\imi \alpha} = \frac{\bar A^{+0}}{A^{+0}}\,.
\end{equation}
This complex ratio cannot be determined from a single measurement, but it is possible to reconstruct
the two isospin triangles corresponding to $CP$-conjugate amplitudes using branching ratios and $CP$ asymmetries for all the modes. Fig.~\ref{fig:triangles} illustrates this construction, which translates the measurement of $\alpha_{\rm eff}$ into a determination of the CKM angle $\alpha$. This procedure is affected by discrete ambiguities, since there are several manners of reconstructing the two isospin triangles. This leads to a fourfold ambiguity for $\sin(2\alpha)$, i.e. an eightfold ambiguity on the solutions of $\alpha$ in $[0,180]^\circ$, in general. These additional solutions are called ``mirror solutions''. If one or both triangles are flat, several mirror solutions  become degenerate, decreasing the number of distinct solutions for $\alpha$.

\section{Determining the weak angle $\alpha$}\label{sec:alpha}
\subsection{Procedure} \label{sec:procedure}

If we consider that each of the five amplitudes $A^{ij}$ $(ij=+-,-+,+0,0+,00)$ receives two complex contributions, tree and penguin, the  hadronic contributions to the generic decay system $B^{i+j}\to h_1^ih_2^j$ can be parametrised with 20 real  parameters, in addition to the weak phase $\alpha$ (one overall phase being irrelevant). In the case of a  Bose-symmetric final state ($h_1=h_2$), the dimension of the parameter space reduces down to 13. Assuming the isospin relations between the amplitudes discussed in Sec.~\ref{sec:isospin},  the three decay systems $B\to\pi\pi$, $B\to\rho\rho$ and $B\to\rho\pi$ provide enough experimental measurements to fully constrain the parameter space, hereafter denoted $\vec p$=($\alpha$,$\vec\mu$), where $\vec\mu$ represents the set of independent hadronic parameters (tree and penguin amplitudes). The actual dimension of the parameter space depends on the decay system and will be discussed in the following subsections. 

The constraints on the parameter $\alpha$ are determined by an  exploration of the $N$-dimensional parameters space through the frequentist statistical approach discussed in detail in Refs.~\cite{Hocker:2001xe,Charles:2004jd,thePapIII,Charles:2016qtt} and in Sec.~\ref{sec:statistics}, but we find it useful to briefly summarise its main features here. The set of experimental observables, denoted $\vec {\cal O}_{exp}$, is measured in terms of likelihoods that can be used to build a \chisq-like test statistic:
\begin{equation}
\chi^2(\vec{p})=-2\log {\mathcal{L}}(\vec {\cal O}_{exp}-\vec {\cal O}(\vec p))\,,
\end{equation}
where $\vec {\cal O}(\vec p)$ represent the theoretical value of the observables  in SM for fixed  parameters $\vec p$.  The test statistic \chisq is first minimised over the whole parameter space, letting all $N$ parameters $\vec p$ free to vary. The absolute minimum value of the test statistic, $\chi^2_{\rm min}$, quantifies the agreement of the data with the theoretical model (assuming  the validity of the SM and \su{2} isospin symmetry in the present case). Converting  $\chi^2_{\rm min}$ into a $p$-value is however not trivial a priori, as one has to interpret $\chi^2(\vec{p})$
as a random variable distributed according to a $\chi^2$ law with a certain number of degrees of freedom. The actual 
number of degrees of freedom of the system can be ill defined in the case where the experimental observables are interdependent (see for instance the related discussion in Sec.~\ref{subsec:rhopi}).
In the case of $M$ independent observables, the number of degrees of freedom of the system is defined as $N_{dof}=M-N$. This occurs in the Gaussian case, but it also can apply in non-Gaussian cases in the limit of large samples, under the conditions of Wilks' theorem \cite{Wilks}. 

It is also possible to perform the metrology of specific parameters of the model, and in particular $\alpha$~\cite{Hocker:2001xe,Charles:2004jd,thePapIII,Charles:2016qtt}. Indeed, if we consider the $\vec\mu$ hadronic parameters as ``nuisance parameters'', 
we can define the test statistic from the \chisq difference:
\begin{equation}\label{eq:deltachisq}
 \Delta\chi^2(\alpha) = \min_{\vec\mu}[\chi^2(\alpha)] - \chi^2_{\rm min}
\end{equation}
where  $\min_{\vec\mu}[\chi^2(\alpha)]$ is the value of  \chisq, minimised with respect to the nuisance parameters for a  fixed $\alpha$ value. This test statistic assesses how a given hypothesis on the true value of $\alpha$ agrees with the data, irrespective of the value of the nuisance parameters. Confidence intervals on $\alpha$ can be derived from the resulting $p$-value, which is computed assuming that $\Delta\chi^2(\alpha)$ is \chisq-distributed with one degree of freedom:
\begin{equation}\label{eq:prob}
 p(\alpha) = {\rm Prob}(\Delta\chi^2(\alpha),N_{dof}=1)\,, \qquad  {\rm Prob}\!\left(\Delta \chi^2,N_{dof}\right)=\frac{\Gamma(N_{dof}/2,\Delta\chi^2/2)}{\Gamma(N_{dof}/2)}\,.
\end{equation}
One can express ${\rm Prob}$ for $N_{dof}=1$ in terms of the complementary error function:
\begin{equation}
{\rm Prob}(\Delta\chi^2(\alpha),N_{dof}=1)={\rm Erfc(\sqrt{\Delta\chi^2(\alpha)/2})}\,.
\end{equation}
Confidence intervals at a given confidence level (CL) are obtained by selecting the values of $\alpha$ with a $p$-value larger than $1-$CL.
The derivation, robustness and coverage of this definition for the $p$-value will be further discussed in Sec.~\ref{sec:statistics}.

Although the relevant information on the  $\alpha$ constraint is fully contained in the $p$-value function $p(\alpha)$, confidence intervals will be derived in the following subsections. It is worth noticing that the $p$-value for $\alpha$ usually presents a highly non-Gaussian profile and that the \su{2} isospin analysis suffers from (pseudo-)mirror ambiguities. Therefore,  the confidence intervals provided must be interpreted with particular care.

\begin{figure}[t]
\begin{center}
  \includegraphics[width=18pc]{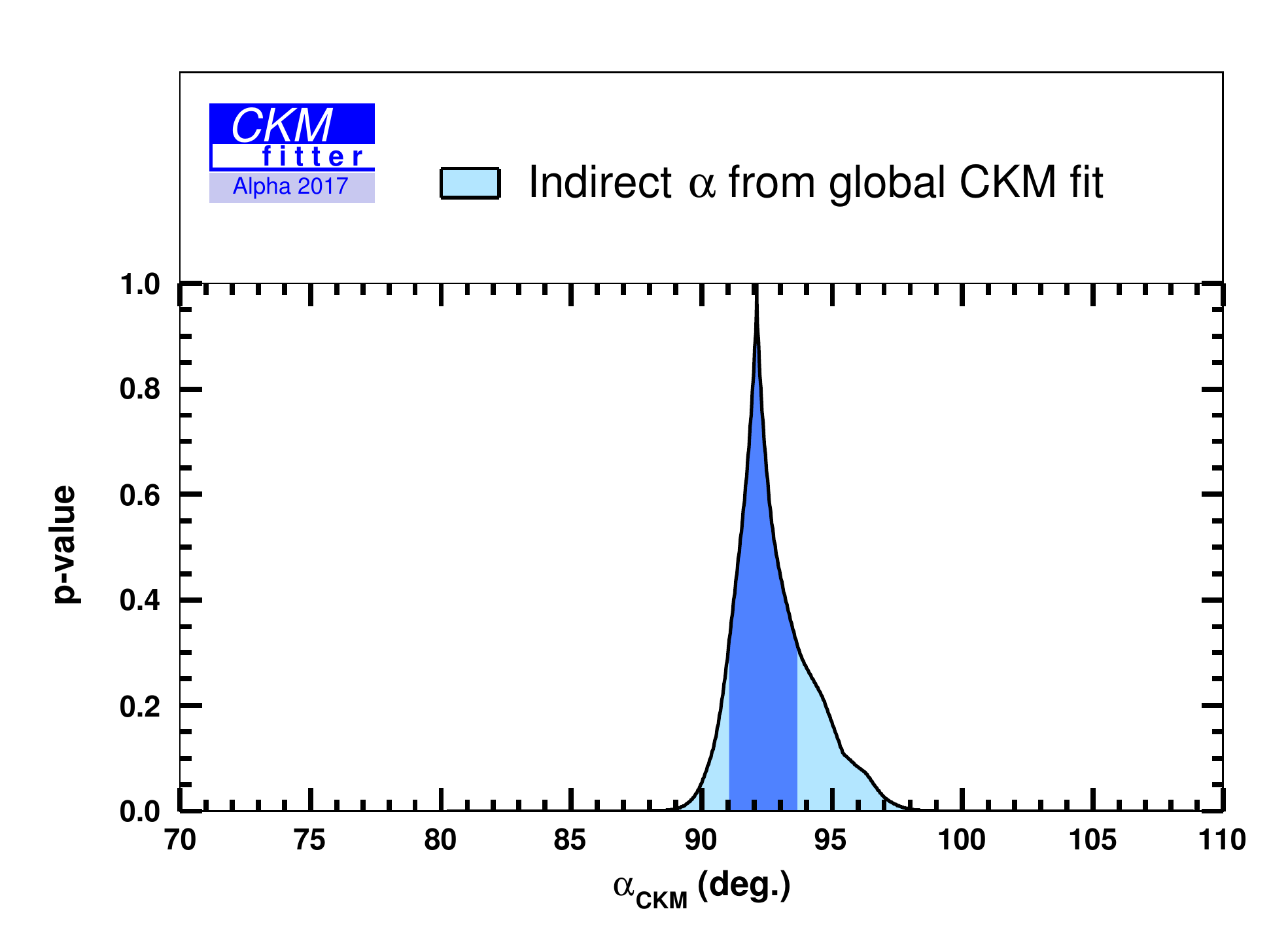}
  \includegraphics[width=18pc]{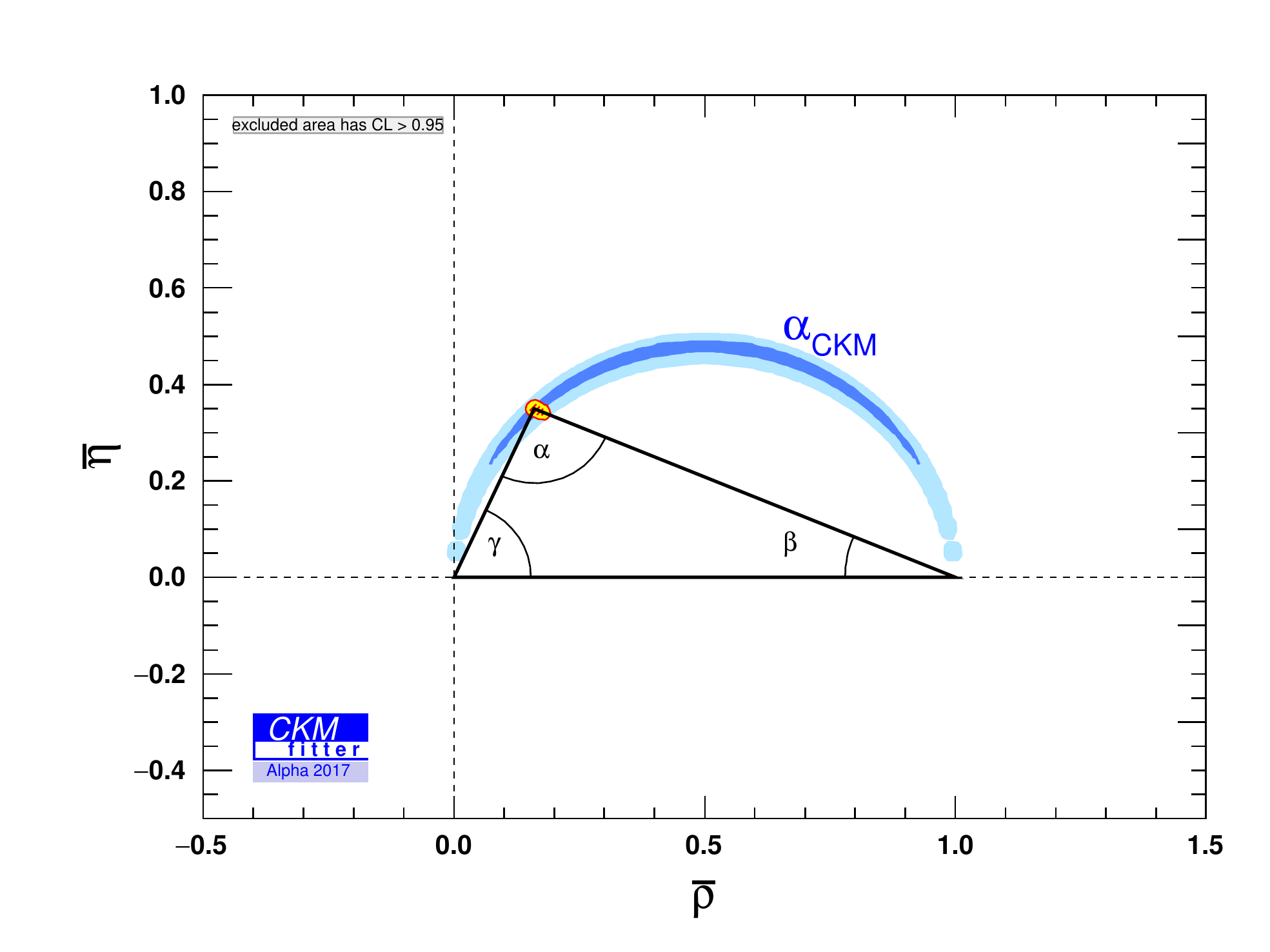}
\caption{\it\small On the left: $p$-value as a function of the assumed value for $\alpha$, extracted from the global CKM fit excluding direct information on $\alpha$ from charmless $B$ decays. On the right: 95\% confidence-level constraint (light blue area) in the $(\bar\rho,\bar\eta)$ plane defining the apex of the $B$-meson unitarity triangle. The corresponding 68\% confidence-level interval, indicated by the dark blue area on the two figures, is given by Eq.~(\ref{eq:alphaInd}).}
\label{fig:alphaInd}
\end{center}       
\end{figure}

We consider here all the experimental inputs available up to Winter 2017 conferences (for each channel, we provide the corresponding references). We compare the results of the isospin-based direct determination of $\alpha$ from these inputs with the indirect result from the global CKM fit obtained in Summer 2016. This indirect determination includes all the quark-flavour constraints described in Refs. \cite{thePapIII,CKMfitterSummer16}, apart from the inputs from the $B\to\pi\pi$, $B\to\rho\rho$ and $B\to \rho\pi$ decays used for the direct determination of $\alpha$. The indirect determination, hereafter denoted $\alpha_{\rm ind}$, is illustrated in Fig.~\ref{fig:alphaInd} and the corresponding 68\% CL interval is given in Eq.~(\ref{eq:alphaInd}).

\subsection{Isospin analysis of the $B\to\pi\pi$ system\label{subsec:pipi}}

The three decays $B^{0,+}\to(\pi\pi)^{0,+}$ depend on 12 hadronic parameters and the weak phase $\alpha$.   One can set one irrelevant global phase to zero, and one can eliminate further parameters using the complex isospin relation Eq.~(\ref{eqn:triangular}) and its $CP$ conjugate (four real constraints) as well as the absence of the penguin contribution to the charged mode amplitude (two real constraints). 
The remaining system of amplitudes features six degrees of freedom. The isospin-related $B\to\pi\pi$ decays (and similarly each of the helicity states of the $B\to\rho\rho$ mode) can  thus be described with six real independent parameters,  including the common weak phase $\alpha$.

The relevant available experimental observables and their current world-average value for the $B\to\pi\pi$ modes are summarised\footnote{The $CP$ asymmetry for the charged mode explicitly vanishes in the parametrisation of the amplitudes  based on exact \su{2} isospin symmetry. 
Therefore the corresponding  observable is not included here. The current experimental measurement, fully consistent with the null hypothesis, would only affect the minimal value $\chisq_{\rm min}$ of the fit, but not the metrology of the parameters.  This additional observable will be useful to study isospin-breaking effects in Sec.~\ref{sec:tests}.} in Tab.~\ref{input_pipi}. Six independent observables are available, allowing us to constrain the six-dimensional parameter space of the \su{2} isospin analysis.
These branching fractions and ${CP}$ asymmetries are related to the decay amplitudes as:
\begin{equation}
\frac{1}{\tau_{B^{i+j}}} {\cal B}^{ij} = \frac{ |A^{ij}|^2 +  |{\bar A}^{ij}|^2}{2}, \qquad
                      {\cal C}^{ij} = \frac{ |A^{ij}|^2 -  |{\bar A}^{ij}|^2}{ |A^{ij}|^2 +  |{\bar A}^{ij}|^2}, \qquad
                      {\cal S}^{ij} = \frac{ 2 {\cal I}m({\bar A}^{ij}A^{ij}{}^{*})}{ |A^{ij}|^2 +  |{\bar A}^{ij}|^2}\,,
\end{equation}
where $\tau_{B^{i+j}}$ is the measured lifetime of the charged ($i+j=1$) or neutral ($i+j=0$) $B$ meson.

\begin{table}[t]
\begin{center}
\begin{tabular}{|l|c|l|}
\hline
Observable & World average & References \\
\hline
\small ${\cal B}^{+-}_{\pi\pi}$ $(\times 10^6)$  &\small \val{5.10}{0.19}        &\small   \cite{Babar_pp_bpm,Belle_pp_bpm,CDF_pp_bpm,Cleo_pp_bpm,LHCb_pp_bpm} \\
\small ${\cal B}^{+0}_{\pi\pi}$ $(\times 10^6)$  &\small \val{5.48}{0.34}        &\small   \cite{Belle_pp_bpm,CDF_pp_bpm,Cleo_pp_bpm,Babar_pp_bp0}             \\
\small ${\cal B}^{00}_{\pi\pi}$ $(\times 10^6)$  &\small \val{1.59}{0.18}        &\small   \cite{Babar_pp_bpm,Julius:2017jso}                                    \\
\small ${\cal C}^{00}_{\pi\pi}$                  &\small \val{-0.34 }{0.22}      &\small   \cite{Babar_pp_bpm,Julius:2017jso}                       \\
\small ${\cal C}^{+-}_{\pi\pi}$                  &\small \val{-0.284}{0.039}                  &\small   \cite{Babar_pp_cpm,Belle_pp_cpm,LHCb_pp_cpm}           \\
\small ${\cal S}^{+-}_{\pi\pi}$                  &\small \val{-0.672}{0.043}                  &\small   \cite{Babar_pp_cpm,Belle_pp_cpm,LHCb_pp_cpm}            \\
\small $\rho(C^{+-}_{\pi\pi},S^{+-}_{\pi\pi})$     &\small $+0.013$                           &\small   \cite{Babar_pp_cpm,Belle_pp_cpm,LHCb_pp_cpm}            \\
\hline
\end{tabular}
\caption{\it\small  World averages for the relevant experimental observables in the $B\to\pi^i\pi^j$ modes: branching fraction ${\cal B}^{ij}_{\pi\pi}$, time-integrated $CP$ asymmetry ${\cal C}^{ij}_{\pi\pi}$, time-dependent asymmetry  ${\cal S}^{ij}_{\pi\pi}$ and correlation ($\rho$).\label{input_pipi}}
\end{center}
\end{table}

Taking into account the isospin relation Eq.~(\ref{eqn:triangular}), the amplitudes can be parametrised in terms of their penguin and tree contributions as
\begin{eqnarray}
 {A}^{+-}= T^{+-} e^{-\imi\alpha}  + P   &,& {\bar A}^{+-}= T^{+-} e^{+\imi\alpha} + P, \nonumber \\
 \sqrt{2}{A}^{00}= T^{00} e^{-\imi\alpha}  - P   &,&  \sqrt{2}{\bar A}^{00}= T^{00} e^{+\imi\alpha} - P\,, \nonumber \\
 \sqrt{2}{A}^{+0}=  ( T^{+-} + T^{00} ) e^{-\imi\alpha} &,&  \sqrt{2}{\bar A}^{+0}= ( T^{+-} + T^{00}) e^{+\imi\alpha},  \label{eq:systemtriangular}
\end{eqnarray}
where $T^{+-}$, $T^{00}$ and $P$ are three complex parameters, among which one can be taken as real to set the global phase convention.

The actual choice of the representation for the amplitudes is irrelevant from the mathematical point of view in the frequentist approach adopted here \cite{Charles:2006vd}. We use the following alternative representation of the amplitude system, which proves more convenient for our purposes \cite{Pivk:2004hq}:
\begin{eqnarray}
 {A}^{+0}= \mu e^{\imi(\Delta-\alpha)}   &,& {\bar A}^{+0}= \mu e^{\imi (\Delta+\alpha)} \,,  \nonumber\\
 {A}^{+-}= \mu a                         &,& {\bar A}^{+-}= \mu {\bar a}  e^{2\imi {\bar\alpha}} \,, \nonumber\\
 {A}^{00}= {A}^{+0}-\frac{{A}^{+-}}{\sqrt{2}}&,&  {\bar A}^{00}= {\bar A}^{+0}-\frac{{\bar A}^{+-}}{\sqrt{2}} \,,
\end{eqnarray}
where the parameters $a$, ${\bar a}$ and $\mu$ are real positive parameters related to the modulus of the decay amplitudes and ${\bar\alpha}$ and $\Delta$ are relative phases. 
$A^{+-}$ is chosen to be a real positive quantity, which sets the phase convention. The weak phase $\alpha$ clearly appears as $2\alpha=\arg({\bar A}^{+0}/A^{+0})$. The parameter $\bar\alpha$, satisfying  $\bar\alpha=2\arg({\bar A}^{+-}/A^{+-})$, 
would coincide with  $\alpha$ in the limit of a vanishing penguin contribution.\footnote{This model is a valid representation of the amplitude system for $\alpha\ne 0$. When $\alpha$ vanishes exactly,  the  constraints $\bar a=a$ and $\bar\alpha=0$ must be added to the system in order to satisfy the equality of the $CP$-conjugate amplitudes $A^{+-}={\bar A^{+-}}$.}

This alternative representation provides a convenient and compact parametrisation enabling a fast and stable exploration of the six-dimensional parameter space. Besides its technical convenience, this representation has a  pedagogical benefit, as it exhibits the discrete ambiguities affecting the determination of the weak phase $\alpha$  in the $B\to hh$ modes in a clear way. 
Indeed, the above experimental $B\to\pi\pi$ observables can be written as a function of the chosen set of parameters as
\begin{eqnarray}
 {\cal B}^{+0}  &=& \tau_{B^{+}}\mu^2, \qquad
 {\cal B}^{+-}  = \tau_{B^{0}}\mu^2\frac{ a^2 +  {\bar a}^2}{2}\,,\nonumber\\
 {\cal B}^{00}  &=& \tau_{B^{0}}\frac{\mu^2}{4}(4+a^2+{\bar a}^2-2\sqrt{2}(ac + {\bar a}.{\bar c})\,,\nonumber\\
 {\cal C}^{+-}  &=& \frac{  a^2 -  {\bar a}^2 }{ a^2 +  {\bar a}^2}, \qquad
 {\cal S}^{+-}  = \frac{  2a{\bar a}}{ a^2 +  {\bar a}^2}\sin(2\bar\alpha)\,,\nonumber\\
 {\cal C}^{00}  &=& \frac{a^2-{\bar a}^2-2\sqrt{2}(ac + {\bar a}{\bar c})}{4+a^2+{\bar a}^2-2\sqrt{2}(ac + {\bar a}{\bar c})}.
\end{eqnarray}
where we define $c=\cos(\alpha-\Delta)$ and ${\bar c}=\cos(\alpha+\Delta-2{\bar\alpha})$. The system exhibits a fourfold trigonometric ambiguity under the phase redefinitions:
\begin{equation}\label{eq:phaseredf1}
  (\alpha,\Delta)\to
(\Delta,\alpha),(2{\bar\alpha}-\alpha,2{\bar\alpha}-\Delta),(2{\bar\alpha}-\Delta,2{\bar\alpha}-\alpha)\,.
\end{equation}
There is also an additional discrete symmetry involving not only $\alpha$ and $\Delta$, but also $\bar\alpha$: indeed,
$c$, $\bar{c}$ and ${\cal S}^{+-}$ are left invariant by the reflection:
\begin{equation} \label{eq:phaseredf2}
({\bar\alpha},\alpha,\Delta)\to\frac{\pi}{2}-({\bar\alpha},\alpha,\Delta)\,.
\end{equation}
 The combination of these two discrete symmetries yields
an heighfold ambiguity in the determination of the $\alpha$ angle in the $[0,180]^\circ$ range. Geometrically speaking, the fourfold ambiguity results from the choice left concerning the position of the apex with respect to the $A^{+0}$ base for each of the two isospin triangles, and a twofold 
ambiguity arises in relation with the relative direction of the two $A^{+0}$ and $\bar{A}^{+0}$ bases.  

In the absence of any additional input, the available $B\to\pi\pi$ observables lead thus to height strictly equivalent mirror solutions for $\alpha$. The degeneracy could be partially lifted with the measurement of the time-dependent $CP$ asymmetry in 
the $B^0\to\pi^0\pi^0$ decay, $S^{00}$.
This observable can be written as:
\begin{eqnarray}\label{eq:S00pipi}
 {\cal S}^{00} &=& \frac{4\sin(2\alpha)+2a{\bar a}\sin(2{\bar\alpha})-2\sqrt{2} ( as+{\bar a}{\bar s})}{4+a^2+{\bar a}^2-2\sqrt{2}(ac + {\bar a}{\bar c})}\,,
\end{eqnarray}
where $s=\sin(\alpha+\Delta)$ and ${\bar s}=\sin(\alpha-\Delta+2{\bar\alpha})$ are not invariant under the phase redefinition of $\alpha$ and $\Delta$ Eq.~(\ref{eq:phaseredf1}), leaving only a twofold ambiguity on $\alpha$. The $B^0\to\pi^0\pi^0$ decay is so far observed only in the 4-photon final state preventing the measurement of the time-dependent decay rate. Future high-luminosity facilities have investigated the feasibility of the  $\S^{00}_{\pi\pi}$ asymmetry measurement, either by considering the rare Dalitz decays of the neutral pions or by exploiting the conversion of photons in the detector material~\cite{Ishino:2007pt,Aushev:2010bq}.

\begin{figure}[t]
\begin{center}
  \includegraphics[width=18pc]{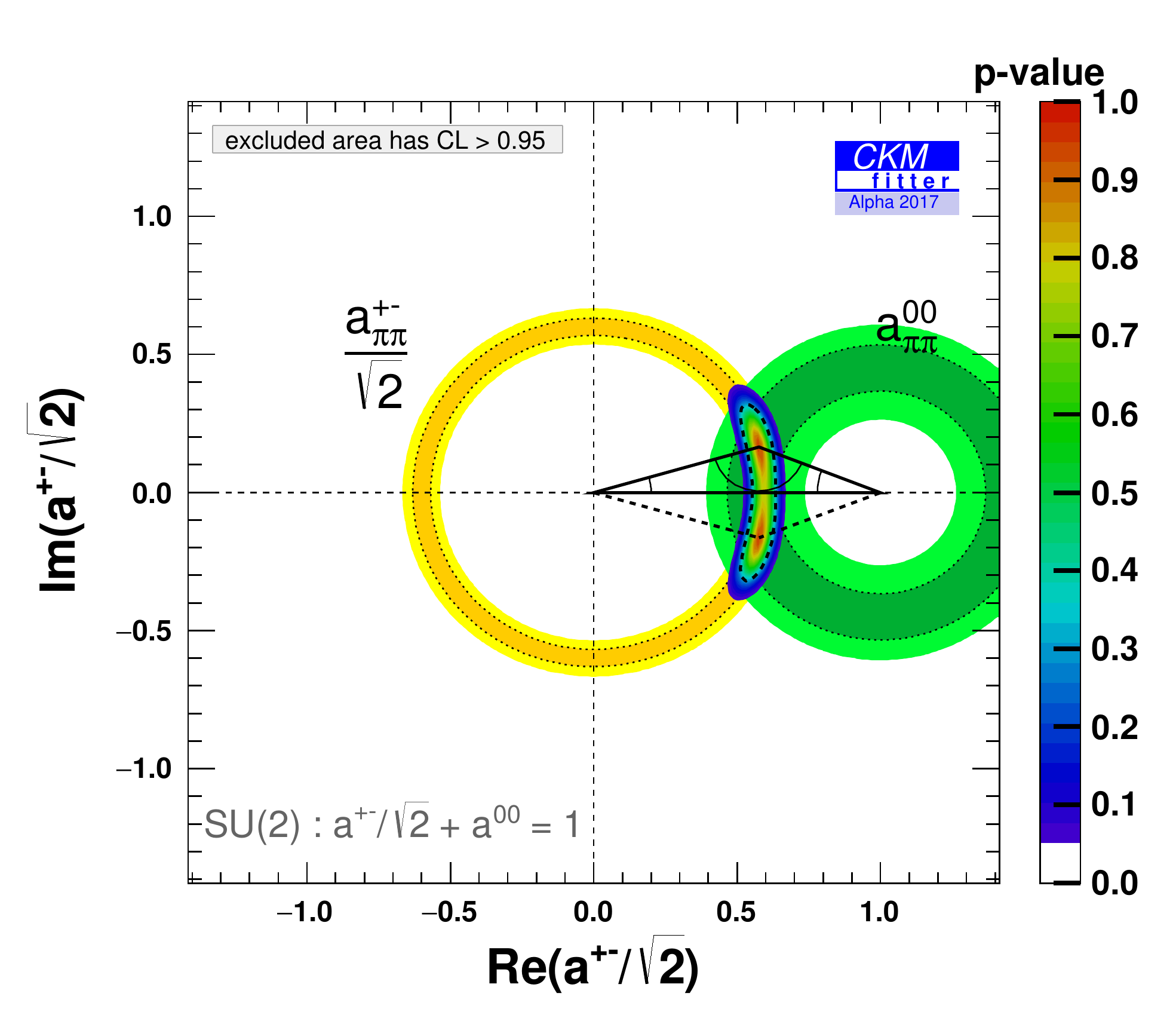}
  \includegraphics[width=18pc]{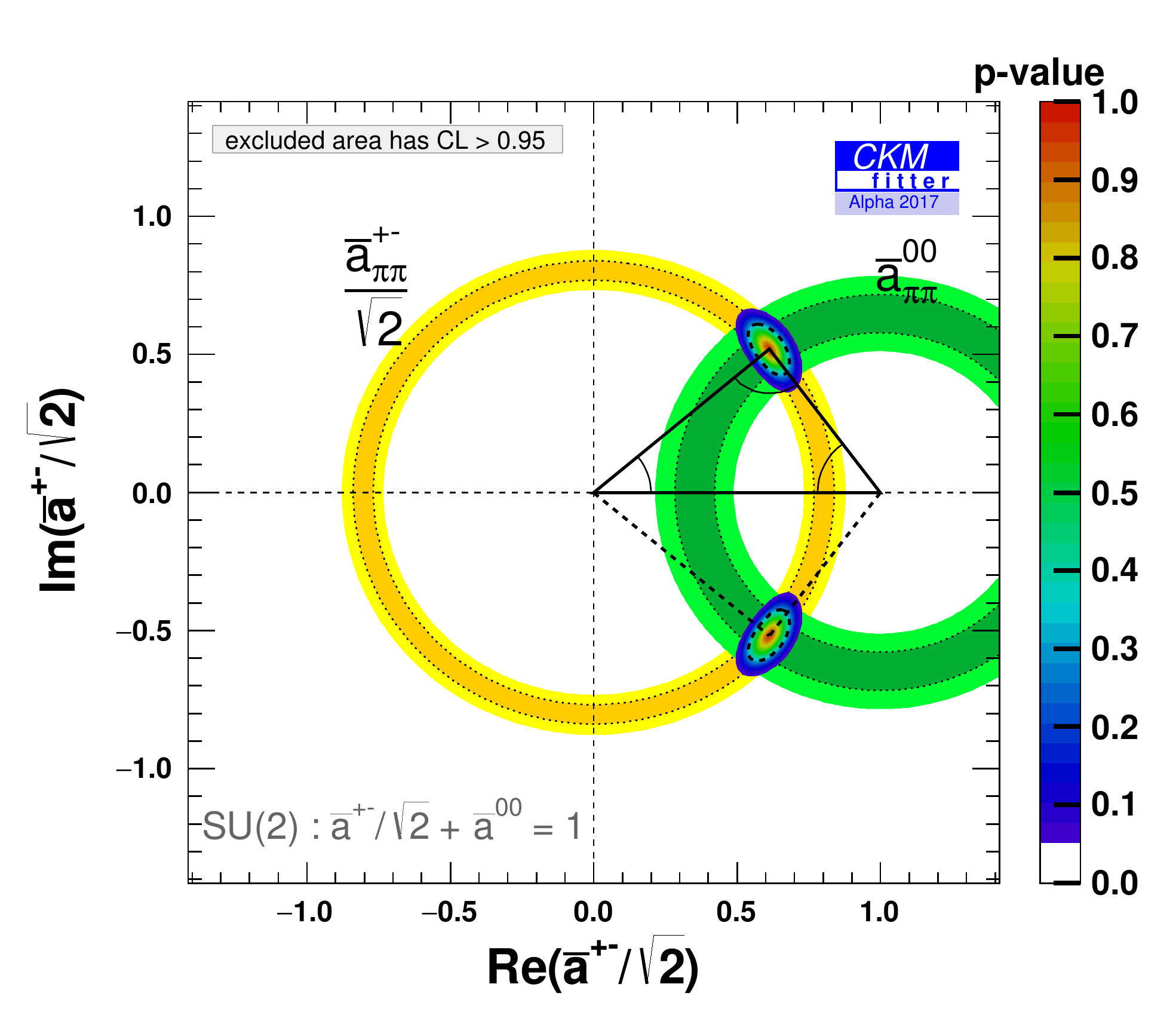}
\caption{\it\small  Constraint on the reduced amplitude $a^{+-}=A^{+-}/A^{+0}$ in the complex plane for the $B\to\pi\pi$ (left) and ${\bar B}\to\pi\pi$ systems (right). The individual constraint from the $B^0(\bar B^0)\to\pi^+\pi^-$ observables and from the $B^0(\bar B^0)\to\pi^0\pi^0$ observables are indicated by the yellow and green circular areas, respectively. The  corresponding isospin triangular relations $a^{00}+{a^{+-}}/{\sqrt{2}}=1$ (and $CP$ conjugate) are represented by the black triangles.}
\label{fig:PiPiSU2}
\end{center}       
\end{figure}

\begin{figure}[t]
\begin{center}
  \includegraphics[width=18pc]{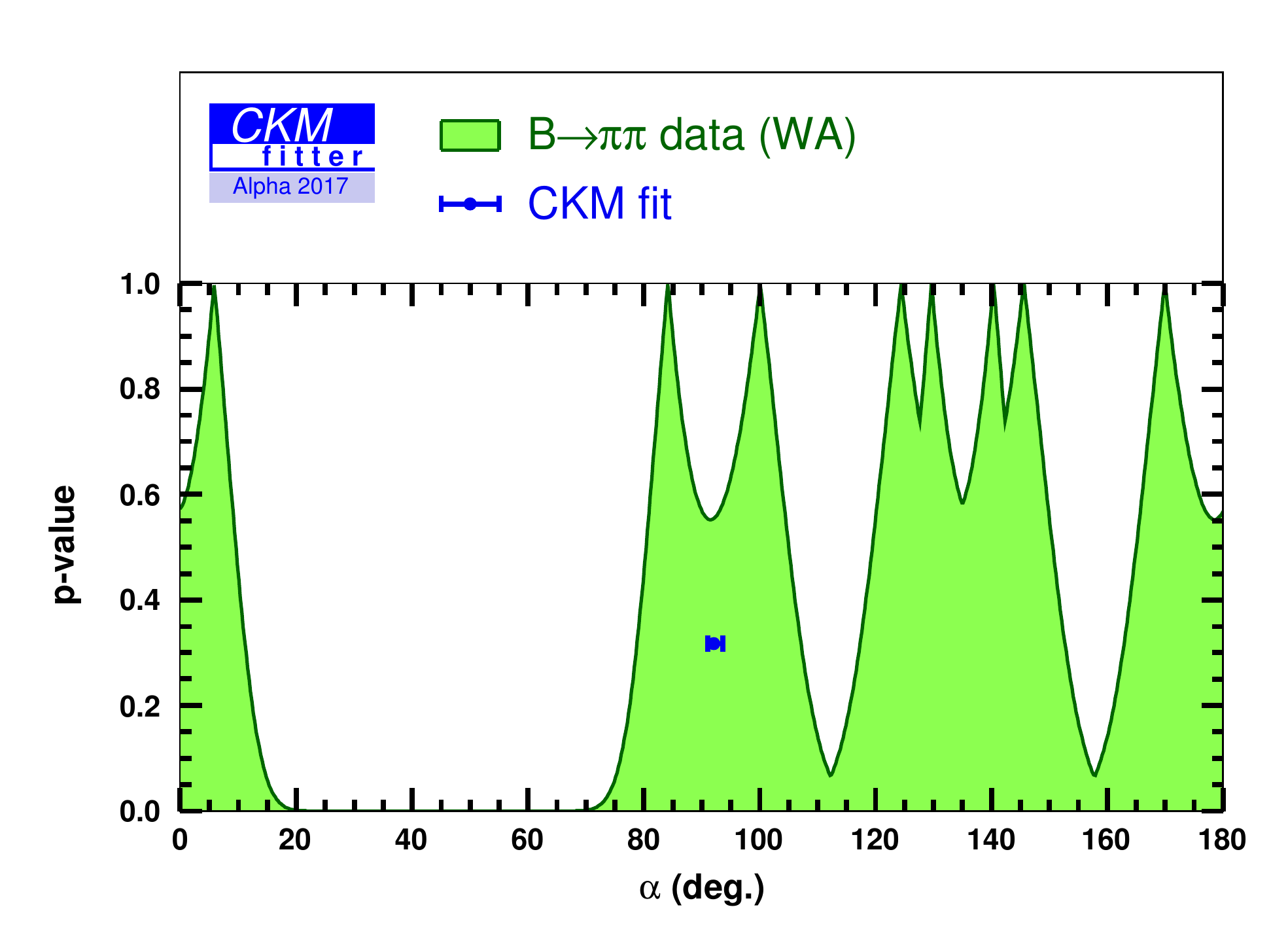}
  \includegraphics[width=18pc]{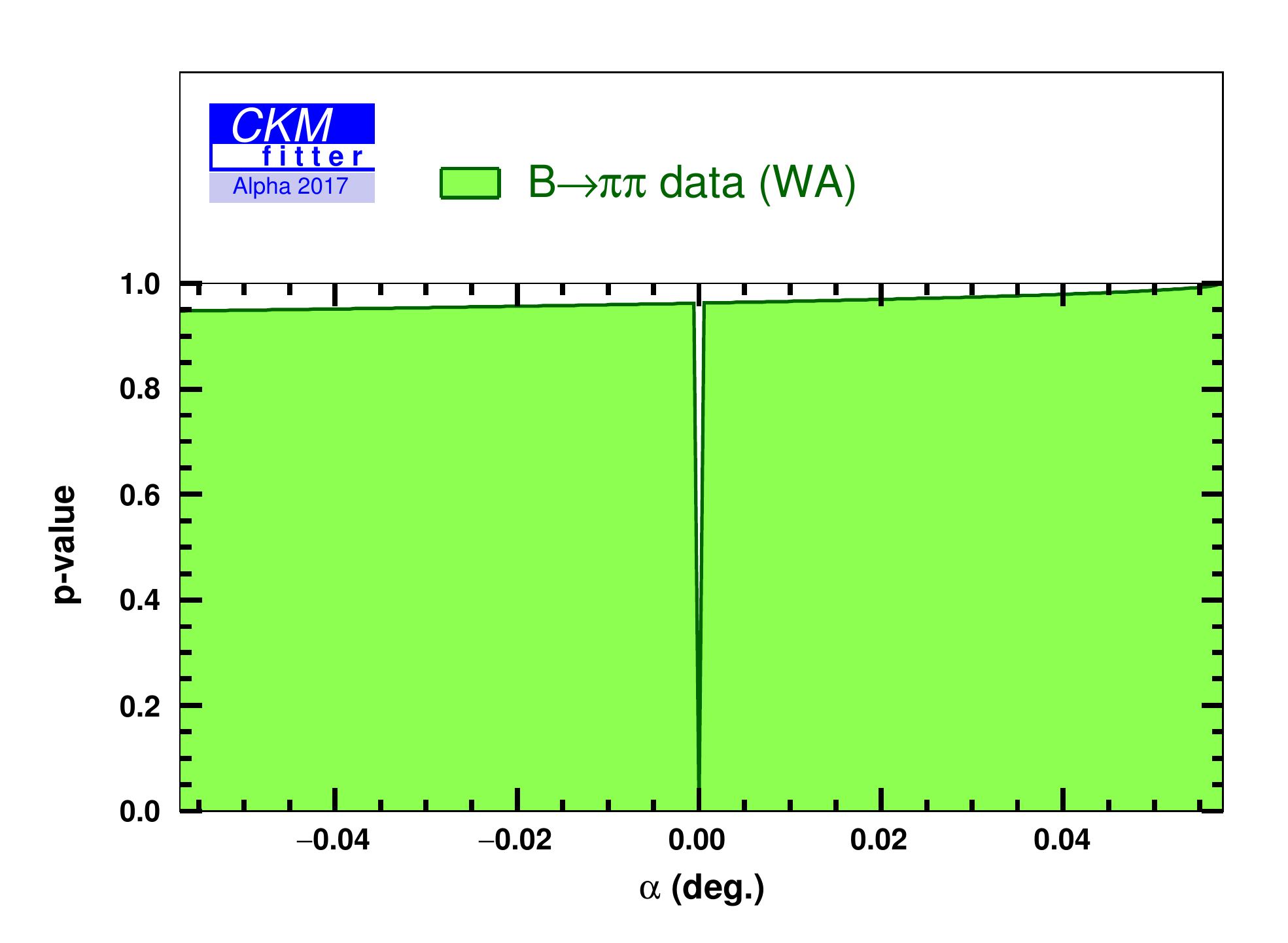}
\caption{\it\small  One-dimensional scan of $\alpha$ in the $[0,180]^\circ$ range (left) from the \su{2} isospin analysis of the $∫B\to\pi\pi$ system. The interval with a dot indicates the indirect $\alpha$ determination introduced in Eq.~(\ref{eq:alphaInd}).
The zoom in the vicinity of $\alpha=0$ (right figure) displays the punctual discontinuity, whose width due to the scan binning has no physical meaning.}
\label{fig:PiPialpha}
\end{center}       
\end{figure}

The consistency of the experimental $B\to\pi\pi$ data   with the triangular isospin relation of Eq.~(\ref{eqn:triangular}) can  be assessed by the means of the two-side sum:
\begin{equation}
t=\frac{|a^{+-}|}{\sqrt{2}}+|a^{00}| 
\end{equation}
where $a^{ij}$ denotes the normalised amplitude $a^{ij}={A^{ij}}/{A^{+0}}$. The sum of the length of two sides of a triangle being greater than the third one \cite{Euclide}, we have $t\geq 1$, the equality occurring for a flat triangle. The values
\begin{equation}
t=1.05\pm 0.09 \qquad \textrm{and} \qquad
{\bar t}=1.45\pm 0.08
\end{equation}
are obtained for the $B\to\pi\pi$ and ${\bar B}\to\pi\pi$ systems and are consistent with an almost-flat and an open triangle, respectively, as illustrated in Fig.~\ref{fig:PiPiSU2}.  Both isospin triangles display the expected mirror symmetry, therefore the $\alpha$ constraint  exhibits height non-degenerate solutions in $[0,180]^\circ$, as illustrated in Fig.~\ref{fig:PiPialpha}.  The minima of the \chisq test statistic over the parameter space ($\chi^2_{\rm min}=0$) are found at  
$\alpha=5.9^\circ$, $\alpha=84.1^\circ$, $\alpha=100.1^\circ$, $\alpha=124.4^\circ$, $\alpha=129.6^\circ$, $\alpha=140.4^\circ$, $\alpha=125.6^\circ$ and $\alpha=169.9^\circ$.
This system of six observables for six independent parameters is not over-constrained and the vanishing value of $\chi^2_{\rm min}$ reflects the closure of both the $\bar B\to\pi\pi$ and the $B\to\pi\pi$  isospin triangles. The corresponding (symmetrised) $\alpha$ intervals at 68\% and 95\% CL are
\begin{eqnarray} \label{eq:alphapipi}
\alpha_{\pi\pi}: &&(\val{93.0}{14.0}   \quad\cup\quad \val{135.0}{17.7} \quad\cup\quad \val{177.1}{14.2})^\circ~~(68\%\ {\rm CL})~~\mathrm{and}~~\nonumber\\
&&(\val{135.3}{60.3})^\circ~~(95\%\ {\rm CL})\,.
\end{eqnarray}
The interval $\alpha = [20.0,71.0]^\circ$ is excluded at more than 3 standard deviations by the \su{2} isospin analysis of the $B\to\pi\pi$ system. 
The solution around $90^\circ$ is in excellent agreement with the indirect determination of $\alpha$, Eq.~(\ref{eq:alphaInd}), from the other constraints of the global CKM fit \cite{CKMfitterSummer16}.  If we include $\alpha_{\rm ind}$  derived from Fig.~\ref{fig:alphaInd} as an additional and independent constraint to the \su{2} isospin fit, the minimal \chisq  increases by 0.3, indicating a consistency at the level of 0.6 standard deviation. 

It is interesting to note the isospin constraints present a discontinuity at $\alpha=0$ (modulo $\pi$) as illustrated on the right-hand side of Fig.~\ref{fig:PiPialpha}. The $\alpha=0$ hypothesis is rejected with a high significance ($\chi^2(\alpha=0)=296$) as expected by the direct $CP$ violation observed in the $B^0\to\pi^+\pi^-$ decay \cite{Babar_pp_cpm,Belle_pp_cpm}.  However, for any finite $\alpha$ in the vicinity of zero, the isospin relation can be satisfied if arbitrary large penguin amplitudes are allowed (the limit case $\alpha\to 0$ happens if $|T^{+-}|$, $|T^{00}|$, $|P|$ are sent to infinity, see Eq.~(\ref{eq:systemtriangular}) and Refs.~\cite{Charles:2006vd,Charles:2007yy}). The limit $\alpha\to 0$  was also discussed in the context of the Bayesian statistical approach in Ref.~\cite{Bona:2007qta}.
External data, e.g., based on \su{3} consideration, could be used to set bounds on the penguin over tree ratio and thus further constrain  $\alpha$ around 0. These bounds stand beyond the pure $\su{2}$ approach adopted here and we will not attempt at including such additional information for the following reasons. On one hand, other charmless modes (such as $\rho\pi$) will provide further information on the  small $\alpha$ region, and on the other hand, we will ultimately use the determination of $\alpha$ in a global fit together with other well-controlled constraints on the CKM parameters \cite{CKMfitterSummer16}: both types of constraints will rule out the region around $\alpha=0$, so that additional theoretical hypotheses on the size of the $B\to\pi\pi$ amplitudes are not necessary for our purposes.

\subsection{Isospin analysis of the $B\to\rho\rho$ system\label{subsec:rhorho}}

A similar \su{2} isospin analysis can be performed on each polarisation state of the $B\to\rho\rho$ system. 
Experimentally, the decays are completely dominated by the longitudinal polarisation of the $\rho$ mesons, and in the following, we will always imply this polarisation for the $\rho\rho$ final state (which will be denoted explicitly as $\rho_L\rho_L$ when we quote experimental measurements).
This final state presents several advantages with respect to $B\to\pi\pi$. The measured branching fractions ${\cal B}^{+-}$ and ${\cal B}^{+0}$ are approximately five times larger than for $B\to\pi\pi$, while  ${\cal B}^{00}$ is of same order of magnitude, indicating a smaller penguin contamination in the $\rho\rho$ case. Moreover, the time-dependent $CP$ asymmetry in the $B^0\to\rho^0\rho^0$ decay, ${\cal S}^{00}$, is experimentally accessible for the final state with four charged pions, potentially lifting some of the discrete ambiguities affecting the determination of $\alpha$. However, the current measurement ${\cal S}^{00}=0.3\pm0.7$ suffers from large uncertainties, leaving pseudo-mirror solutions in $\alpha$. The available experimental observables\footnote{As can be seen from Tab.~\ref{input_rhorho}, the longitudinal polarisation $f_L$ is always used as an input in combination with branching ratios ${\cal B}$, and both are defined as $CP$-averaged quantities.} and their current world averages are summarised in Tab.~\ref{input_rhorho}. Under the \su{2} isospin hypothesis, the direct $CP$ asymmetry in $B^+\to \rho^+\rho^0$ vanishes and we will not take into account the experimental measurement of this quantity (which is consistent with our hypothesis and will be used to test this assumption in Sec.~\ref{sec:tests}).
Seven independent observables are available for the longitudinal helicity state of the $\rho$ mesons, allowing us to over-constrain the six-dimensional  parameter space of the \su{2} isospin analysis. 

\begin{table}[t]
\begin{center}
\begin{tabular}{|l|c|l|}
\hline
Observable & World average & References \\
\hline
\small ${\cal B}^{+-}_{\rho\rho}~ \times~ f_L^{+-}$ $(\times 10^6)$ &\small $(27.76\pm 1.84)$ $\times$ $(0.990\pm 0.020)$ &\small \cite{Babar_rr_bpm,Belle_rr_pm} \\ 
\small ${\cal B}^{+0}_{\rho\rho}$ $\times$ $f_L^{+0}$   $(\times 10^6)$ &\small $(24.9\pm 1.9)$ $\times$  $(0.950\pm 0.016)$       &\small \cite{Babar_rr_bp0,Belle_rr_bp0} \\
\small ${\cal B}^{00}_{\rho\rho}$ $\times$ $f_L^{00}$ $(\times 10^6)$ &\small $(0.93\pm 0.14)$ $\times$ $(0.71\pm 0.06)$        &\small \cite{Babar_rr_b00,Belle_rr_b00,LHCb_rr_b00} \\
\small ${\cal C}^{+-}_{\rho_L\rho_L}$                &\small $-0.00\pm 0.09$                      &\small \cite{Babar_rr_bpm,Belle_rr_pm} \\
\small ${\cal S}^{+-}_{\rho_L\rho_L}$                &\small $-0.15\pm 0.13$                      &\small \cite{Babar_rr_bpm,Belle_rr_pm} \\
\small $\rho(C^{+-}_{\rho_L\rho_L},S^{+-}_{\rho_L\rho_L})$ &\small $+0.0002$                         &\small \cite{Babar_rr_bpm,Belle_rr_pm} \\
\small ${\cal C}^{00}_{\rho_L\rho_L}$                &\small $0.2\pm 0.9$                         &\small \cite{Babar_rr_b00,Belle_rr_b00,LHCb_rr_b00} \\
\small ${\cal S}^{00}_{\rho_L\rho_L}$                &\small $0.3\pm 0.7$                         &\small \cite{Babar_rr_b00,Belle_rr_b00,LHCb_rr_b00} \\
\hline
\end{tabular}
\caption{\it\small  World averages for the relevant experimental observables in the $B\to\rho^i\rho^j$ modes: branching fraction ${\cal B}^{ij}_{\rho\rho}$, fraction of longitudinal polarisation $f_L^{ij}$, time-integrated $CP$ asymmetry ${\cal C}^{ij}_{\rho\rho}$, time-dependent asymmetry  ${\cal S}^{ij}_{\rho\rho}$ and correlation ($\rho$).\label{input_rhorho}}
\end{center}
\end{table}

\begin{figure}[t]
\begin{center}
  \includegraphics[width=18pc]{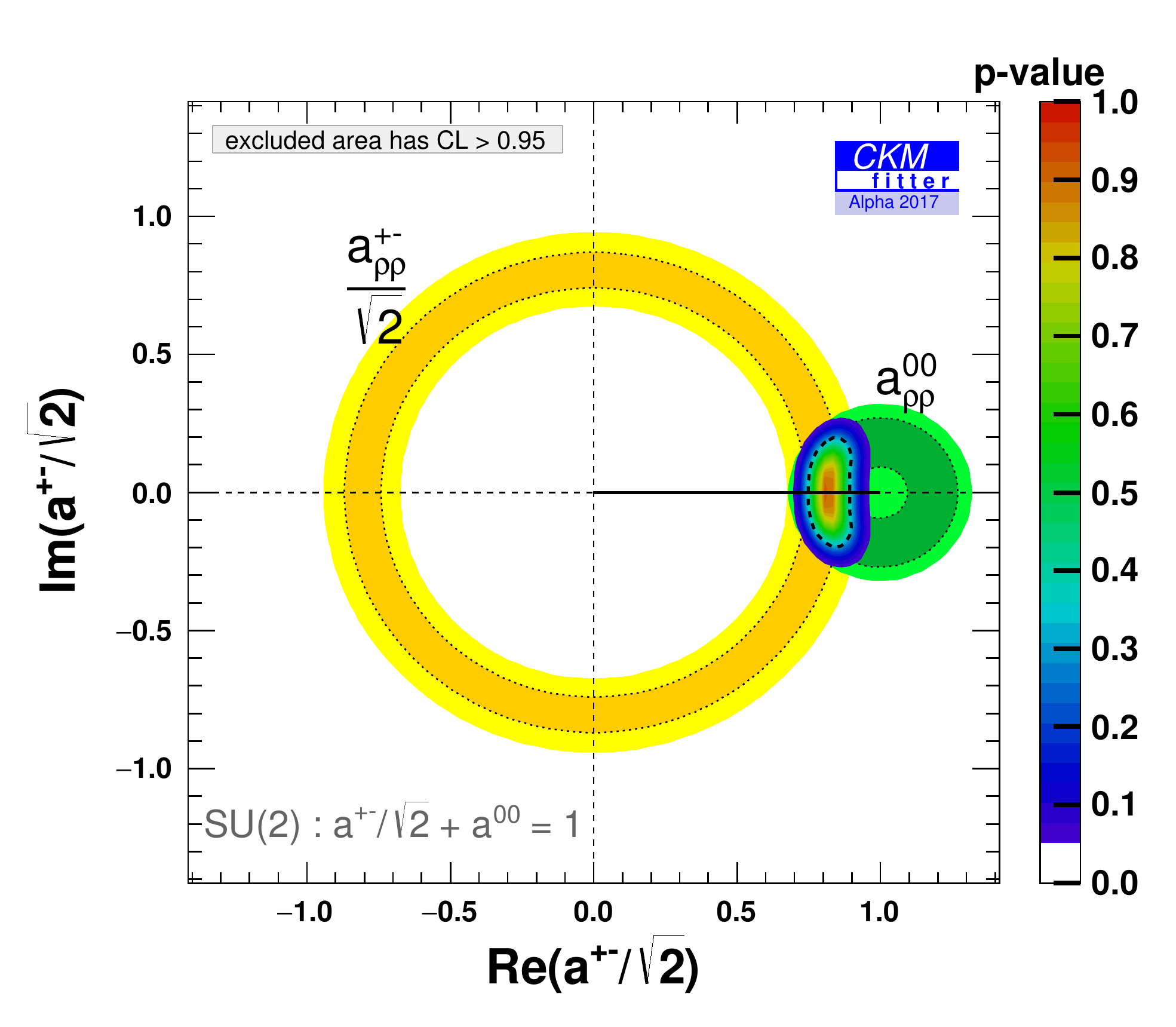}
  \includegraphics[width=18pc]{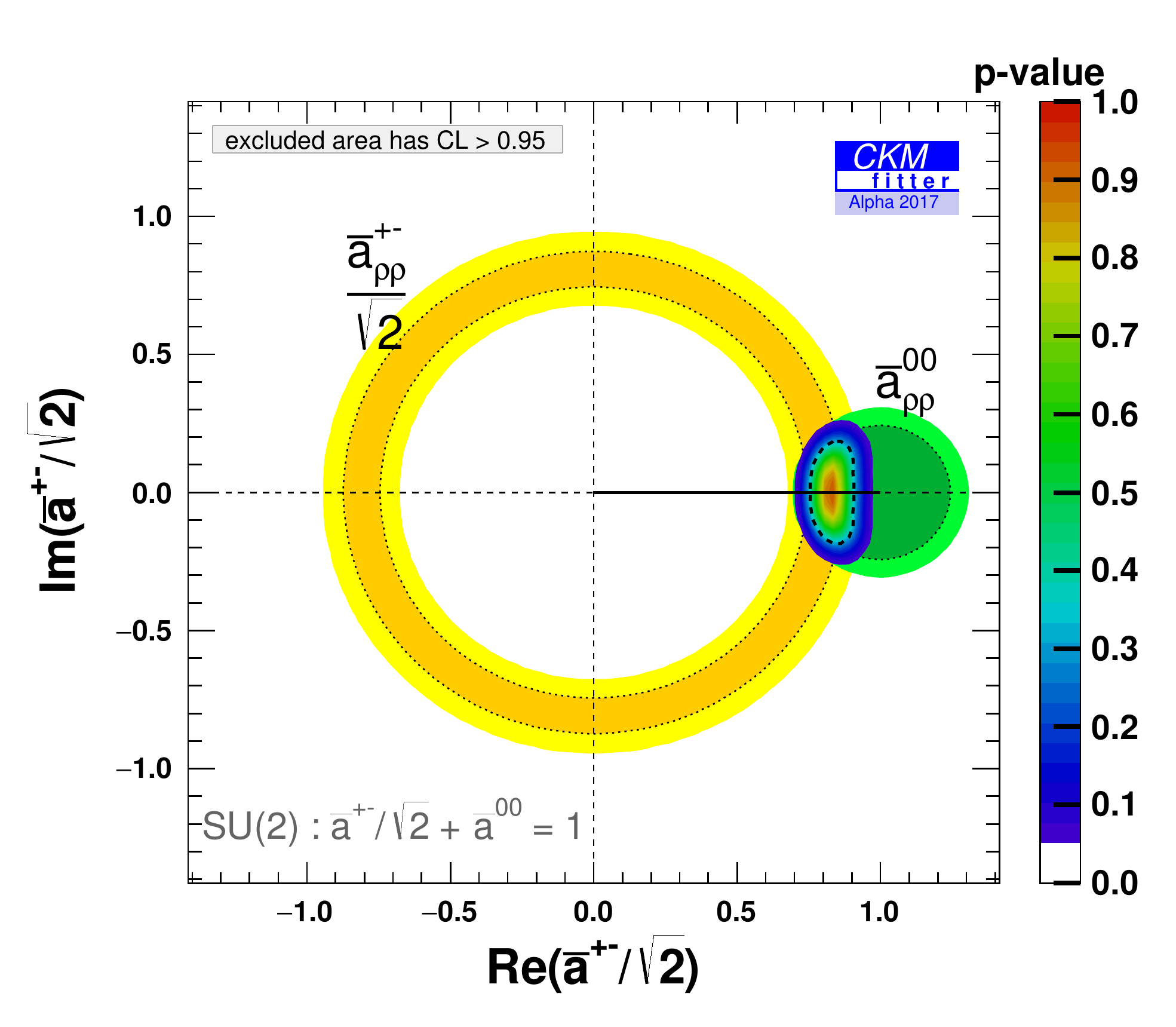}
\caption{\it\small Constraint on the reduced amplitude $a^{+-}=A^{+-}/A^{+0}$ in the complex plane for the $B\to\rho\rho$ (left) and ${\bar B}\to\rho\rho$ systems (right). The individual constraint from the $B^0(\bar B^0)\to\rho^+\rho^-$ observables and from the $B^0(\bar B^0)\to\rho^0\rho^0$ observables are indicated by the yellow and green circular areas, respectively. The  corresponding isospin triangular relations $a^{00}+{a^{+-}}/{\sqrt{2}}=1$ (and $CP$ conjugate) are represented by the black line. Both $B\to\rho\rho$ and ${\bar B}\to\rho\rho$ isospin triangles are flat.}
\label{fig:RhoRhoSU2}
\end{center}       
\end{figure}

\begin{figure}[t]
\begin{center}
  \includegraphics[width=18pc]{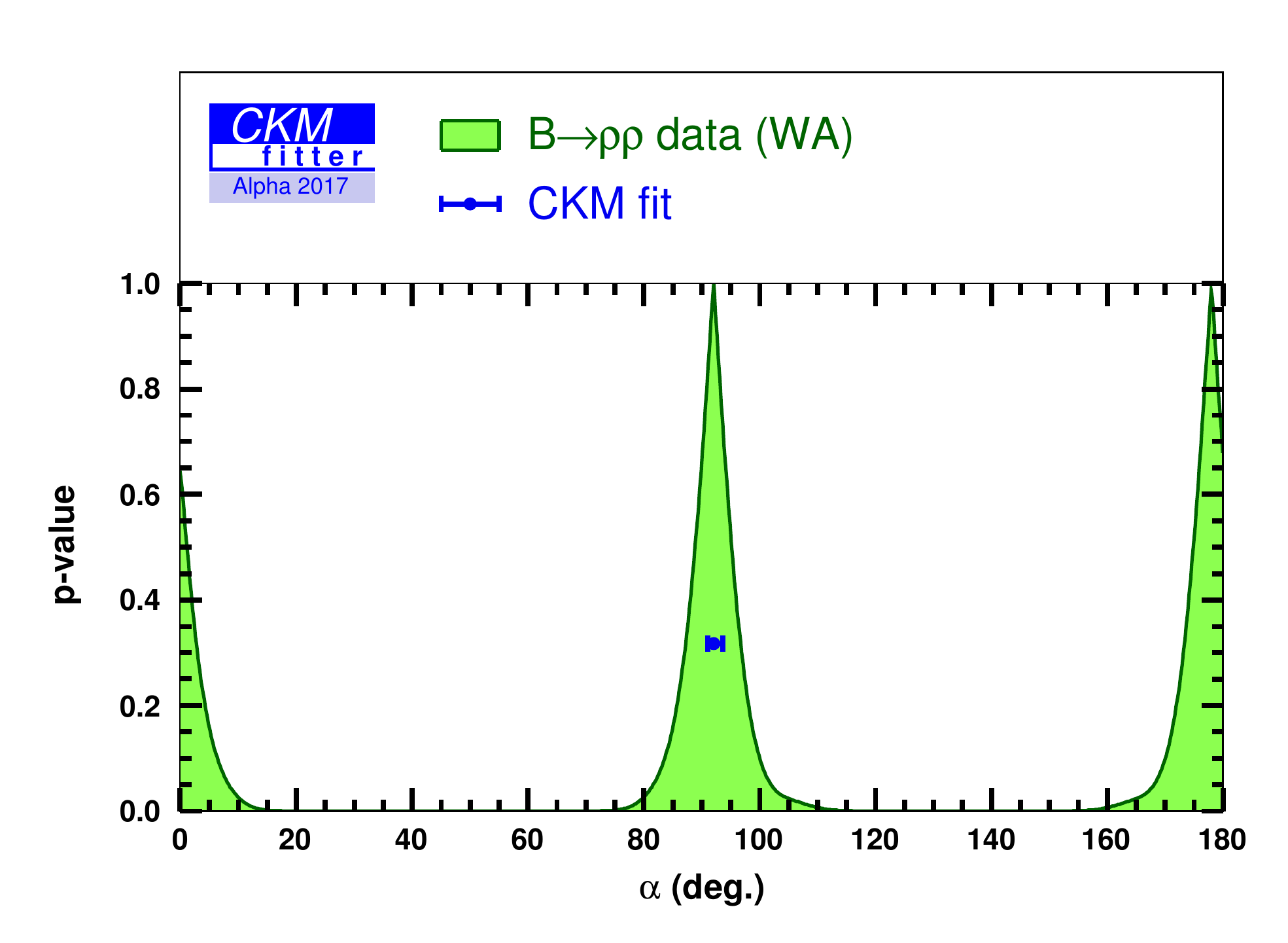}
  \includegraphics[width=18pc]{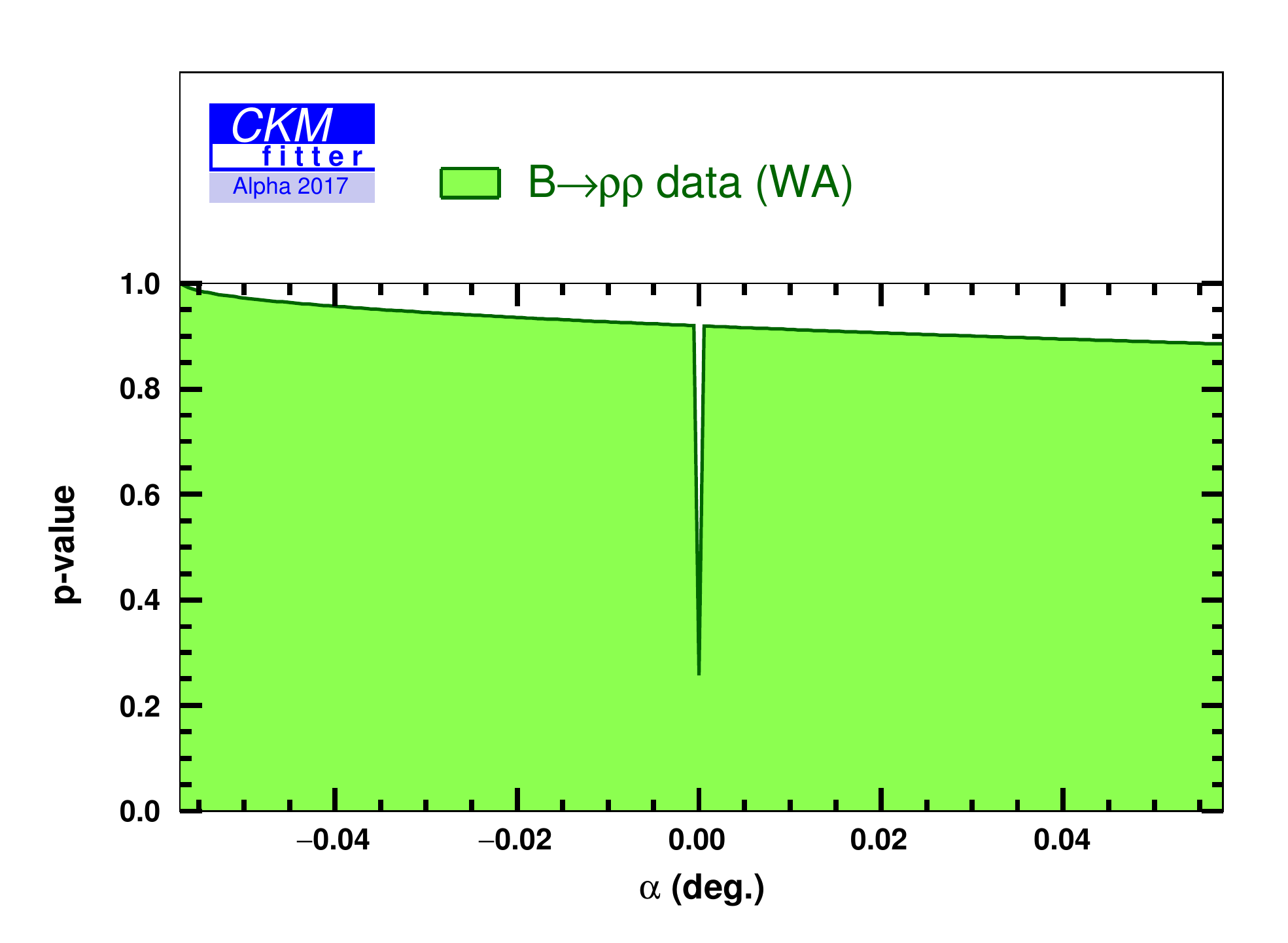}
\caption{\it\small  One-dimensional scan of $\alpha$ in the $[0,180]^\circ$ range (left) from the \su{2} isospin analysis of the $B\to\rho\rho$ system. The interval with a dot indicates the indirect $\alpha$ determination introduced in Eq.~(\ref{eq:alphaInd}).
The zoom in the vicinity of $\alpha=0$ (right) displays the punctual discontinuity, whose width, due to the scan binning, has no physical meaning.}
\label{fig:RhoRhoalpha}
\end{center}       
\end{figure}

The situation of the $B\to\rho\rho$ system is illustrated in Fig.~\ref{fig:RhoRhoSU2}.
The following two-side sums are consistent with flat triangles for both $B\to\rho\rho$ and ${\bar B}\to\rho\rho$ systems:
\begin{equation}
t=1.00\pm 0.10 \qquad {\rm and}\qquad {\bar t}=0.97\pm 0.11\,.
\end{equation}
 This further reduces the degeneracy of the (pseudo) mirror solutions in $\alpha$ leaving a twofold ambiguity; see Fig.~\ref{fig:RhoRhoalpha}. The minima of the \chisq test statistic over the parameter space ($\chi^2_{\rm min}=0.14$) are found at  $\alpha=92.1^\circ$ and $178.0^\circ$. The corresponding 68 and 95\% CL intervals on $\alpha$ are
\begin{eqnarray}\label{eq:alpharhorho}
\alpha_{\rho\rho} : && (\val{92.0}{+4.7}{-4.8}   \quad\cup\quad  \val{177.9}{+4.9}{-4.6})^\circ ~~(68\%\ {\rm CL})~~\mathrm{and} \nonumber\\
&& (\val{92.0}{+10.0}{-11.0}   \quad\cup\quad \val{177.9}{+10.7}{-10.0})^\circ ~~(95\%\ {\rm CL})\,.
\end{eqnarray}
The solution around $90^\circ$ is in excellent agreement (less than 0.1 standard deviation) with the indirect determination of $\alpha$, Eq.~(\ref{eq:alphaInd}), from the other constraints of the global CKM fit \cite{CKMfitterSummer16}, and is more tightly constrained than in the $B\to \pi\pi$ case, in relation with the smaller penguin contamination.

The region $\alpha = [14.0,76.0]^\circ \cup [112.0,158.0]^\circ$ is excluded at more than 3 standard deviations by the \su{2} isospin analysis of the $B\to\rho\rho$ system. On the right hand side of Fig.~\ref{fig:RhoRhoalpha}, a zoom around $\alpha=0$ exhibits a small discontinuity, corresponding to $\chi^2(\alpha=0)=1.61$ and indicating that this hypothesis is mildly disfavoured by the data, consistently with the absence of large direct CP asymmetries in $B\to\rho\rho$ decays.

\subsection{Isospin analysis of the $B\to\rho\pi$ system\label{subsec:rhopi}}

The three neutral and two charged  $B^{0,+}\to(\rho\pi)^{0,+}$ decays can be described with 10 complex (tree and penguin) amplitudes and one weak phase $\alpha$, i.e., 21 real parameters. Assuming the  pentagonal isospin relation Eq.~(\ref{eqn:pentagonal}) that leaves only two independent complex penguin contributions, the number of degrees of freedom of the system is reduced to 12 after setting an irrelevant global phase to zero. The dimension of the system gets down to 10 if we consider the three neutral decays only.  The time-dependent Dalitz analysis of the neutral  $B^{0}\to(\rho^\pm\pi^\mp,\rho^0\pi^0)\to\pi^+\pi^-\pi^{0}$ transitions has been shown to carry enough information to fully constrain the isospin-related $B\to\rho\pi$ system,  thanks to the finite width of the intermediate $\rho$-mesons that yields a richer interference pattern of the three-body decay \cite{Gronau:1990ka,Lipkin:1991st}. For instance, 11 independent phenomenological observables can be defined from the flavour-tagged time-dependent Dalitz distribution of the three-pion decay, e.g., the  branching fractions of the three intermediate $\rho^i\pi^j$ transitions, the corresponding direct and time-dependent $CP$ asymmetries and  two relative phases between their amplitudes.  

The extraction of the parameter $\alpha$ through the  $\su{2}$ analysis of the neutral modes will be referred to as  the ``Dalitz  analysis'' and will be considered first in Sec.~\ref{subsec:dalitz}. An  extended analysis  (referred to as  the ``pentagonal analysis''), including the information coming from the charged decays  $B^+\to\rho^+\pi^0$ and $B^+\to\rho^0\pi^+$,  will be discussed in Sec.~\ref{subsec:pentagonal}.

\subsubsection{Dalitz  analysis of the neutral $B^0\to(\rho\pi)^0$ modes\label{subsec:dalitz}}

Considering the neutral modes only, the time-independent amplitude of the $\pi^+\pi^-\pi^0$ final state can be written as
\begin{eqnarray}
 A_{3\pi}&=&  f_+ A^{+} + f_- A^{-} + f_0 A^{0}\,,\\
 {\bar A}_{3\pi}&=&  f_+ {\bar A}^{+} + f_- {\bar A}^{-} + f_0 {\bar A}^{0}\,, \nonumber
\end{eqnarray}
where $A^{i}$ (${\bar A}^{i}$) is the amplitude of the $B^0({\bar B}^0)\to\rho^i\pi^j$ transition\footnote{With this convention the $A^i$ (${\bar A}^i$) amplitude carries a superscript referring to the electric charge of the $\rho^i$ meson in both $B^0$ and ${\bar B}^0$ decays, i.e.,  $A^{+}=A^{+-}$, $A^{-}=A^{-+}$ and ${\bar A}^{+}={\bar A}^{-+}$,  ${\bar A}^{-}={\bar A}^{+-}$  where the $A^{ij}$ (${\bar A}^{ij}$) amplitude is defined in Sec.~\ref{sec:amplitudes}.}  and $f_i$ ($i=-,+,0$) is the form factor accounting for the $\rho^i$ line-shape.
Neglecting the tiny $B^0$ width difference $\Delta\Gamma_d$, the time-dependent decay rate can be written as a function of three combinations of these amplitudes:
\begin{equation}
\Gamma(t)_{{B^0}/{\bar B^0}\to\pi^+\pi^-\pi^0} \propto e^{-t/\tau_{B^0}}\left[ \Gamma_{0} \mp  \Gamma_{\cal C}\cos(\Delta m_d t ) \pm 2\Gamma_{\cal S}\sin(\Delta m_d t )\right], 
\end{equation}
with   $\Gamma_0=(|A_{3\pi}|^2 + |{\bar A}_{3\pi}|^2)$,  $\Gamma_{\cal C}=(|A_{3\pi}|^2 - |{\bar A}_{3\pi}|^2)$ and $\Gamma_{\cal S}={\cal I}m\left(\frac{q}{p} {\bar A}_{3\pi}A_{3\pi}^*\right)$.

These amplitudes are described phenomenologically in terms of form factors $f_i$ describing the decay of the $\rho$ meson into two pions, multiplied by coefficients, denoted $\cal U$ and $\cal I$, related to the $B\to\rho\pi$ dynamics:
\begin{eqnarray}
  |A_{3\pi}|^2\pm |{\bar A}_{3\pi}|^2              &=& \sum\limits_{i=+,-0} {\cal U}^\pm_i |f_i|^2    +2\sum\limits_{j<i}  {\cal U}^{\pm,{\cal R}e}_{ij} {\cal R}e( f_i f_j^*) -2\sum\limits_{j<i}  {\cal U}^{\pm,{\cal I}m}_{ij} {\cal I}m( f_i f_j^*)  \nonumber \\
  {\cal I}m\left(\frac{q}{p} {\bar A}_{3\pi}A_{3\pi}^*\right) &=& \sum\limits_{i=+,-0} {\cal I}_i |f_i|^2       + \sum\limits_{j<i}  {\cal I}^{{\cal R}e}_{ij} {\cal R}e( f_i f_j^*)     + \sum\limits_{j<i}  {\cal I}^{{\cal I}m}_{ij} {\cal I}m( f_i f_j^*)
\end{eqnarray}
with 
\begin{eqnarray}
{\cal U}^\pm_{i} =  |A^{i}|^2 \pm |{\bar A}^{i}|^2\,, && 
{\cal I}_{i} =  {\cal I}m( {\bar A}^{i}A^{i^*})\,, \nonumber \\
{\cal U}^{\pm,{\cal R}e}_{ij} = {\cal R}e(A^{i} A^{j^{*}}\pm {\bar A}^{i}{\bar A}^{j^*})\,, &&
{\cal I}^{{\cal R}e}_{ij} =  {\cal R}e( {\bar A}^{i}A^{j^*} - {\bar A}^{i} A^{j^*})\,, \nonumber \\
{\cal U}^{\pm,{\cal I}m}_{ij} = {\cal I}m(A^{i} A^{j^*}\pm {\bar A}^{i}{\bar A}^{j^*})\,, &&
{\cal I}^{{\cal I}m}_{ij} = {\cal I}m( {\bar A}^{i}A^{j^*} + {\bar A}^{i} A^{j^*})\,.   \label{eq:UandI}
\end{eqnarray}
For given form factors $f_i$, these 27 coefficients  $\cal U$ and $\cal I$ fully parametrise the time-dependent Dalitz distribution of the $B\to\pi^+\pi^-\pi^0$ decay. The \U parameters are related to the branching fractions and the direct $CP$ asymmetries in the  $B\to\rho^i\pi^j$ intermediate states while the \I coefficients parametrise the mixing-induced $CP$ asymmetries.

\begin{table}[t]
\begin{center}
\includegraphics[width=24pc]{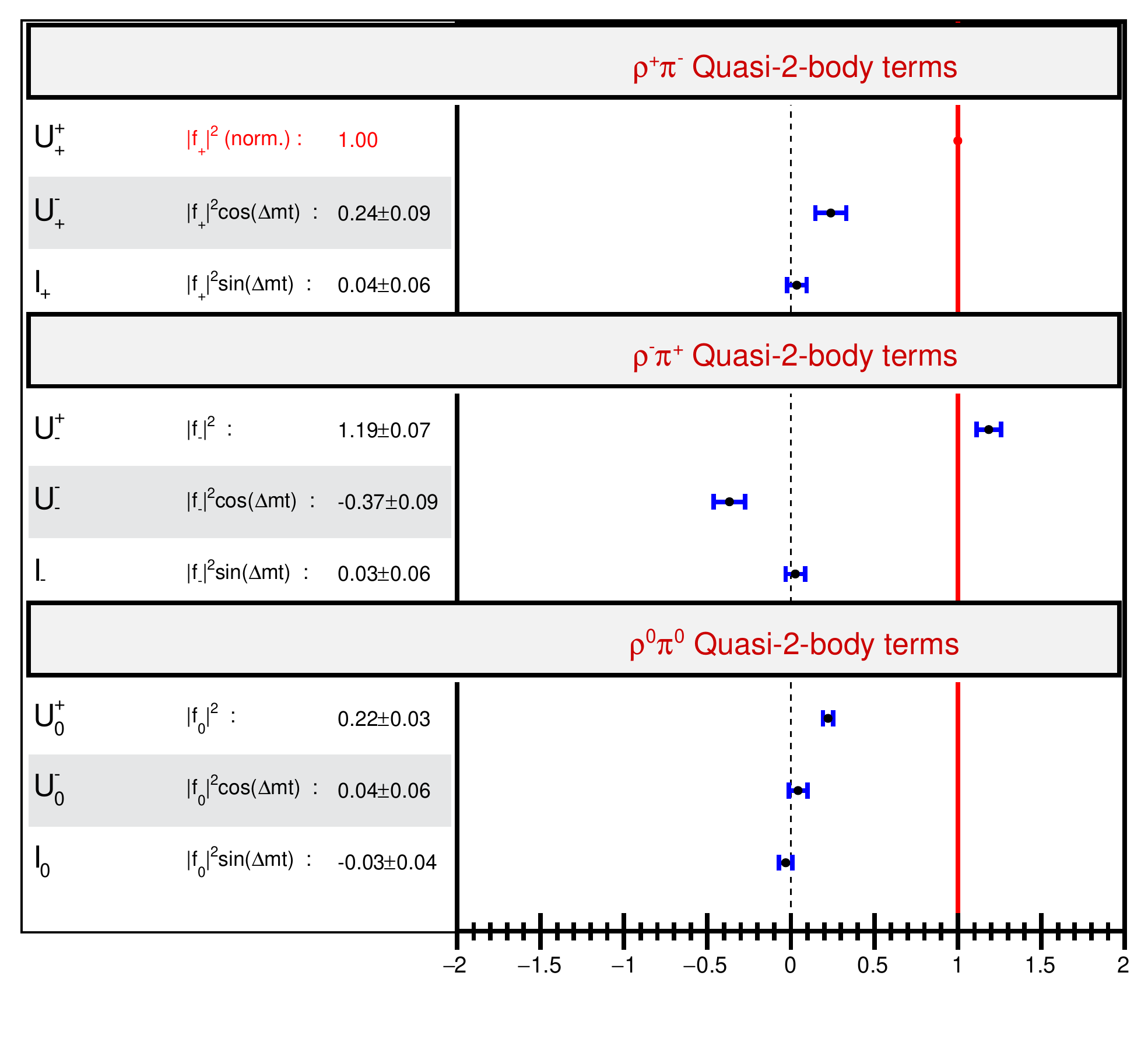}
\caption{\it \small Relative value of the quasi-two-body related $\cal U$ and $\cal I$ coefficients extracted from the time-dependent Dalitz analysis of the $B^0\to(\rho\pi)^0$  decay \cite{Babar_rp,Belle_rp}. The corresponding form-factor bilinear is indicated for each coefficient. The red vertical line indicates the overall normalisation, defined by  $U^+_+=1$.}\label{input_rhopi1}
\end{center}       
\end{table}

\begin{table}[t]
\begin{center}
\includegraphics[width=24pc]{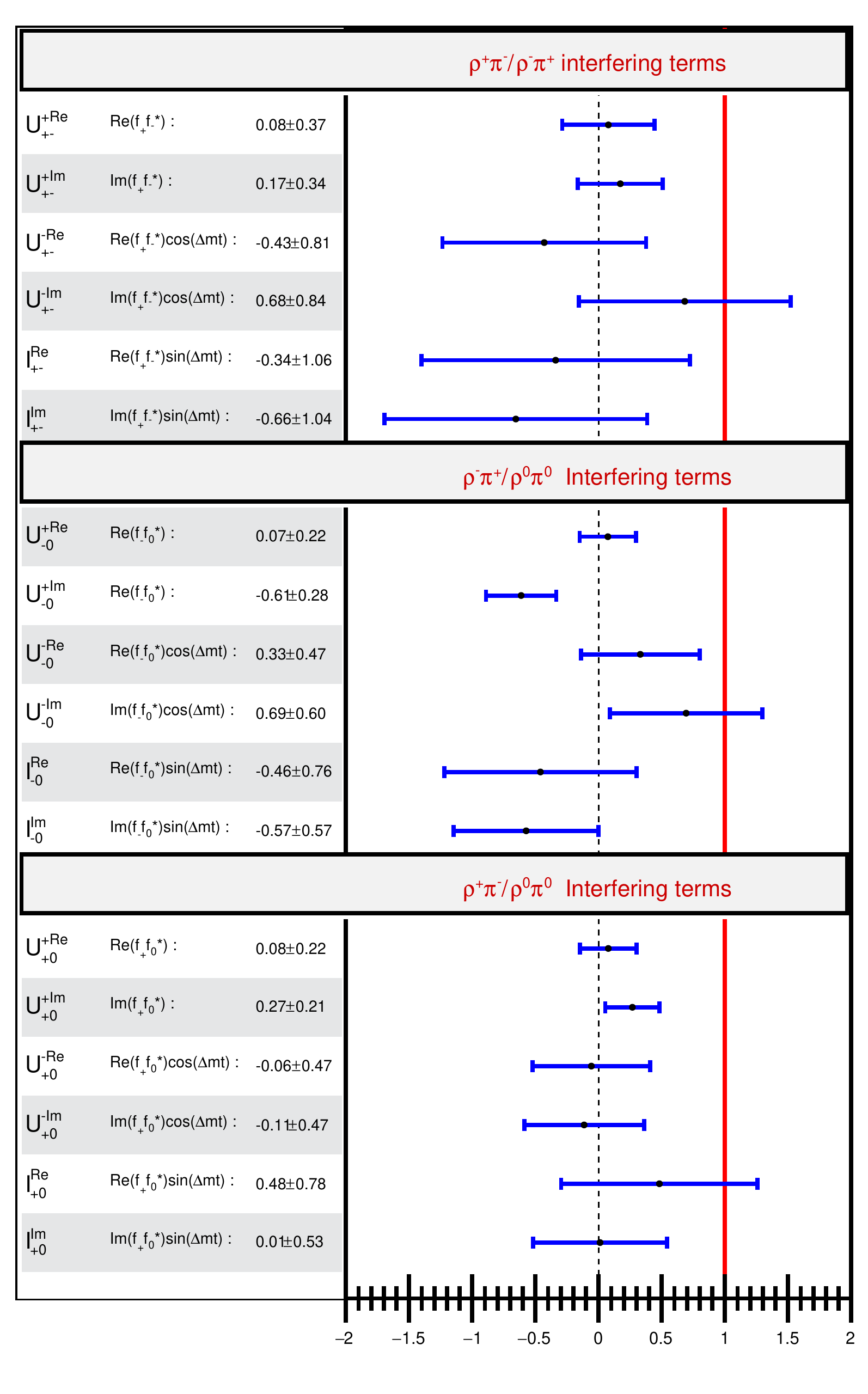}
\caption{\it \small Relative value of the interference-related $\cal U$ and $\cal I$ coefficients extracted from the time-dependent Dalitz analysis of the $B^0\to(\rho\pi)^0$  decay \cite{Babar_rp,Belle_rp}. The corresponding form-factor bilinear is indicated for each coefficient. The red vertical line indicates the overall normalisation, defined by  $U^+_+=1$.}\label{input_rhopi2}
\end{center}       
\end{table}

\begin{table}[t]
\begin{center}
\includegraphics[width=30pc]{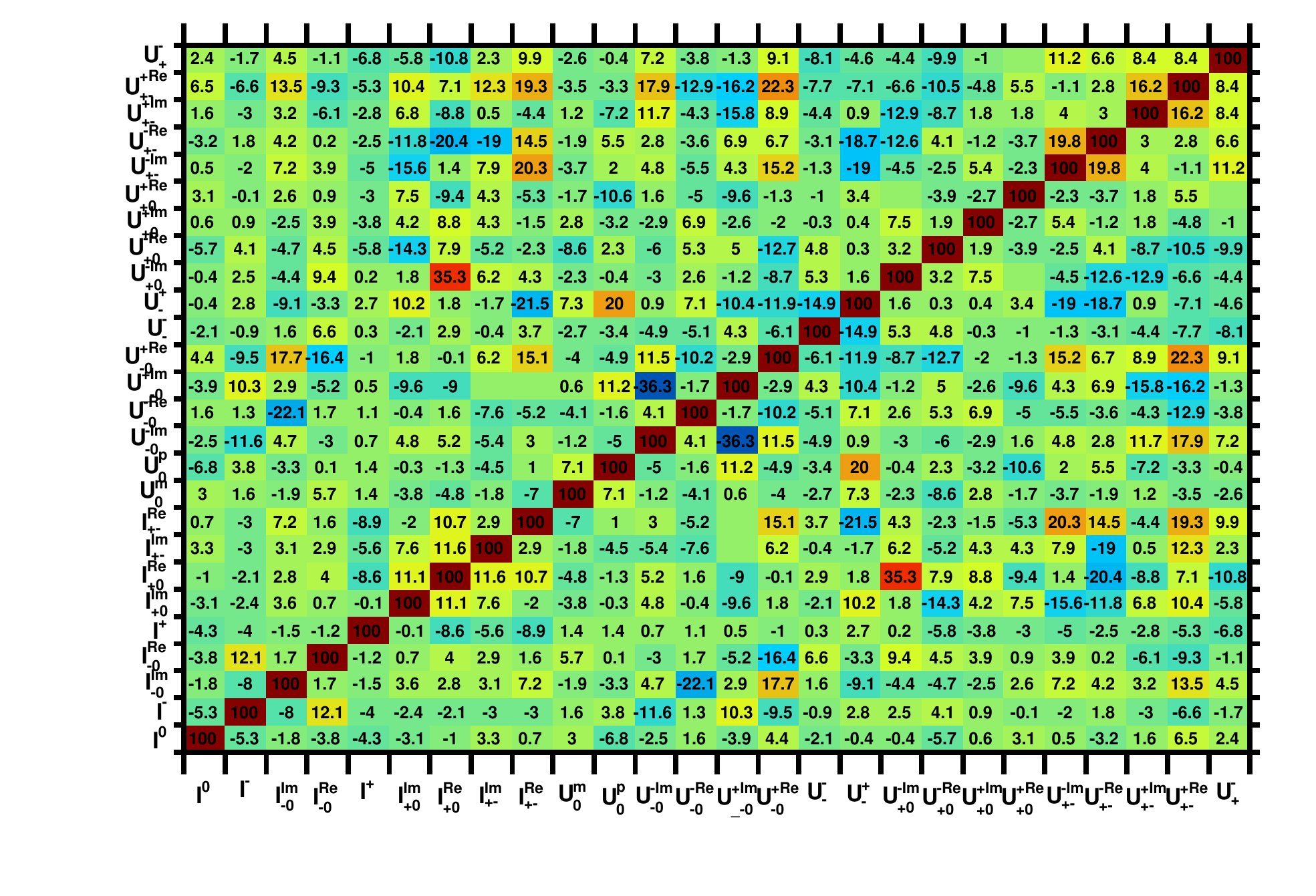}
\caption{\it \small
Correlation matrix of correlations, in percent, for the $\cal U$ and $\cal I$ coefficients extracted from the time-dependent Dalitz analysis of the $B^0\to(\rho\pi)^0$  decay \cite{Babar_rp,Belle_rp}.
}\label{input_rhopi_corr}
\end{center}       
\end{table}

\begin{figure}[t]
\begin{center}
  \includegraphics[width=18pc]{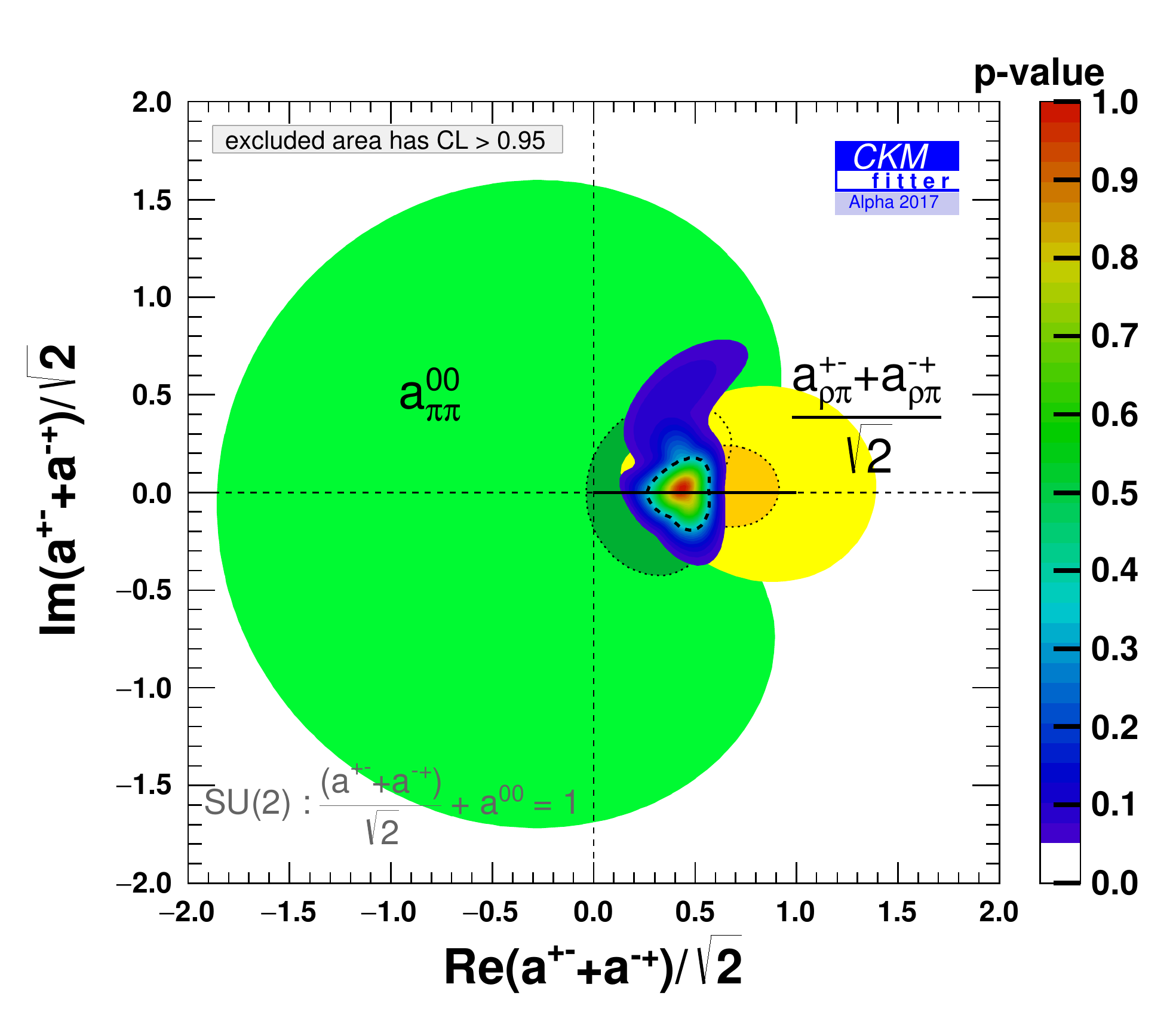}
  \includegraphics[width=18pc]{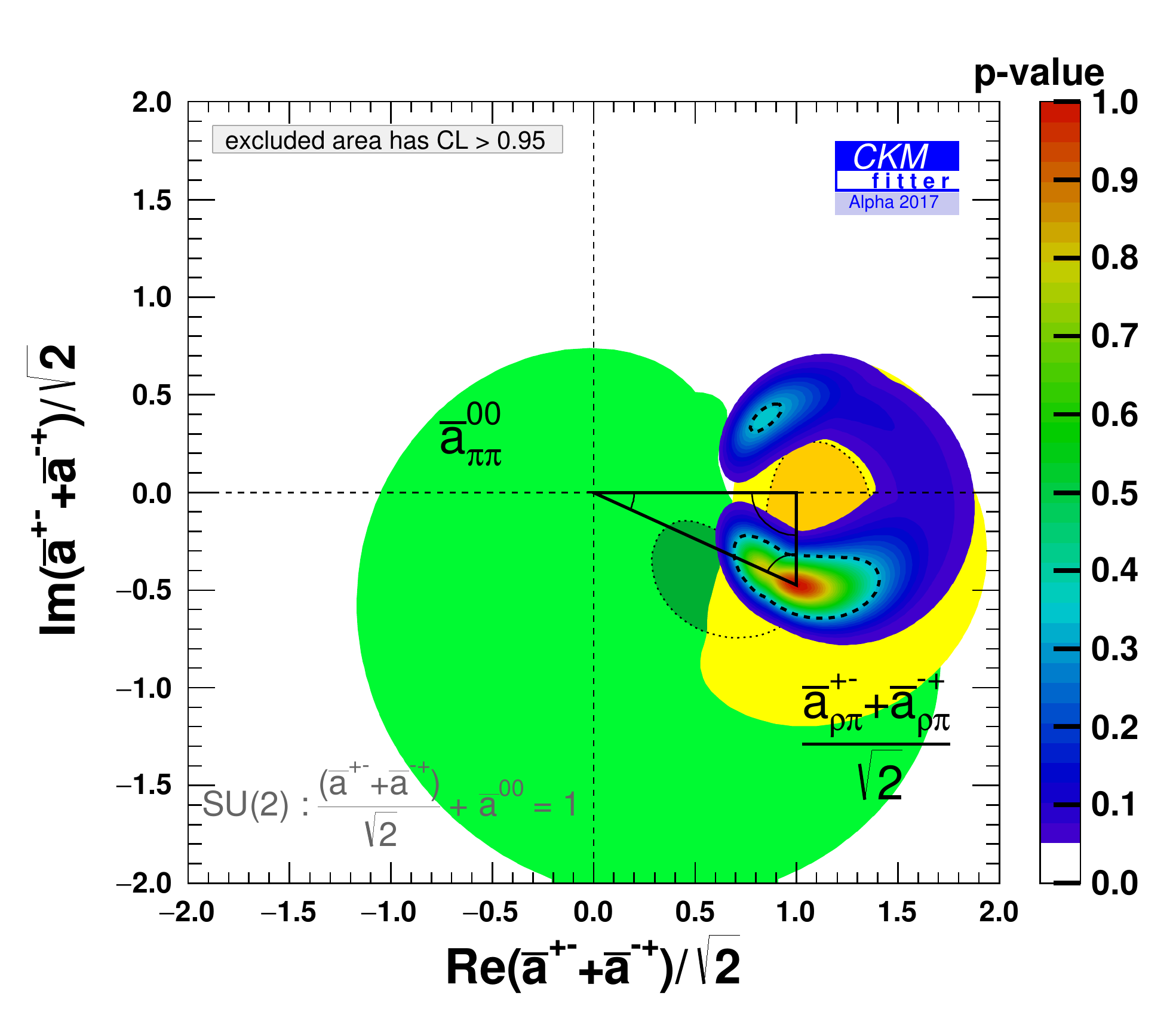}
\caption{\it\small Constraint on the reduced amplitude $(a^{+-}+a^{-+})=(A^{+-}+A^{-+})/(A^{+0}+A^{0+})$ in the complex plane for the $B^0\to(\rho\pi)^0$ (left) and ${\bar B}^0\to(\rho\pi)^0$ systems (right). 
The individual 95\% CL constraint from the $B^0(\bar B^0)\to\rho^\pm\pi^\mp$ observables and from the $B^0(\bar B^0)\to\rho^0\pi^0$ observables are indicated by the yellow and green areas, respectively, whose non-trivial shapes are due to the correlations between the two modes. The  corresponding isospin triangular relations $a^{00}+{a^{+-}}/{\sqrt{2}}=1$ (and $CP$ conjugate) are represented by the black line.}
\label{fig:RhoPiSU2}
\end{center}       
\end{figure}

\begin{figure}[t]
\begin{center}
  \includegraphics[width=18pc]{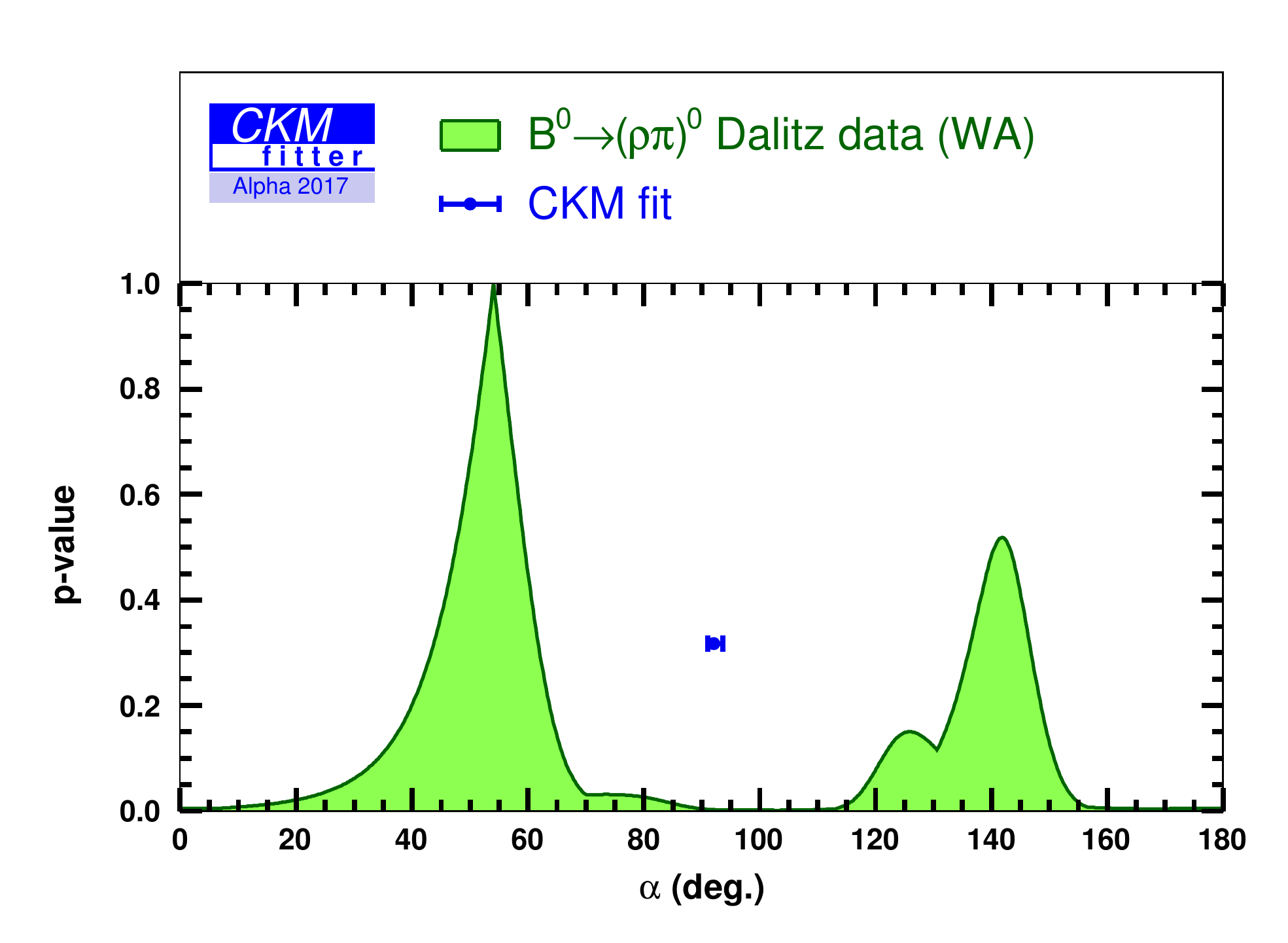}
  \includegraphics[width=18pc]{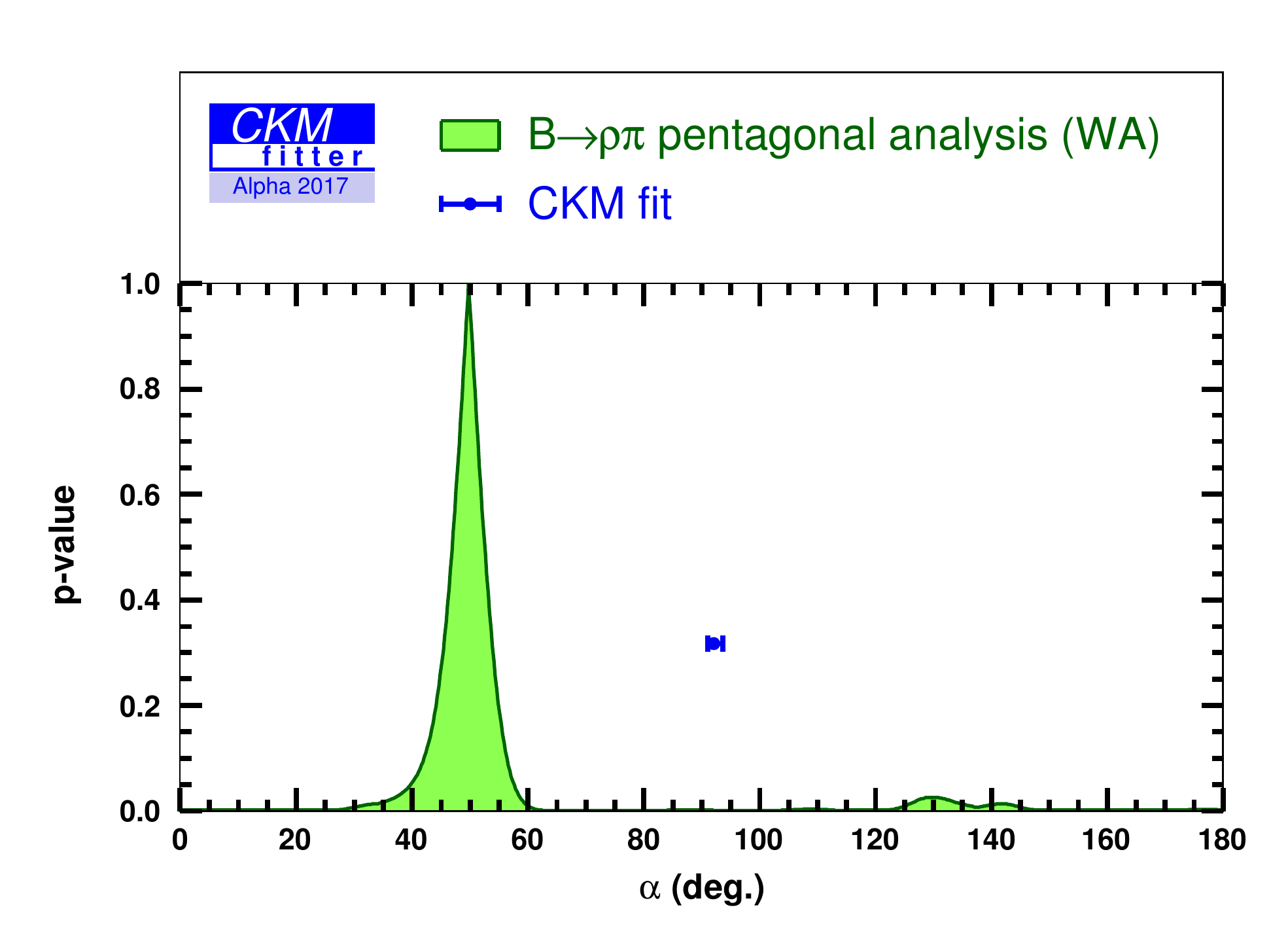}
\caption{\it\small  One-dimensional scan of $\alpha$ in the $[0,180]^\circ$ range from the ``Dalitz'' \su{2} isospin analysis of the $B^0\to(\rho\pi)^0$ system (left) and the ``pentagonal'' analysis (right). The interval with a dot indicates the indirect $\alpha$ determination introduced in Eq.~(\ref{eq:alphaInd}).}\label{fig:RhoPiDalitzalpha}
\end{center}       
\end{figure}

As long as only the neutral modes are considered, it is worth noticing that the absolute normalisation of these coefficients is irrelevant for the \su{2} isospin analysis. The measured values and correlations \cite{Babar_rp,Belle_rp} of the 26 remaining $\cal U$ and $\cal I$  coefficients (normalised with respect to ${\cal U^+_+}$ that is set to unity) are collected in Tabs.~\ref{input_rhopi1}, \ref{input_rhopi2}, and \ref{input_rhopi_corr}. The following compact representation  of the amplitude system for the neutral transitions can be chosen:
\begin{eqnarray}\label{eq:amplsystemneutral}
  {A}^{+}= \cos(\Theta^{+})         &,&   ~~~ {\bar A}^{+}= \sin(\Theta^{+})e^{\imi \Psi^{+}}\,, \nonumber\\
  {A}^{-}= \mu^{-}e^{\imi\phi^{-}}  &,&   ~~~ {\bar A}^{-}= \nu^{-}e^{\imi\Psi^{-}} \,, \nonumber\\
  {A}^{0}= \mu^{0}e^{\imi\phi^{0}}  &,&   ~~~ 2{\bar A}^{0}=e^{2\imi\alpha}(2A^{0}+A^{+}+A^{-})  - {\bar A}^{+}-{\bar A}^{-}\,.
\end{eqnarray}
This  modelisation explicitly embeds the isospin relations Eq.~(\ref{eqn:pentagonal}) as well as the normalisation choice ${\cal U^+_+} = |A^{+}|^2 + |{\bar A}^{+}|^2 =1$. We thus obtain a system depending on 9 real parameters, consisting in four relative amplitudes ($\cos(\Theta^{+})$, $\mu^{-}$, $\mu^{0}$, $\nu^{-}$) and 5 phases  ($\alpha$, $\phi^{-}$, $\phi^{0}$, $\Psi^{+}$, $\Psi^{-}$) (including the weak phase $\alpha$).  
It is worth stressing again that the actual parametrisation of the amplitudes has no impact on the results obtained in our frequentist framework \cite{Charles:2006vd}. The parametrisation Eq.~(\ref{eq:amplsystemneutral}) is mostly based on technical considerations: this minimal representation ensures a fast and stable exploration of the nine-dimensional parameter space constrained by the 26 correlated observables.  The isospin relations for both $CP$-conjugate modes are illustrated on Fig.~\ref{fig:RhoPiSU2}. The resulting  constraint on  $\alpha$  is shown on the left-hand side of Fig.~\ref{fig:RhoPiDalitzalpha}.

Using the 26 correlated $\cal U$ and $\cal I$ observables, the minimum of the \chisq test statistic over the parameter spaces ($\chi^2_{\rm min}=8.6$) is found at $\alpha=54.1^\circ$.
The corresponding 68\% and 95\% CL intervals are
\begin{eqnarray}\label{eq:alpharhopi}
\alpha_{(\rho\pi)^0} ~ ({\rm Dalitz}): &&(\val{54.1}{+7.7}{-10.3} \quad\cup\quad  \val{141.8}{+4.8}{-5.4})^\circ ~~(68\%\ {\rm CL})~~ \textrm{and} \nonumber\\
  &&(\val{54.1}{+15.0}{-27.0} \quad\cup\quad  \val{141.8}{+11.2}{-23.8})^\circ ~~(95\%\ {\rm CL})\,.
\end{eqnarray}
The interval $\alpha = [92.0,112.0]^\circ$ is excluded at more than 3 standard deviations by the \su{2} isospin analysis of the $B^0\to(\rho\pi)^0$ system. These results do not show a good agreement with the indirect determination of $\alpha$, Eq.~(\ref{eq:alphaInd}), based on the other constraints of the global CKM fit \cite{CKMfitterSummer16}. Adding the indirect determination as an additional constraint to the \su{2} isospin fit, the minimal \chisq  increases by 9.1 units, corresponding in the Gaussian limit to a 3.0 $\sigma$ discrepancy. In front of this uncomfortable discrepancy, we should remember that this result will be combined with constraints from $\rho\rho$ and $\pi\pi$ systems. Without a dedicated statistical study, it is difficult to determine the effective number of degrees of freedom associated with the set of 26 interdependent observables for the $\rho\pi$ system, but it remains interesting to notice that the $\chi^2_{\rm min}$ for $\rho\pi$ appears less significant (statistically) than in the case of  $\rho\rho$ (which dominates the combination) and $\pi\pi$ (which contributes significantly). The latter constraints will dominate, leading to a combined determination remaining in good agreement with the indirect value of $\alpha$ obtained from the other constraints of the global fit.  A detailed study of the actual significance of this \chisq offset in the $\rho\pi$ case is discussed in Sec.~\ref{sec:statistics}. In addition, the $\rho\pi$ Dalitz observables have been scrutinized to understand more precisely the origin of the tension, as reported in section~\ref{sec:rhopiObs}.

\subsubsection{``Pentagonal'' analysis of the  $B\to\rho\pi$ system\label{subsec:pentagonal}}

The amplitudes of the two charged decay modes $B^+\to\rho^+\pi^0$ and $B^+\to\rho^0\pi^+$ are related to the amplitudes of the three neutral decays $B^0\to(\rho\pi)^0$ through the pentagonal relation Eq.~(\ref{eqn:pentagonal}). The measurement of the branching ratios and ${CP}$-asymmetries of the charged modes may provide additional constraints to the isospin system.  Considering simultaneously the charged and the neutral modes, the $\cal U$ and $\cal I$ observables that describe the relative amplitudes of the neutral decays must be supplemented with an absolute normalisation. This is achieved by identifying the sum of  branching ratios, ${\cal B}^{+-}_{\rho\pi}+{\cal B}^{-+}_{\rho\pi}$,  with the scaled amplitude $\mu^{+}({\cal U}^+_+ + {\cal U}^+_-){\tau_{B^0}}/{2}$, where $\mu^{+}$ is the absolute normalisation of the relative $\cal U$ and $\cal I$ coefficients. The experimental  measurements of the  branching fractions for the charged and neutral $B\to\rho\pi$ modes and  the ${CP}$ asymmetries for the charged modes are listed in Tab.~\ref{input_rhopiC}. 

At this stage, we would like to comment on some of these inputs. The measurement of the  branching fractions for the  neutral intermediate states $B^0\to\rho^i\pi^j$  
is experimentally tricky. Indeed, the three $\rho$ mesons, charged and neutral, contribute to the neutral modes, whereas only one $\rho$ state contributes to each of the charged modes. A quasi-two-body analysis was performed by the  \babar and  CLEO experiments \cite{Cleo_rp_pm0,Babar_rp_pm,Babar_rp_00}  assuming that the interferences are negligible in the Dalitz region where the measurements were performed. The Belle measurement \cite{Belle_rp_pm0} extracts the partial branching ratio $\B^{\pm\mp}_{\rho\pi}$  and $\B^{00}_{\rho\pi}$ simultaneously from a global $B^0\to\pi^+\pi^-\pi^0$ Dalitz analysis, rescaling the  \U coefficients to the overall three-body branching fractions. The reliability of the measurements on the  quasi-two-body intermediate states, $B^0\to\rho^\pm\pi^\mp$ and $B^0\to\rho^0\pi^0$, is thus questionable and this may bias the pentagonal constraint. Since the Dalitz-based \su{2} isospin analysis already provides a complete and self-consistent description of the $B^0\to(\rho\pi)^0$ amplitude system, we will use the Dalitz analysis rather than the pentagonal one in the combined direct determination of the $\alpha$ angle. However, for completeness, we briefly discuss the results of the pentagonal analysis, interpreting the $\B^{\pm\mp}_{\rho\pi}$  measurement as an independent constraint on ${\cal B}^{+-}_{\rho\pi}+{\cal B}^{-+}_{\rho\pi}$. Further discussions can be found  in Sec.~\ref{sec:pentagonalObservables}, where predictions for these observables can be found.

\begin{table}[t]
\begin{center}
\begin{tabular}{|l|c|l|}
\hline
Observable & World average & References \\
\hline 
\small ${\cal B}^{+0}_{\rho\pi}$ $(\times 10^6)$ &\small $10.9\pm 1.4$  &\small \cite{Cleo_rp_pm0,Babar_rp_p0,Belle_rp_p0} \\ 
\small ${\cal C}^{+0}_{\rho\pi}$                &\small $-0.02\pm 0.11$ &\small \cite{Cleo_rp_pm0,Babar_rp_p0,Belle_rp_p0} \\
\small ${\cal B}^{0+}_{\rho\pi}$ $(\times 10^6)$ &\small $8.3\pm 1.2$   &\small \cite{Cleo_rp_pm0,Babar_rp_0p,Belle_rp_0p} \\ 
\small ${\cal C}^{0+}_{\rho\pi}$                &\small $-0.18^{+0.17}_{-0.09}$&\small \cite{Cleo_rp_pm0,Babar_rp_0p,Belle_rp_0p} \\ 
\hline
\small ${\cal B}^{\pm\mp}_{\rho\pi}$ $(\times 10^6)$ &\small $23.0\pm 2.3$   &\small \cite{Cleo_rp_pm0,Babar_rp_pm,Belle_rp_pm0} \\ 
\small ${\cal B}^{00}_{\rho\pi}$ $(\times 10^6)$ &\small $2.0\pm 0.5$   &\small \cite{Cleo_rp_pm0,Babar_rp_00,Belle_rp_pm0} \\ 
\hline
\end{tabular}
\caption{\it\small World averages for the branching fractions ${\cal B}^{ij}_{\rho\pi}$ and  time-integrated $CP$ asymmetries ${\cal C}^{ij}_{\rho\pi}$ for the charged (top) and neutral (bottom) $B\to\rho\pi$ decay modes.\label{input_rhopiC}}
\end{center}
\end{table}

Including the charged decays, the amplitude model for the neutral modes discussed in Sec.~\ref{subsec:dalitz} can be completed with two additional real parameters ($\mu^{+0}$,$\phi^{+0}$):
\begin{equation}
  {A}^{+0}= \mu^{+0}e^{\imi \phi^{+0}}\,,\qquad
  {A}^{0+}= -{A}^{+0} + \frac{2A^{0}+A^{+}+A^{-}}{\sqrt{2}},
\end{equation}
where the second identity uses the pentagonal relation Eq.~(\ref{eqn:pentagonal}) explicitly.
Using Eq.~(\ref{eqn:alpha3}), the $CP$-conjugate amplitudes are defined as
\begin{eqnarray}
  {\bar A}^{+0}&=&(A^{+0}-\frac{A^{+}-A^{-}}{\sqrt{2}})e^{2\imi\alpha}+\frac{{\bar A}^{+}-{\bar A}^{-}}{\sqrt{2}}\,,\nonumber\\
  {\bar A}^{0+}&=&(\sqrt{2}(A^{+}+A^{0})-A^{0+})e^{2\imi\alpha}-\frac{{\bar A}^{+}-{\bar A}^{-}}{\sqrt{2}}\,.
\end{eqnarray}
Adding the normalisation factor $\mu^{+0}$, the whole $B\to\rho\pi$ dynamics is described through a 12-parameter system that is over-constrained by the set of observables for neutral and charged modes given in Tabs.~\ref{input_rhopi1},~\ref{input_rhopi2}, \ref{input_rhopi_corr} and \ref{input_rhopiC}. 

The minimum of the \chisq test statistic over the parameter space ($\chi^2_{\rm min}=12.1$) is found at $\alpha=49.8^\circ$. The following 68\%  and 95\% CL intervals on $\alpha$ are obtained:
\begin{eqnarray}
\alpha_{\rho\pi}\ ({\rm pentagonal}): && \val{49.8}{+4.1}{-4.3}{$^\circ$}~~(68\%\ {\rm CL})~~{\rm and}  \nonumber \\
 && \val{49.8}{+8.0}{-10.4}{$^\circ$}~~(95\%\ {\rm CL})\,.   
\end{eqnarray}
The compatibility with the indirect $\alpha$ determination is estimated at 3.2 standard deviations, slightly larger than for the Dalitz analysis of the neutral modes. The constraint on $\alpha$  is displayed on the right-hand side of Fig.~\ref{fig:RhoPiDalitzalpha}.  The same comments as in Sec.~\ref{subsec:dalitz} apply also here concerning the relative statistical significance of this constraint compared to $B\to \pi\pi$ and $B\to \rho\rho$ channels.  

\subsection{Combined result \label{subsec:alpha}}
\begin{figure}[t]
\begin{center}
  \includegraphics[width=25pc]{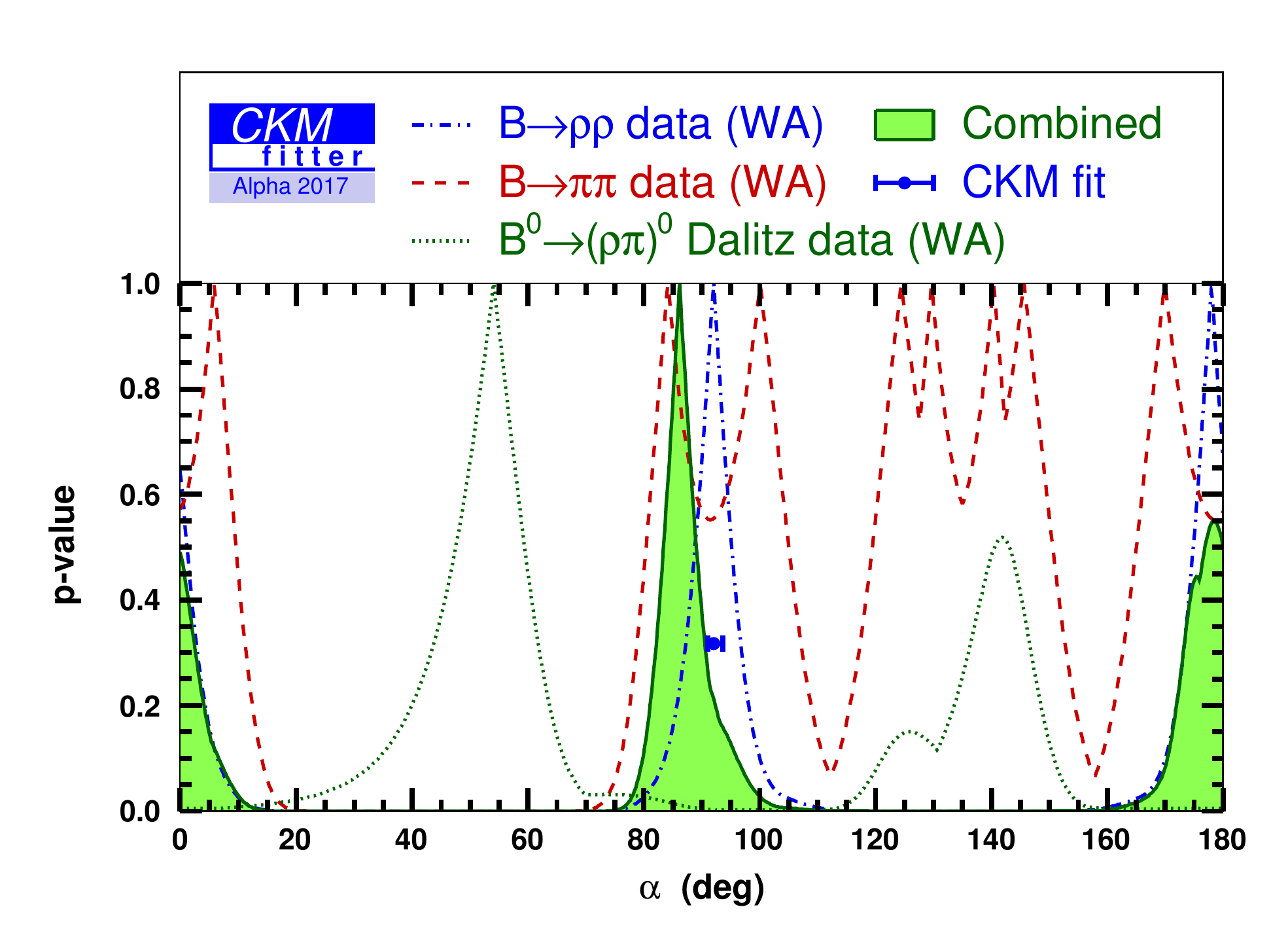}
\caption{\it\small \su{2} constraint on $\alpha$ (green shaded area) from the combination of $B\to\pi\pi$ (dashed line), $B\to\rho\rho$ (dashed-dotted line) and $B^0\to\pi^+\pi^-\pi^0$  Dalitz analyses (dotted line) compared to the indirect value (dot) obtained from the global CKM fit~\cite{CKMfitterSummer16}. 
}
\label{fig:alpha}
\end{center}       
\end{figure}

The \su{2} isospin analyses of $B\to\pi\pi$, $B\to\rho\rho$ and $B\to\rho\pi$ provide three independent constraints on $\alpha$. A combined analysis can be performed by summing the individual $\chi^2(\alpha)$ curves. Using the Dalitz analysis for the $B^0\to(\rho\pi)^0$ mode together with $B\to\pi\pi$ and $B\to\rho\rho$, the minimum  of the \chisq test statistic ($\chi^2_{\rm min}=17.1$) is obtained for $\alpha=86.2^\circ$, as illustrated on Fig.~\ref{fig:alpha}. A slightly disfavoured second solution peaks at $\alpha=178.4^\circ$  with $\Delta\chi^2=\chi^2(\alpha=178.4^\circ)-\chi^2_{\rm min}=0.4$.  The  corresponding 68\% and 95\% CL intervals are
\begin{eqnarray}
\alpha_{\rm dir} : && (\val{86.2}{+4.4}{-4.0} \quad\cup\quad \val{178.4}{+3.9}{-5.1})^\circ~~(68\%\ {\rm CL})~~ \textrm{and}\nonumber \\
  && (\val{86.2}{+12.5}{-7.5}\quad\cup\quad \val{178.4}{+10.5}{-9.9})^\circ ~~(95\%\ {\rm CL})\,.
\end{eqnarray}
The preferred solution near $90^\circ$ is consistent with the indirect determination  $\alpha_{\rm ind.}$  given in Eq.~(\ref{eq:alphaInd})   at 1.3 standard deviations. As discussed earlier, the combined constraint is dominated by $B\to \rho\rho$ and to a lesser extent, by $B\to \pi\pi$, whereas $B\to \rho\pi$ plays only a limited role.  The  constraint resulting of the partial combination based on  $B\to \pi\pi$ and $B\to \rho\rho$ only is reported in Tab.~\ref{tab:summary}.

\begin{figure}[t]
\begin{center}
  \includegraphics[width=25pc]{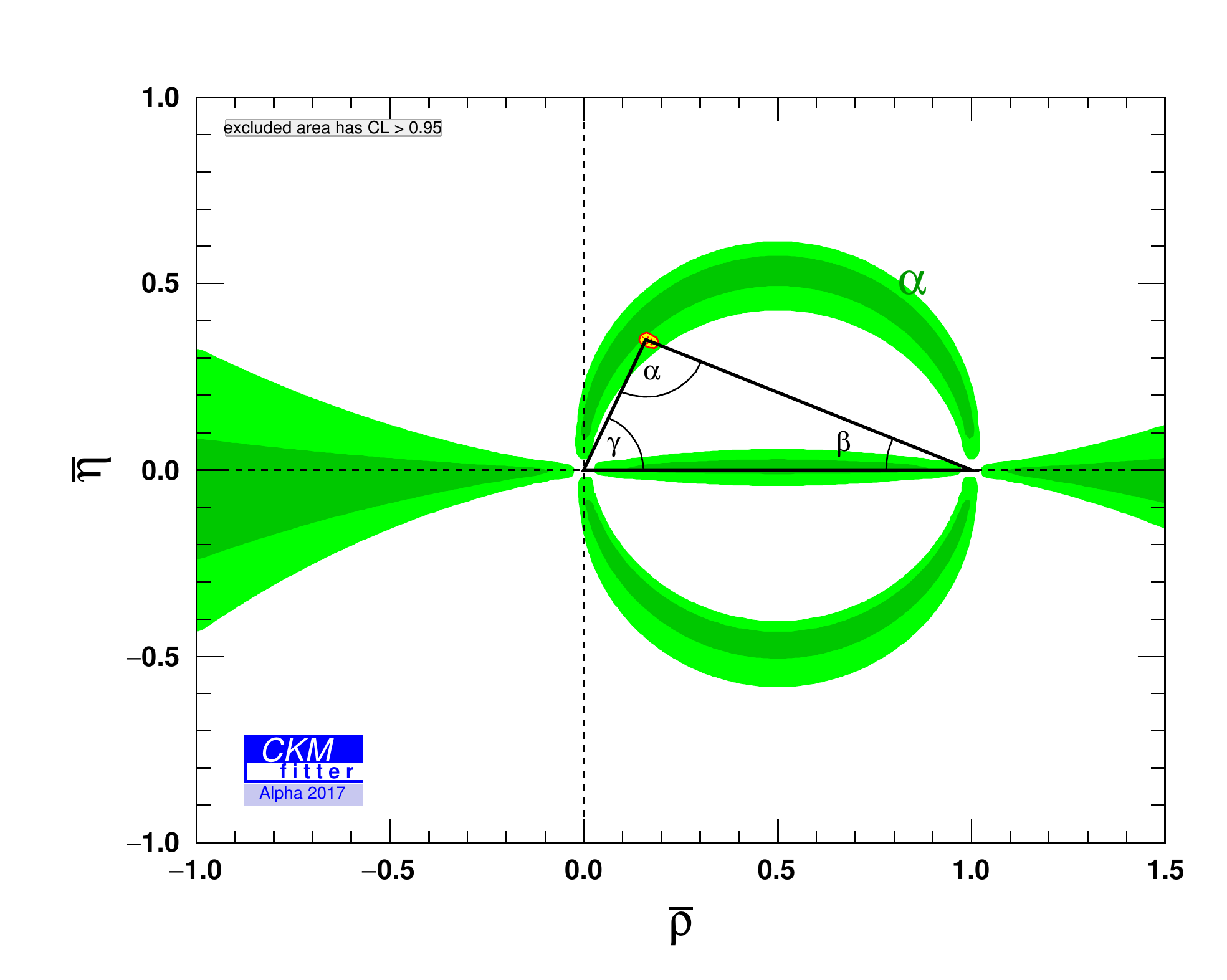}
  \caption{\it\small 95\% CL (dark green) and 68\% (light green) constraints on $\alpha$ in the ($\bar\rho,\bar\eta)$ plane from the combined \su{2} isospin analysis of $B\to\pi\pi$, $\B\to\rho\rho$ and $B\to\rho\pi$ (Dalitz) decays. The small yellow area indicates the 95\% CL region for the apex of the $B$-meson unitarity triangle (solid black lines) from the global CKM fit excluding the charmless $B\to hh$ data used in the direct $\alpha$ determination~\cite{CKMfitterSummer16}.}
\label{fig:rhoeta}
\end{center}       
\end{figure}

The one-dimensional constraint on $\alpha$ can be recast in a constraint on 
the  $(\bar\rho,\bar\eta)$ Wolfenstein parameters of the CKM matrix representing the apex of the $B$-meson unitarity triangle \cite{thePapIII,CKMfitterSummer16}. The following relation can be derived:
\begin{equation}
\left({\bar\eta}-\frac{{\rm cotan}(\alpha)}{2}\right)^2+\left({\bar\rho}-\frac{1}{2}\right)^2 = \frac{1}{4\sin^2(\alpha)}\,,
\end{equation}
so that the curves at fixed $\alpha$ consist in circles centred on the point $(\bar\rho,\bar\eta)=(1/2,{\rm cotan}(\alpha)/2)$. In particular, the curve at $\alpha$=$90^\circ$  is a circle of radius $1/2$ centered on $(\bar\rho,\bar\eta)=(1/2,0)$, while $\alpha$=$0^\circ$ amounts to a circle of an infinite radius tangent to the axis $\bar\eta=0$. All the  curves of constant $\alpha$ meet at the points $(0,0)$ and $(1,0)$. Fig.~\ref{fig:rhoeta} displays the constraint resulting from the direct determination of $\alpha$ in the $(\bar\rho,\bar\eta)$ plane.

In this section, we have used world averages for all the observables, combining information from the \babar, Belle and LHCb experiments. Since both $B$-factory experiments, \babar and Belle, have measured all these observables independently, it is also possible to perform separate \su{2} isospin analyses for each of the three decay channels and each experiment. The corresponding constraints are discussed in App.~\ref{sec:experiments}.

\section{Additional uncertainties in the extraction of $\alpha$} \label{sec:syst}

In this section, we are going to test the limits of the assumptions made to extract $\alpha$ from the data: the breakdown of isospin symmetry and the statistical approach used to build confidence intervals.

\subsection{Testing the \su{2} isospin framework} \label{sec:tests}

As discussed in Sec.~\ref{sec:isospin}, the extraction of the weak phase $\alpha$ relies on isospin symmetry, which is used at different levels \cite{Gardner:1998gz,Zupan:2004hv}:
\begin{itemize}
\item The charges of the $u$ and $d$ quarks are taken as identical. The  $\Delta I=\ofrac{3}{2}$  contribution induced by the electroweak penguins topology to the $B^+\to h^+h^0$ decay is considered negligible. As the gluonic penguins only yield a $\Delta I=\ofrac{1}{2}$ isospin contribution,  the pure $\Delta I=\ofrac{3}{2}$   $B^+\to h^+h^0$ decay only receives tree contributions in the absence of electroweak penguins. In this limit,  the weak phase $\alpha$ can be identified as half the phase difference between the amplitude of the charged mode and its $CP$ conjugate.
\item The masses of the $u$ and $d$ quarks are taken as identical. Isospin symmetry is assumed to be exact  in the strong hadronisation process following the weak transition $\bar b(q)\to \bar u u\bar d(q)$ where $q$ represents the light spectator quark $u$ or $d$. This assumption allows one to relate the decay amplitudes of the charged ($\bar bu$) meson to the decay amplitudes of its isospin-related neutral state $\bar bd$  according to the pentagonal (triangular) identity given in Eq.~(\ref{eqn:pentagonal}) (respectively Eq.~(\ref{eqn:triangular})).
\item The $\rho\rho$ final state is supposed to obey Bose--Einstein statistics, and thus the analysis for the $\pi\pi$ and the $\rho\rho$ systems are supposed to follow the same isospin decomposition (for a given $\rho$ polarisation). However, the $\rho$ mesons cannot be distinguished only in the limit of a vanishing width. Once the finite $\rho$ width is taken into account, additional terms (forbidden by Bose symmetry fin the limit $\Gamma_\rho=0$) must be taken into account in the isospin decomposition of the amplitudes.
\end{itemize}
All these hypotheses are valid a priori to a very good approximation, but the accuracy reached in the determination of $\alpha$ in Sec.~\ref{subsec:alpha} is an incentive to investigate their limitations in more detail.

\subsubsection{$\Delta I=\ofrac{3}{2}$ electroweak penguins}\label{sec:ewpeng}

While preserving the isospin relations between charged and neutral decay amplitudes, the electroweak penguin topology induces a $\Delta I=\ofrac{3}{2}$ contribution to the charged modes $B^+\to h^+h^0$. With such a contribution, the system of amplitudes in the $\pi\pi$ and $\rho\rho$ cases still obeys the triangular relation
Eq.~(\ref{eqn:triangular}), but Eq.~(\ref{eq:systemtriangular}) has to be rewritten as
\begin{eqnarray}
 {A}^{+-}        &=& T^{+-} e^{-\imi \alpha}                  +  P^{+-} \,,      \nonumber\\
 \sqrt{2}{A}^{00}&=& T^{00} e^{-\imi \alpha}                  - P^{+-}   +  P^{+0}_\textsc{ew}\,, \nonumber\\
 \sqrt{2}{A}^{+0}&=&  ( T^{+-} + T^{00} ) e^{-\imi \alpha}                 + P^{+0}_\textsc{ew}\,,
\end{eqnarray}
where the amplitude $P^{+0}_\textsc{ew}$ accounts for the $\Delta I=\ofrac{3}{2}$ contribution from  electroweak penguins to the charged mode while $P^{+-}$ is redefined to absorb the contributions from both gluonic and electroweak penguins to the  $B^0\to h^+h^-$ neutral decay. 

The electroweak penguin contribution can be related to the tree amplitude in a model-independent way by performing the Fierz transformation of  the relevant current-current operators in the effective Hamiltonian for $B\to \pi\pi$ decays \cite{Buras:1998rb,Neubert:1998jq,Neubert:1998pt}. This leads to  an estimate of the relative contribution of the electroweak penguin, $P^{+0}_\textsc{ew}$, compared to the tree amplitude, $T^{+0}=(T^{00}+T^{+-})$, for the charged decay.
Neglecting the penguins with internal light quarks $u$ and $c$ as well as the electroweak operators ${\cal O}_7$ and ${\cal O}_8$ (suppressed by tiny Wilson coefficients), the amplitude ratio:
\begin{equation}
r_{\ewp}=\frac{P_\textsc{ew}^{+0}}{T^{+0}} \simeq -\frac{3}{2}\left(\frac{{\cal C}_9+{\cal C}_{10}}{{\cal C}_1+{\cal C}_2}\right)\frac{V_{td}V^*_{tb}}{V_{ud}V^*_{ub}}e^{-\imi\alpha}
\end{equation}
can be computed in terms of the short-distance Wilson coefficients ${\cal C}_i$ associated to the effective operators ${\cal O}_i$ describing the dominant electroweak penguin processes ($i=9,10$) and the current-current tree processes ($i=1,2$).  Following this estimate, the electroweak penguin amplitude does exhibit no strong phase difference compared to the tree amplitude, preserving  the charge symmetry $|A^{+0}|=|\bar A^{+0}|$  in the charged $B$ decay. Consequently, the impact of the electroweak penguin can be accounted for introducing the single real parameter $r_{\ewp}={P_\textsc{ew}^{+0}}/{T^{+0}}$ in the isospin analysis of the $B\to hh$ systems. 

\begin{figure}[t]
\begin{center}
  \includegraphics[width=18pc]{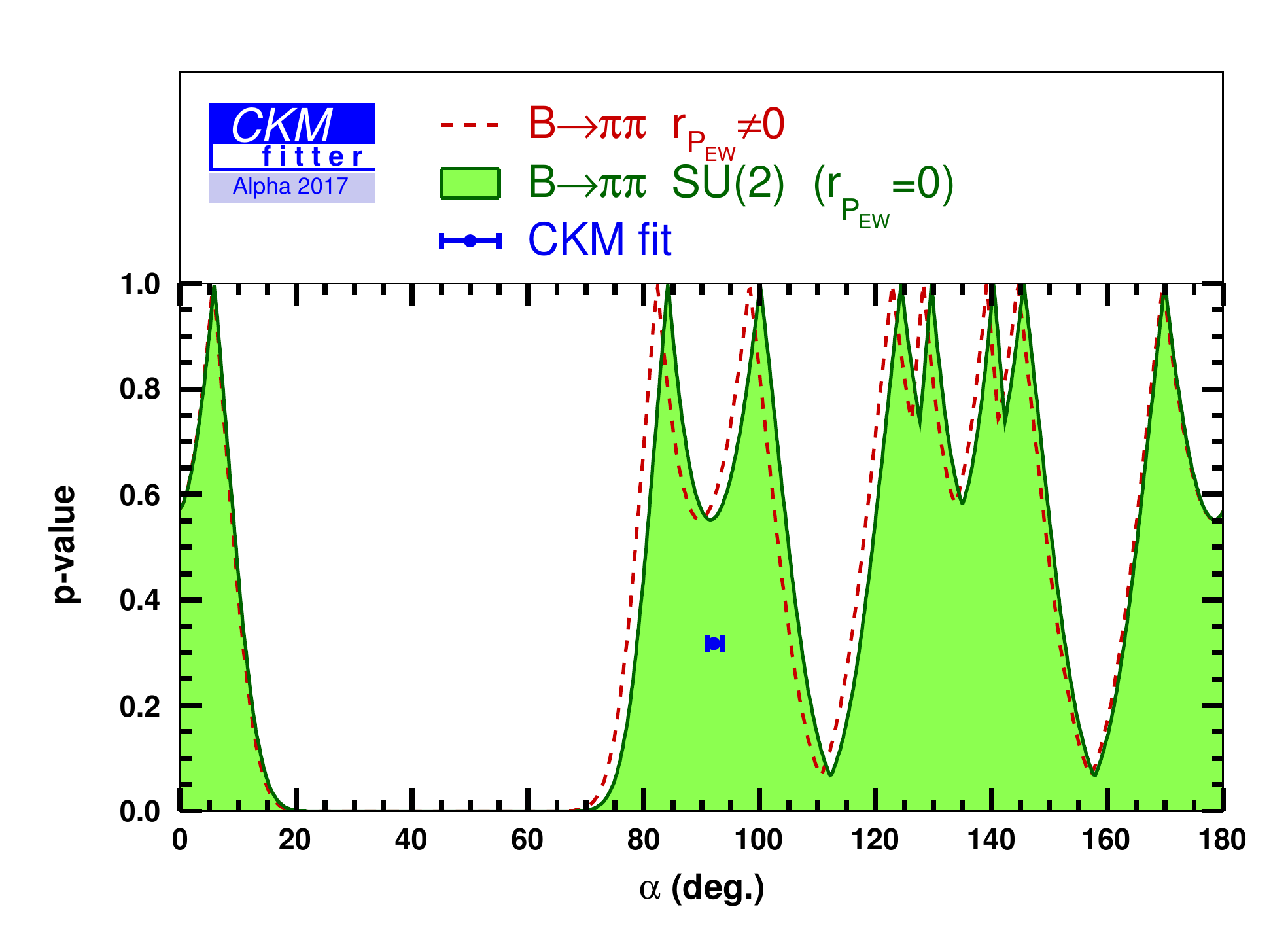}
  \includegraphics[width=18pc]{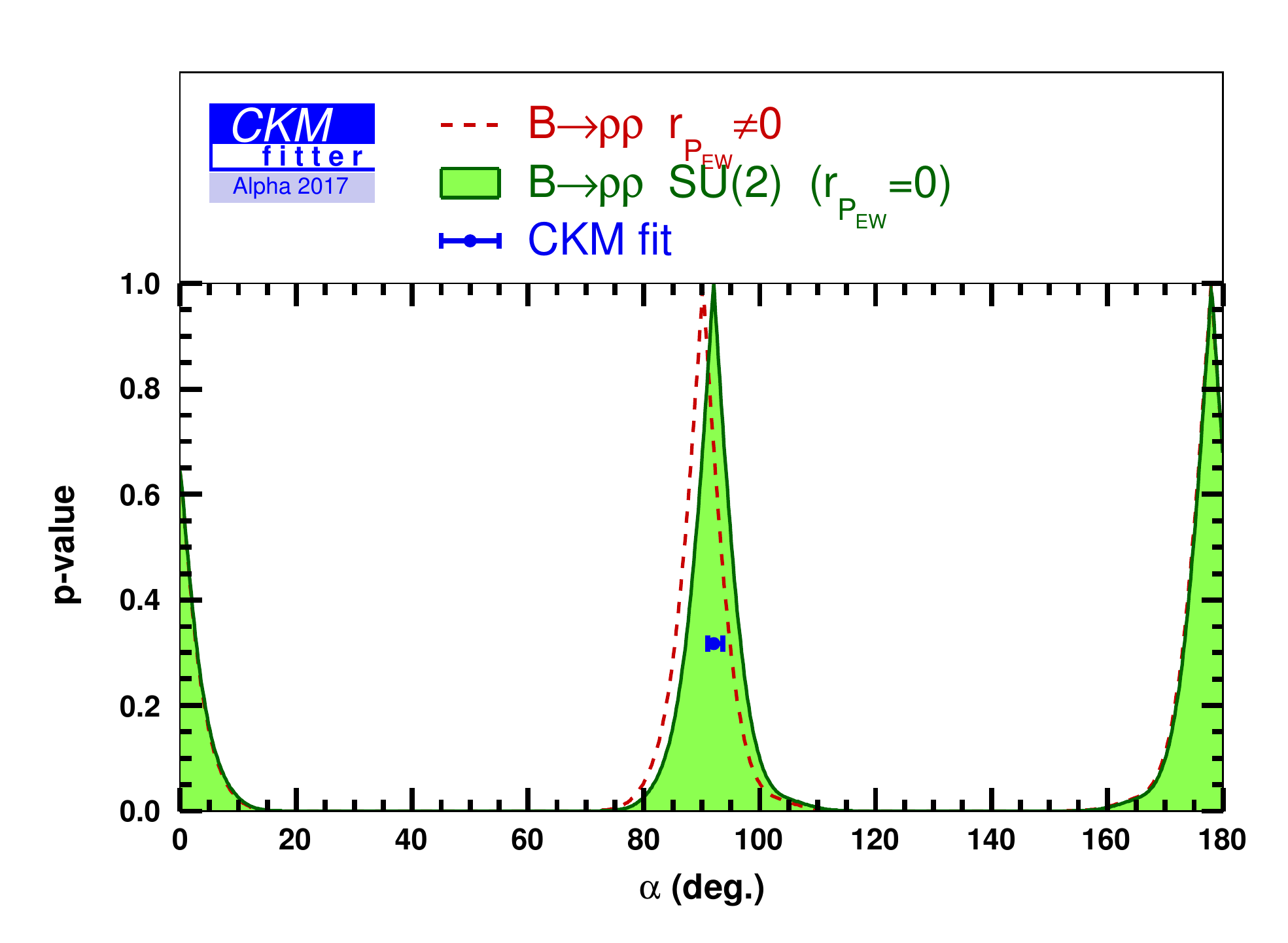}
\caption{\it\small  \su{2} constraint on $\alpha$ including an electroweak penguin contribution given by Eq.~(\ref{eqn:ewp}) (red dashed line) and neglecting this penguin contribution (green shaded area) for $B\to\pi\pi$ (left) and $B\to\rho\rho$ (right).}\label{fig:alphaEWP}
\end{center}       
\end{figure}

The above modification of the amplitude $A^{+0}$ affects the determination of $\alpha$, which is the argument of  $A^{+0}$ in the absence of $\Delta I=\ofrac{3}{2}$ penguin contributions.
The CKM phase measured in $B\to hh$ systems under this hypothesis is an effective phase $\alpha_0$, which can be related to the true phase $\alpha$ through the relation:
\begin{equation}
  \tan\alpha_0 = \frac{\sin\alpha}{r_{\ewp}+\cos\alpha}\,,
\end{equation}
and the relative electroweak contribution $r_{\ewp}$ induces the shift:
\begin{equation}
  \delta\alpha = \alpha-\alpha_0={\rm arcsin}[r_{\ewp} \sin\alpha_0]\,.
\end{equation}
on the determination of $\alpha$. 
A numerical evaluation yields~\cite{Charles:2004jd,Buras:1998rb,Neubert:1998jq,Neubert:1998pt}:
\begin{equation}
  r_{\ewp}=\frac{P_\textsc{ew}^{+0}}{T^{+0}} = -(1.35\pm0.12)\times 10^{-2}\times \left|\frac{V_{td}V^*_{tb}}{V_{ud}V^*_{ub}}\right| = -(3.23\pm0.30)\times 10^{-2}\,,\label{eqn:ewp}
\end{equation}
using the CKM matrix parameters constrained from the global analysis of the flavour constraints \cite{CKMfitterSummer16}.
A maximal shift $\delta\alpha=-1.9^\circ$ occurs at $\alpha_0=90^\circ$, in agreement with the numerical results shown in Fig.~\ref{fig:alphaEWP}. 
The resulting 68\% CL intervals are
\begin{eqnarray}
\alpha_{\pi\pi}^\textrm{\scriptsize \ewp}   : && (\val{91.2}{14.2} ~~\and~ \val{133.8}{18.1} ~~\and~~ \val{177.1}{14.8})^\circ,\\\nonumber
\alpha_{\rho\rho}^\textrm{\scriptsize \ewp} : && (\val{90.1}{+4.7}{-4.8} ~~\and~ \val{177.9}{+4.9}{-4.8})^\circ.\nonumber
\end{eqnarray}

\begin{figure}[htpb]
\begin{center}
  \includegraphics[width=18pc]{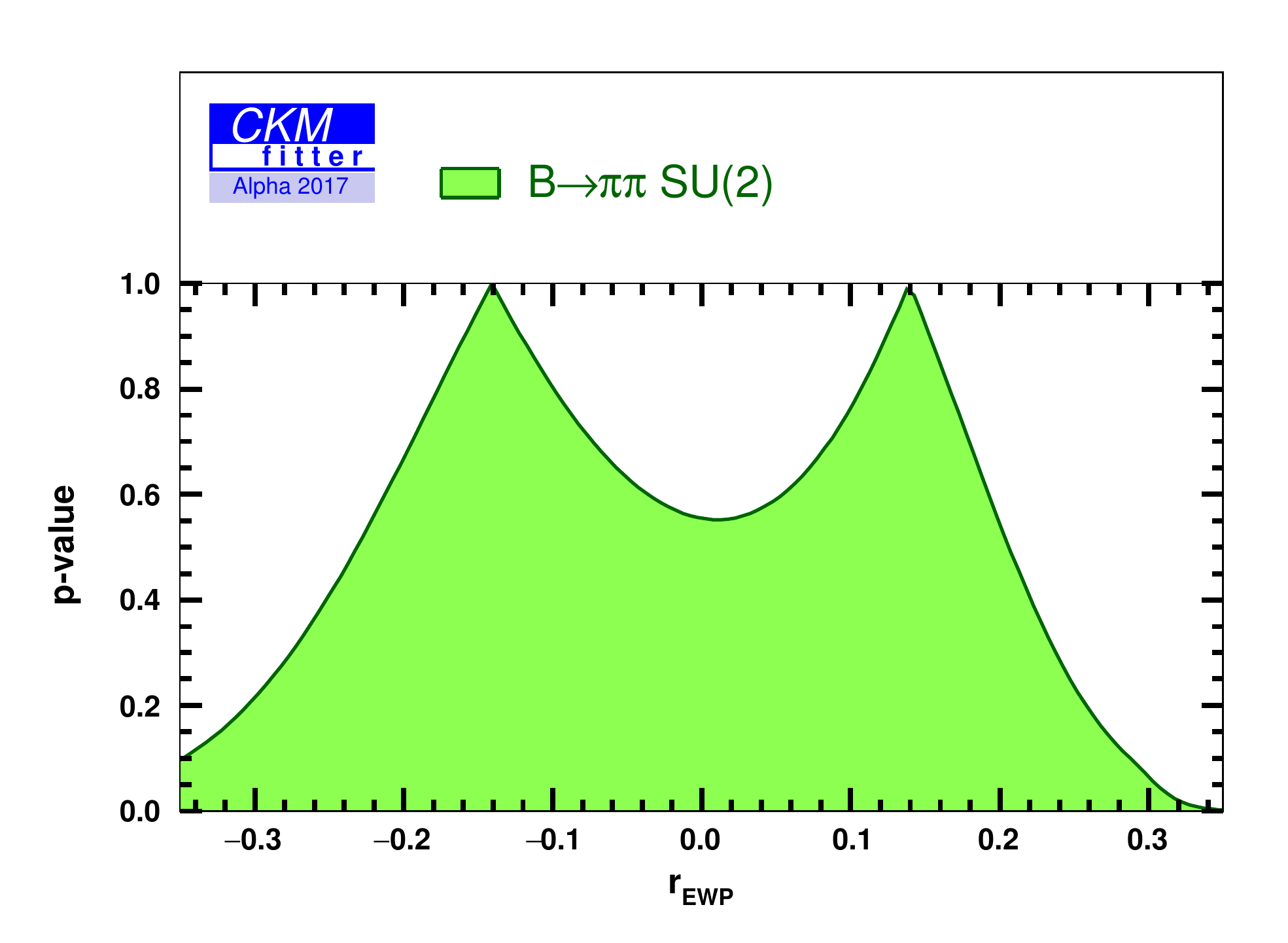}
  \includegraphics[width=18pc]{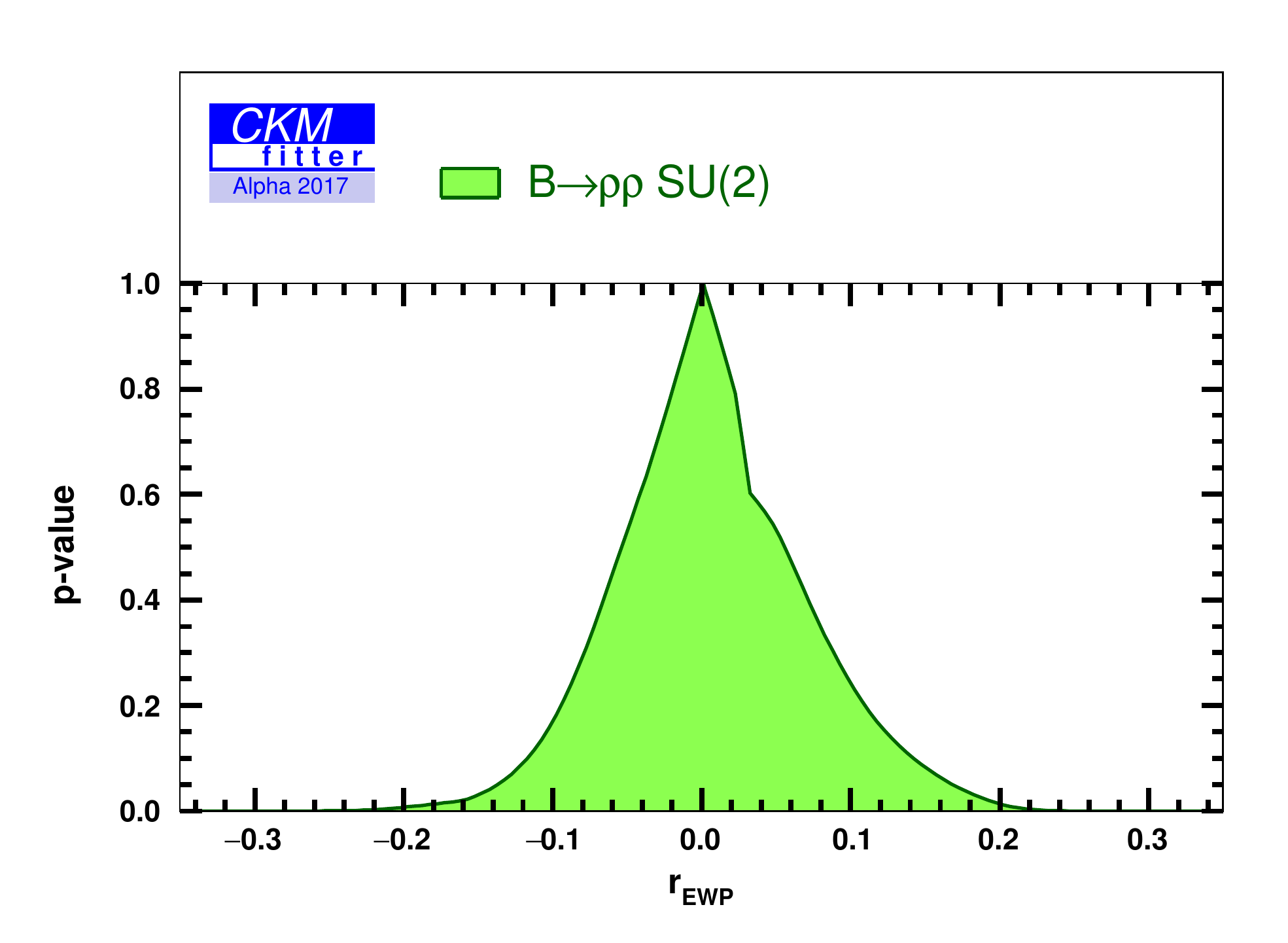}
\caption{\it\small  Constraint on the relative contamination from $\Delta I=\ofrac{3}{2}$ electroweak penguin $r_{\ewp}=P_\textsc{ew}^{+0}/T^{+0}$ from the \su{2} analysis of $B\to\pi\pi$ (left) and $B\to\rho\rho$ (right), 
obtained by constraining the weak phase $\alpha$ with the indirect determination provided by the global CKM fit \cite{CKMfitterSummer16}. An additional constraint on the strong and electroweak penguin hierarchy has been implemented ($|P_\textsc{ew}^{+0}| < |P^{+-}|$) in order to single out the solution for $\alpha$ compatible with the indirect determination.
}\label{fig:EWP}
\end{center}       
\end{figure}

Conversely, neglecting all other isospin-breaking effects, we can set a limit on the electroweak penguin contamination in the $B\to\pi\pi$ and $B\to\rho\rho$ systems. For this purpose, we require the modified isospin analysis 
to agree with the indirect prediction of $\alpha$, Eq.~(\ref{eq:alphaInd}), determined from all the other constraints used in the global CKM fit~\cite{CKMfitterSummer16}. Fig.~\ref{fig:EWP} shows the resulting constraints  on the amplitude ratio $r_{\ewp}$  for $B\to\pi\pi$ (left panel) and $B\to\rho\rho$ (right panel) systems. We have rejected
the mirror solutions that would yield very large, unphysical, differences $\delta\alpha$ and thus exceedingly large electroweak penguin contributions (see the discussion at the end of Sec.~\ref{subsec:pipi}). This is achieved by applying an additional constraint on the strong and electroweak penguin hierarchy $|P_\textsc{ew}^{+0}| < |P^{+-}|$ for this specific study.\footnote{While eliminating mirror solutions, this constraint does not distort the $p$-value for $\alpha$ in the vicinity of the solution compatible with the indirect determination. We stress that we do not use such a constraint anywhere else in this article.} The following 68\% CL intervals are obtained:
\begin{eqnarray}
  r_{\ewp}&=& (\val{-2}{26}) \times 10^{-2} \textrm{ for the $B\to\pi\pi$ system, and }\nonumber\\
  r_{\ewp}&=& \val{0.1}{+8.4}{-7.8}{$~\times~ 10^{-2}$} \textrm{ for the $B\to\rho\rho$ system}\,,
\end{eqnarray}
showing results consistent with an electroweak penguin contribution of only a few percents, in agreement with our theoretical expectations.

The contribution of the electroweak penguin  can be added in a similar way to the  $B\to\rho\pi$ system \cite{Gronau:2005pq}:
\begin{eqnarray}
A^{+-}+A^{-+}+2A^{00}&=&\sqrt{2}(A^{+0}+A^{0+})=T^{+}e^{-\imi\alpha}+P_\textsc{ew}^{+}\,,\\\nonumber \label{eq:rhopiPEW}
{\bar A}^{+-}+{\bar A}^{-+}+2{\bar A}^{00}&=&\sqrt{2}({\bar A}^{+0}+{\bar A}^{0+})=T^{+}e^{+\imi\alpha}+P_\textsc{ew}^{+}\,,
\end{eqnarray}
where $T^{+}$  and  $P_\textsc{ew}^{+}$ denote the  penguin and tree  contributions to the sum of the charged amplitudes.
Considering only the Dalitz analysis of the neutral modes, the penguin triangular relation Eq.~(\ref{eq:Prhopi1}) is violated:
\begin{equation}
P^{+-} + P^{-+} + 2P^{00} = P_\textsc{ew}^{+}.
\end{equation}
The weak phase $\alpha$ derived from
\begin{equation}
e^{2\imi \alpha}=\frac{\bar A^{+-}+\bar A^{-+}+2\bar A^{00}-P_\textsc{ew}^{+}}{A^{+-}+A^{-+}+2A^{00}-P_\textsc{ew}^{+}}\,,
\end{equation}
can still be extracted by constraining the electroweak penguin-to-tree ratio:
\begin{equation}
r_{\ewp}(\rho\pi)=\frac{P_\textsc{ew}^{+}}{A^{+-}+A^{-+}+2A^{00}-P_\textsc{ew}^{+}}=\frac{P_\textsc{ew}^{+}}{T^{+}}\,.
\end{equation}
As discussed in Refs.~\cite{Ciuchini:2006kv,Gronau:2006qn}, the numerical expectation given in Eq.~(\ref{eqn:ewp}) does not apply to the $\rho\pi$ system:  there are additional contributions in this case compared to $\pi\pi$ and $\rho\rho$, coming from  matrix elements of tree operators that do not cancel anymore since the final state does not have a symmetric wave function. Following Refs.~\cite{Ciuchini:2006kv,Gronau:2006qn}, we parametrise the penguin-to-tree ratio for the $B\to\rho\pi$ transition as
\begin{equation}
r_{\ewp}(\rho\pi)=r_{\ewp}\frac{1-|r_a|e^{\imi\delta_a}}{1+|r_a|e^{\imi\delta_a}}\,,
 \quad r_a=\frac{{\cal C}_1-{\cal C}_2}{{\cal C}_1+{\cal C}_2}\times \frac{\langle (\rho\pi)^+|{\cal O}_1^u-{\cal O}_2^u|B^+\rangle}{\langle (\rho\pi)^+|{\cal O}_1^u+{\cal O}_2^u|B^+\rangle}\,,
\end{equation}
$r_{\ewp}$ is the ratio given by  Eq.~(\ref{eqn:ewp}) for the symmetrical $B\to hh$ transition where the Bose-Einstein symmetry applies. On the other hand, $r_a=|r_a|e^{\imi\delta_a}$ is a correction parameter accounting for the non-vanishing $I_f=1$ contribution, with the notation $\langle (\rho\pi)^+|=\langle \rho^+\pi^0|+\langle \rho^0\pi^+|$.
We let the correction parameter free to vary up to $|r_a|<0.3$: this rather conventional rule of thumb seems fairly conservative compared to estimates based on naive factorisation \cite{Ciuchini:2006kv,Gronau:2006qn}, but data-based extractions in the case of $B\to K^*\pi$ decays suggest that it is the appropriate order of magnitude \cite{Perez}. The unknown phase $\delta_a$, which can generate a charge asymmetry in $B^+\to(\rho\pi)^+$ decays, is unconstrained in the $B^0\to(\rho\pi)^0$ Dalitz analysis considered here.

\begin{figure}[t]
\begin{center}
  \includegraphics[width=18pc]{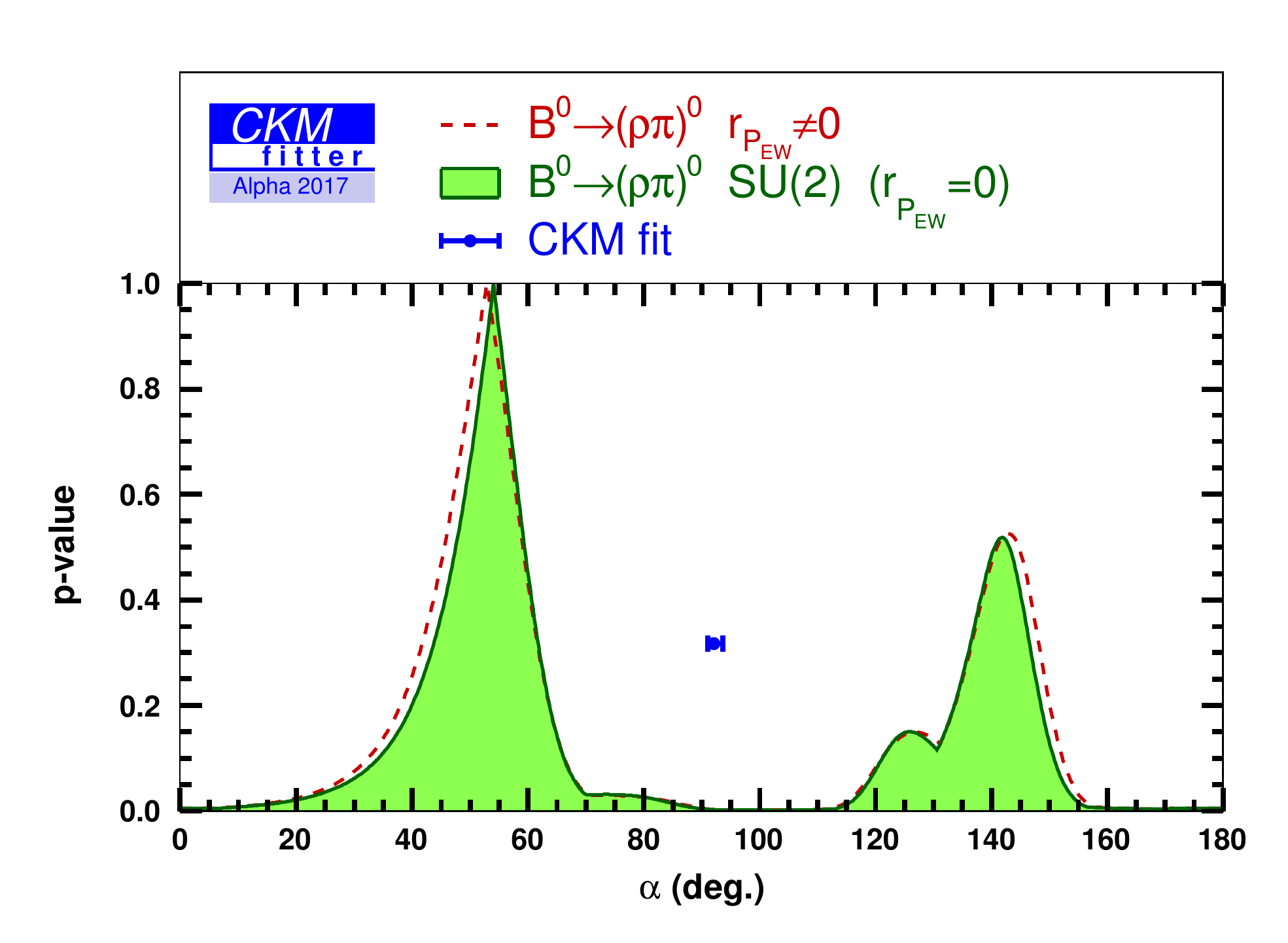}
  \includegraphics[width=18pc]{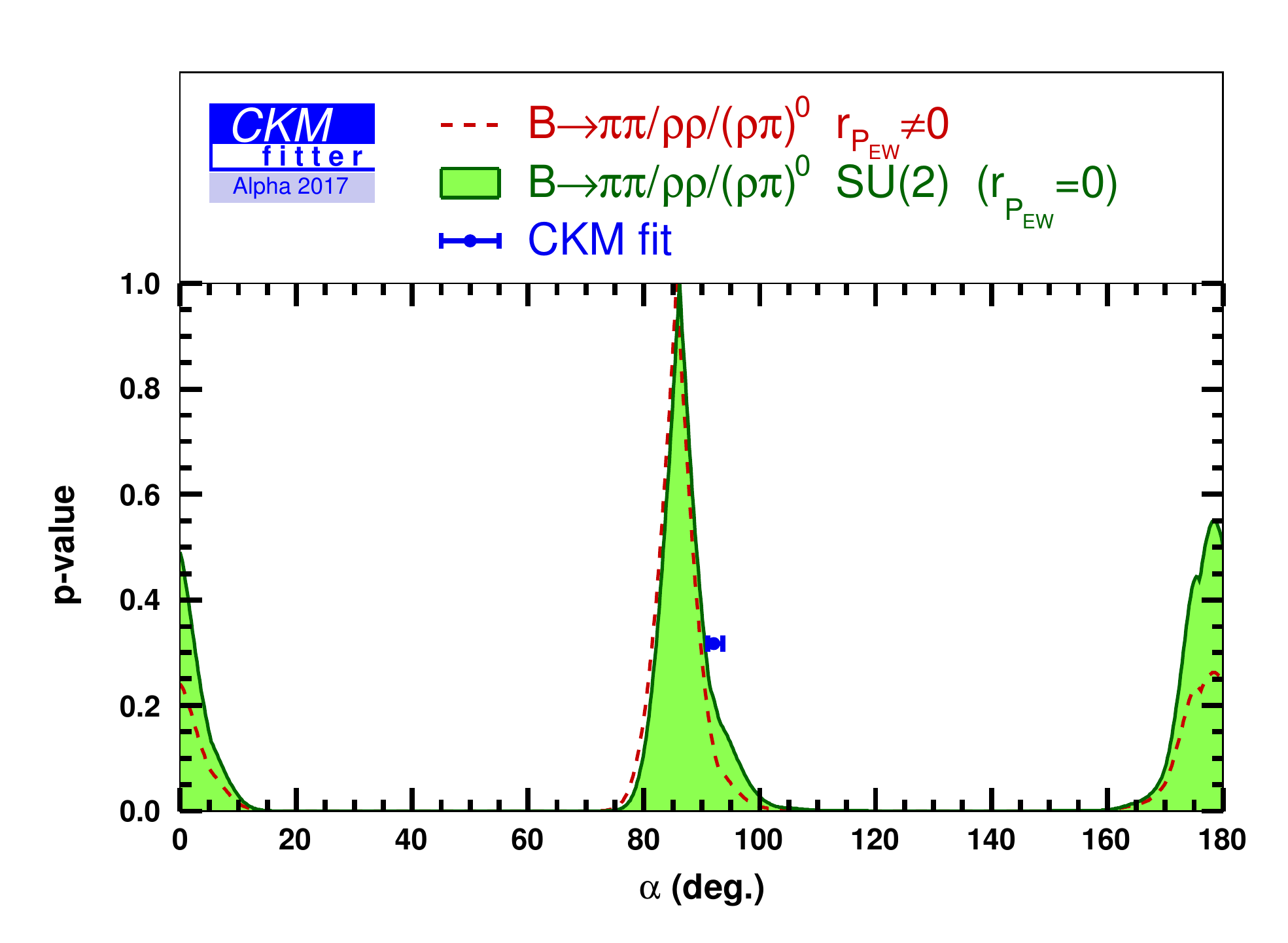}
\caption{\it\small  \su{2} constraint on $\alpha$ including an electroweak penguin contribution given by Eq.~(\ref{eqn:ewp}) (red dashed line) and neglecting this penguin contribution (green shaded area) for the $B\to\rho\pi$ system alone (left) and the combination of the three charmless systems (right).}\label{fig:alphaEWP2}
\end{center}       
\end{figure}

As illustrated on the left panel of Fig.~\ref{fig:alphaEWP2}, including the estimate of $\Delta I=\ofrac{3}{2}$ electroweak penguin in the $B^0\to(\rho\pi)^0$ isospin analysis induces a small shift of $\delta\alpha= -1.2^\circ$ for the preferred value for $\alpha$:
\begin{eqnarray}
\alpha_{(\rho\pi)^0}^\textrm{\scriptsize \ewp}&:& (\val{52.9}{+8.7}{-11.1} ~~\and~ \val{142.9}{+5.3}{-6.3})^\circ\,,
\end{eqnarray}
indicating that even a larger bound on $|r_a|$ would have only a limited impact.

 The right panel of Fig.~\ref{fig:alphaEWP2} shows the combination of the three channels in the presence of the $\Delta I=\ofrac{3}{2}$ electroweak penguin  that produces an overall shift of $\delta\alpha= -0.7^\circ$ on the solution near $90^\circ$.  The corresponding 68\% CL interval is:
\begin{eqnarray}
\alpha_{\su{2}}^\textrm{\scriptsize \ewp}& :&  \val{85.6}{+4.1}{-4.2}{$^\circ$}\,,
\end{eqnarray}
 in agreement with the indirect value $\alpha_{\rm ind}$ given in Eq.~(\ref{eq:alphaInd})  at 1.5 standard deviations.
 
 Assuming the same amplitudes ratio, $r_{\ewp}$, hold for the three decay systems (modulo the correction term included for the asymmetrical $B\to\rho\pi$ decay), the combined 68\% CL interval:
\begin{equation}
  r_{\ewp}= (\val{8.1}{+4.3}{-4.8})\times 10^{-2}
\end{equation}
is obtained when $\alpha$ is constrained to the indirect determination from Eq.~(\ref{eq:alphaInd}) and  any other \su{2}-breaking effect is neglected, in good agreement with the theoretical expectation Eq.~(\ref{eqn:ewp}).

\subsubsection{Isospin-breaking effects due to mixing in the $\pi\pi$ system}\label{sec:breaking}

We have worked up to now under the assumption that isospin symmetry was exact for the hadronic part of the $B$-meson decay. Even though isospin-breaking effects are known to be tiny, due to the very small mass difference between the $u$ and $d$ quarks, it is interesting to assess more precisely how it could affect our analysis.

In the $B\to\pi\pi$ system, the breaking of isospin symmetry due to the difference of quark masses triggers
a mixing between light pseudoscalar mesons. This affects in particular the $\pi^0$ meson.
Following Refs.~\cite{Gardner:1998gz,Zupan:2004hv}, at leading order of isospin breaking, the $\pi^0$ state can be written as
\begin{eqnarray}
  \ket{\pi^0} = \ket{\pi_3}+\epsilon_{\eta}\ket{\eta}+\epsilon_{\eta'}\ket{\eta'} \qquad \qquad(\epsilon_{\eta},\epsilon_{\eta'} \ll 1)\,,
\end{eqnarray}
 where the flavour states are defined as follows:  $\ket{\pi_3}$  represents the $I=0$ component of the $\su{2}$ triplet  and $\ket{\eta}$, $\ket{\eta'}$ are the states resulting from the mixing of the \su{3} pseudoscalar meson octet and singlet components (according to the flavour decomposition $3\otimes 3=8\oplus 1$).
The triangular isospin relation that applies to the isospin amplitudes:
\begin{equation}
  \frac{A^{+-}}{\sqrt{2}}+A^{33}=A^{+3}
\end{equation}
 can be translated into mass-eigenstate amplitudes upon introducing the shifts $\delta A^{i0}$:
\begin{equation}
  A^{33} = A^{00} - \delta A^{00}\,, \qquad\qquad
  A^{+3} = A^{+0} - \delta A^{+0}\,.
\end{equation}
The mixing among light pseudoscalar mesons thus generates a slight modification of the triangular relation in the $B\to\pi\pi$ amplitudes:
\begin{equation}
  \frac{A^{+-}}{\sqrt{2}} + A^{00} =A^{+0}(1-e)\,,\qquad\qquad\label{eq:modSU2}
  e = \frac{\delta A^{+0} - \delta A^{00}}{ A^{+0}}\,,
\end{equation}
with a similar equation for the $CP$-conjugate amplitudes.

\begin{table}[t]
\begin{center}
\begin{tabular}{|l|c|}
\hline
Observable & World average \\
\hline
\small ${\cal B}^{\pi^0\eta}$ $(\times 10^6)$    &\small \val{0.41}{+0.18}{-0.17}        \\
\small ${\cal B}^{\pi^0\eta'}$ $(\times 10^6)$  &\small  \val{1.2}{0.4}        \\
\small ${\cal B}^{\pi^+\eta}$ $(\times 10^6)$    &\small \val{4.02}{0.27}     \\
\small ${\cal B}^{\pi^+\eta'}$ $(\times 10^6)$   &\small \val{2.7}{+0.5}{-0.4}\\
\hline
\small ${\cal C}^{\pi^+\eta}$    &\small \val{0.14}{0.05}        \\
\small ${\cal C}^{\pi^+\eta'}$   &\small \val{-0.06}{0.15}        \\
\hline
\end{tabular}
\caption{\it\small  World averages for  the branching ratios and direct $CP$ asymmetries of the $B^{0,+}\to\pi^{0,+}\eta^{(')}$ modes~\cite{Amhis:2016xyh}.\label{tab:input_eta}}
\end{center}
\end{table}

At leading order in $\epsilon_{\eta^{(')}}$,  the amplitude shifts $\delta A^{i0}$ can be written in terms of the $B\to\pi\eta^{(')}$ amplitudes $A^{\pi\eta^{(')}}$ as 
\begin{equation}
  \delta A^{00}=\sqrt{2}(\epsilon_{\eta} A^{\pi^0\eta}+ \epsilon_{\eta'} A^{\pi^0\eta'})\,,\qquad\qquad
  \delta A^{+0}=\epsilon_{\eta} A^{\pi^+\eta} + \epsilon_{\eta'} A^{\pi^+\eta'}\,,
\end{equation}
where the factor $\sqrt{2}$ in the first equation accounts for the Bose--Einstein symmetry in the symmetric $A^{00}-A^{33}$ amplitude.  Both neutral and charged $B\to\pi\eta^{(')}$ decay amplitudes can be  constrained using the experimental branching ratios and \CP-asymmetries reported in Tab.~\ref{tab:input_eta}. Contrary to the
$\Delta I=\ofrac{3}{2}$ contribution discussed in Sec.~\ref{sec:ewpeng}, the a priori unknown strong phase affecting $\delta A^{+0}$ may generate a charge asymmetry $|A^{+0}|\neq|\bar A^{+0}|$ in the charged $B$ meson decay. Consequently, the amplitude system is further constrained by introducing the measured  $CP$-asymmetry in the  $B^+\to\pi^+\pi^0$ mode, see Tab.~\ref{tab:input_CPcharged}, which was not considered in the isospin-symmetric analysis. 

\begin{table}[t]
\begin{center}
\begin{tabular}{|l|c|l|}
\hline
Observable & World average   & Reference \\
\hline
\small ${\cal C}^{+0}_{\pi\pi}$                   &\small $(-2.6\pm 3.9)10^{-2}$                  &\small   \cite{Babar_pp_bp0,Belle_pp_bpm}                       \\
\small ${\cal C}^{+0}_{\rho_L\rho_L}$                   &\small $(-5.1\pm 5.4)10^{-2}$                  &\small   \cite{Babar_rr_bp0,Belle_rr_cp0}                       \\
\hline
\end{tabular}
\caption{\it\small  World averages for the $C^{+0}$ direct $CP$ asymmetries for the $B\to\pi^+\pi^0$  decay and for the longitudinally polarised state in $B^+\to\rho^+\rho^0$ decay. } \label{tab:input_CPcharged}
\end{center}
\end{table}

\begin{figure}[t]
\begin{center}
  \includegraphics[width=21pc]{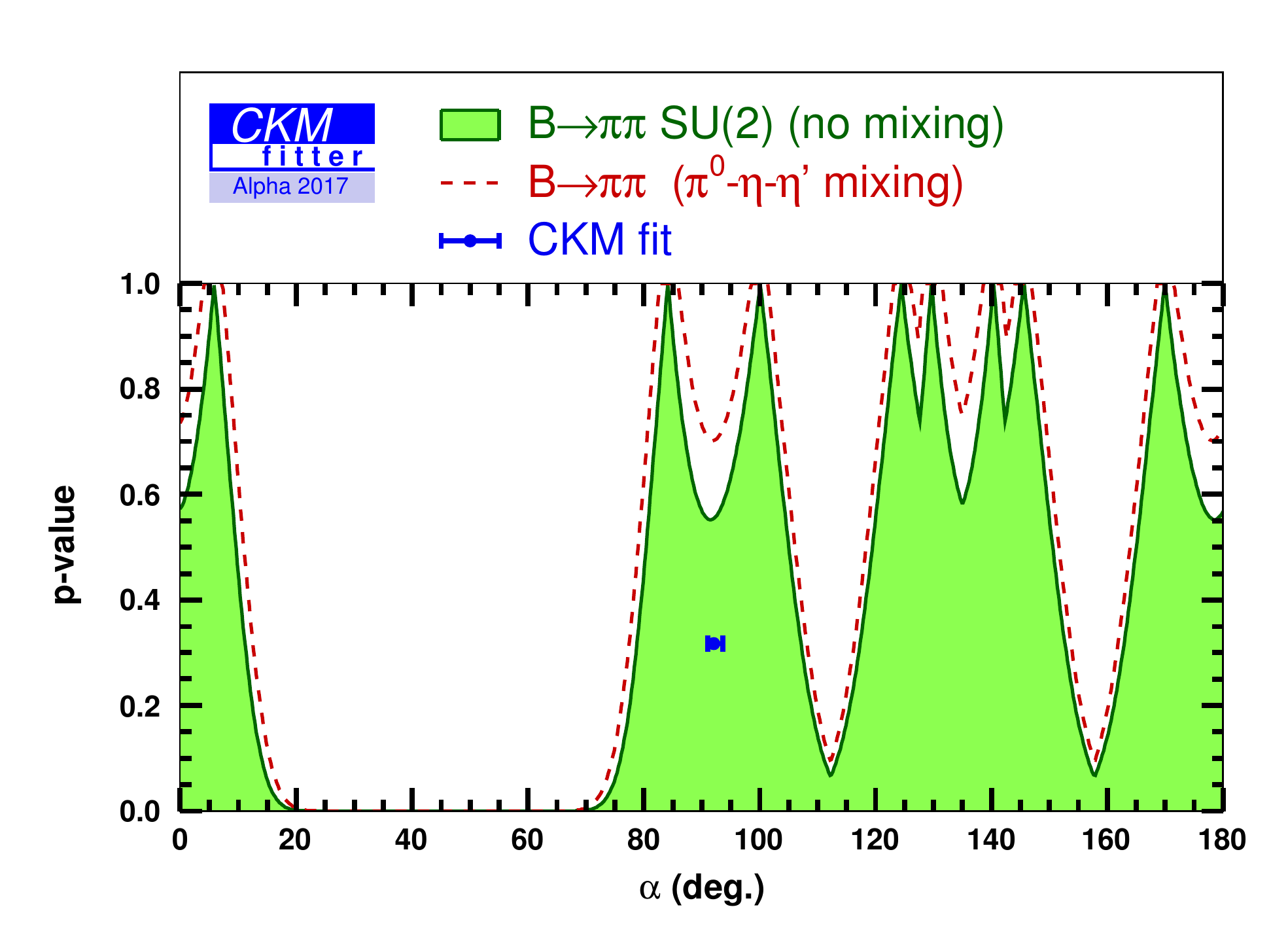}
\caption{\it\small  Determination of $\alpha$ from the $B\to\pi\pi$ system  including $\pi^0-\eta-\eta'$ mixing  (red dashed line) and neglecting this mixing (green shaded area).}\label{fig:alphaMix}
\end{center}       
\end{figure}

We must add eight complex amplitudes, $A^{\pi^+\eta}$, $A^{\pi^+\eta'}$,  $A^{\pi^0\eta}$,  $A^{\pi^0\eta'}$ and their $CP$ conjugates, to the $B\to\pi\pi$ system. We use Ref.~\cite{Kroll:2004rs} for the numerical estimates of the mixing parameters:
\begin{equation}
\epsilon_{\eta} = (1.7\pm 0.3)\times 10^{-2}\,,\qquad\qquad
\epsilon_{\eta'} = (0.4\pm 0.1)\times 10^{-2}\,,
\end{equation}
leading to a slightly modified constraint on the weak phase $\alpha$, shown in Fig.~\ref{fig:alphaMix}:
\begin{eqnarray}
\alpha_{\pi\pi}^\textrm{\scriptsize mixing}   &:& (\val{92.5}{15.5} ~~\and~ \val{135.0}{19.0} ~~\and~~ \val{172.2}{10.4})^\circ, 
\end{eqnarray}
at 68\% CL. We obtain thus a global shift for $\alpha$ of $\delta\alpha=-0.5^\circ$. The linear increase of the 68\% interval, namely $\pm 1.5^\circ$, is mostly due to the limited experimental resolution on the $B\to\pi\eta^{(')}$ data and it must be considered as an upper limit of the mixing-induced breaking effect on $\alpha$.

It is possible to go one step further using the admittedly naive simplification that there is no gluonic component to the $\eta^{(')}$ states. Additional relations between the $B\to\pi\pi$ and the $B\to\pi\eta^{(')}$ amplitudes can then be derived, unaffected by first-order \su{3} breaking \cite{{Dighe:1995gq}}:
\begin{eqnarray}
A^{+\eta}&=&\cos(\phi_P) A^{+0}+ \sqrt{2} A^{0\eta}\,,\\
A^{+\eta'}&=&\sin(\phi_P) A^{+0} + \sqrt{2} A^{0\eta'}\,,
\end{eqnarray}
where $\phi_P$ is the mixing angle for the isospin singlets \ket{\eta_q} = ${(\ket{u{\bar u}}+\ket{d{\bar d}})}/{\sqrt{2}}$ and \ket{\eta_s}=\ket{s{\bar s}} forming the $\eta^{(')}$ states. With these additional constraints, only four additional complex parameters are needed to account for the $B\to\pi\eta^{(')}$  amplitudes. In that case, the deviation for the triangular isospin relation Eq.~(\ref{eq:modSU2}) is given by:
\begin{equation}
e=\cos(\phi_P)\ \epsilon_{\eta} + \sin(\phi_P)\ \epsilon_{\eta'}.
\end{equation}
Using this relation and without further assumption on the $\eta$-$\eta'$ mixing angle, the isospin-breaking effect on the $B\to\pi\pi$ determination of $\alpha$ is found to be $\delta\alpha=-1^\circ$ with a slightly reduced resolution  $\pm 1.0 ^\circ$. Assuming the  physical $\eta^{(')}$ states are well described by the flavour combinations:
\begin{equation}
\ket{\eta} =\frac{\sqrt{2}\ket{\eta_q}-\ket{\eta_s}}{\sqrt{3}}\,,\qquad \ket{\eta'}= \frac{\ket{\eta_q}+\sqrt{2}\ket{\eta_s}}{\sqrt{3}}\,,
\end{equation}
corresponding to an $\eta-\eta'$ mixing angle $\tan(\phi_P)=1/\sqrt{2}$, a deviation  $\delta\alpha=(-0.3\pm1.4)^\circ$ is obtained for the $B\to\pi\pi$ system, in the same ball park as our previous estimate.

\subsubsection{Additional isospin-breaking effects for the $\rho\rho$ and $\rho\pi$ systems}

In the limit of a vanishing $\rho$ meson width, the isospin analysis of each $B\to\rho\rho$  polarisation state is identical to the  $B\to\pi\pi$ analysis, with a dominance of the longitudinal polarisation. In parallel with the previous section, isospin breaking manifests itself though the $\rho-\omega-\phi$ mixing:
\begin{equation}
\ket{\rho^0}\sim\ket{\rho_3}+\epsilon_\omega\ket{\omega}+\epsilon_\phi\ket{\phi}\,,
\end{equation}
where the mixing term $\epsilon_\omega$ is of  ${\cal O}(1\%)$ and $\epsilon_\phi$ is  negligible
as there is an almost ideal mixing in the case of the $\phi$ meson.
Additional sources of isospin breaking can occur in the $\rho\to\pi\pi$  decay  (see Ref.~\cite{Gronau:2005pq} for a detailed review):
\begin{itemize}
\item The isospin breaking due to the $\pi^0-\eta-\eta'$ mixing may affect the $\rho^+\to\pi^+\pi^0$ decay but turns out negligible, as the leading term in $\epsilon_\eta$ is suppressed by the small $\rho^+\to\pi^+\eta$ decay rate: ${\Gamma(\rho^+\to\pi^+\eta)}/{\Gamma(\rho^+\to\pi^+\pi^0)}<0.6\%$~\cite{Olive:2016xmw}.
\item Differences in the di-pion couplings for the neutral and  charged $\rho$ mesons are experimentally limited to less than 1\%: indeed, $1-({\Gamma_{\rho^+}}/{\Gamma_{\rho^0}})=(0.2\pm 0.9)\%$~\cite{Olive:2016xmw}
\item Isospin breaking affecting the $\rho^0,\omega\to\pi^-\pi^+$  interference is restricted to a small window in the $\pi\pi$ mass spectrum: this effect, integrated over the whole $\pi\pi$ range, is estimated at the order of 2\% \cite{Gronau:2005pq}.
\item The dominant source of isospin breaking is actually due to the large decay width $\Gamma_\rho$ that makes the two final-state mesons distinguishable in the $B\to\rho_1(\pi\pi)\rho_2(\pi\pi)$ decay. This results in a
residual  $I_f=1$ amplitude contribution, forbidden by Bose--Einstein symmetry in the limit $\Gamma_\rho=0$, but potentially as large as $({\Gamma_\rho}/{m_\rho})^2\sim4\%$. It is a slight abuse of language to call this effect an isospin-breaking contribution, as it does not vanish in the limit $m_u=m_d$ (but it does in the limit of a vanishing decay width $\Gamma_\rho\sim (m_{(\pi\pi)_{1}}-m_{(\pi\pi)_{2}})=0$).
\end{itemize}
 \begin{figure}[t]
\begin{center}
  \includegraphics[width=21pc]{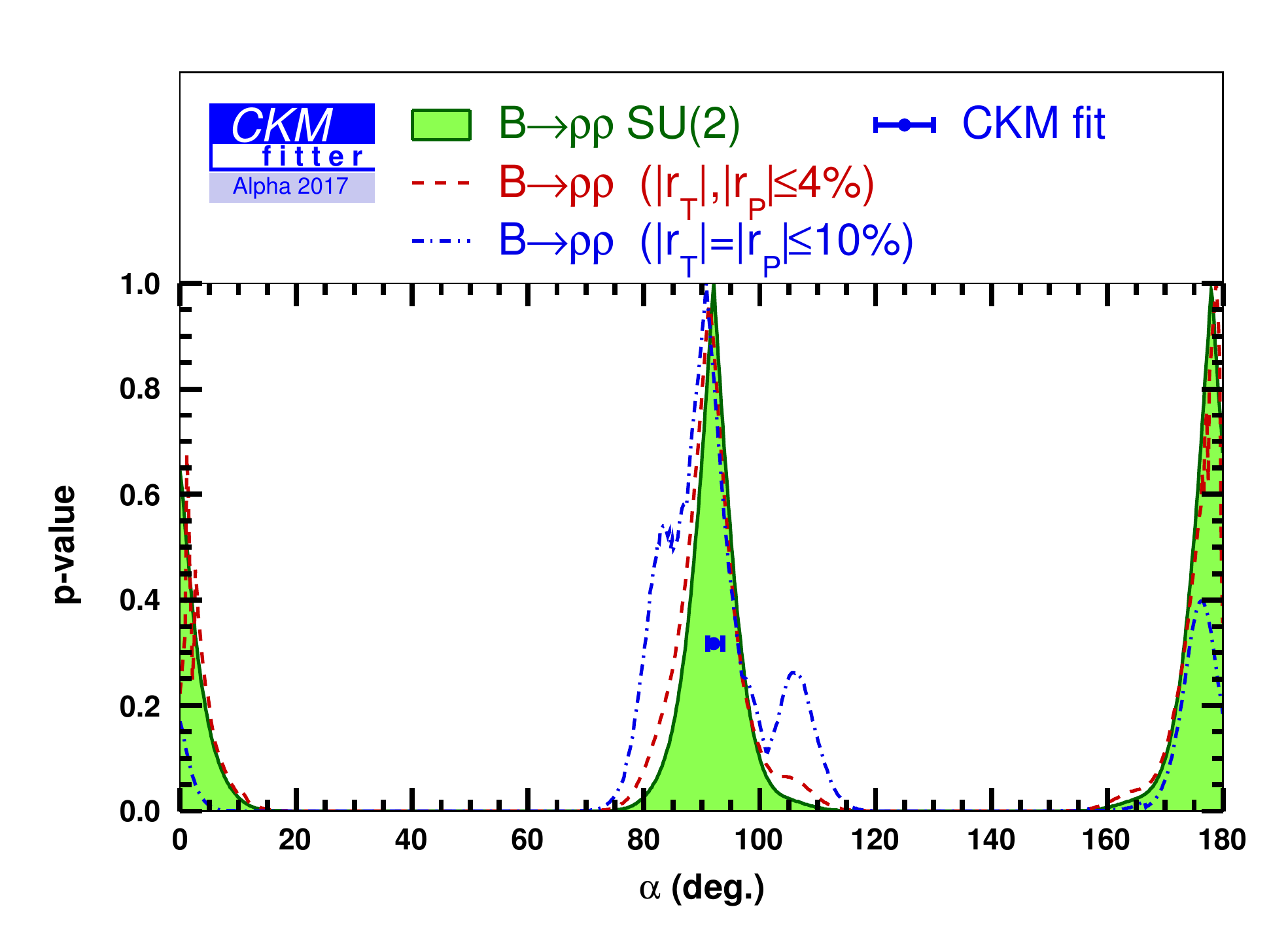}
\caption{\it\small  Determination of $\alpha$ from the $B\to\rho\rho$ system  including \su{2}-breaking amplitudes limited to 4\% (red dashed line), limited to 10\% (blue dot-dashed line) and neglecting this contributions (green shaded area).}\label{fig:rhorhoBreak}
\end{center}       
\end{figure}

The above isospin-breaking effects can be accounted for by 
modifying the isospin triangular relation in the same way as in Eq.~(\ref{eq:modSU2}):
\begin{equation}
\frac{A^{+-}}{\sqrt{2}}+A^{00}=A^{+0} - \Delta^{+0}\,.
\end{equation}
The unknown amplitude $\Delta^{+0}$ is parametrised in a generic way:
\begin{equation}
\Delta^{+0}= r_T T^{+-} + r_P P^{+-}\,,
\end{equation}
where $r_T$ and $r_P$ are two independent complex parameters accounting for isospin breaking in the tree and penguin contributions, respectively. The arbitrary  strong phase affecting these parameters can generate $CP$ violation in the charged decays. The measured  $CP$ asymmetry in the  $B^+\to\rho^+\rho^0$ mode given in Tab.~\ref{tab:input_CPcharged} is thus included as an additional constraint in the presence of these isospin-breaking terms. Fig.~\ref{fig:rhorhoBreak} illustrates the determination of the weak phase $\alpha$ in the case of isospin-breaking contributions  smaller than $|r_T|,|r_P|<4\%$ and the very conservative bound  $|r_T|,|r_P|<10\%$. A small shift on the preferred $\alpha$ value near $90^\circ$, $\delta\alpha=-0.6^\circ$ $(-1.2^\circ)$, results from isospin-breaking contributions limited to 4\% (10\%) respectively. The linear increase of the 68\% interval, $({}^{+0.2}_{-1.7})^\circ$ for $|r_T|,|r_P|<4\%$, has to be understood as an upper limit of the impact of these contributions on the determination of $\alpha$.

Isospin breaking  arises  similarly in the $B\to\rho\pi$ system through mixing in the light pseudoscalar and vector sectors. However, as long as the weak phase determination is limited to the neutral $B^0\to(\rho\pi)^0$ Dalitz analysis, the isospin constraints reduces to the triangular penguin relation Eq.~(\ref{eq:Prhopi1}). Any isospin breaking in the tree amplitudes  can be absorbed by a redefinition of the unconstrained charged amplitudes $T^{+0}$, $T^{0+}$. Corrections to the penguin amplitude relation Eq.~(\ref{eq:Prhopi1}) in the $B\to\rho\pi$ system will be suppressed by the  small penguin-to-tree ratios. For instance, the impact of  the $\pi^0-\eta-\eta'$ mixing on the weak phase determination in the $B^0\to(\rho\pi)^0$  analysis is suppressed by factors like
$\epsilon_{\eta^{(')}}\left|{P_{\rho\eta^{(')}}}/{T^+}\right|$,
 where $P_{\rho\eta^{(')}}$ represents the penguin contribution to the $B^+\to\rho^+\eta^{(')}$ transition and $T^+$ is defined in Eq.~(\ref{eq:rhopiPEW}). The corresponding deviation is found to be negligible, $\delta\alpha<0.1^\circ$~\cite{Gronau:2005pq}. Therefore,
apart for the electroweak penguin contribution discussed in section \ref{sec:ewpeng}, no further isospin breaking will be considered for the $B^0\to(\rho\pi)^0$  analysis.

\subsection{Impact of the statistical treatment \label{sec:statistics}}

\subsubsection{$p$-values based on the bootstrap approach and Wilks' theorem}

In Sec.~\ref{sec:procedure}, we have outlined our framework to determine confidence intervals on $\alpha$, starting from
a test statistic $\Delta \chisq(\alpha)$, which is converted into a $p$-value following Eq.~(\ref{eq:prob}). This procedure can be proven to be exact in the simple case where the observables obey Gaussian probability distribution functions and they are linearly related to the parameters of interest. 
Following Wilks' theorem, this can be extended to more general cases, at least asymptotically, if the data sample is large and the observables resolutions are small enough to consider the problem as locally linear \cite{Wilks}. In such case, the construction in Sec.~\ref{sec:procedure} ensures exact coverage: if one repeated the determination of $\alpha$ using  independent data sets from many identical experiments, the  68\% CL interval for $\alpha$ would encompass the true value of $\alpha$ in 68\% of the cases.

However, many effects can alter this picture. Even in the exact Gaussian case, the $p$-value can be distorted 
when the observables have a nonlinear dependence on the fundamental parameters of interest. An example consists in
Eq.~(\ref{eq:alphaeff}), where the reexpression of the $CP$-asymmetries in terms of $\alpha_{\rm eff}$ is nonlinear and implements 
the trigonometric boundary on $CP$-asymmetries $\sqrt{{\cal C}^2+{\cal S}^2}<1$~\cite{Charles:2004jd}. More generally,
one may wonder if we stand close to the hypotheses of Wilks' theorem with the current set of data. Otherwise, the construction
given in Sec.~\ref{sec:procedure} might suffer from under- or over-coverage.
  In this section, we will assess the finite-size errors associated with our statistical framework by considering a different construction of the $p$-value that takes into account some of the effects deemed subleading in the Wilks-based approach. 

We start by recalling some elements related to the construction of $p$-values in our context, as discussed in Refs.~\cite{Charles:2016qtt,James:2006zz,Kendall-Stuart,Cowan:2013pha}. We want to assess how
much the data is compatible with the hypothesis that the true value of the weak phase $\alpha$, denoted $\alpha_t$, is equal to some fixed value $\alpha$, i.e.,
${\mathcal H}_{\alpha}: \alpha_t = \alpha$.
This hypothesis is composite, as it sets the value of some of the theoretical parameters, but not all of them. Indeed, the hadronic parameters (tree and penguin amplitudes) are also theoretical parameters required in our isospin analysis but are not set in ${\mathcal H}_{\alpha}$. These hadronic parameters are nuisance parameters, denoted collectively as $\vec\mu$. 

The test statistic $\Delta\chisq(\alpha)$ defined in Eq.~(\ref{eq:deltachisq}) can be seen as the maximal likelihood ratio comparing the most 
plausible configuration under ${\mathcal H}_\alpha$ with the most plausible one in general. It is a definite positive function 
chosen in a way that large values  indicate that the data present evidence against ${\mathcal H}_\alpha$: in the following, we will state explicitly its dependence on the data $X$ by using the notation $\Delta\chisq(X;\alpha)$.
A $p$-value is built by calculating the probability
to obtain a value for the test statistic at least as large as the one that was
actually observed, assuming that the hypothesis ${\mathcal H}_{\alpha}$ is true:
\begin{eqnarray}\label{pvalue}
1-p(X_0;\alpha,\vec\mu) &=& \int^{\Delta\chisq(X_0;\alpha)}_{0} d\Delta\chisq\, h(\Delta\chisq|\alpha,\vec\mu)
 ={\mathcal P}[\Delta\chisq<\Delta\chisq(X_0;\alpha)]\,,
\end{eqnarray}
where the probability distribution function (PDF) $h$ of the test statistic is obtained from the PDF $g$ of
the data
as
\begin{equation}\label{eq:pdft}
h(\Delta\chisq|\alpha,\vec\mu) = \int dX\, \delta\left[\Delta\chisq-\Delta\chisq(X;\alpha)\right] g(X;\alpha,\vec\mu)\,.
\end{equation}
 A small value of the $p$-value  thus provides evidence against the hypothesis ${\mathcal H}_{\alpha}$. We notice that in general the $p$-value Eq.~(\ref{pvalue}) exhibits a dependence on the nuisance parameters through the PDF $h$, even though the test statistic $\Delta\chisq$ itself is independent of $\vec\mu$. 

For linear models, in which the observables $X$ depend linearly on the parameter $\alpha_t$, $\Delta\chisq(X;\alpha)$ is
a sum of standard normal random variables, and is distributed as a $\chisq$ with $N_{dof}=1$. Under the conditions of Wilks' theorem \cite{Wilks}, this property can be extended to non-Gaussian cases, the distribution of $\Delta\chisq(X;\alpha)$ will converge to a $\chisq$ law
depending only on the number of parameters tested. The $p$-value becomes independent of the nuisance parameters $\vec\mu$ and can be still interpreted as coming from a $\chisq$ law with $N_{dof}=1$. This is the rationale for the statistical framework presented in Sec.~\ref{sec:procedure}.

We compare this Wilks-based approach with the bootstrap one based on the plug-in principle \cite{Efron2}. It consists here in defining
\begin{equation}\label{pvaluebootstrap}
p^{\rm bootstrap}(X_0;\alpha) = p(X_0;\alpha,\hat{\vec\mu}(\alpha))\,, \qquad \chisq(\alpha,\hat{\vec\mu}(\alpha))=\min_{\vec\mu}\chisq(\alpha,\vec\mu)\,.
\end{equation}
We perform the evaluation of the $p$-value for a given value of $\alpha$ by setting the nuisance parameters to the value minimising $\chisq(\alpha,\vec\mu)$. This approach assesses
the role played by nuisance parameters by replacing them with an estimator $\hat{\vec\mu}(\alpha)$ that depends on $\alpha$. 
Other constructions could have been considered, with more conservative statistical properties (supremum, constrained supremum, etc.)~\cite{Trabelsi}, but this would go beyond the scope of our study. We focus here on the bootstrap approach: it is relatively simple to implement, it exhibits good coverage properties in the examples considered here and it provides a first glimpse of the role played by nuisance parameters in coverage, which is neglected in the Wilks-based approach used up to now.

\begin{figure}[t]
\begin{center}
  \includegraphics[width=18pc]{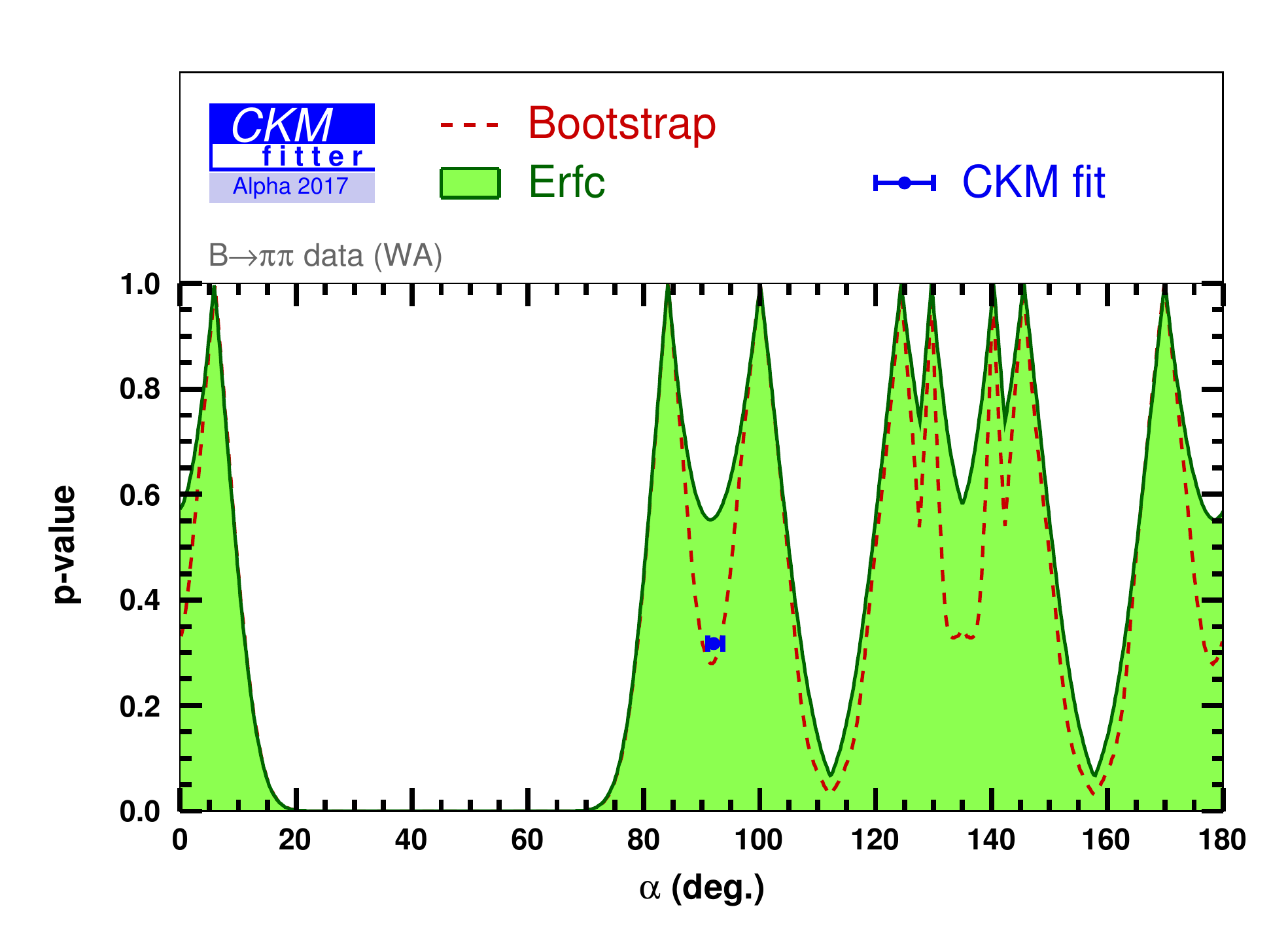}
  \includegraphics[width=18pc]{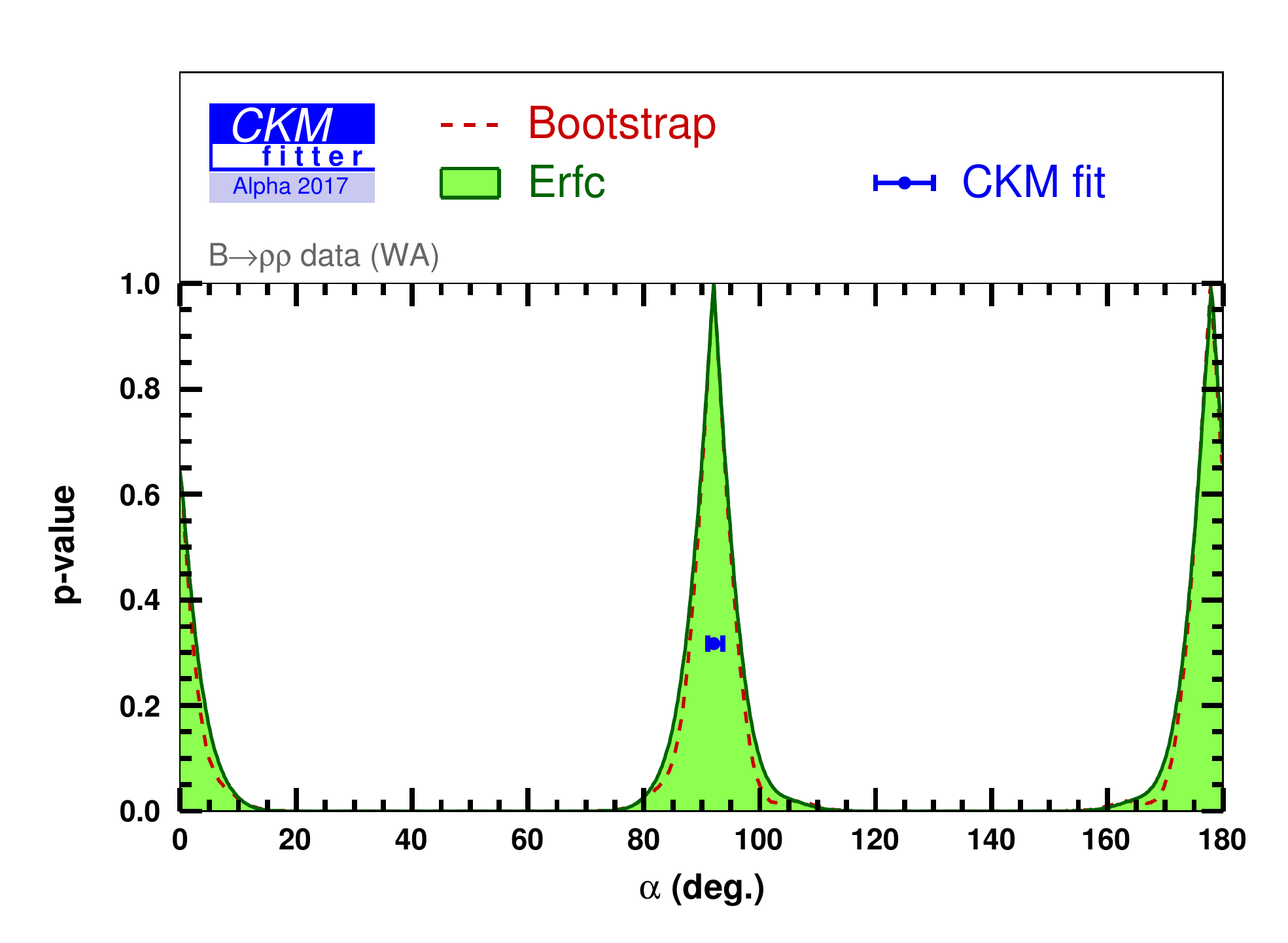}
  \includegraphics[width=18pc]{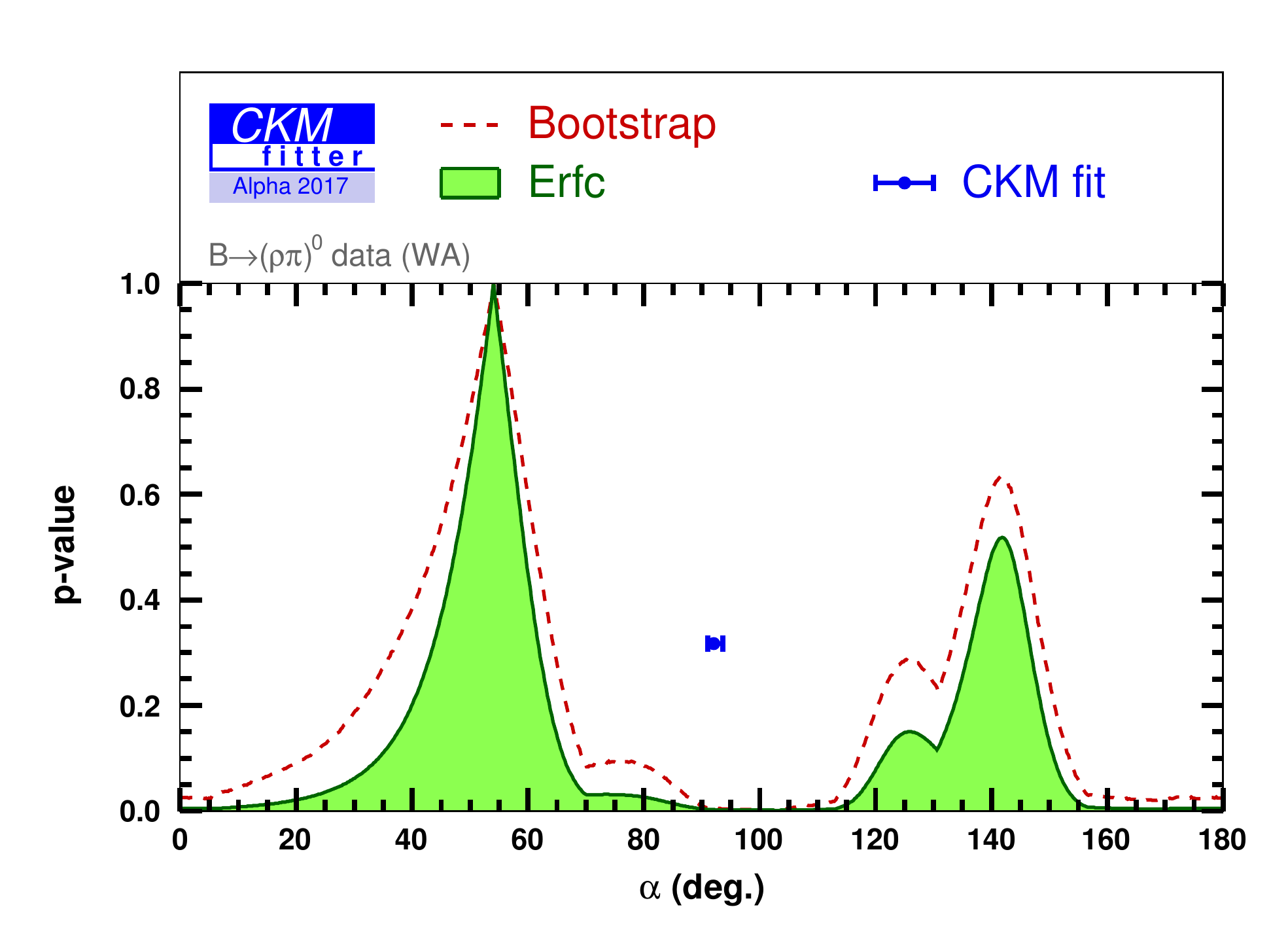}
\caption{\it\small  p-value for $\alpha$ comparing the Wilks-based approach  (denoted ${\rm Erfc}$) and the full frequentist approach based on the bootstrap method for $B\to\pi\pi$ (top left), $B\to\rho\rho$ (top right), $B\to\rho\pi$ (bottom).}\label{fig:stat_B2hh}
\end{center}       
\end{figure}

\subsubsection{Comparison of $p$-values for the extraction of $\alpha$}\label{sec:pvaluecompar}

The comparison between the two approaches may be performed by examining their coverage properties.
Once computed, a $p$-value can be used to determine confidence-level intervals for the parameter of interest $\alpha$. These intervals have a correct frequentist interpretation if the $p$-value exhibits exact coverage, i.e., for any $\beta$ between 0 and 1, $P(p\leq \beta)=\beta$. Over-coverage $P>\beta$ corresponding to a conservative $p$-value and too wide a CL interval, under-coverage $P<\beta$ corresponding to a liberal $p$-value and too narrow a CL interval. In general, an exact $p$-value, or if not possible, a reasonably conservative $p$-value, is desirable, at least for the confidence levels of interest. This conservative approach is generally adopted in high-energy physics in order to avoid rejecting a hypothesis (such as ``the SM is true'') too hastily.

We have already discussed the extraction of the $p$-value using the Wilks-based approach.
In the case of the bootstrap approach, the statistical coverage of the $\alpha$ intervals  has been studied through a full frequentist exploration of the space of nuisance parameters. A complete analysis with toy Monte Carlo simulations was carried out in order to compute the PDF $h$ and thus the $p$-value for the angle $\alpha$. The individual constraints on $\alpha$ from the $B\to\pi\pi$, $B\to\rho\rho$ and $B^0\to\pi^+\pi^-\pi^0$ systems, are displayed in Fig.~\ref{fig:stat_B2hh}, comparing the bootstrap and Wilks-based approaches. Considering the 68\% CL interval on $\alpha$, the Wilks-based approach is slightly more conservative than the bootstrap one for both $B\to\pi\pi$ and $B\to\rho\rho$ systems, whereas the situation is reversed for the $B^0\to(\rho\pi)^0$ analysis. 
The  68\% CL intervals obtained with the bootstrap method for the three systems are
\begin{eqnarray}
\alpha_{\pi\pi}^{\rm bootstrap} &:& \val{84.1}{+6.0}{-5.3}{$^\circ$} ~\and~ \val{100.1}{+6.1}{-6.9}{$^\circ$} ~\and~ (\val{135.0}{17.0})^\circ ~\and~ \val{169.9}{+6.9}{-6.1}{$^\circ$} ~\and~ \val{5.9}{+5.3}{-6.0}{$^\circ$}\,,  \nonumber\\
\alpha_{\rho\rho}^{\rm bootstrap} &:&\val{92.0}{+4.2}{-4.1}{$^\circ$} ~\and~ \val{177.9}{+4.2}{-4.1}{$^\circ$} \,, \nonumber\\
\alpha_{\rho\pi}^{\rm bootstrap} &:&\val{54.0}{+10.0}{-17.0}{$^\circ$} ~\and~ \val{142.0}{+6.8}{-8.7}{$^\circ$}\,.
\end{eqnarray}
For the latter system, the bootstrap approach slightly reduces the tension with the indirect determination $\alpha_{\rm ind}$ to 2.7 standard deviations.

\begin{figure}[t]
\begin{center}
  \includegraphics[width=18pc]{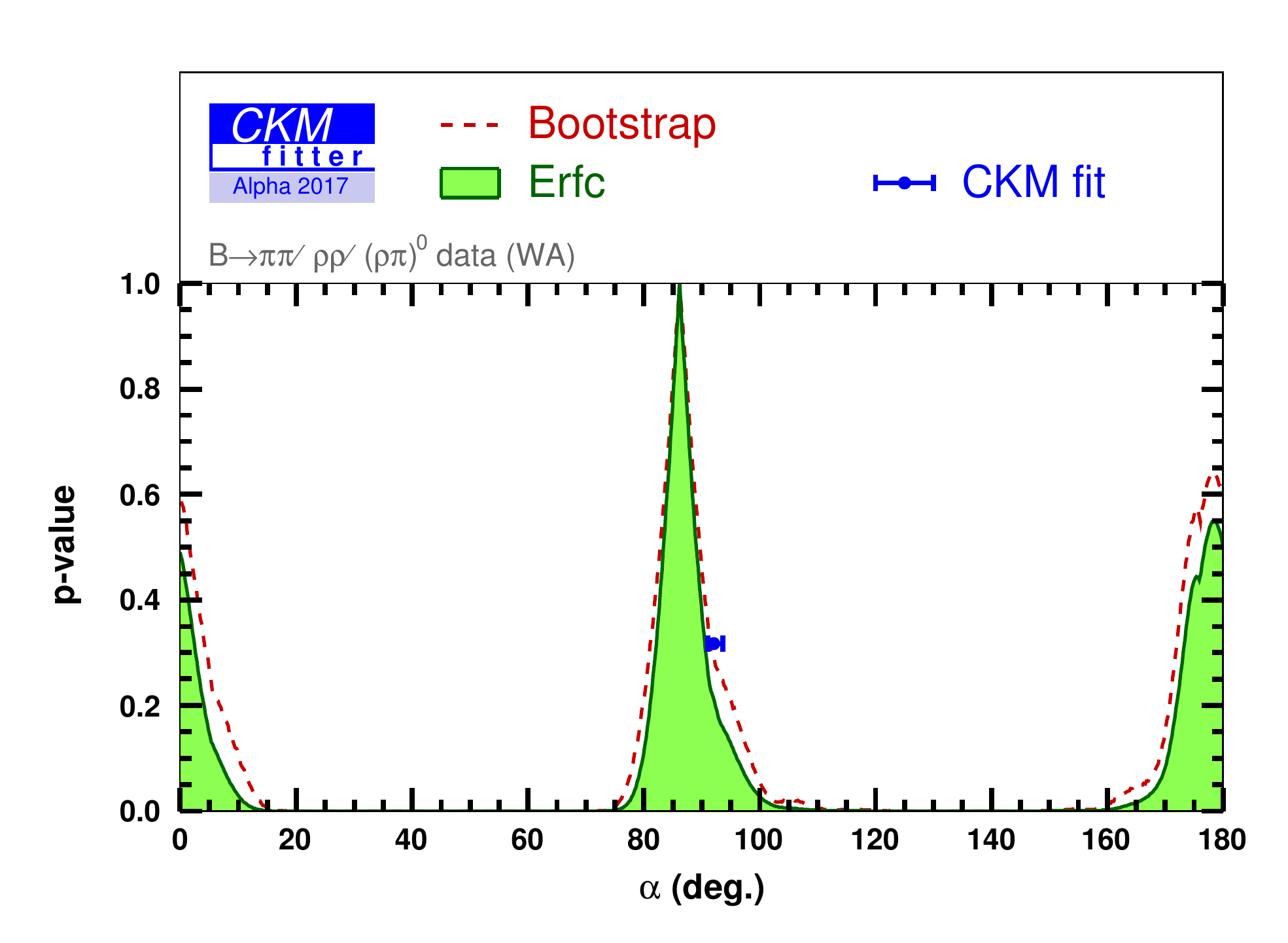}
  \includegraphics[width=18pc]{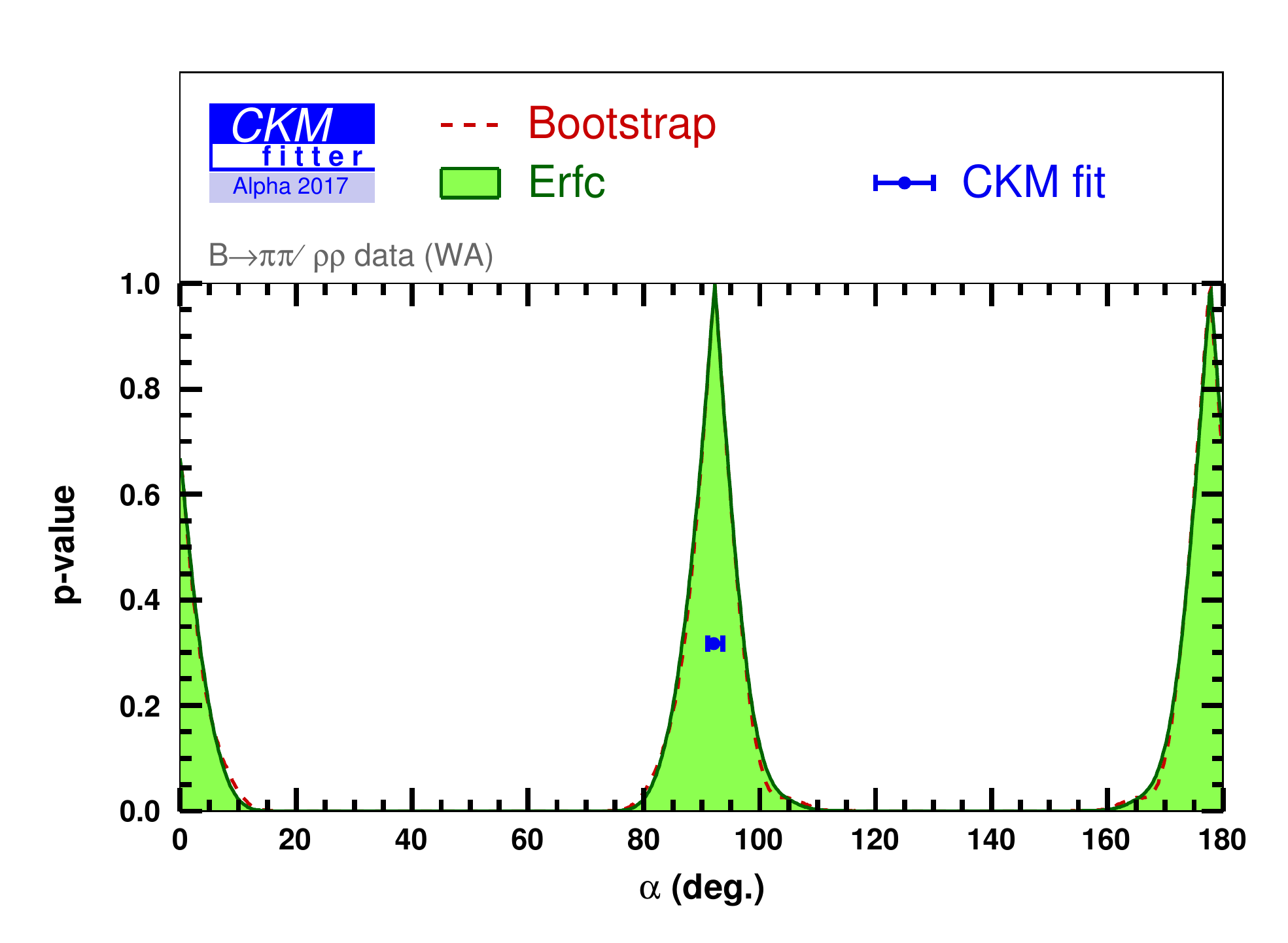}
\caption{\it\small  $p$-value for $\alpha$ comparing the Wilks-based approach (denoted ${\rm Erfc}$)  and the full frequentist approach based on the bootstrap method, for the combination of the three $B\to hh$ decay modes (left) and for the partial combination based on  $B\to\pi\pi$ and  $B\to\rho\rho$ only.}\label{fig:stat_B2UU}
\end{center}       
\end{figure}

As shown on Fig.~\ref{fig:stat_B2UU}, when combining the three analyses for the 68\% CL interval, the over-coverage for $B\to\pi\pi$ and $B\to\rho\rho$ almost compensates the under-coverage for $B\to\rho\pi$.
The  68\% CL interval  for the preferred solution close to $90^\circ$ increases by about 1$^\circ$, which can be considered as an estimate of the finite-size errors associated to the statistical framework. The second  solution near $180^\circ$ is slightly less disfavoured,  increasing the corresponding 68\% CL interval by $\pm 1.4^\circ$ with respect to the Wilks-based approach.  The overall combination results in
\begin{eqnarray}
\alpha^{\rm bootstrap}_{\rm dir} : &&\val{86.1}{+5.3}{-5.0}{$^\circ$} ~\and~  \val{178.5}{+5.5}{-6.4}{$^\circ$} ~~\textrm{(68\% CL)}~~{\rm and}\nonumber\\
                               &&\val{86.1}{+14.3}{-9.3}{$^\circ$} ~\and~  \val{178.5}{+13.5}{-12.5}{$^\circ$} ~~\textrm{(95\% CL)}\,.
\end{eqnarray}
This direct $\alpha$ measurement within the bootstrap approach is consistent with the indirect CKM determination $\alpha_{\rm ind}$ at the level of 1.1 standard deviation.

Going one step further, we can compare the coverage of the two methods. This can be done by studying the PDF of the $p$-values to check whether they are exact, conservative or liberal.
Even though $p^\mathrm{bootstrap}$ does not depend on the true value of the nuisance parameters explicitly, its distribution does in general. For illustrative purposes, we choose here to compute the distribution of both tests assuming as true values $\alpha = 92.5^\circ$ and $\vec{\mu} = \hat{\vec{\mu}}(\alpha = 92.5^\circ)$ (we do not attempt to investigate other values of $\alpha$ or the nuisance parameters). The computation of the PDF of $p^\mathrm{bootstrap}$ requires a twofold recursion of the bootstrap procedure (double bootstrap)\footnote{This computation is a CPU-consuming exercise which was carried on at the CC-IN2P3 computing farm (scoring $49 \cdot 10^3\ \mathrm{HS06 \cdot hours}$, i.e. approximatively 200 CPUs over one day).}.  The $p$-value for the $\Delta\chi^2$ test is 
obtained assuming that it obeys Wilks' theorem while the bootstrap test is assumed to be a true $p$-value, i.e., uniformly distributed over $[0,1]$.

\begin{figure}[t]
\begin{center}
  \includegraphics[width=18pc]{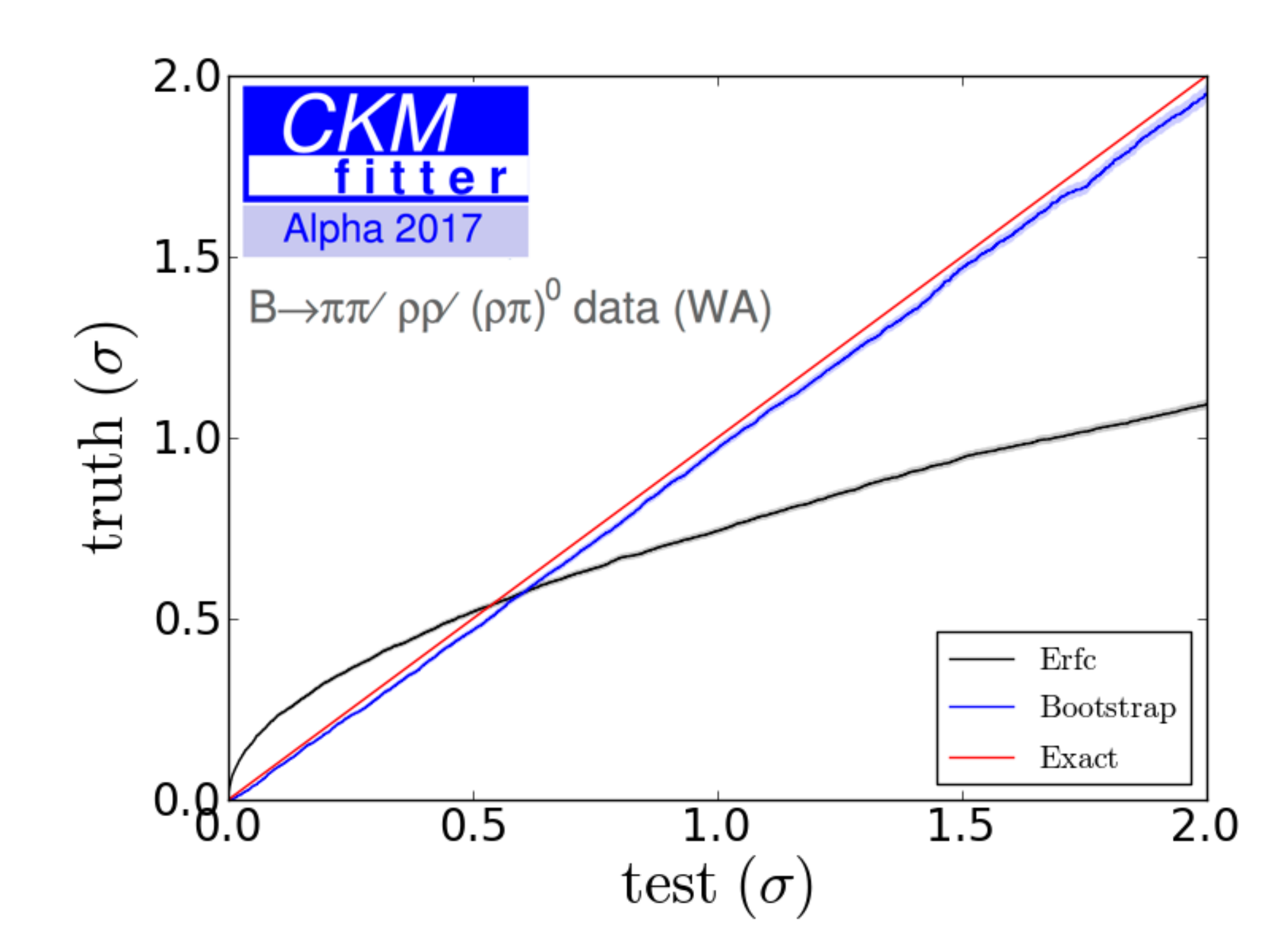}
  \includegraphics[width=18pc]{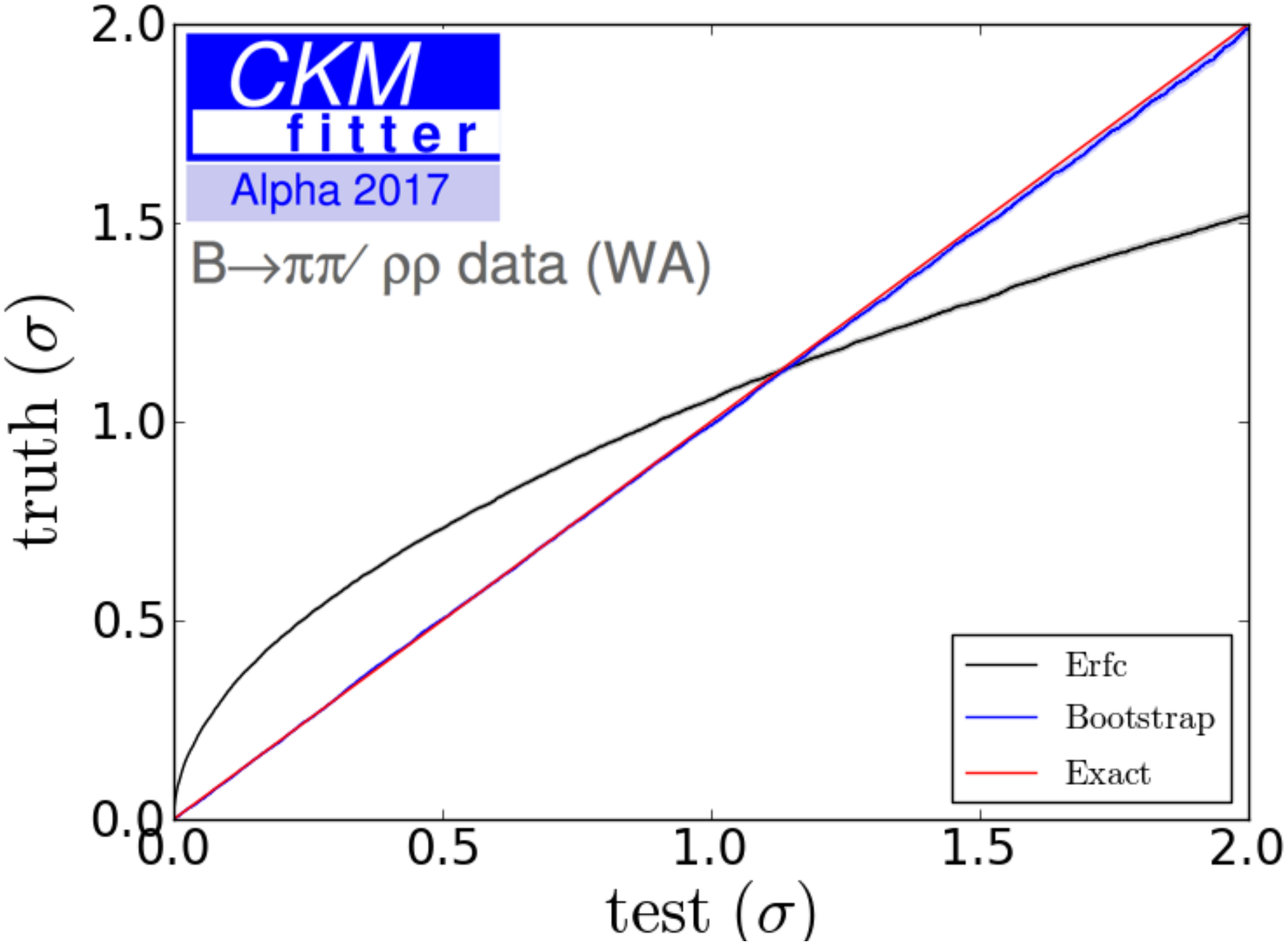}
\caption{\it\small  Coverage test displaying the true significance as a function of the test significance using the bootstrap and Wilks-based approaches (the latter being denoted ${\rm Erfc}$) for the combined extraction of $\alpha$ (left) and for the partial combination based on $B\to\pi\pi$ and $B\to\rho\rho$ only (right).}\label{fig:stat_coverage}
\end{center}       
\end{figure}

The results are shown in Fig.~\ref{fig:stat_coverage}. Rather than showing the PDF of the two $p$-values as a function of $p$, we have expressed $p$ in units of $\sigma$ (assumed test significance) and expressed the cumulative distribution function (integral of the PDF from 0 to $p$) once again in units of $\sigma$ (true test significance). A $p$-value with exact coverage corresponds to a diagonal straight line, over-coverage (under-coverage) happens if the curve is above (below) this diagonal.
 In our particular case, one can clearly see the improved coverage from the bootstrap test with respect to the $\Delta\chi^2$ case, assuming Wilks' theorem. The bootstrap distribution has a flat distribution at least up to the $2\sigma$ level, as shown by the curve in Fig.~\ref{fig:stat_coverage}, which is close to a diagonal straight line (this is expected since we set the true values of the nuisance parameters to their plug-in values). On the other hand, the $\Delta\chi^2$ test combined with Wilks' theorem is too aggressive (under-covers) for confidence levels above $0.6\sigma$ for the global combination of the three $B\to hh$ decays modes and above  $1.2\sigma$ when considering $B\to\pi\pi$ and $B\to\rho\rho$ only. 
Due to the highly CPU-consuming nature of this computation, we have not tried to perform the same computation for other true values of $\alpha$ and the nuisance parameters, but one may expect that the results of these coverage tests should hold for values of $\alpha$ in the vicinity of the solution compatible with its indirect determination. 
   
The difference in the 68\% CL intervals between the bootstrap and Wilks-based approaches could be taken as a linear correction to the direct extraction of the parameter $\alpha$, i.e. focusing only on the solution around $90^\circ$:
\begin{equation}\label{eq:alphadirectwithsyst}
\alpha_{\rm dir}= \left(86.2 ~~\err{+4.4}{-4.0}~{\rm\small (stat.)} ~\oplus~ \err{+0.9}{-1.0}~{\rm\small (bootstrap)}\right)^\circ\,.
\end{equation}
It is important to notice that the size of the bootstrap correction is an indication of the non-asymptotic regime only for the direct determination of the weak phase based on the current data on charmless decays.
This estimate is likely to change when including additional or improved observables. In particular, the observed under-coverage for the Wilks-based approach is likely to be reduced in a global fit of the CKM matrix, where the direct $\alpha$ measurement Eq.~(\ref{eq:alphadirectwithsyst}) is combined with the more precise indirect determination Eq.~(\ref{eq:alphaInd}). The bootstrap interval correction has been computed here for the direct measurement of $\alpha$ alone, for a particular set of observables. Therefore this correction
cannot be considered as an absolute uncertainty to be included automatically when combined with another independent measurement. 

\subsection{Summary for the direct determination of $\alpha$}

\begin{table}[t]
\begin{center}\footnotesize
\begin{tabular}{|c|c|c||l|l||l|l||l|}
\hline
\!\!Iso\!\! &  \!\!\ewp \!\!    &\!\! Stat\!\! & $B\to\rho\rho$   & $B\to\pi\pi$    & $B\to\pi\pi$ + $\rho\rho$  & $B^0\to(\rho\pi)^0$ & All combined \\
\hline
\footnotesize Y  & N    & W       
  \! & \!\val{92.0}{+4.7}{-4.8}  \pull{0.03} \! & \!\val{93.0}{14.0}  \pull{0.6} \!& \!\val{92.1}{+5.2}{-5.5} \pull{0.03} \! & \!\val{54.1}{+7.7}{-10.3} \pull{3.0} \! & \!\val{86.2}{+4.4}{-4.0} \pull{1.3}  \\
\footnotesize Y & N    & B
  \! & \!\val{92.0}{+4.2}{-4.0}  \pull{0.03} \! & \!\val{92.5}{13.5}  \pull{1.1} \!&\!\val{92.2}{+4.9}{-5.3} \pull{0.03} \! & \!\val{54.0}{+10.0}{-17.0} \pull{2.7}  \! & \!\val{86.1}{+5.3}{-5.0} \pull{1.1} \\
\hline
\footnotesize  Y & Y    & W        
  \! & \!\val{90.1}{+4.7}{-4.8}  \pull{0.4} \! & \!\val{91.2}{14.2}  \pull{0.5}\!&\!\val{90.1}{+5.1}{-5.6} \pull{0.3} \! & \!\val{52.9}{+8.7}{-11.1} \pull{3.0} \! & \!\val{85.6}{+4.1}{-4.2} \pull{1.5}  \\
\footnotesize N & N    & W     
  \! & \!\val{91.4}{+4.9}{-5.7}  \pull{0.2} \! & \!\val{92.5}{15.5}  \pull{0.4} \!&\!\val{91.4}{+5.4}{-6.0} \pull{0.2} \! & \!\val{54.1}{+7.7}{-10.3} \pull{3.0}  \! & \!\val{84.4}{+5.2}{-4.3}  \pull{1.5}   \\
\footnotesize N & Y     & W       
  \! & \!\val{89.8}{+4.6}{-4.9}  \pull{0.5} \! & \!\val{91.0}{15.0}  \pull{0.3}  \!&\!\val{89.8}{+4.9}{-5.3} \pull{0.5} \! & \!\val{52.9}{+8.7}{-11.1} \pull{3.0}  \! & \!\val{83.3}{+6.1}{-3.1} \pull{1.6}  \\
\hline
\end{tabular}
\caption{\it\small 68\% CL intervals for the weak phase $\alpha$ for different theoretical hypotheses, statistical approaches and channels.
In the ``Iso'' column, Y indicates that the analysis is performed assuming isospin symmetry, wheras N denotes the inclusion of 
 isospin breaking, namely, the effect of $\pi^0-\eta-\eta'$ mixing in $B\to\pi\pi$ and the breakdown of isospin triangle relations up to $|r_T|,|r_P|<4\%$ in $B\to\rho\rho$, as discussed in Sec.~\ref{sec:breaking} (isospin-breaking effects are neglected in the $B^0\to(\rho\pi)^0$ Dalitz analysis). 
In the ``\ewp'' column, Y indicates that a contamination from $\Delta I=3/2$ electroweak penguins is included as discussed in Sec.~\ref{sec:ewpeng}, whereas N corresponds to setting these penguin contributions to zero. In the ``Stat'' column, B corresponds to the bootstrap approach and W to the Wilks-based one. For each channel or combination of channels, the deviation with respect to the indirect determination $\alpha_{\rm ind}$ is indicated within brackets.}\label{tab:summary}
\end{center}
\end{table}

Our results for the direct determination of the weak phase $\alpha$ are summarized in Tab.~\ref{tab:summary} for the different model hypotheses and the different
statistical approaches considered up to now. The 68\%~CL interval (only for the solution near $90^\circ$~\footnote{For $B\to\pi\pi$, the bootstrap approach 
(second row in Tab.~\ref{tab:summary}) provides two peaks that are just separated at 68\%~CL, as can be seen in Fig.~\ref{fig:stat_B2hh}. However, in
 Tab.~\ref{tab:summary}, a single range for $\alpha$ is given for this method, which is obtained by merging the intervals corresponding to the two peaks. Even though this range is not the exact outcome of the bootstrap analysis for $B\to\pi\pi$ at 68\% CL, it allows us to perform more meaningful comparisons with the extractions of $\alpha$ using the Wilks-based approach, for which the two peaks are not distinguished at this level of significance and a single range  is obtained for $\alpha$.}) is reported as well as the compatibility with 
the indirect $\alpha_{\rm ind}$ determination Eq.~(\ref{eq:alphaInd}). We see that depending on the approach, the central value for the combination shifts by $2^\circ$ or less, remaining thus within the error quoted in Eq.~(\ref{eq:alphadir}). The uncertainty remains between $4^\circ$ and $5^\circ$ when isospin-breaking effects are allowed. On the other hand, as discussed in Sec.~\ref{sec:pvaluecompar}, the comparison between the bootstrap and Wilks-based approaches suggests a sizeable uncertainty attached to the statistical framework, but this uncertainty is attached to the direct extraction of $\alpha$ from the three $B\to\pi\pi$, $B\to \rho\rho$ and $B\to \rho\pi$ modes and cannot be used as such when it is combined with other constraints, e.g., within global fits. It is likely that one would get a smaller uncertainty if a similar analysis was performed with an extended set of observables leading to a more accurate determination of $\alpha$.

These various arguments lead us to keep Eq.~(\ref{eq:alphadir}) as our final answer for the direct extraction of $\alpha$ from charmless $B$ decays to be used in latter analyses.

\section{Hadronic amplitudes\label{sec:nuisance}}

In addition to the CKM angle $\alpha$, our study of the $B\to\pi\pi$, $B\to \rho\pi$ and $B\to\rho\rho$ systems provides constraints on the hadronic amplitudes that cannot be computed in QCD directly. We can thus determine some  features of hadronisation from the data, to be compared with the theoretical approaches proposed to describe these decays. 

In order to sharpen the constraints on these parameters, we will not only consider the experimental measurements discussed previously, but we will also  take as an additional constraint  the indirect prediction $\alpha_{\rm ind}$  derived from Fig.~\ref{fig:alphaInd} and Eq.~(\ref{eq:alphaInd}). This constraint is obtained excluding $B\to\pi\pi$, $B\to\rho\rho$ and $B\to\rho\pi$ data from the global CKM fit, and therefore provides an independent constraint on the isospin analysis.

\begin{figure}[t]
\begin{center}
  \includegraphics[width=18pc]{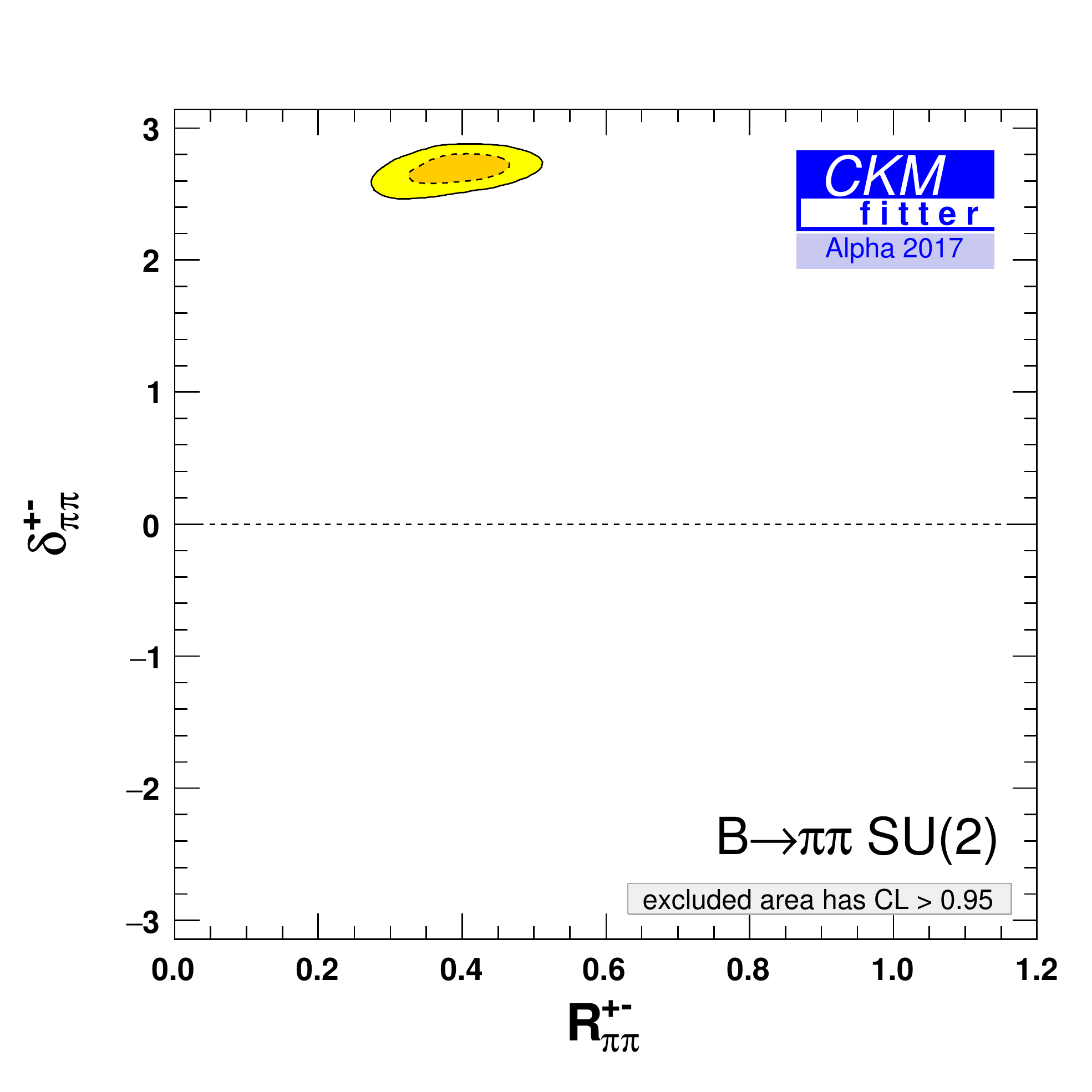}
  \includegraphics[width=18pc]{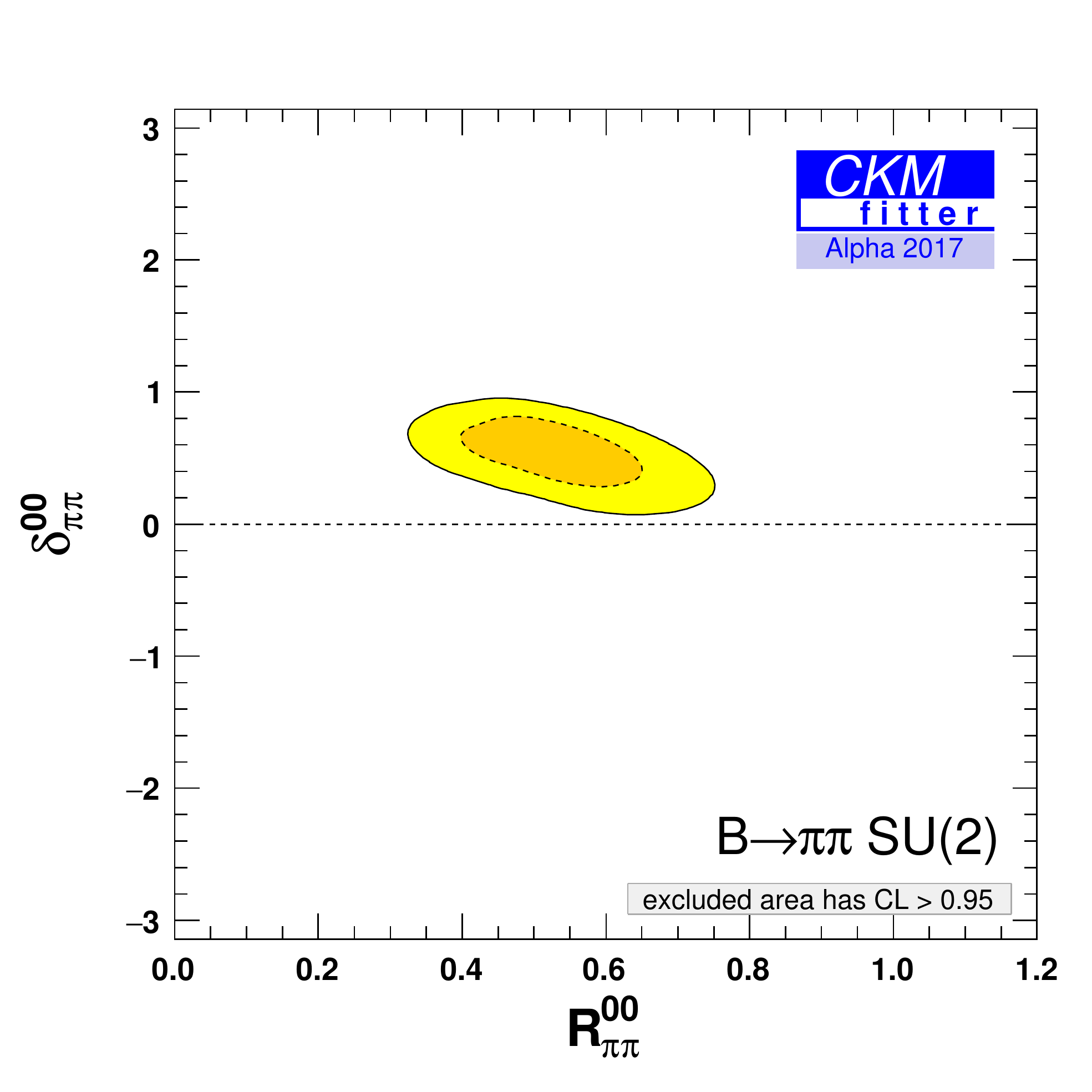}
\caption{\it\small  68\% (dark area) and 95\% CL (light  area)  constraints  on the   modulus and phase for the penguin-to-tree ratio $P^{ij}/ T^{ij}=R^{ij} e^{\imi\delta^{ij}}$  for $B^0\to\pi^+\pi^-$ (left) and  $B^0\to\pi^0\pi^0$ (right).}\label{fig:pOverT_pipi}
\end{center}       
\end{figure}

\subsection{Penguin-to-tree ratios}

Following the ${\cal C}$-convention defined in Sec.~\ref{sec:amplitudes}, the penguin-to-tree ratio is given by:
\begin{equation}
\tilde R^{ij}=\frac{P^{ij}}{T^{ij}} = \frac{R_t}{R_u}\times\frac{{\cal P}^{ij}}{{\cal T}^{ij}}\,,
\qquad\qquad \frac{R_t}{R_u} = \left|\frac{V_{td}V^*_{tb}}{V_{ud}V^*_{ub}}\right|\,.
\end{equation}
The isospin analysis assumes that there are no penguin contributions to the charged decays in the $B\to\pi\pi$, $B\to\rho\rho$ and $B^{0}\to(\rho\pi)^0$ systems. This hypothesis allows us to extract
 $\alpha$ and the tree and penguin contributions simultaneously in each decay mode. We start with the (complex) penguin-to-tree ratio: 
\begin{equation}
\tilde R^{ij}=R^{ij}e^{\imi \delta^{ij}}=\frac{P^{ij}}{T^{ij}}\,,
\end{equation}
 in the $B^0\to (h_1^{i}h_2^{j})^0$ neutral modes. 
The isospin analysis of the $B\to\pi\pi$ system constrains  the ratio of moduli $ R^{ij}_{\pi\pi}$ for both colour-allowed ($B\to\pi^+\pi^-$) and colour-suppressed ($B^0\to\pi^0\pi^0$) neutral modes  at a  precision level of around $20\%$. The two-dimensional constraint  in the (complex)  $\tilde R^{ij}$  plane for these two modes is shown in Fig.~\ref{fig:pOverT_pipi}. A large penguin contamination is observed in both modes. The  one-dimensional extraction of the modulus and the phase gives the following 68\%  (respectively 95\%) CL intervals:
\begin{eqnarray}
  R^{+-}_{\pi\pi}=\val{0.40}{0.04} \quad(\err{+0.08}{-0.10})~   ~~&,&~~   \delta^{+-}_{\pi\pi} = [\val{154.7}{3.6} \quad (\err{+6.9}{-9.2})]^\circ \,,\nonumber\\
  R^{00}_{\pi\pi}=\val{0.52}{+0.08}{-0.07} \quad(\err{+0.18}{-0.16})~    ~~&,&~~   \delta^{00}_{\pi\pi} = [\val{32.7}{+9.2}{-10.3} \quad(\err{+17.2}{-22.3})]^\circ \,,
\end{eqnarray}
We observe that values of the phases close to 0 or $\pi$ are favoured, in agreement with QCD factorisation expectations \cite{Beneke:2000ry,Beneke:2001ev}, which also predicts value for $R^{+-}_{\pi\pi}$ around 0.5~\cite{Beneke:2003zv,Beneke:2006hg}. Recent fits of two-body non-leptonic $B$ decays into light pseudoscalar mesons based on \su{3} flavour symmetry yield smaller values for these ratios \cite{Cheng:2014rfa}.

\begin{figure}[t]
\begin{center}
  \includegraphics[width=18pc]{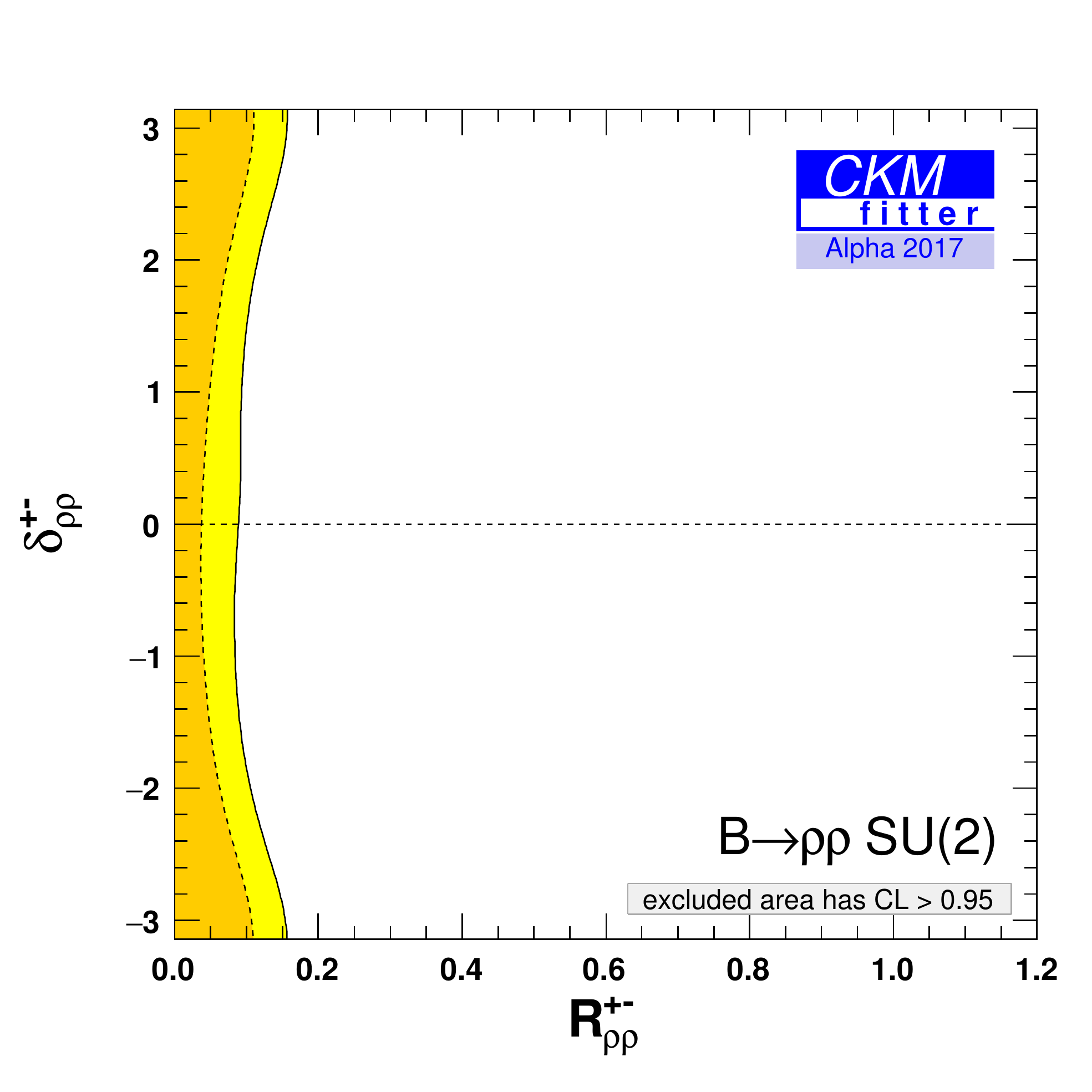}
  \includegraphics[width=18pc]{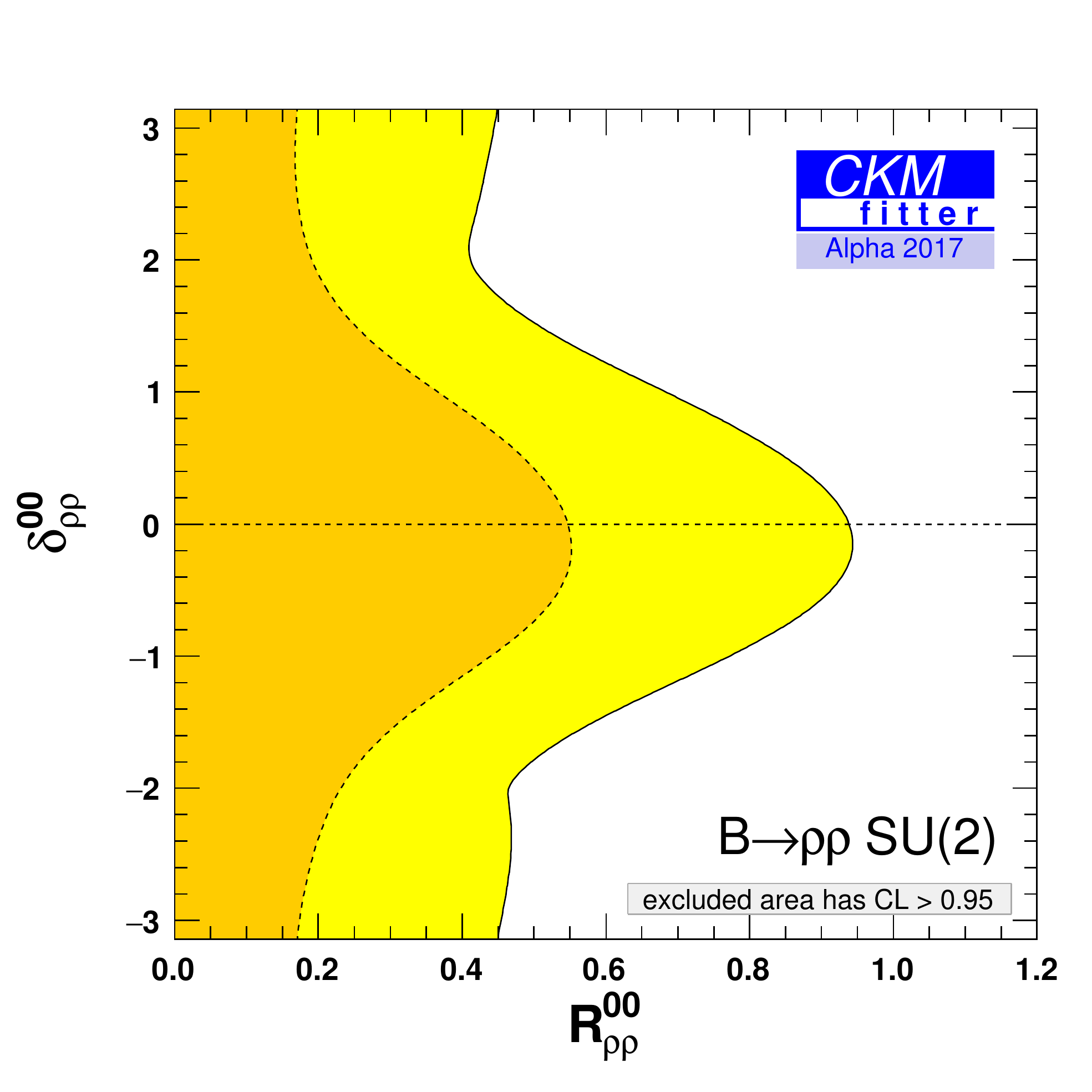}
\caption{\it\small 68\% (dark area) and 95\% CL (light  area)  constraints  on the  modulus and phase of the penguin-to-tree ratio $P^{ij}/T^{ij}=R^{ij} e^{\imi\delta^{ij}}$ for $B^0\to\rho^+\rho^-$ (left) and  $B^0\to\rho^0\rho^0$ (right).}\label{fig:pOverT_rhorho}
\end{center}       
\end{figure}

The penguin contamination to the colour-allowed $B^0\to\rho^+\rho^-$ decay is significantly smaller than for its $B\to\pi\pi$ counterpart, as expected from the  flatness of the $\rho\rho$ isospin triangles discussed in Sec.~\ref{subsec:rhorho}. The two-dimensional constraint in the (complex)  $\tilde R^{ij}_{\rho\rho}$  plane is shown in Fig.~ \ref{fig:pOverT_rhorho}. 
A tight constraint on the penguin contamination to the $B^0\to\rho^+\rho^-$ decay is obtained with a 68\% (respectively 95\%) CL upper limit:
\begin{equation}
 R^{+-}_{\rho\rho} < 0.08~~ (0.13)\,,
 \end{equation}
on the edge of the expectations from QCD factorisation, which predicts this ratio to be around 0.12 \cite{Beneke:2006hg}.
 Due to the small overall amplitudes contributing to the $B^0\to\rho^0\rho^0$ mode, a looser constraint is found:
\begin{equation}
 R^{00}_{\rho\rho}<0.40~~ (0.72). 
\end{equation}
No constraint is obtained on the phase of the ratio for either $B^0\to(\rho\rho)^0$ decays.

\begin{figure}[t]
\begin{center}
  \includegraphics[width=18pc]{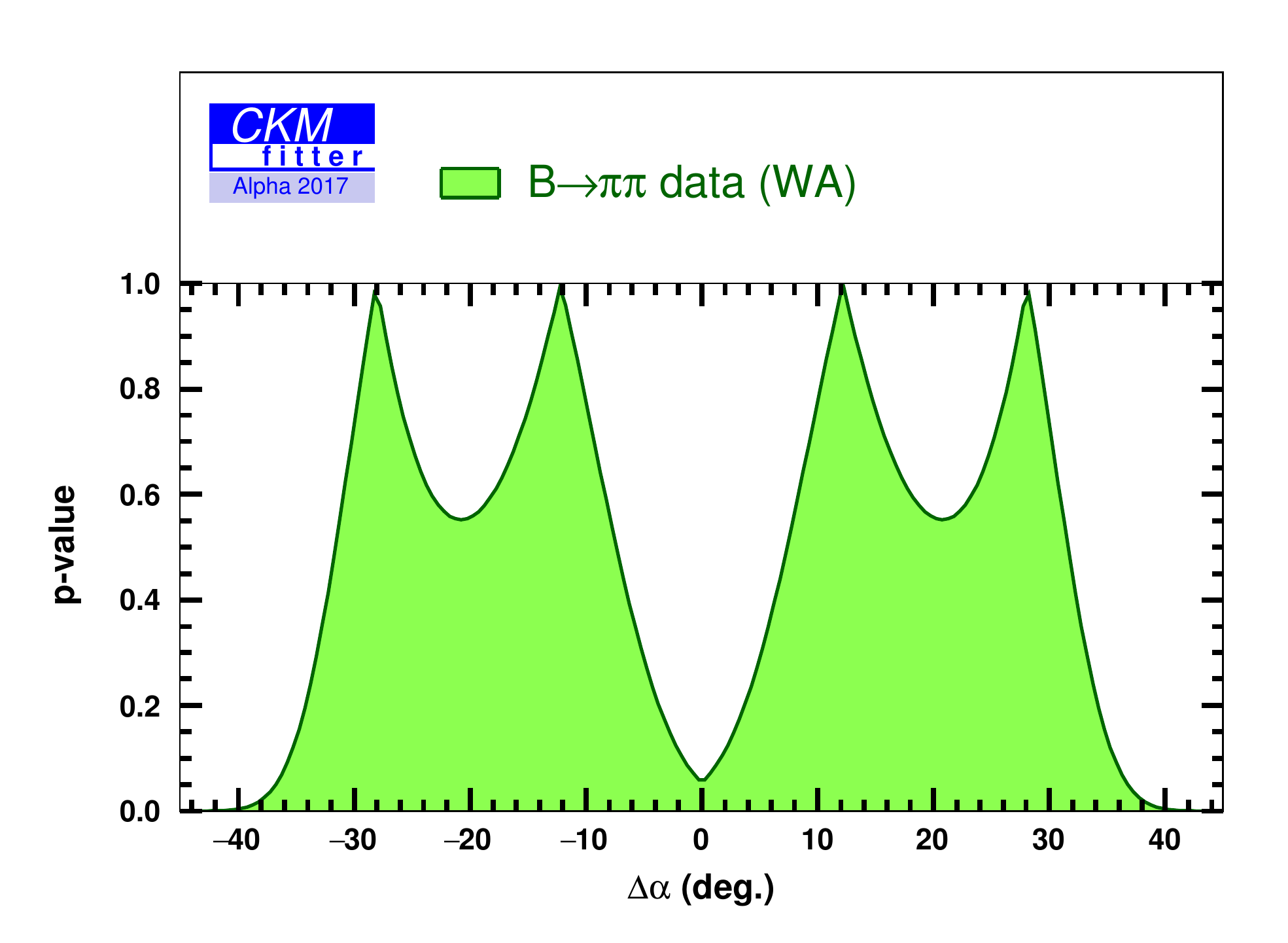}
  \includegraphics[width=18pc]{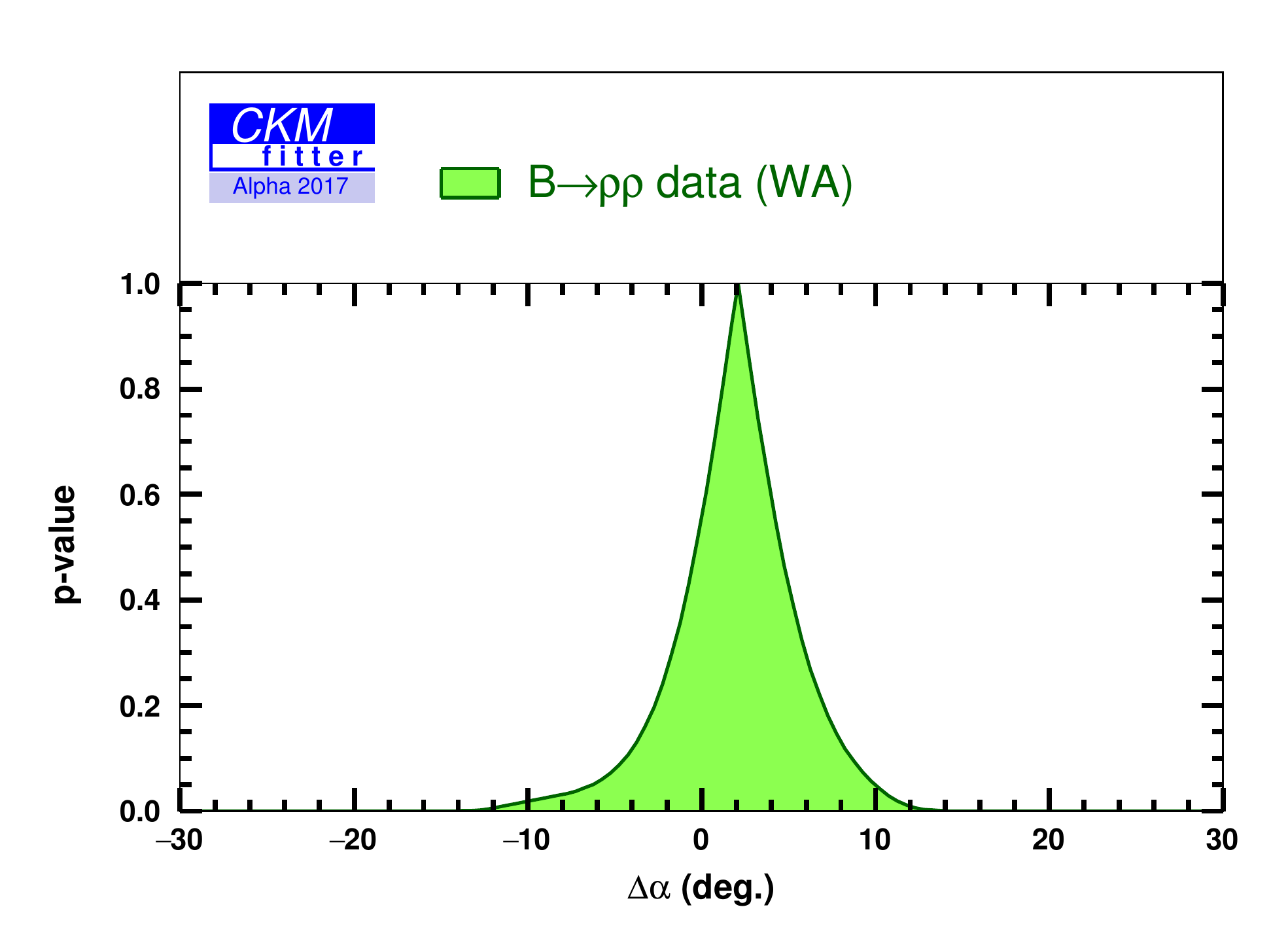}
\caption{\it\small   68\%  CL constraint on the phase shift  $\Delta\alpha=(\alpha-\alpha_{\rm eff})$  for  $B\to\pi\pi$ (left) and  $B\to\rho\rho$ (right).\label{fig:deltaAlpha}}
\end{center}       
\end{figure}
Another representation of the penguin  hierarchy  between the $B^0\to\pi^+\pi^-$ and $B^0\to\rho^+\rho^-$ decays is provided by the phase shift $\Delta\alpha=(\alpha-\alpha_{\rm eff})$ where the effective weak phase $\alpha_{\rm eff}$, introduced in Eq.~(\ref{eq:alphaeff}), describes the $B^0\to h^+h^-$ transition. In the limit of a vanishing penguin contribution, the mixing phase  $\alpha_{\rm eff}$ coincides with the weak phase $\alpha$.
The following 68\% CL intervals on $\Delta\alpha$, derived from Fig.~\ref{fig:deltaAlpha}, are obtained
\begin{equation}
\Delta\alpha_{\pi\pi}=\pm (\val{23.8}{13.9})^\circ\,,\qquad\qquad
\Delta\alpha_{\rho\rho}=\val{2.1}{+3.7}{-3.6}{$^\circ$}\,.
 \end{equation}
The phase shift is consistent with the absence of penguin in the  $B^0\to\rho^+\rho^-$ decay while the sizeable shift obtained in $B^0\to\pi\pi$  excludes the $\Delta\alpha=0$ hypothesis at 1.9~\SIG in this mode, confirming the difference of size of penguin contributions for the two decay channels.
A similar quantity can be defined for the asymmetric $B^0\to\rho^\pm\pi^\mp$ decay and is further discussed in Sec.~\ref{sec:rhopiphase}.

\begin{figure}[t]
\begin{center}
  \includegraphics[width=18pc]{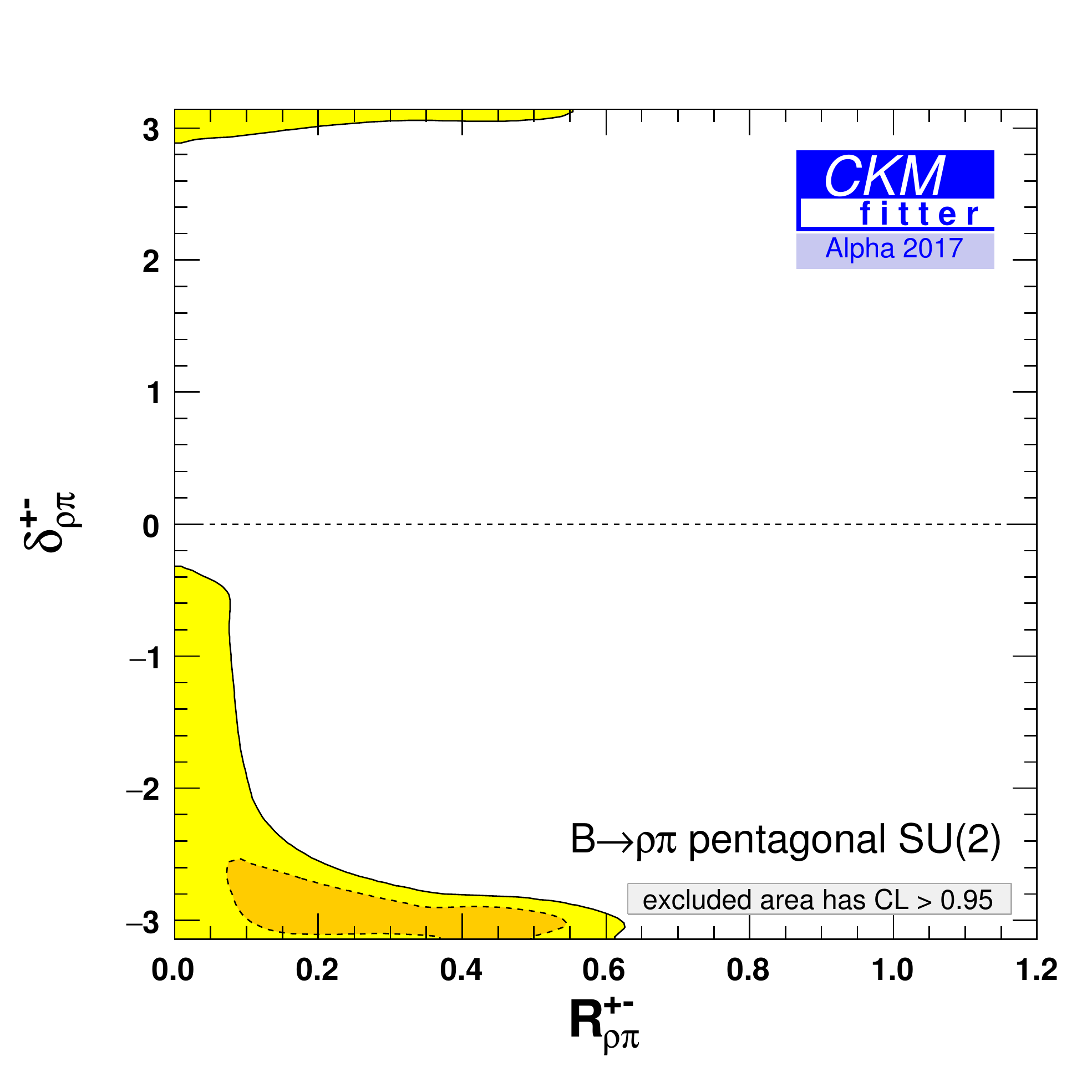}
  \includegraphics[width=18pc]{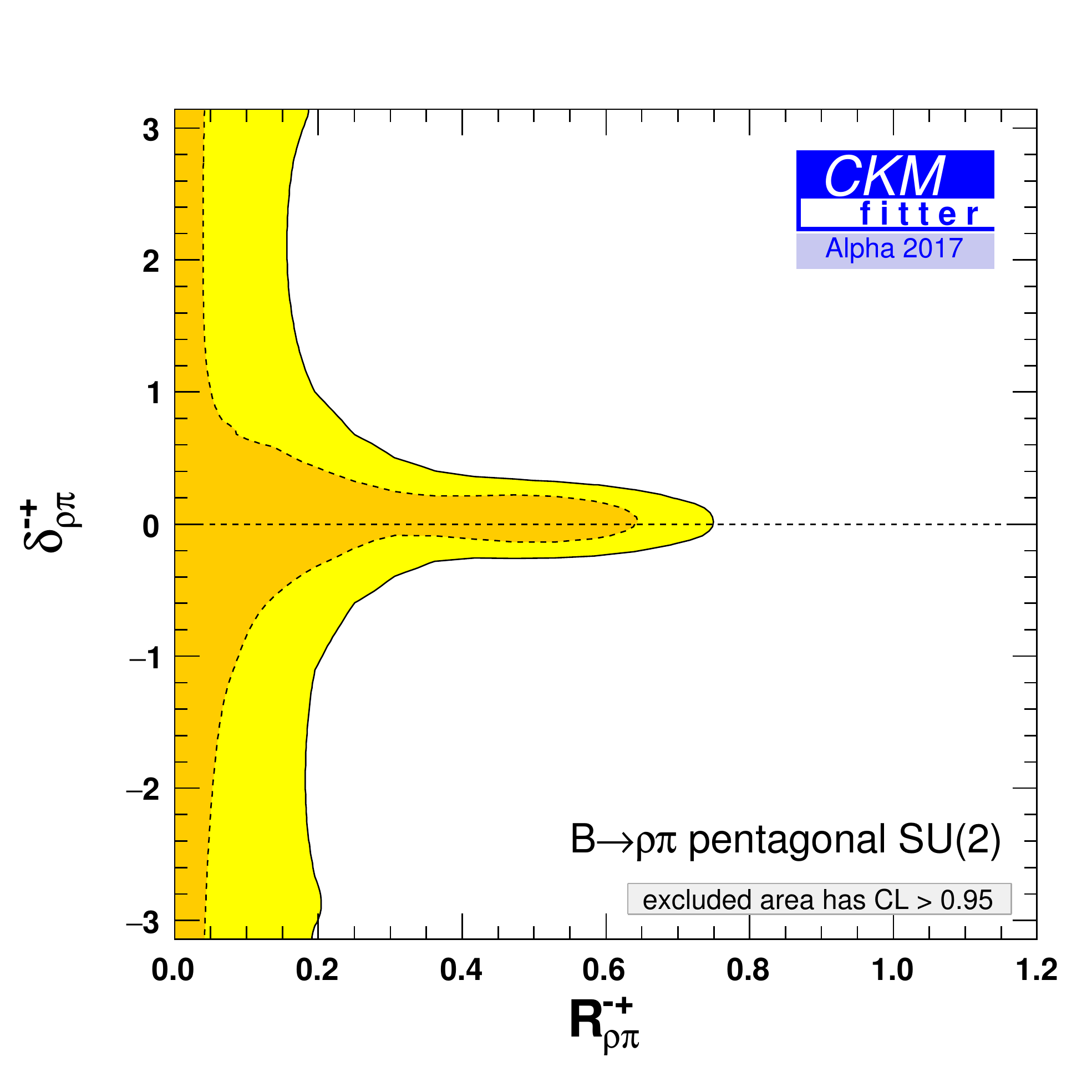}
  \includegraphics[width=18pc]{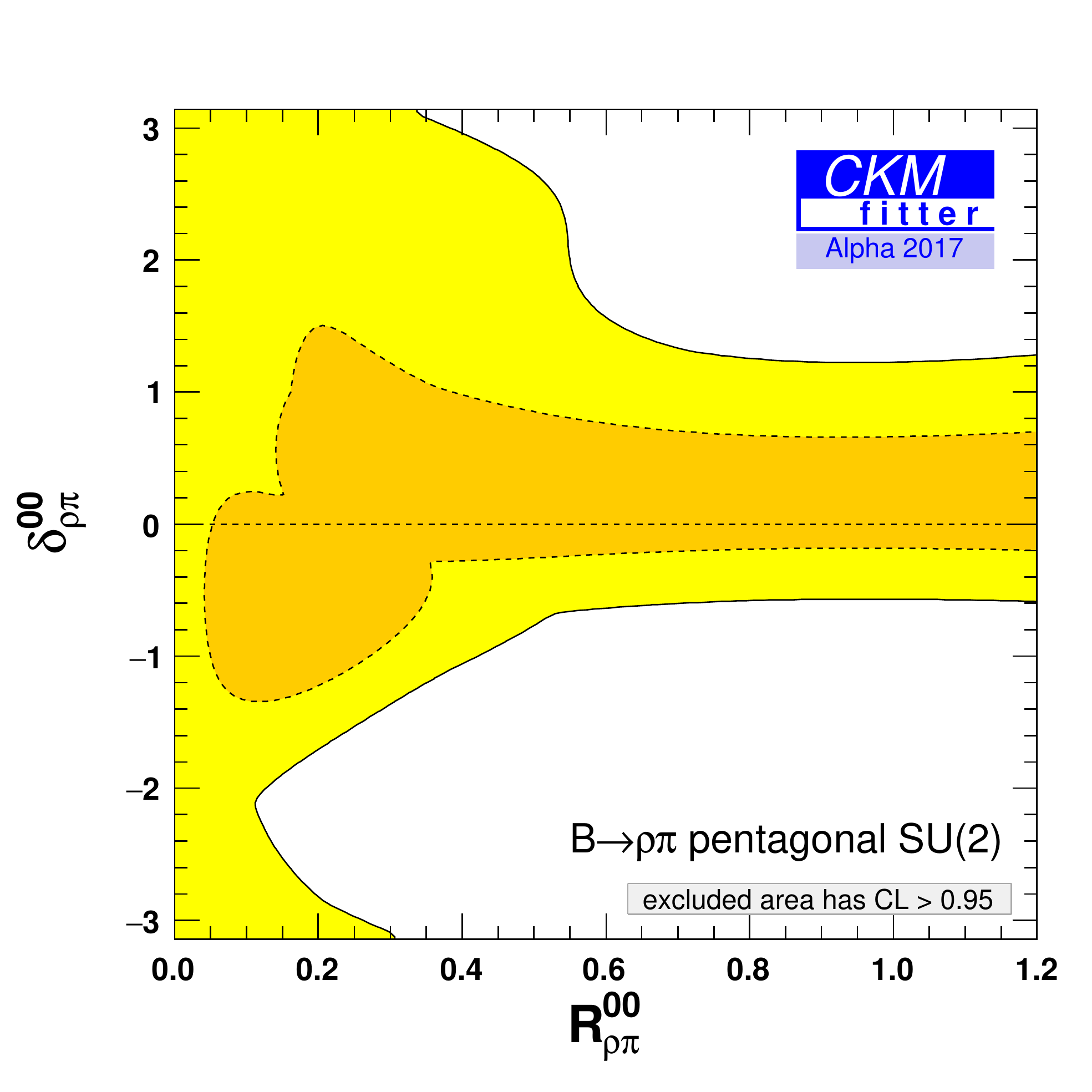}
\caption{\it\small   68\% (dark area) and 95\% CL (light  area)  two-dimensional constraints on the modulus and phase of the penguin-to-tree ratio $P^{ij}/ T^{ij}=R^{ij} e^{\imi\delta^{ij}}$ for $B^0\to\rho^+\pi^-$ (top left), $B^0\to\rho^-\pi^+$ (top right) and  $B^0\to\rho^0\pi^0$ (bottom). These results are obtained using the pentagonal  isospin analysis, including the experimental inputs from the charged $B^+\to (\rho\pi)^+$ decays.\label{fig:pOverT_rhopi}}
\end{center}       
\end{figure}

\begin{figure}[t]
\begin{center}
  \includegraphics[width=40pc]{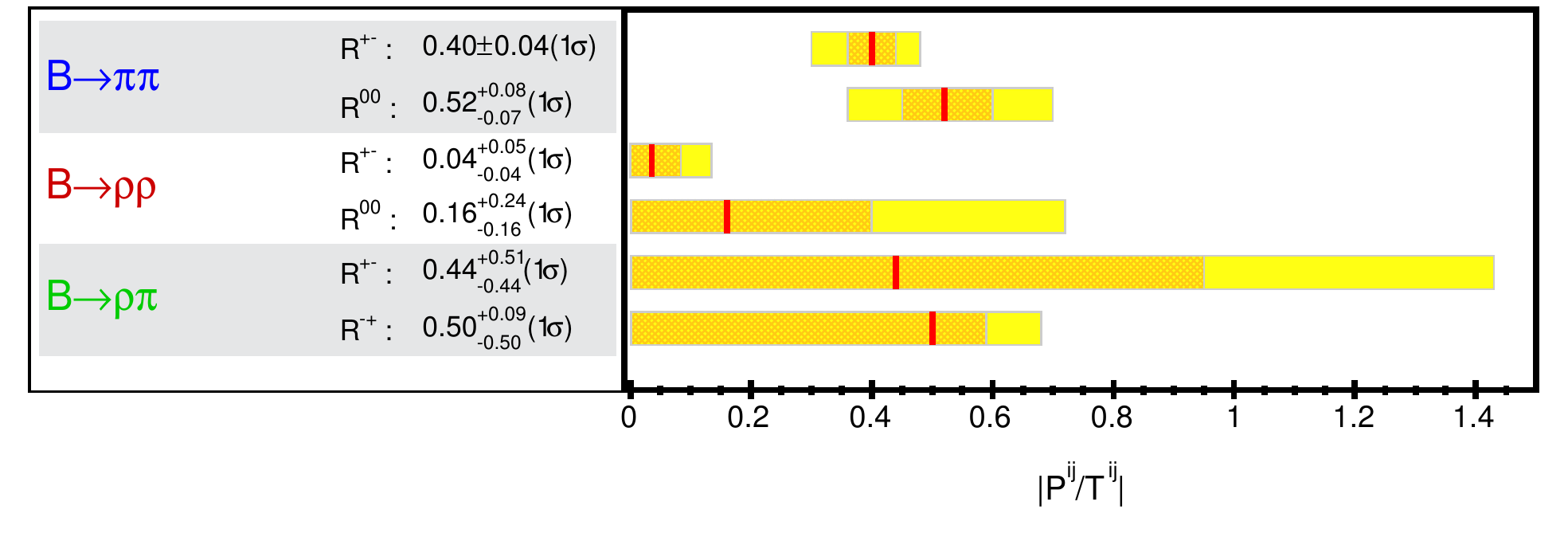}
\caption{\it\small 68\% (dark area) and 95\% CL (light  area) 1D intervals on the  $|P^{ij}/ T^{ij}|$ ratios for the different $B^{0}\to h^{i}h^{j}$ charmless decays.\label{tab:pOverT}}
\end{center}       
\end{figure}

Due to the limited experimental statistics currently available, the Dalitz isospin analysis of the three-pion decay does not lead to a significant constraint on the penguin contribution in any of the three $B^0\to(\rho\pi)^0$ decays. The pentagonal approach, involving further constraints from the charged modes, only provides a weak constraint on $R^{+-}$ and  $R^{-+}$, see Fig.~\ref{fig:pOverT_rhopi}. Phases close to 0 or $\pi$ are mildly favoured, in agreement with the expectations from QCD factorisation.
No constraint can be derived for the $B^0\to\rho^0\pi^0$  mode. For comparison, the constraints on the modulus of the penguin contribution to the various decays modes can be found in Fig.~\ref{tab:pOverT}.

\subsection{Testing colour suppression}

The hadronisation of the ${\bar b}d\to {\bar u}d(u \bar d)$ final state into a pair of charged isovector mesons, $h^+_1h^-_2$,  is colour allowed while the hadronisation of this tree transition into a neutral pair, $h^0_1h^0_2$, is colour suppressed. This colour suppression can be probed through the ratio of the corresponding tree amplitudes\footnote{We identify this ratio of ``tree'' amplitudes to a colour-suppressed ratio from the analysis of tree diagram topologies. In the $\cal C$-convention used here, this identification is correct only in the limit of vanishing $u$- and $c$-penguin topological amplitudes, as discussed in Sec.~\ref{sec:amplitudes}. Moreover, the $W$-exchange topology that provides a colour suppressed contribution that can be absorbed in the tree amplitudes for both $B\to h^+h^-$ and  $B\to h^0h^0$ transitions are neglected here. A similar statement holds for the penguin-to-tree ratio discussed previously.}:
\begin{equation}
  \tilde R_{\cal C}=R_{\cal C}e^{\imi  \delta_{\cal C}}=\frac{T^{00}}{T^{+-}},
\end{equation}
where $R_{\cal C}$ and $\delta_{\cal C}$ represent the modulus and the phase of the complex ratio, respectively. 

\begin{figure}[t]
\begin{center}
  \includegraphics[width=18pc]{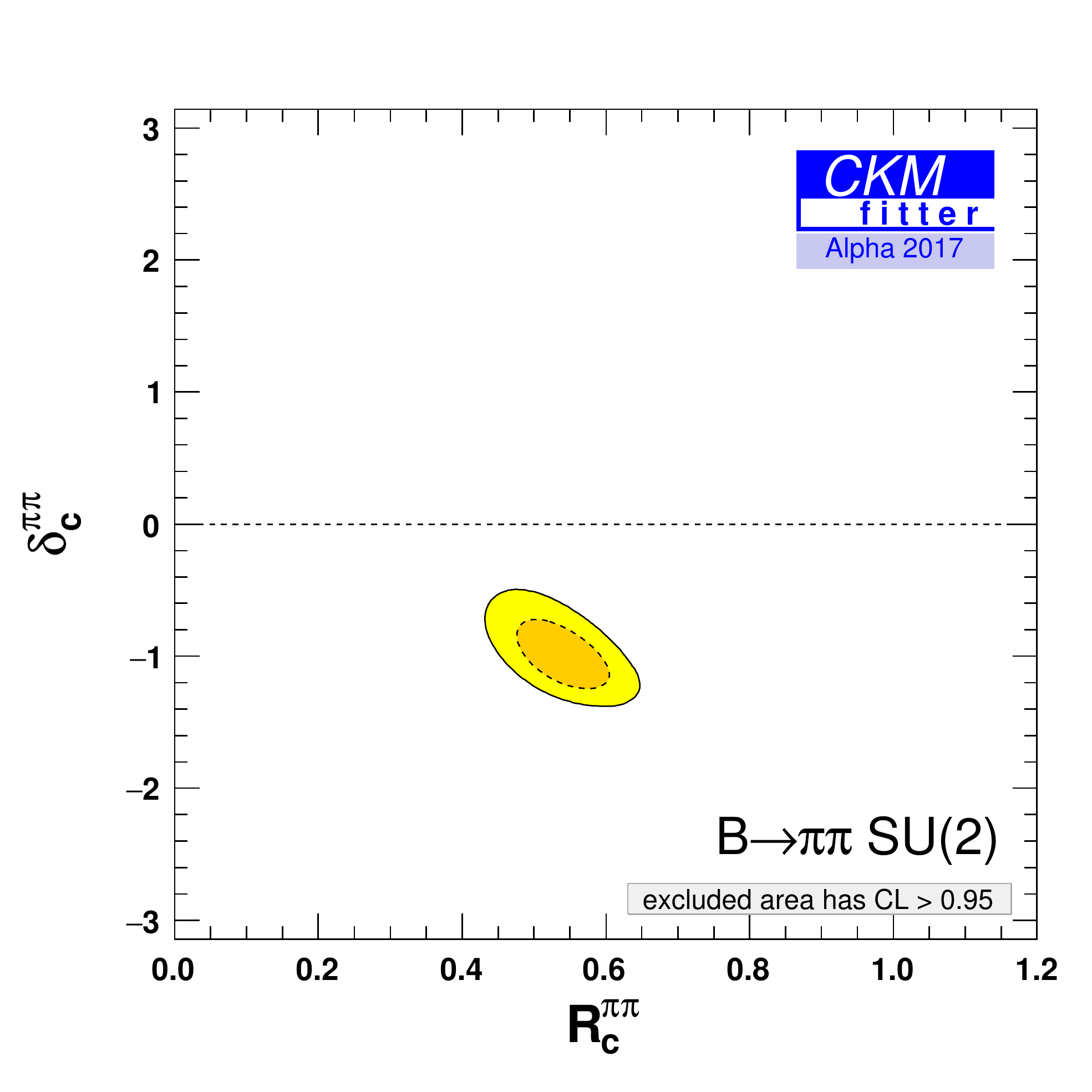}
  \includegraphics[width=18pc]{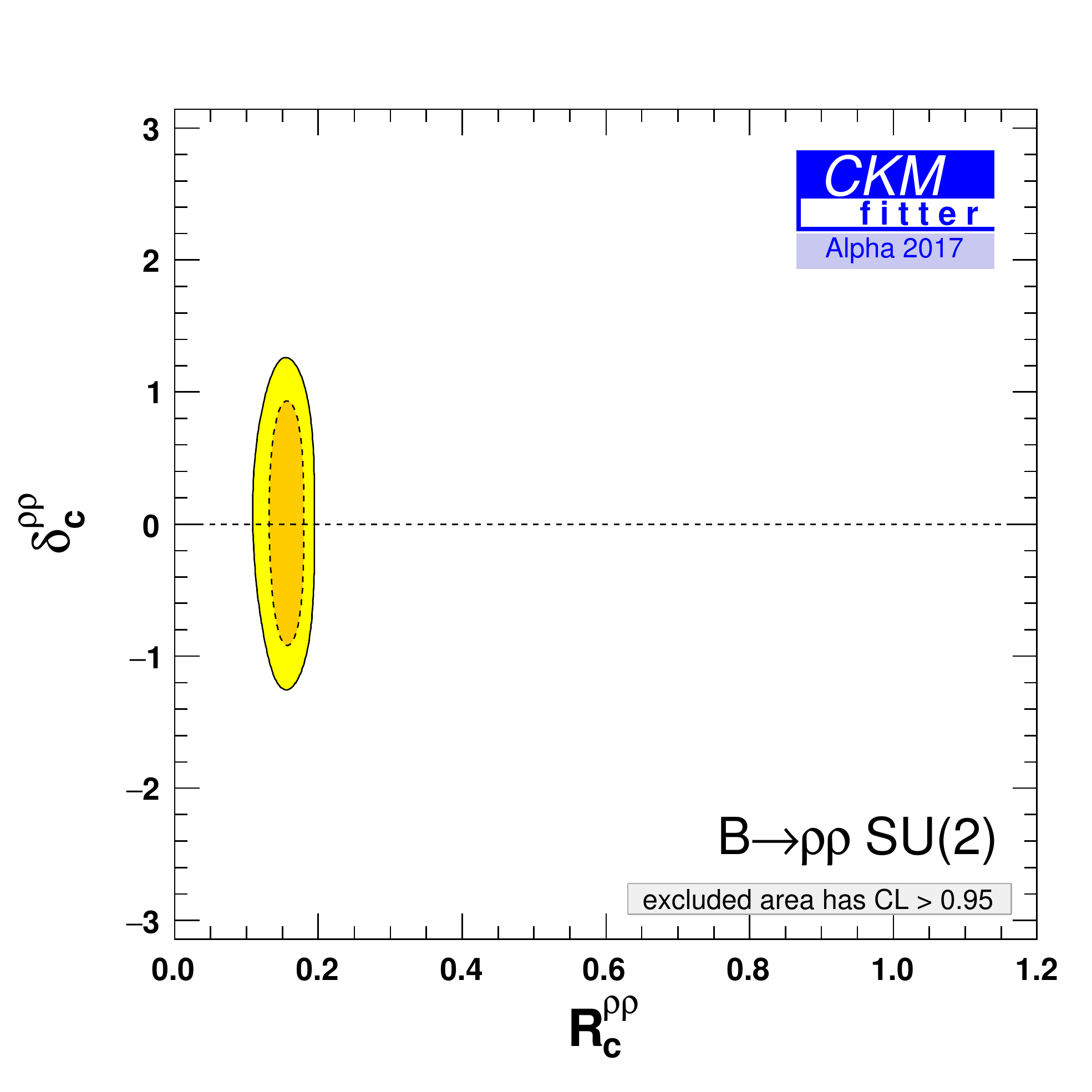}
  \includegraphics[width=18pc]{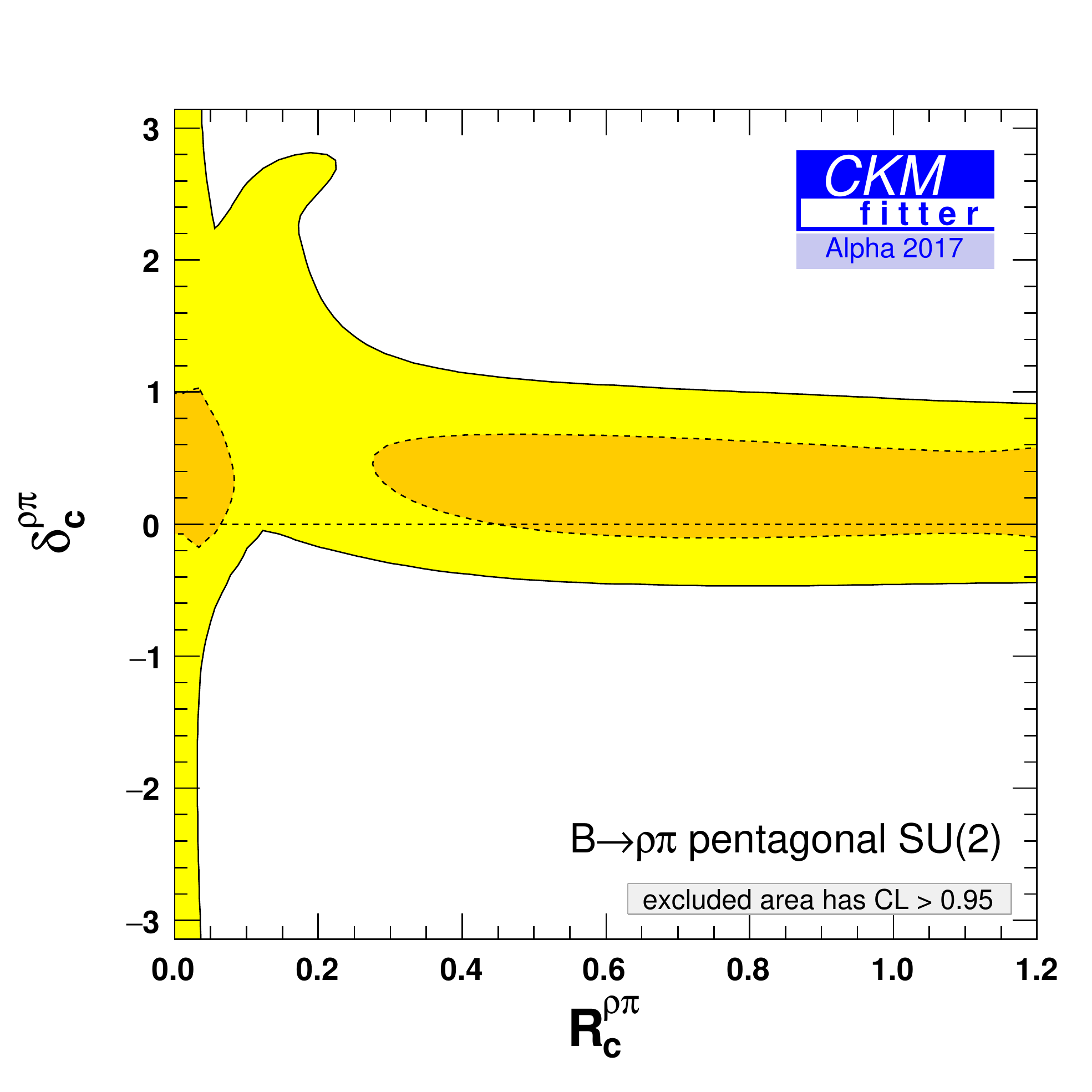}
\caption{\it\small Two-dimensional constraints on the modulus and phase of the colour-suppressed ratio $T^{00}/ T^{+-}=R_{c} e^{\imi\delta_{c}}$ for $B\to\pi\pi$ (top left), $B\to\rho\pi$ (top right) and  $B^{0}\to\rho\pi$ (bottom).}
\label{fig:Rcolour2D}
\end{center}       
\end{figure}

The two-dimensional constraint on the phase and modulus of the colour-suppression ratio is shown in Fig.~\ref{fig:Rcolour2D} for $B\to\pi\pi$, $B\to\rho\rho$ and $B\to\rho\pi$ channels. As expected, no constraint is obtained for that latter due to the limited statistics.  The modulus of the colour-suppression ratio is determined with a relative resolution of about $10\%$ for the $B\to\pi\pi$ and $B\to\rho\rho$ modes.  
The corresponding one-dimensional 68\%  (95\%) CL intervals on the modulus and the phase  are:
\begin{eqnarray}
  R_{\cal C}({\pi\pi})=\val{0.54}{0.04}\quad(\err{+0.08}{-0.09})~   ~~&,&~~   \delta_{\cal C}({\pi\pi}) = [\val{-58.4}{+10.3}{-8.6}\quad(\err{+22.9}{-16.6})]^\circ\,,  \nonumber\\
  R_{\cal C}({\rho\rho})=\val{0.16}{0.02}\quad(\err{+0.03}{-0.04})~   ~~&,&~~   \delta_{\cal C}({\rho\rho}) = [\val{2.9}{+38.4}{-41.8}\quad(\err{63.0})]^\circ\,,
\end{eqnarray}
 indicating that the colour suppression is much more effective for $B\to\rho\rho$ than for $B\to\pi\pi$. In studies based on QCD factorisation \cite{Beneke:2001ev,Beneke:2003zv,Beneke:2006hg}, similar values for these moduli are found, apart from
 the modulus of the ratio between colour-allowed and colour-suppressed tree amplitudes for $\pi\pi$ modes,
$R_{\cal C}({\pi\pi})$, smaller and around 0.2. Similarly, the relative phase for the latter ratio is significantly different from 0 and $\pi$, contrary to general expectations from leading-order QCD factorisation. This may explain the difficulties experienced with this approach to reproduce the $B\to\pi^0\pi^0$ branching ratio \cite{Beneke:2009ek} and may point towards significant power-suppressed contributions for some of these modes. Recent fits of two-body non-leptonic $B$ decays into light pseudoscalar mesons based on \su{3} yield similar values for $R_{\cal C}({\pi\pi})$, indicating that  large phase and modulus of the colour-suppressed tree contribution $T^{00}$ are required not only in the $\pi\pi$ system, but also in other $B\to PP$ decays \cite{Cheng:2014rfa}.

\section{Prediction of  observables \label{sec:observables}}

\begin{figure}[t]
\begin{center}
  \includegraphics[width=18pc]{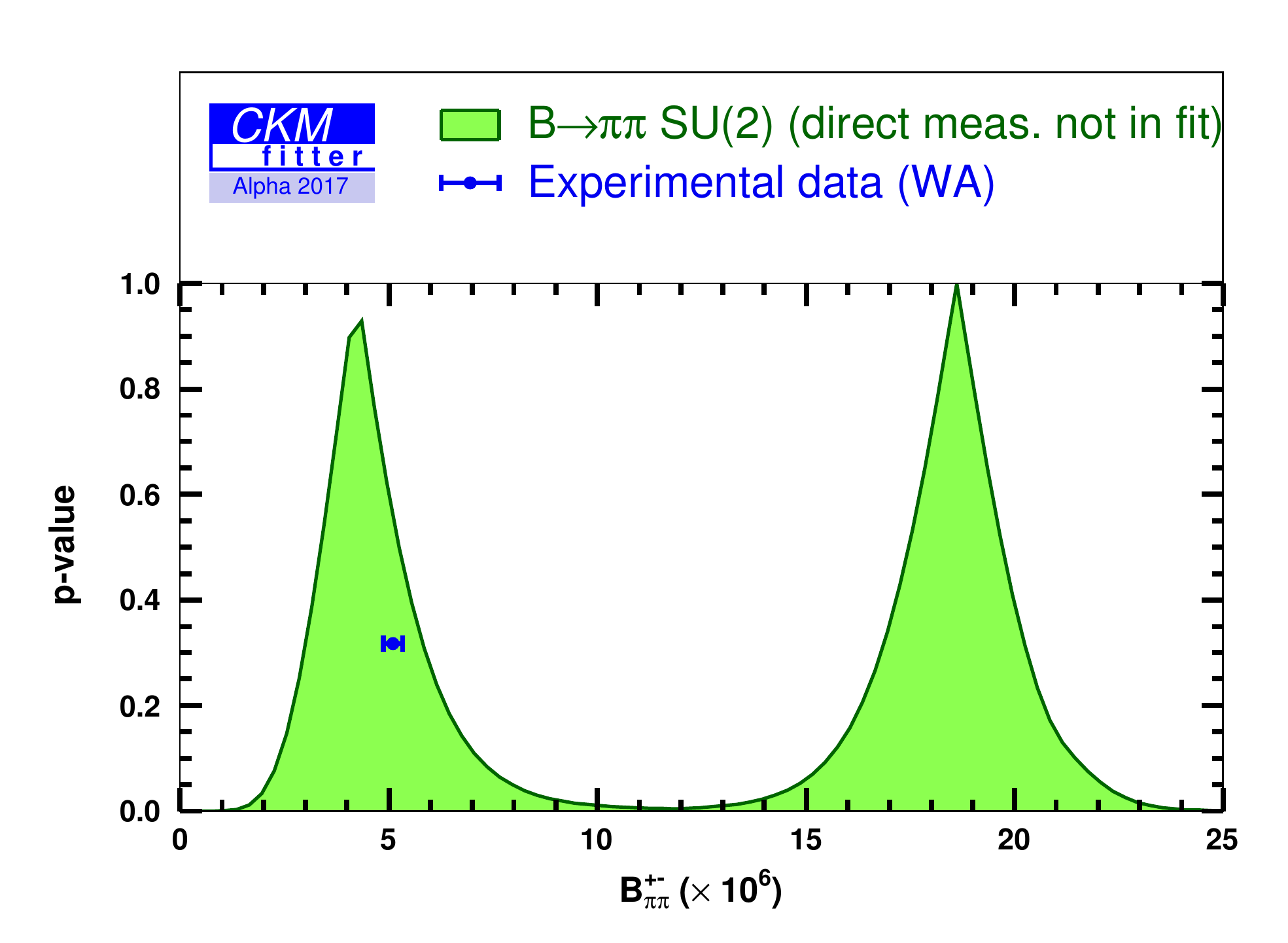}
  \includegraphics[width=18pc]{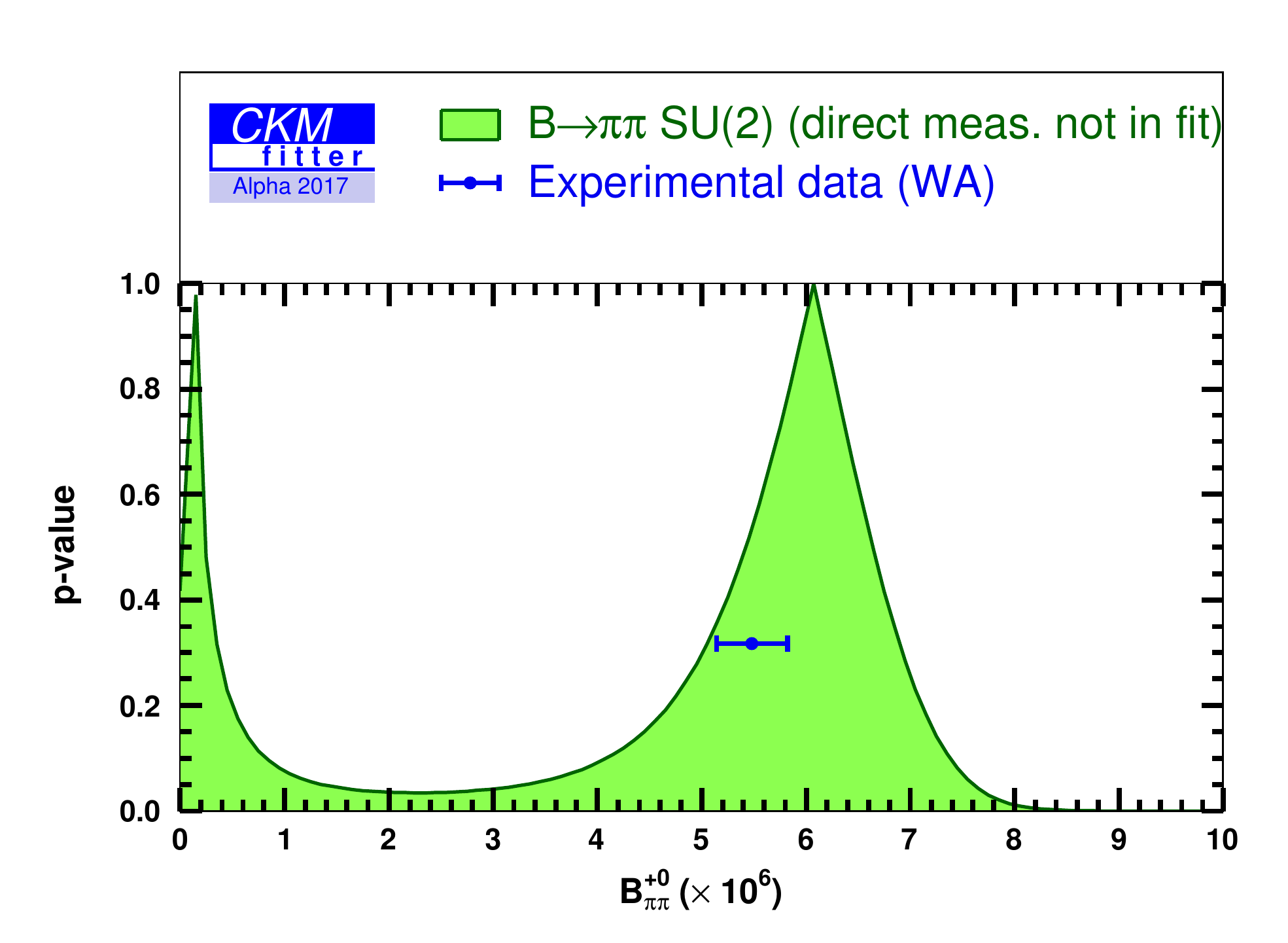}
  \includegraphics[width=18pc]{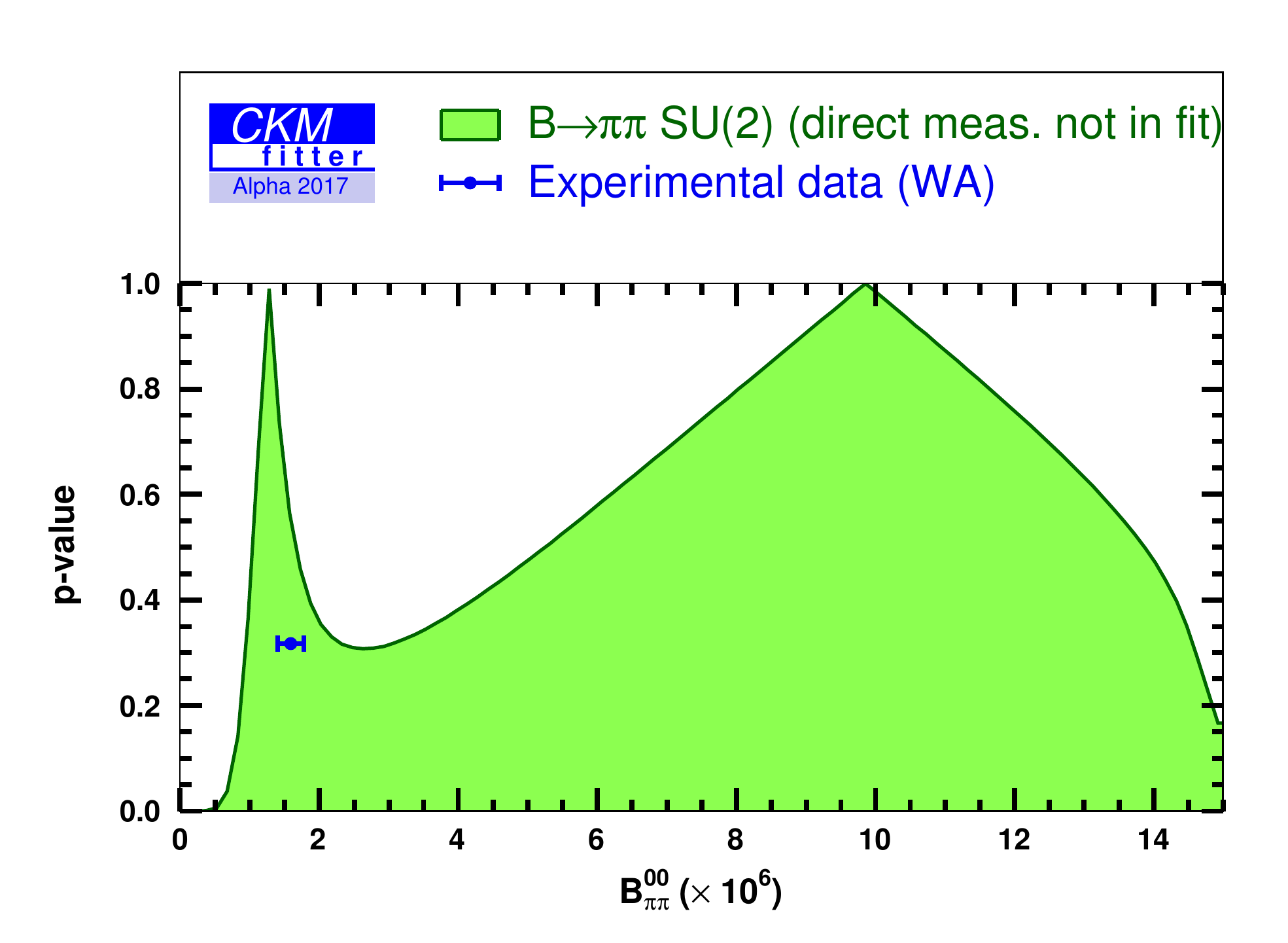}
\caption{\it\small  Constraint on the \decay{B}{\pi\pi} branching fractions from the \su{2} isospin analysis (from left to right, and from top to bottom: $\pi^+\pi^-$, $\pi^+\pi^0$ and $\pi^0\pi^0$). The direct measurement, not included in the fit, is indicated for each mode with an interval with a dot. }
\label{fig:BR_PiPi}
\end{center}       
\end{figure}

If we use the indirect determination of the weak phase $\alpha_{\rm ind}$, Eq.~(\ref{eq:alphaInd}), as an additional input, the system of isospin-related amplitudes becomes over-constrained for the three charmless decay modes of interest. The isospin analysis can then be used to perform an indirect determination for each of the experimental observables introduced in Sec.~\ref{sec:alpha} and to make  predictions for yet unmeasured quantities. For each observable $\O_i$, this indirect determination is obtained by performing a fit to the experimental data excluding the measurement of the observable considered. The compatibility of the indirect determination with the measurement is quantified by comparing the minimal \chisq values reached by the fit when excluding and including the direct measurement. The difference between these two quantities is distributed as a \chisq law with $N_{dof}=1$, whose interpretation in terms of standard deviations yields the compatibility pull defined as:
\begin{equation}\label{eq:pull}
\Pull(\O_i) = \sqrt{\chisq_{\rm min}(\alpha_{\rm ind},\vec\O) - \chisq_{\rm min}(\alpha_{\rm ind},\vec\O_{! i})}\,,
\end{equation}
where $\vec\O=\{\vec\O_{! i},\O_i\}$ represents the full data set of experimental measurements, and $\vec\O_{! i}$ the subset out of which the specific observable $\O_i$ has been excluded.
The pull quantifies the compatibility of the measurement of the observable $\O_i$ with all the other experimental observables under the isospin hypothesis. Let us add that
the individual pulls are not in general independent from each other, 
so that one should refrain from giving a statistical interpretation of their distribution in each of the \decay{B}{\pi\pi},  \decay{B}{\rho\pi},  \decay{B}{\rho\rho} systems.

\subsection{\decay{B}{\pi\pi} observables}

As already indicated in Sec.~\ref{subsec:pipi},  the number of independent amplitudes in the \decay{B}{\pi\pi} system matches the number of experimental measurements available. Including the indirect determination of $\alpha$ as an additional constraint, the minimal \chisq  increases only slightly, reflecting the excellent agreement between $\alpha_{\rm ind}$ and \decay{B}{\pi\pi} data, at the level of 0.6 standard deviation.

\begin{figure}[t]
\begin{center}
  \includegraphics[width=18pc]{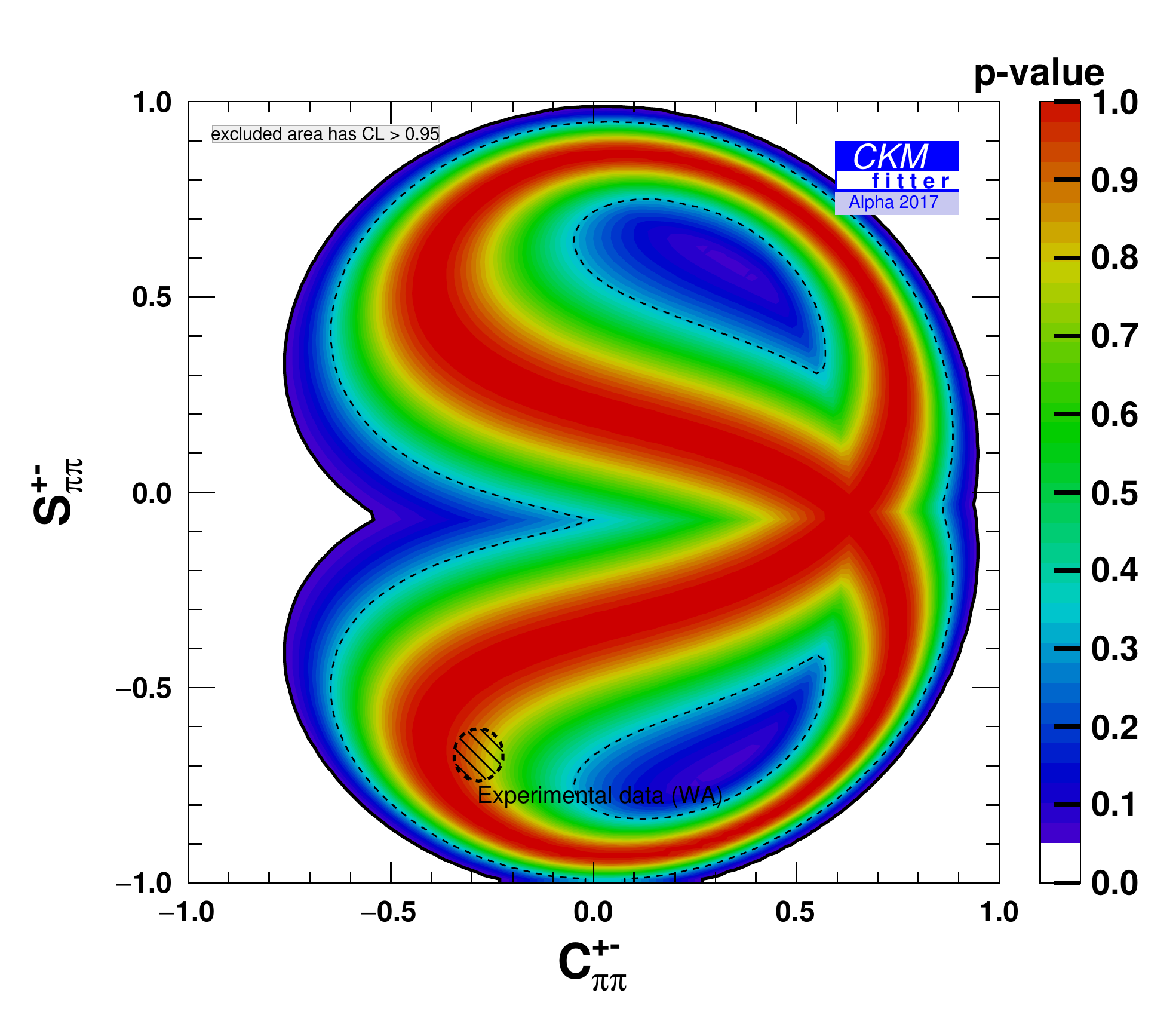}
  \includegraphics[width=18pc]{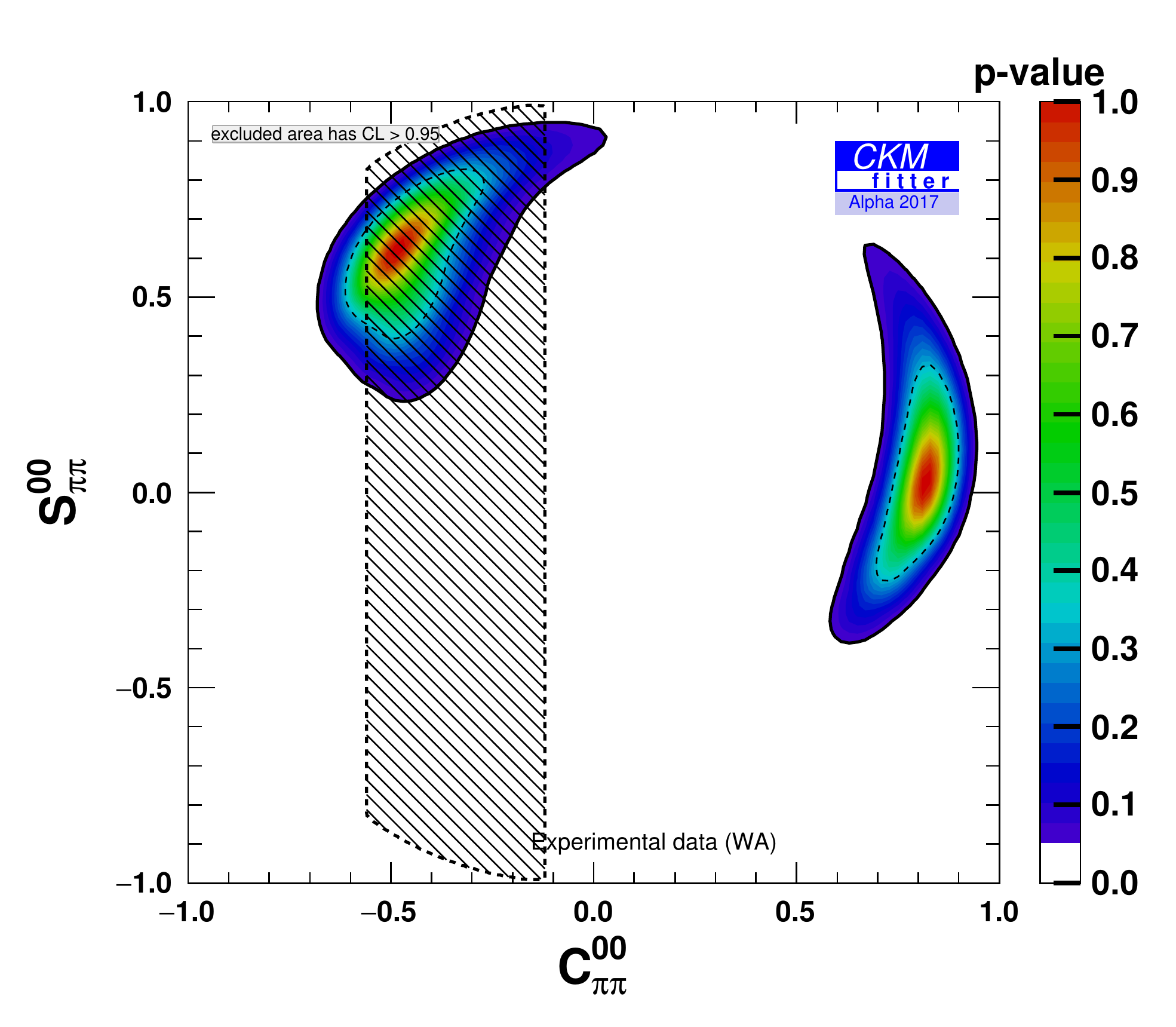}
\caption{\it\small  Two-dimensional \su{2} isospin constraint on the $CP$ asymmetries $\C_{\pi\pi}$ and $\S_{\pi\pi}$ for the  \decay{B^0}{\pi^+\pi^-} decay (left) and the \decay{B^0}{\pi^0\pi^0} decay (right). The direct measurement, not included in the fit, is indicated by the shaded area.  Only the direct $CP$ asymmetry  $C^{00}$ is measured in the $B^0\to\pi^0\pi^0$ decay and we represent the direct constraint in the  $(\C^{00},\S^{00})$ plane taking the theoretical inequality  $\C^2+\S^2 < 1$ into account.}
\label{fig:CS_PiPi}
\end{center}       
\end{figure}

\begin{figure}[t]
\begin{center}
  \includegraphics[width=20pc]{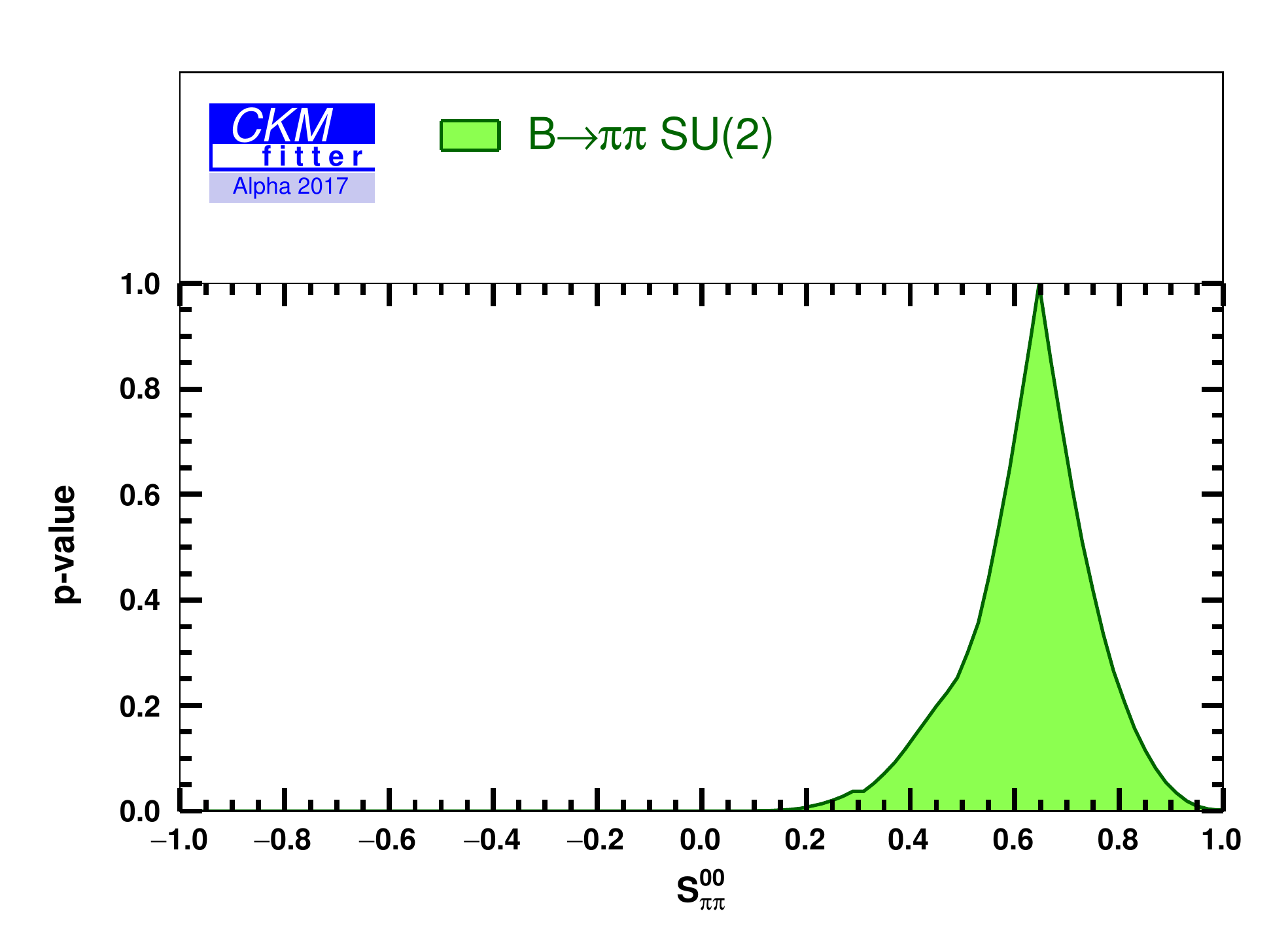}
\caption{\it\small  \su{2} isospin constraint on the unmeasured time-dependent $CP$ asymmetries in the  \decay{B^0}{\pi^0\pi^0} decay.}
\label{fig:S00_PiPi}
\end{center}       
\end{figure}

The indirect constraints on  the three branching fractions $B^{ij}_{\pi\pi}$ are shown in Fig.~\ref{fig:BR_PiPi} while Fig.~\ref{fig:CS_PiPi} displays the two-dimensional constraint on the $CP$ asymmetries ($\C_{\pi\pi},\S_{\pi\pi}$) for the neutral modes \decay{B^0}{\pi^+\pi^-} and \decay{B^0}{\pi^0\pi^0}. 
When excluding any of the experimental measurements, either for the branching ratios $\B^{ij}_{\pi\pi}$ or for the direct $CP$ asymmetries $\C^{ij}_{\pi\pi}$, the \decay{B}{\pi\pi} amplitudes system is no longer over-constrained.
As a consequence, the  corresponding $\chisq_{\rm min}(\alpha_{\rm ind},\vec\O_{! i})$ vanishes and the pull associated to these observables saturates its maximal value: $\Pull(B^{ij}_{\pi\pi},C^{ij}_{\pi\pi})=\sqrt{\chisq_{\rm min}(\alpha_{\rm ind},\vec\O)}$ which turns out to vanish thanks to the closure of both isospin triangles with the current world-average data. 
The yet unmeasured time-dependent asymmetry, $\S^{00}_{\pi\pi}$, is predicted to be:
\begin{equation}
\S^{00}_{\pi\pi}  = \val{0.65}{0.13}\,, \label{eq:s00_prediction}
\end{equation}
with a 68\% confidence level, and larger than $0.33$ at 95\% confidence level. The corresponding one-dimensional projection of the constraint is displayed in Fig.~\ref{fig:S00_PiPi}.

The experimental values, confidence intervals and  pulls for the measured observables are reported in Tab.~\ref{tab:BR_PiPi} in App.~\ref{sec:numerics}.

\subsection{\decay{B}{\rho\rho} observables}

\begin{figure}[t]
\begin{center}
  \includegraphics[width=18pc]{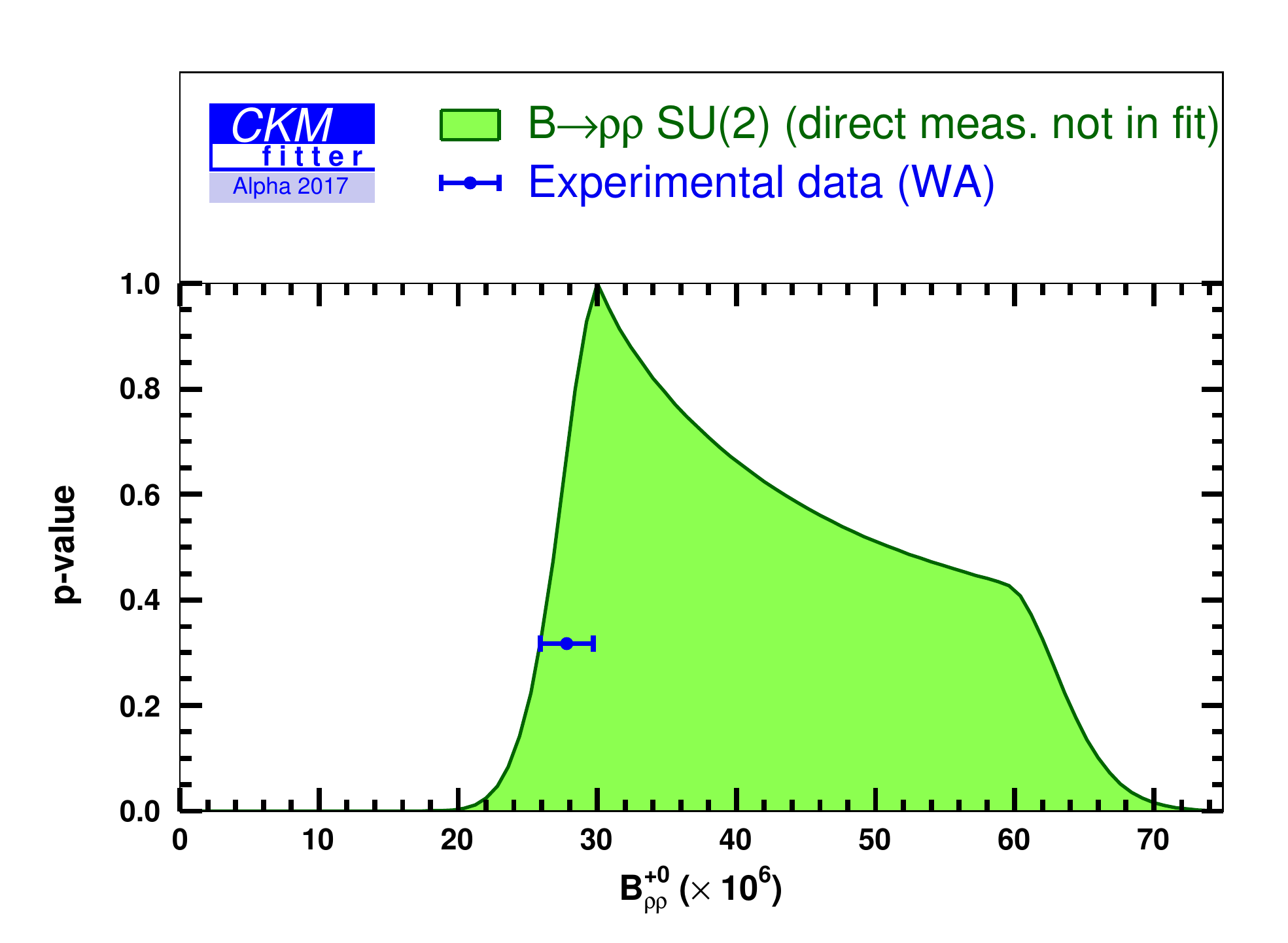}
  \includegraphics[width=18pc]{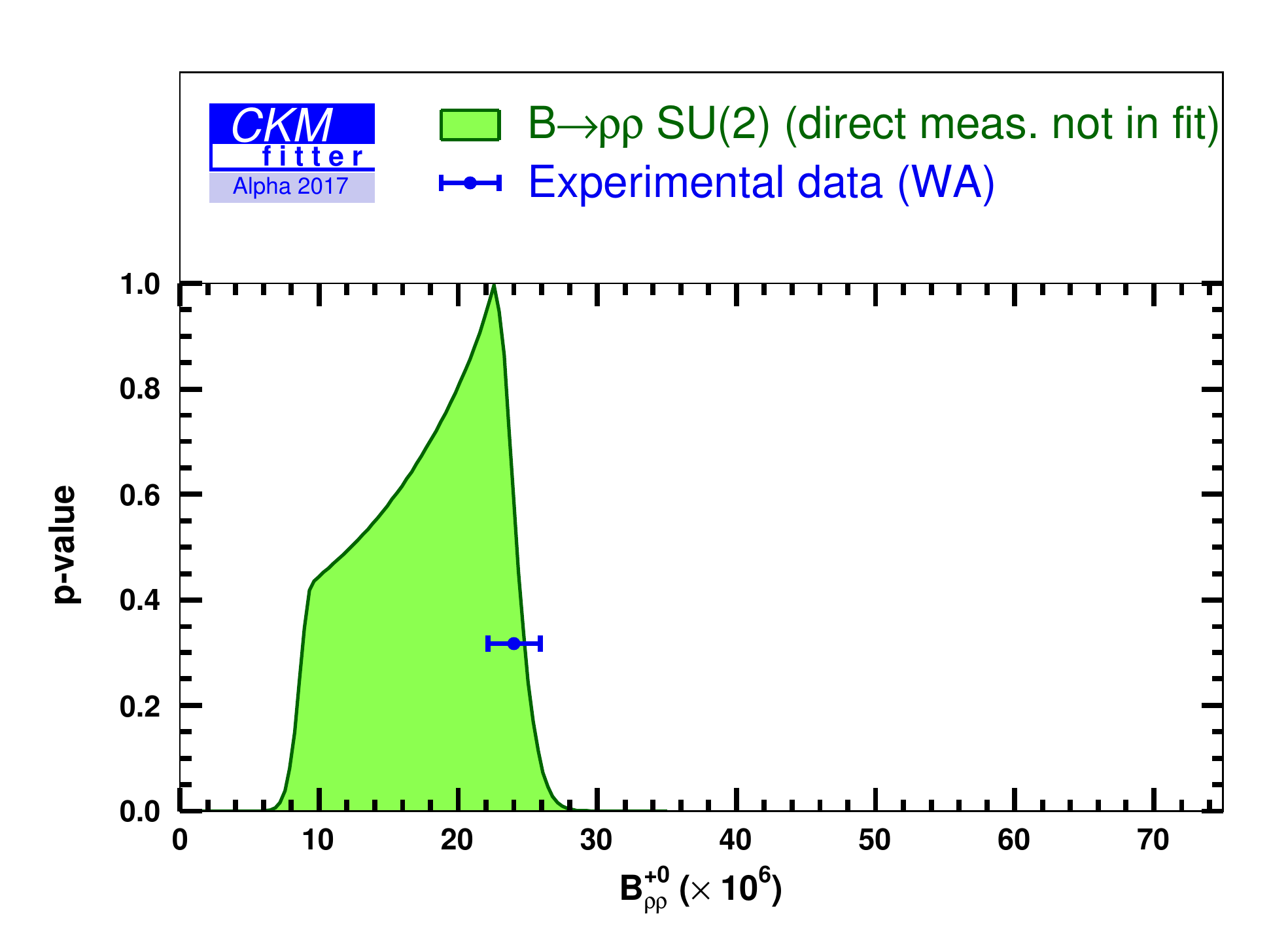}
  \includegraphics[width=18pc]{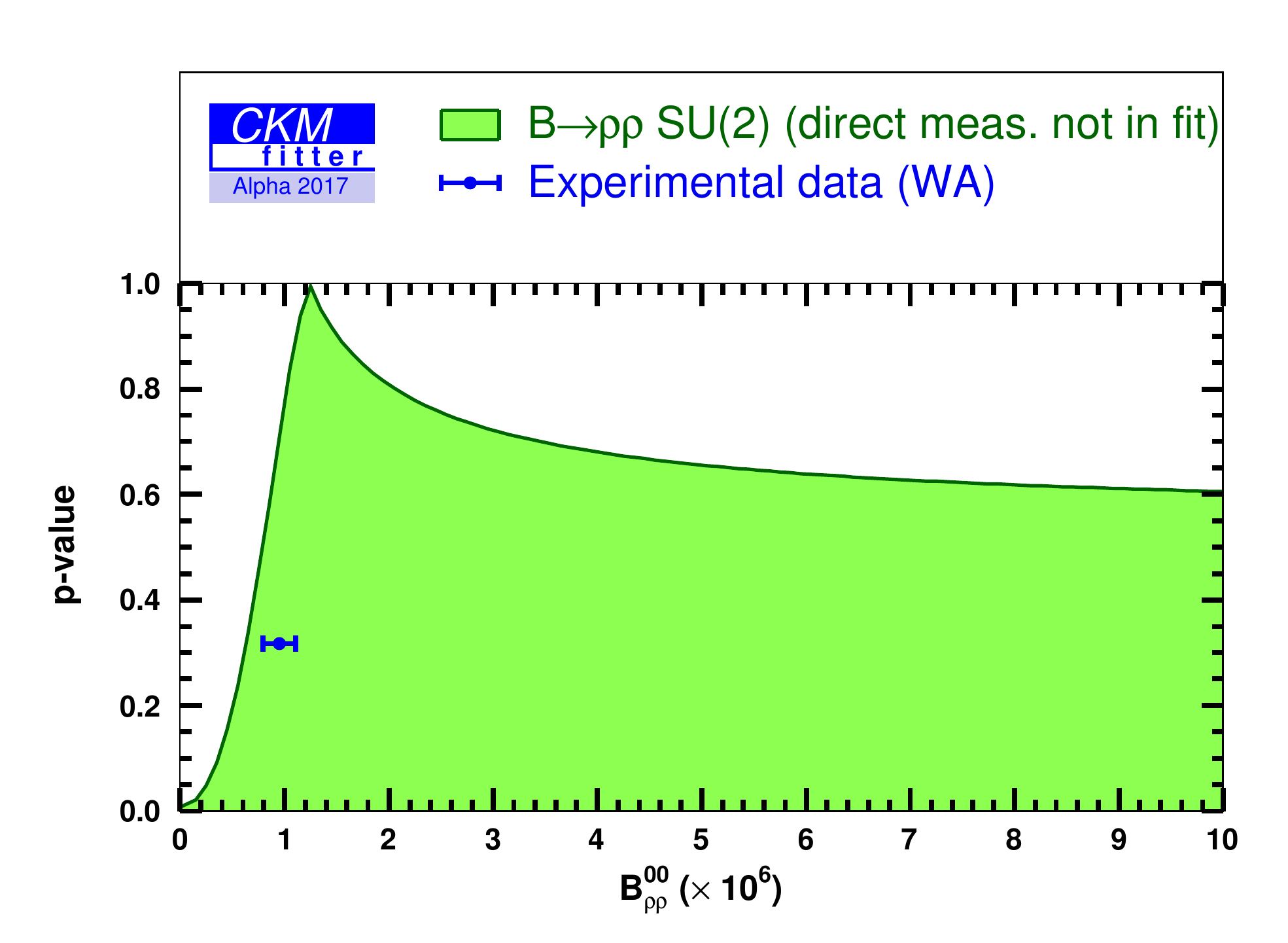}
  \includegraphics[width=18pc]{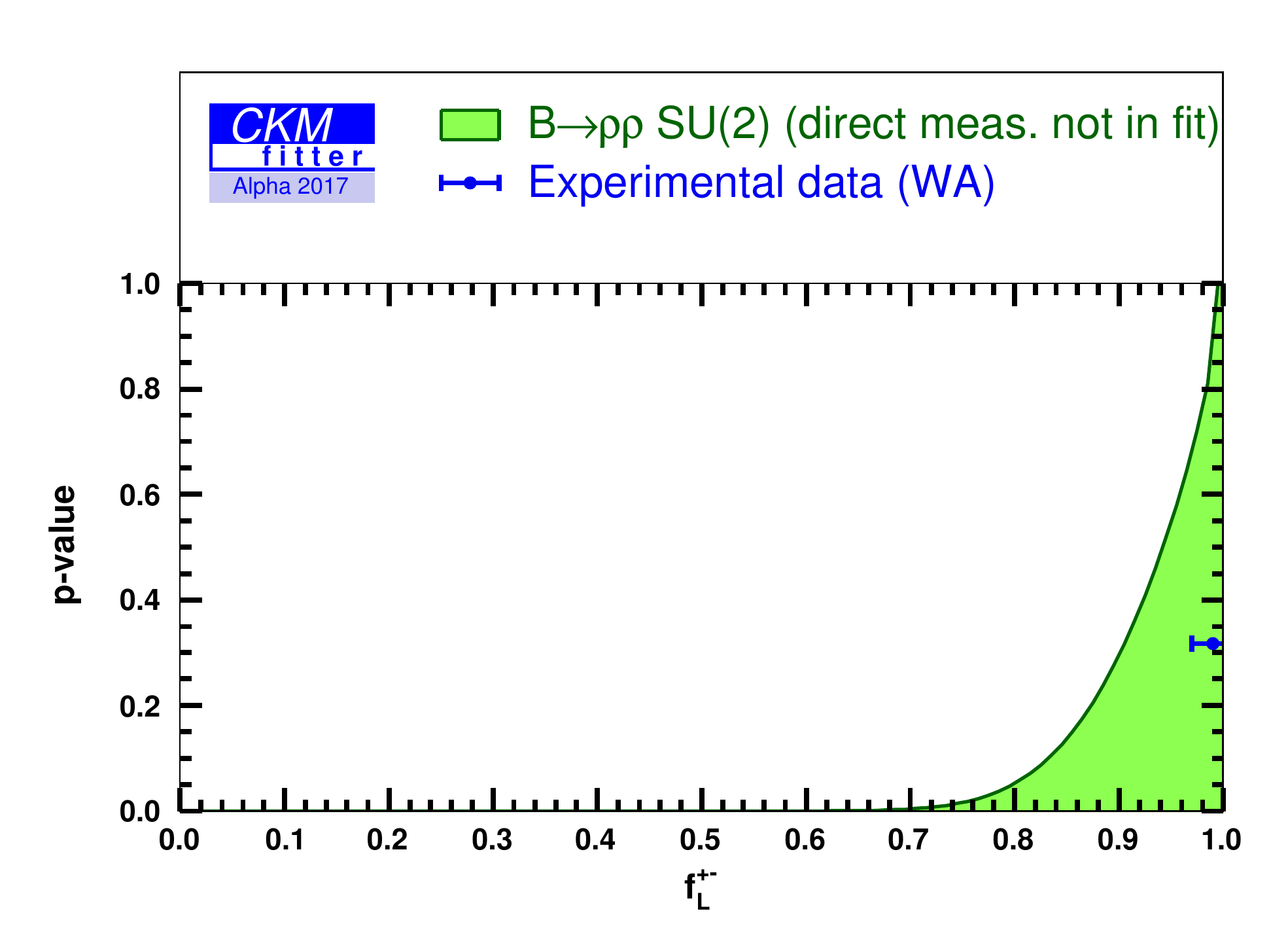}
  \includegraphics[width=18pc]{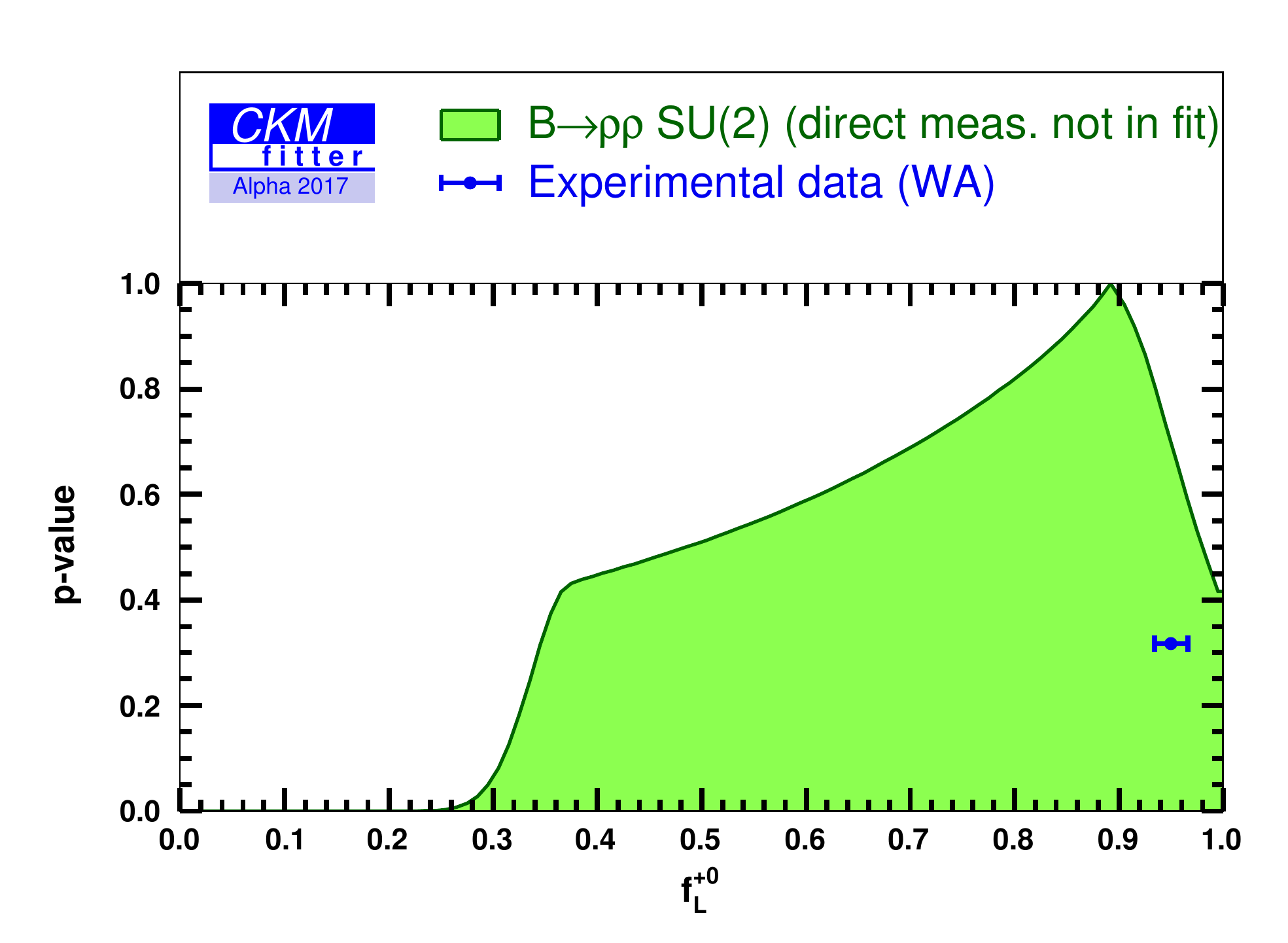}
  \includegraphics[width=18pc]{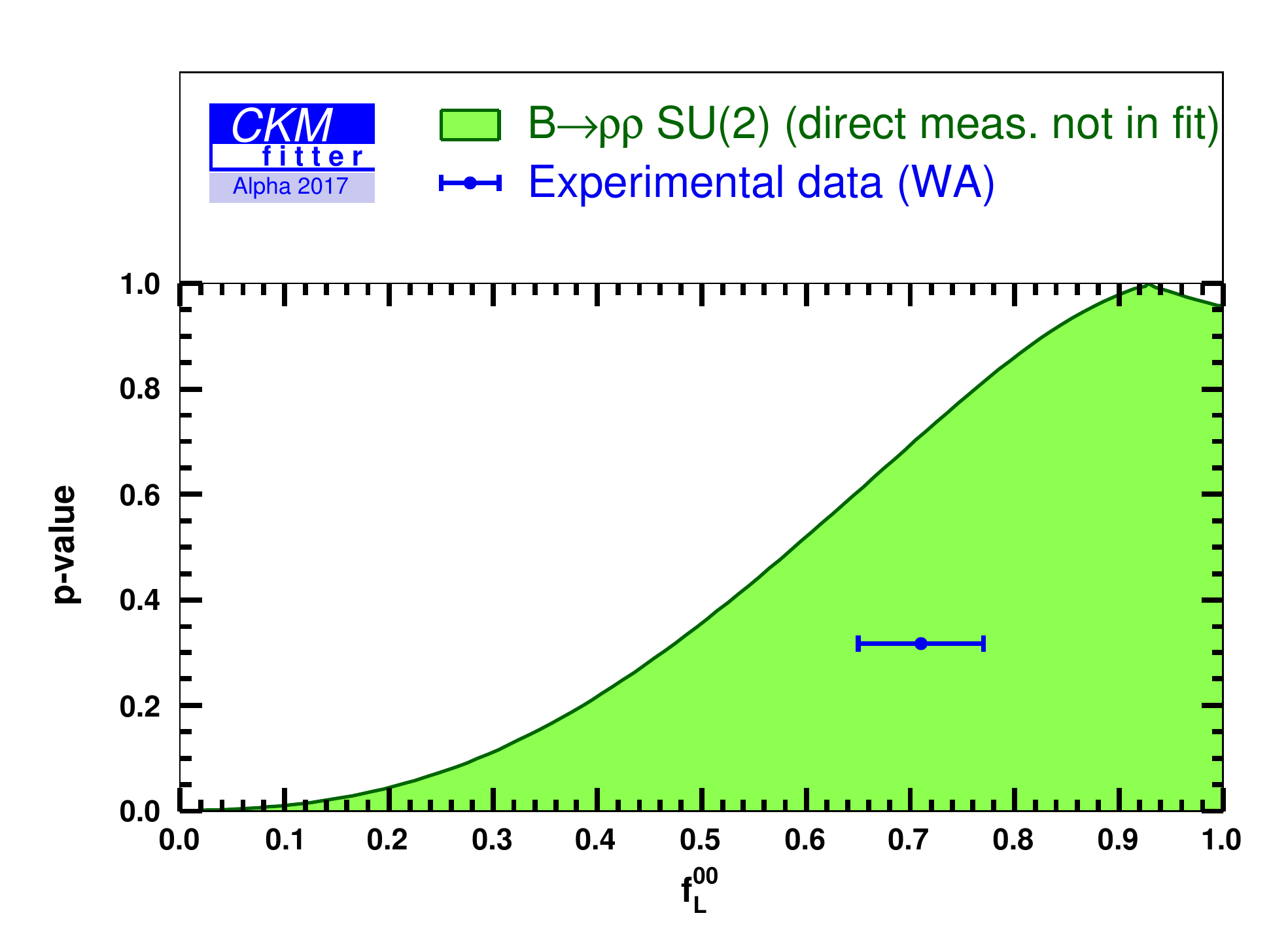}
\caption{\it\small  \su{2} isospin constraint on the \decay{B}{\rho\rho} branching fractions (top and middle left) and the fractions of longitudinal polarisation (middle right and bottom). For each observable, the direct measurement, not included in the fit, is indicated by an interval with a dot.}
\label{fig:BR_RhoRho}
\end{center}        
\end{figure}

As shown in Sec.~\ref{subsec:rhorho}, the   \decay{B}{\rho\rho} data is in very good agreement with the indirect determination of the weak phase $\alpha_{\rm ind}$.
The \su{2} isospin constraint on the $\B^{ij}_{\rho\rho}$ branching ratios and the corresponding fractions of longitudinal polarisation are displayed in Fig.~\ref{fig:BR_RhoRho}. As expected theoretically \cite{Beneke:2006hg}, the isospin fit to the  \decay{B}{\rho\rho} data favours large fractions of longitudinal polarisation in each of the three decay modes. The most precise constraint is obtained for the neutral \decay{B^0}{\rho^+\rho^-} decay with an indirect determination $f_{\rho_L\rho_L;{\rm ind}}^{+-}>0.91$ (0.79) at 68\% (95\%) confidence level, in very good agreement with the experimental measurement $f_{\rho_L\rho_L;{\rm dir}}^{+-}=0.990\pm 0.020$.

\begin{figure}[t]
\begin{center}
  \includegraphics[width=18pc]{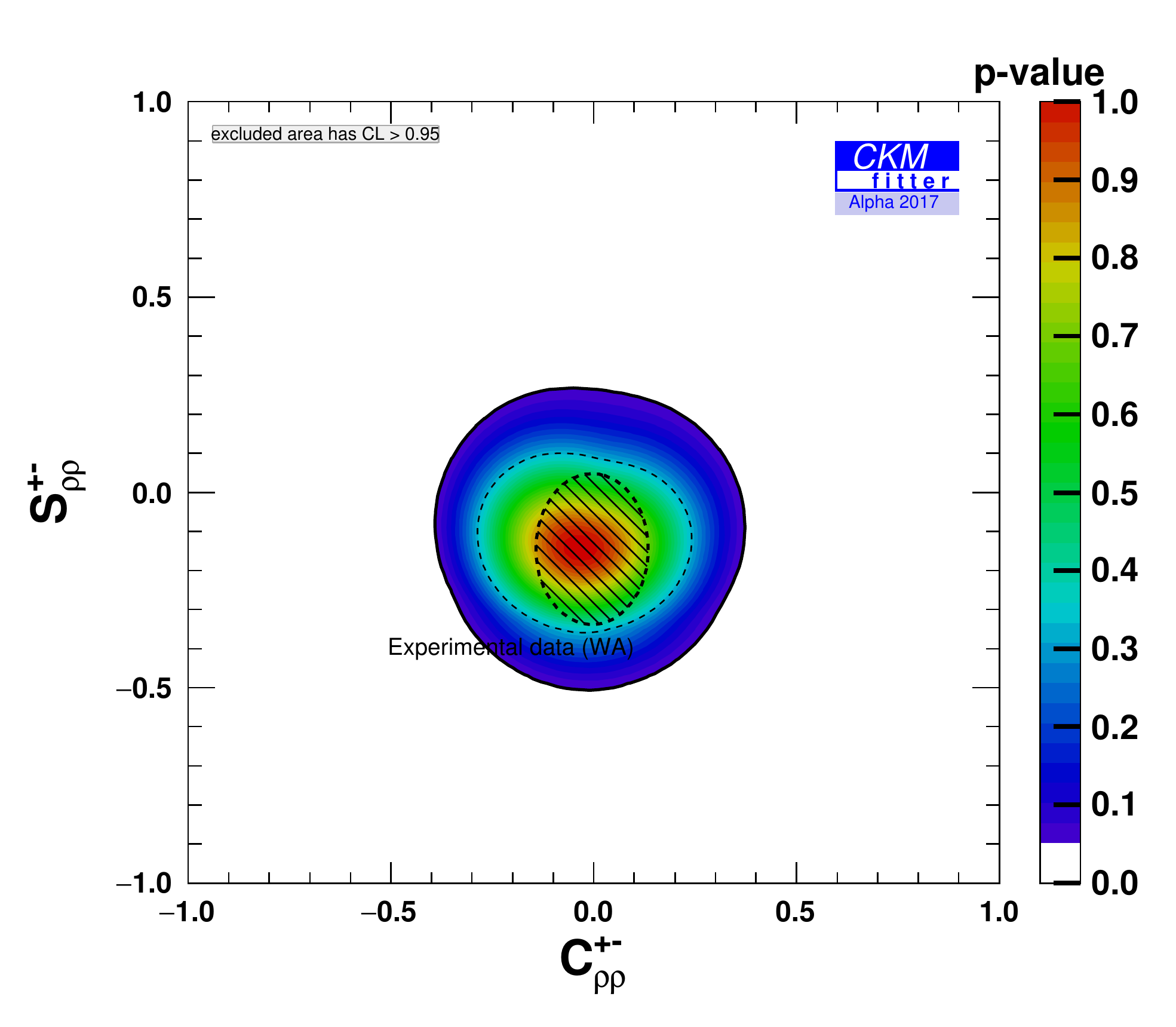}
  \includegraphics[width=18pc]{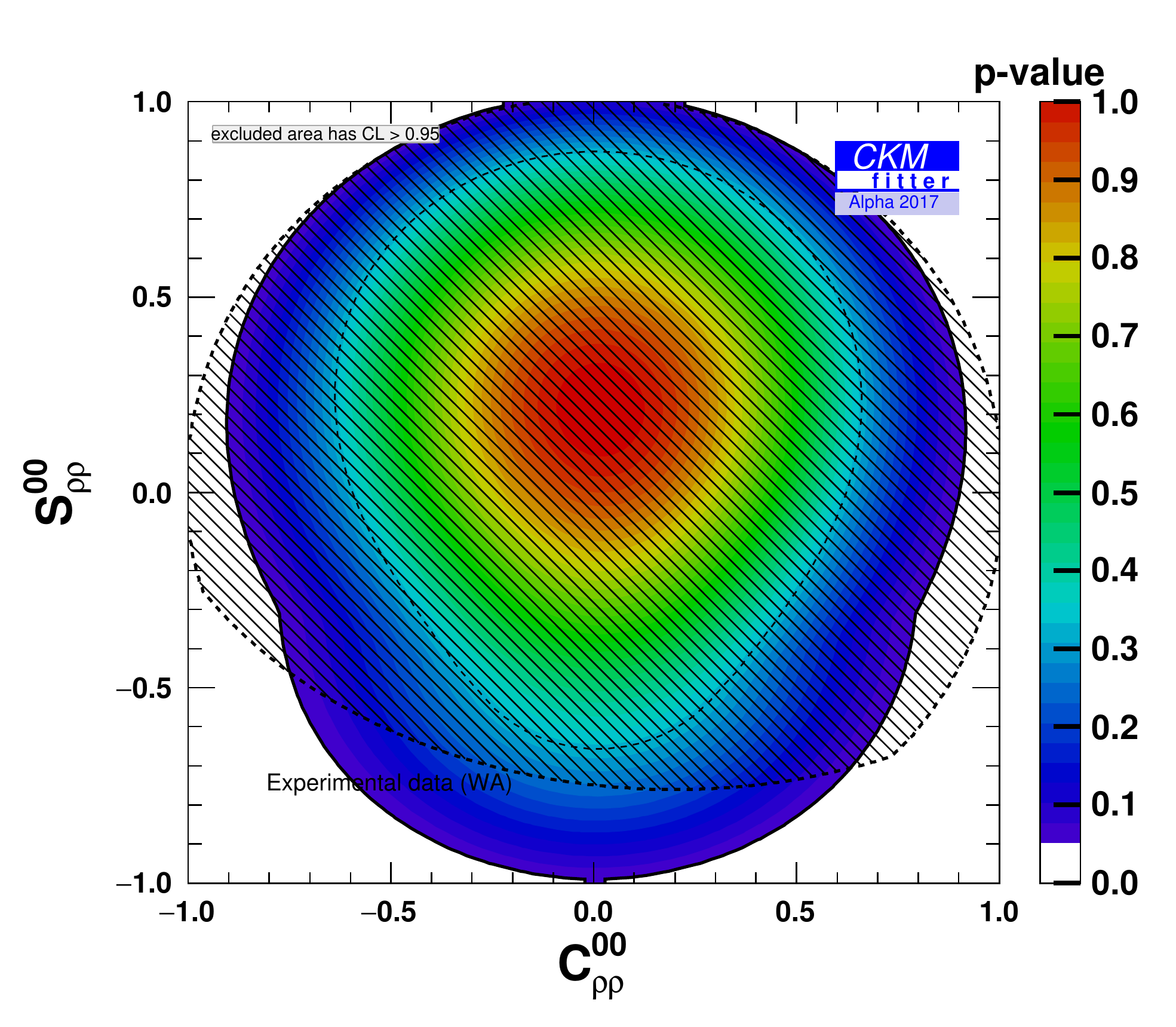}
\caption{\it\small  Two-dimensional constraint on the \decay{B^0}{(\rho\rho)^0} $CP$ asymmetries in the (\C,\S) plane. The direct measurement, not included in the fit, is indicated for each mode as a shaded area.}
\label{fig:CS_RhoRho}
\end{center}       
\end{figure}

Weaker constraints  are obtained on the $CP$ asymmetries ($\C_{\rho\rho},\S_{\rho\rho}$)  for both neutral modes \decay{B^0}{\rho^+\rho^-} and \decay{B^0}{\rho^0\rho^0} as shown in Fig.~\ref{fig:CS_RhoRho}. Numerical values, confidence intervals and pulls for the measured observables are reported on Tab.~\ref{tab:BR_RhoRho} in App.~\ref{sec:numerics}.

\subsection{\decay{B}{\rho\pi} observables  \label{sec:rhopiObs}}

\subsubsection{\decay{B^0}{\pi^{+}\pi^{-}\pi^{0}} Dalitz analysis}

The normalised \U and \I observables provide a complete description of the relative \decay{B^0}{\pi^+\pi^-\pi^0} decay amplitudes.
As discussed in Sec.~\ref{subsec:rhopi}, the data from \decay{B^0}{(\rho\pi)^0\to\pi^{+}\pi^{-}\pi^{0}} Dalitz analyses disagrees with the indirect determination of $\alpha$ at the level of 3.0 standard deviations. This disagreement is reflected through the indirect determination of several of the form-factor coefficients \U and \I as listed in Tab.~\ref{tab:UI_rhopi} in App.~\ref{sec:numerics}, in particular
$\U{+}{-}$, $\U{-}{+}$, $\I{-}$, $\U{+0}{+\Re}$, $\U{-0}{+\Re}$ with pulls above 2 $\sigma$, and $\U{+-}{+\Re}$, $\U{+-}{+\Im}$ with pulls near or above 3 $\sigma$.

\begin{figure}[t]
  \begin{center}
    \includegraphics[width=12pc,height=12pc]{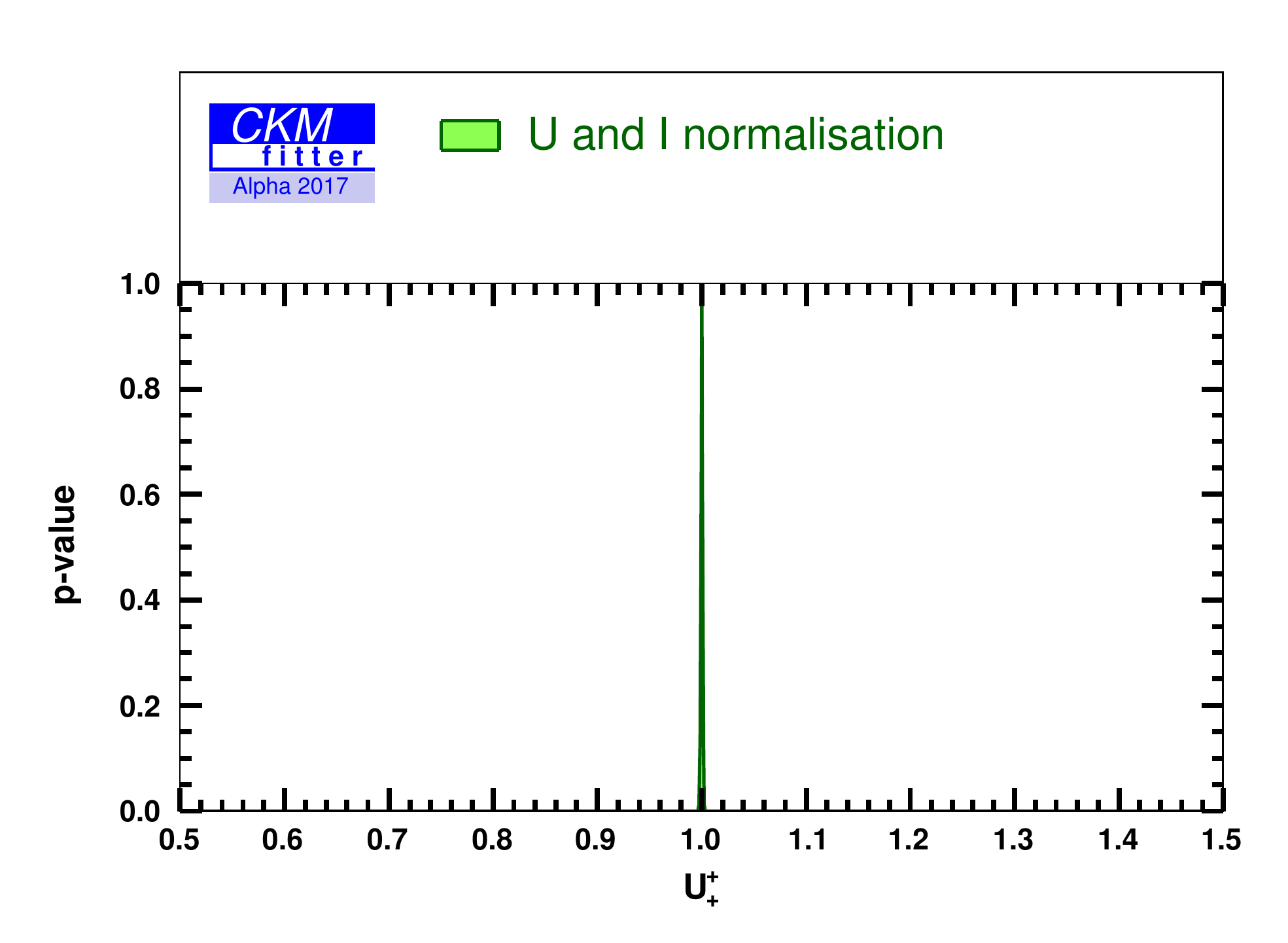} 
    \includegraphics[width=12pc,height=12pc]{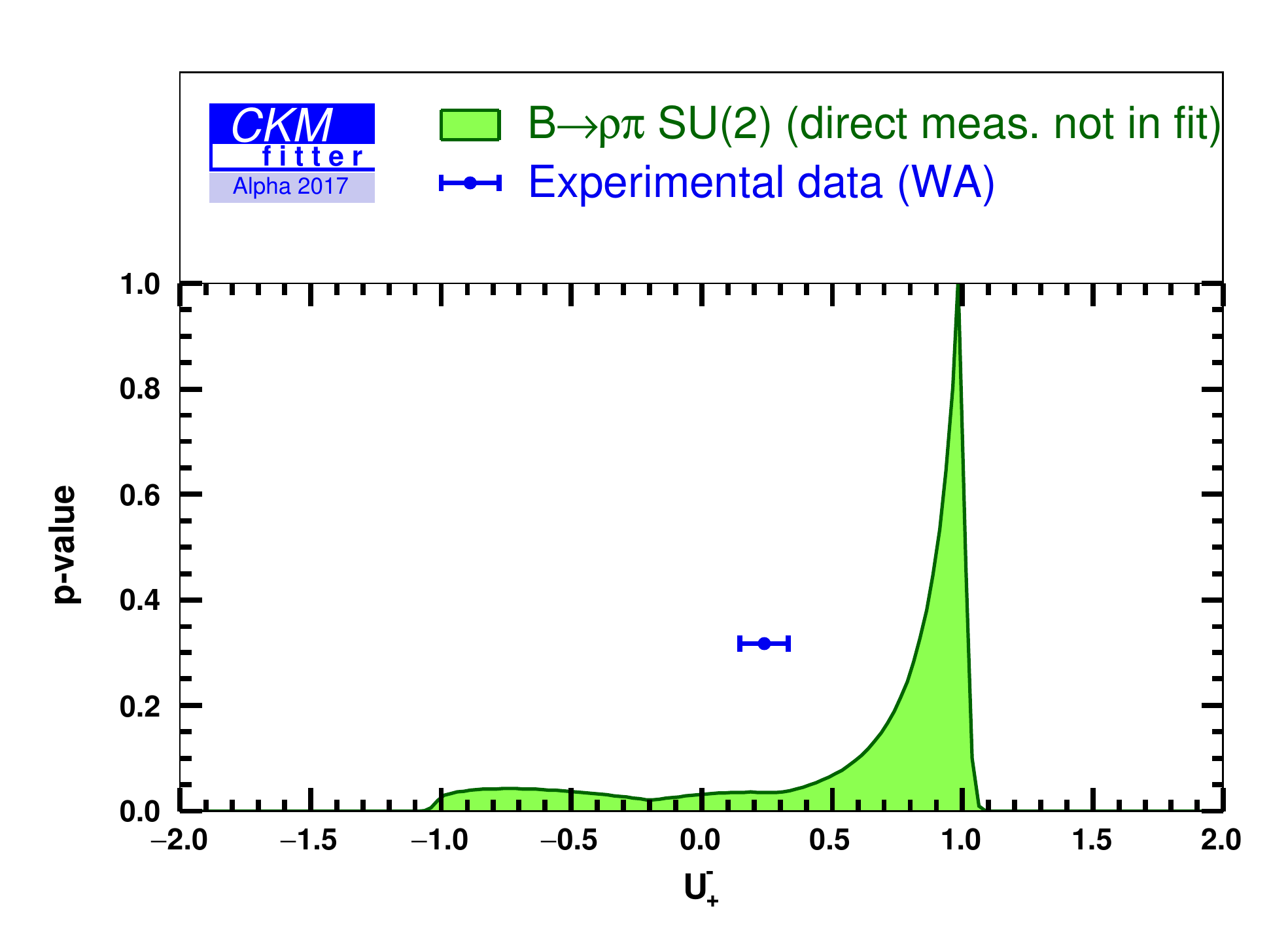}
    \includegraphics[width=12pc,height=12pc]{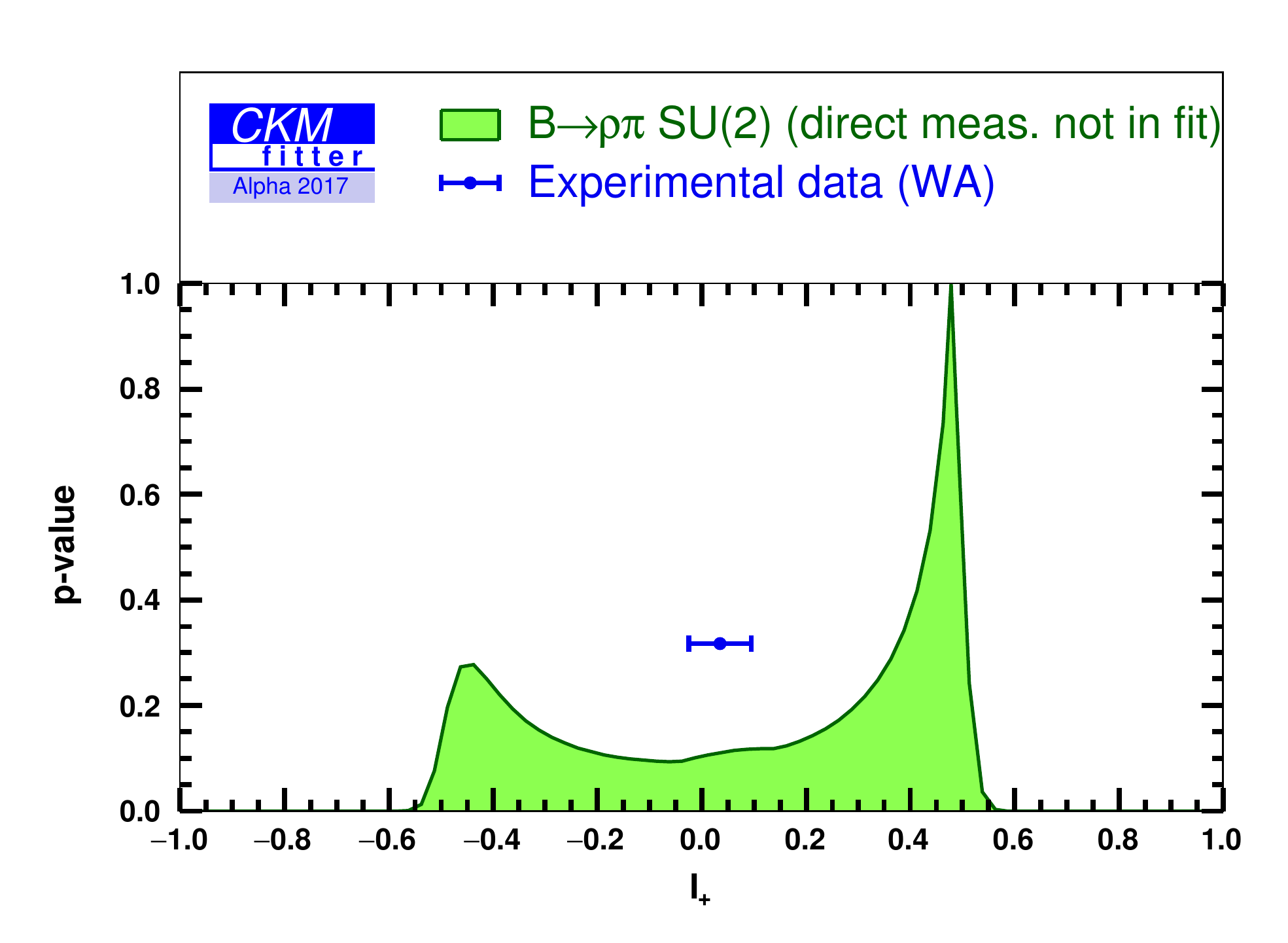}
    \includegraphics[width=12pc,height=12pc]{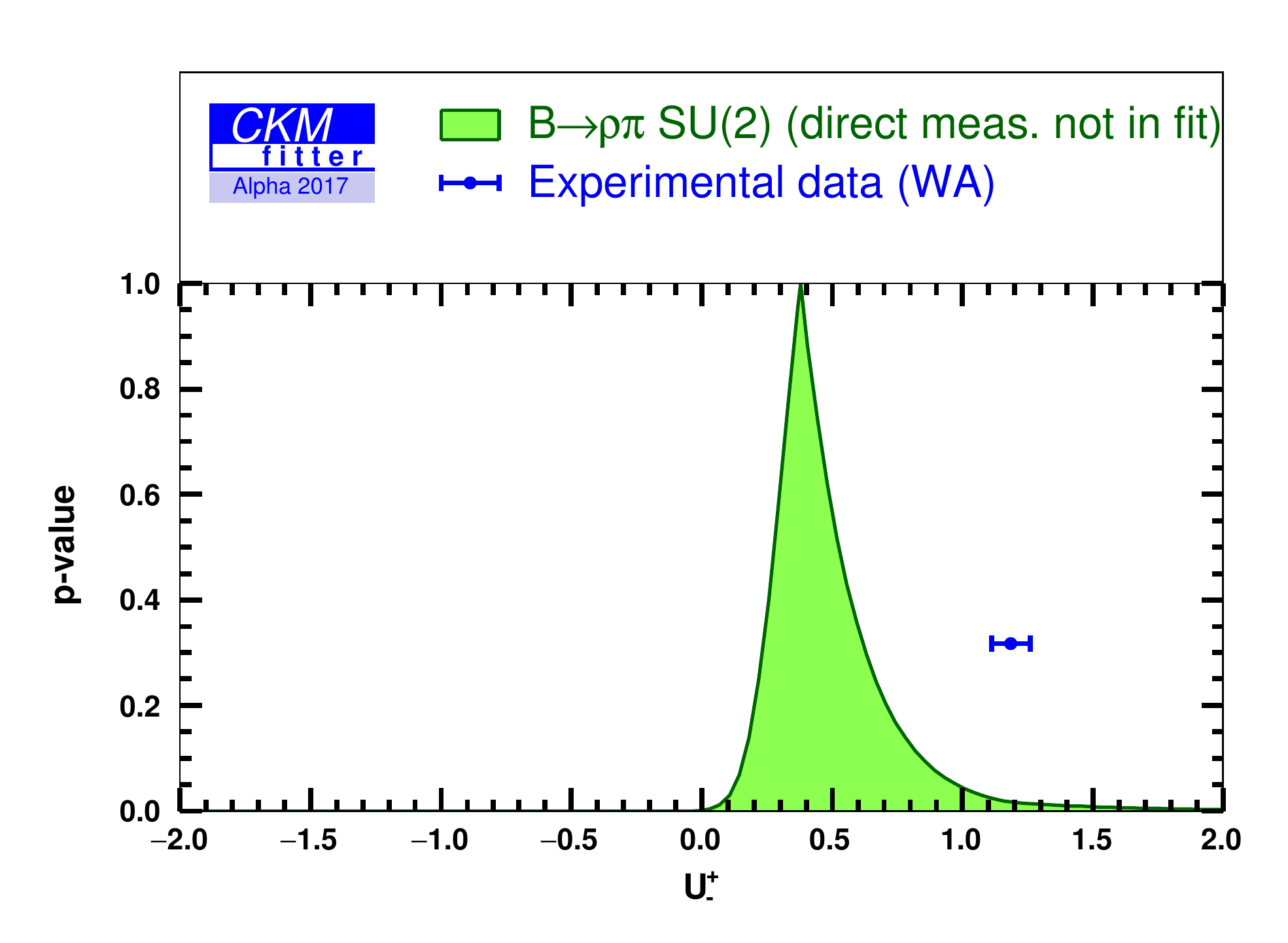}
    \includegraphics[width=12pc,height=12pc]{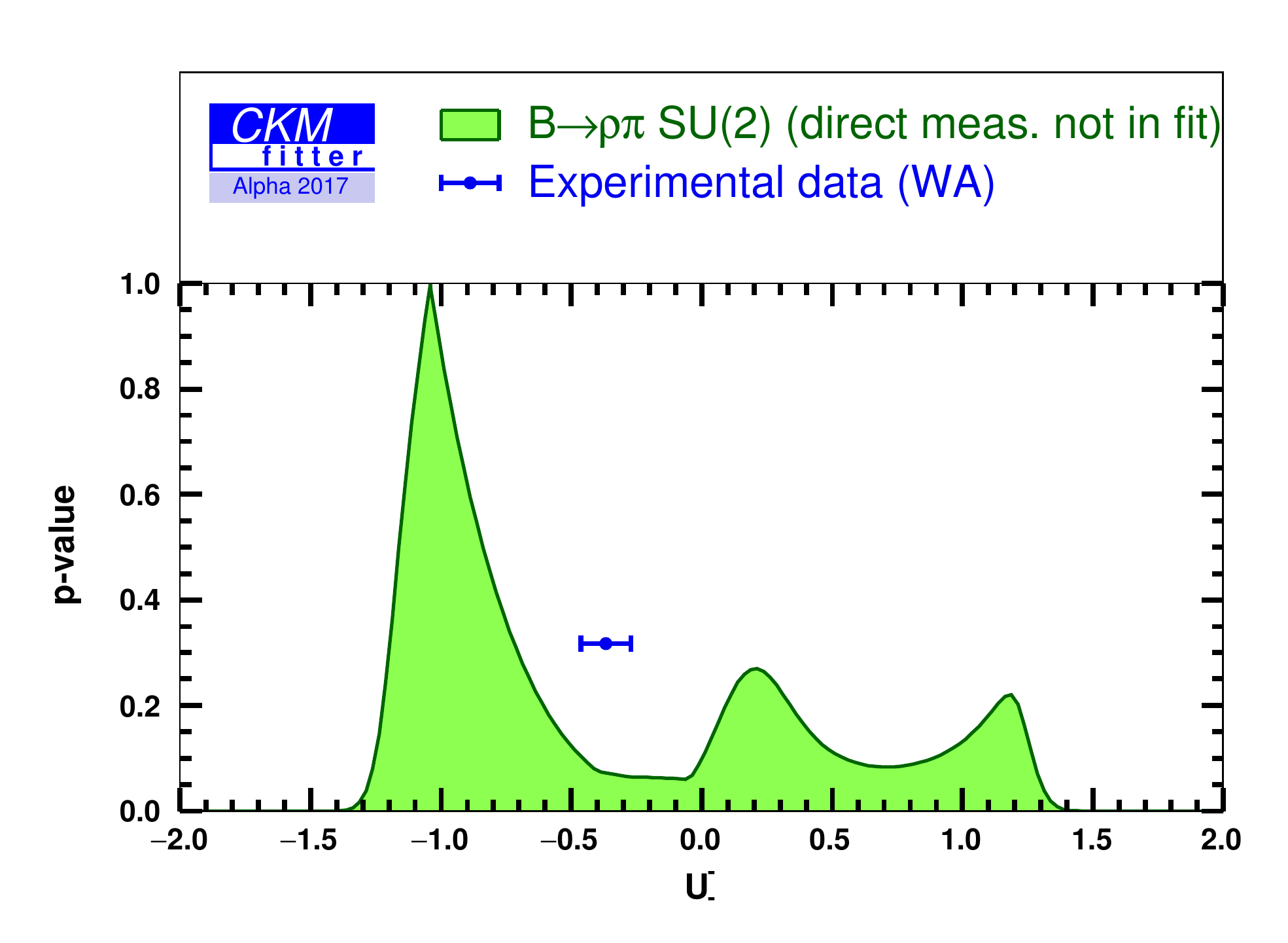}
    \includegraphics[width=12pc,height=12pc]{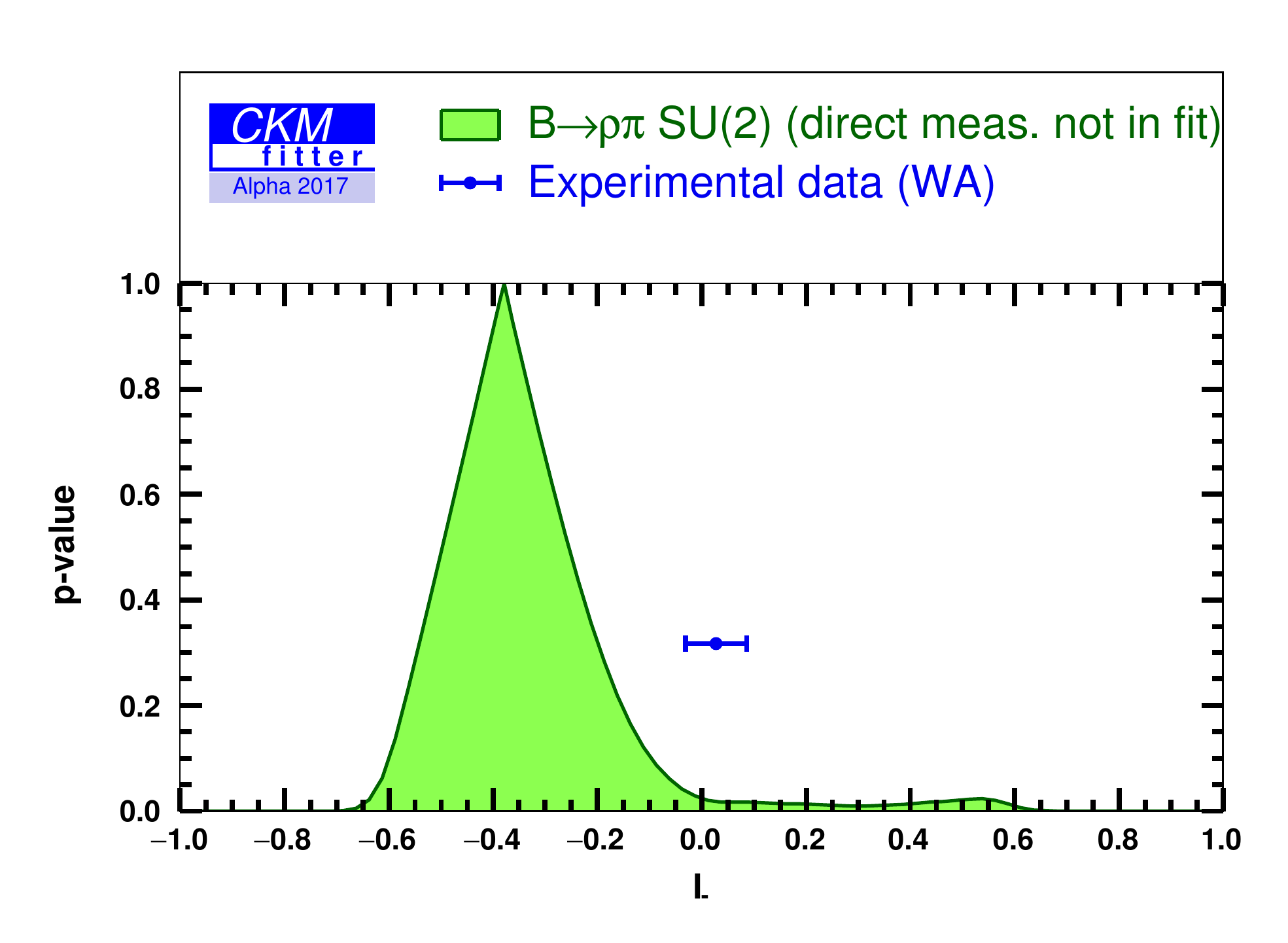}
    \includegraphics[width=12pc,height=12pc]{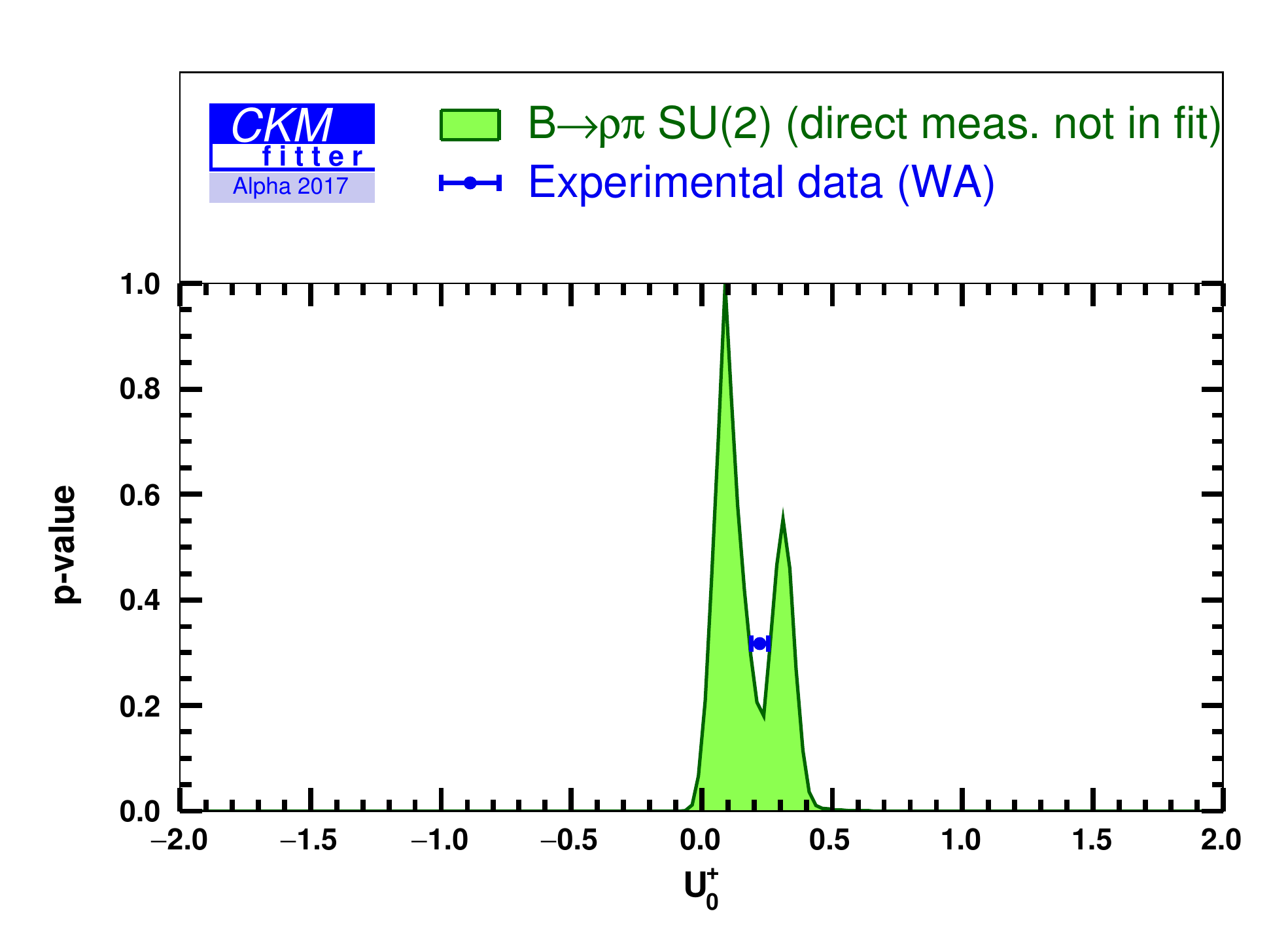}    
    \includegraphics[width=12pc,height=12pc]{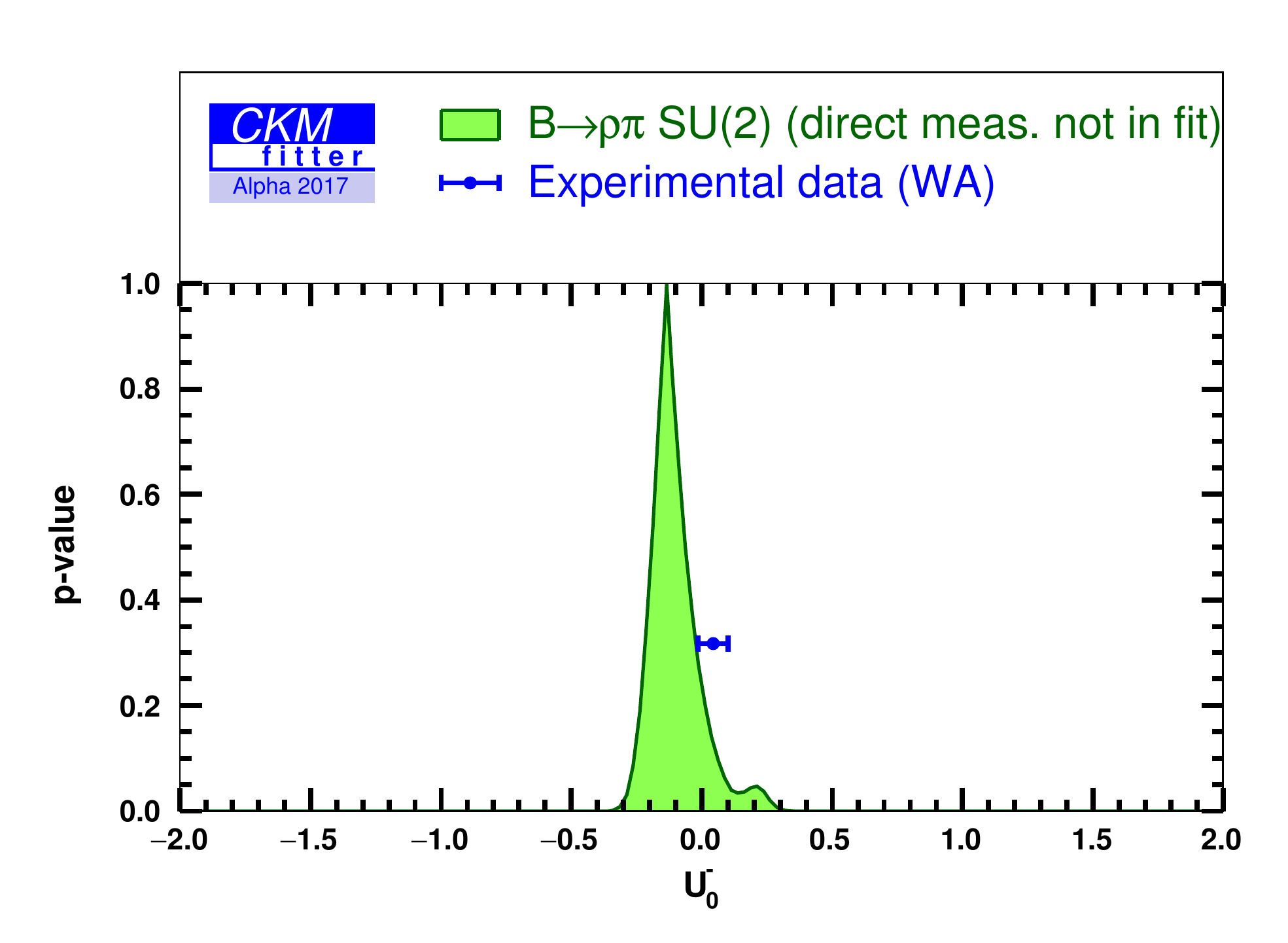}
    \includegraphics[width=12pc,height=12pc]{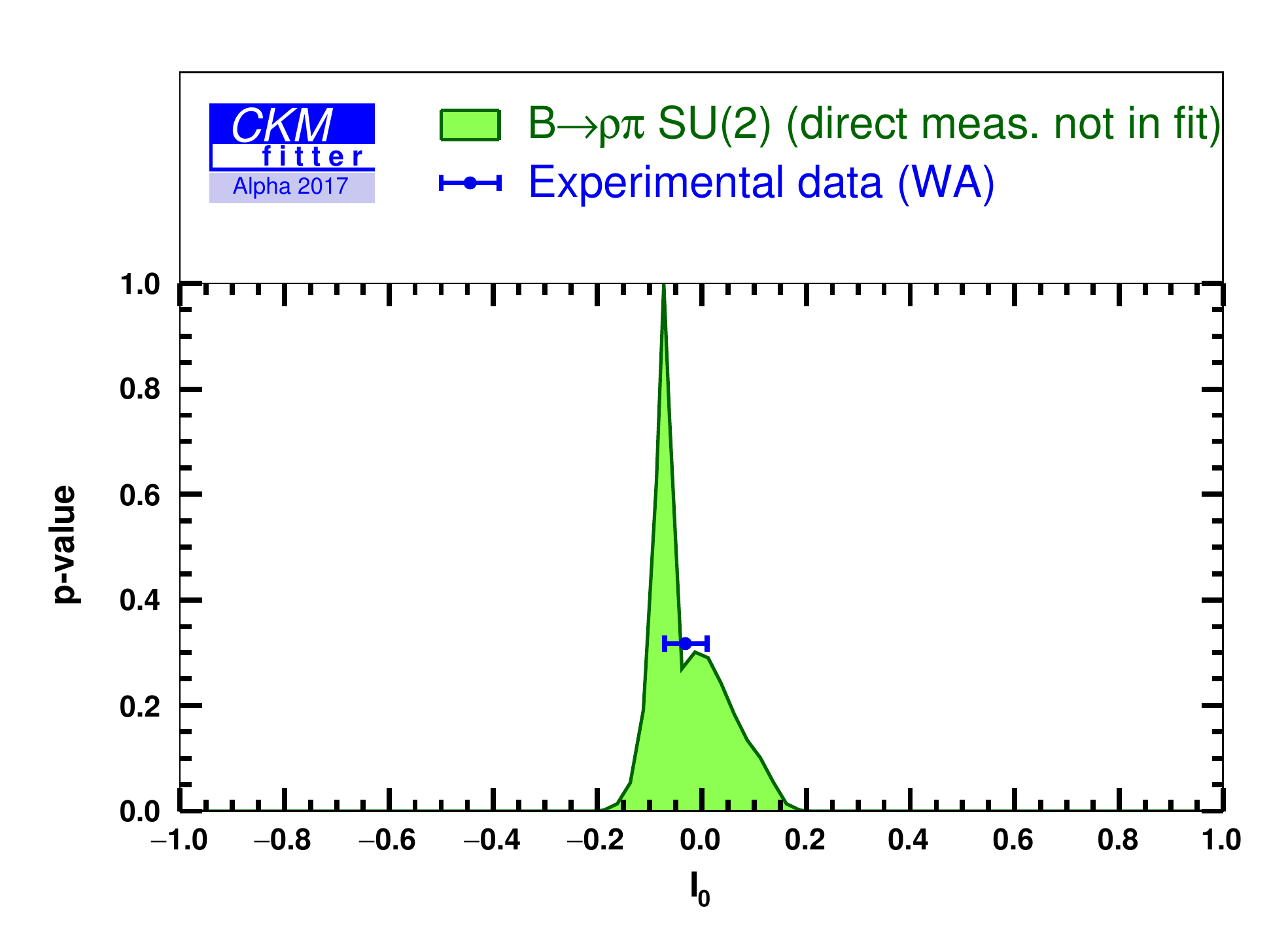}
\caption{\it\small  \su{2} isospin constraint on the Q2B-related coefficients \U and \I describing the \decay{B^0}{(\rho\pi)^{0}} intermediate states. The upper left figure corresponds to the overall  normalisation, \U{+}{+}=1. For each observable, the direct measurement, not included in the fit, is indicated by an interval with a dot.}
\label{fig:UI_Q2B_RhoPi}
\end{center}       
\end{figure}

\begin{figure}[t]
\begin{center}
    \includegraphics[width=12pc,height=12pc]{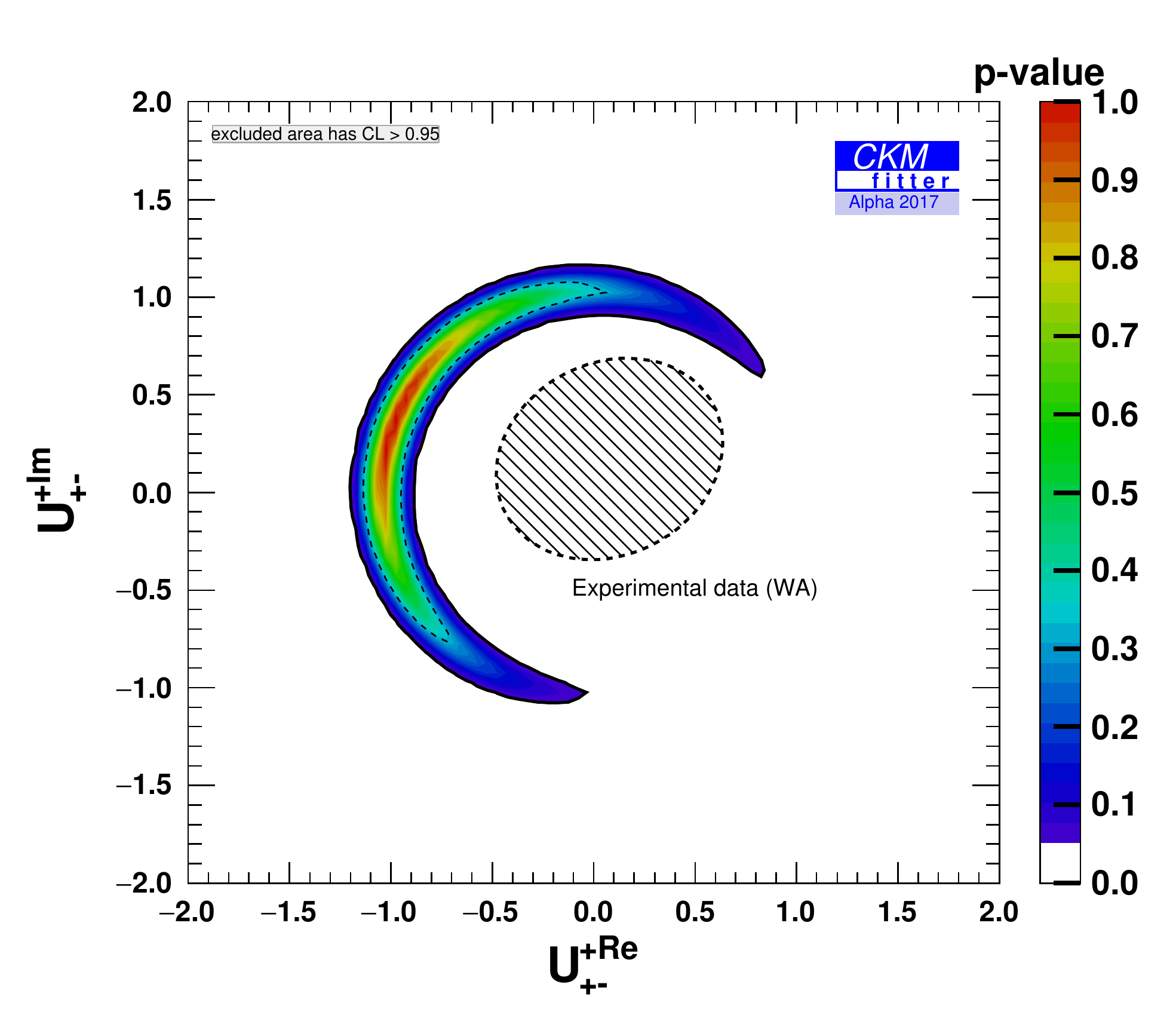}
    \includegraphics[width=12pc,height=12pc]{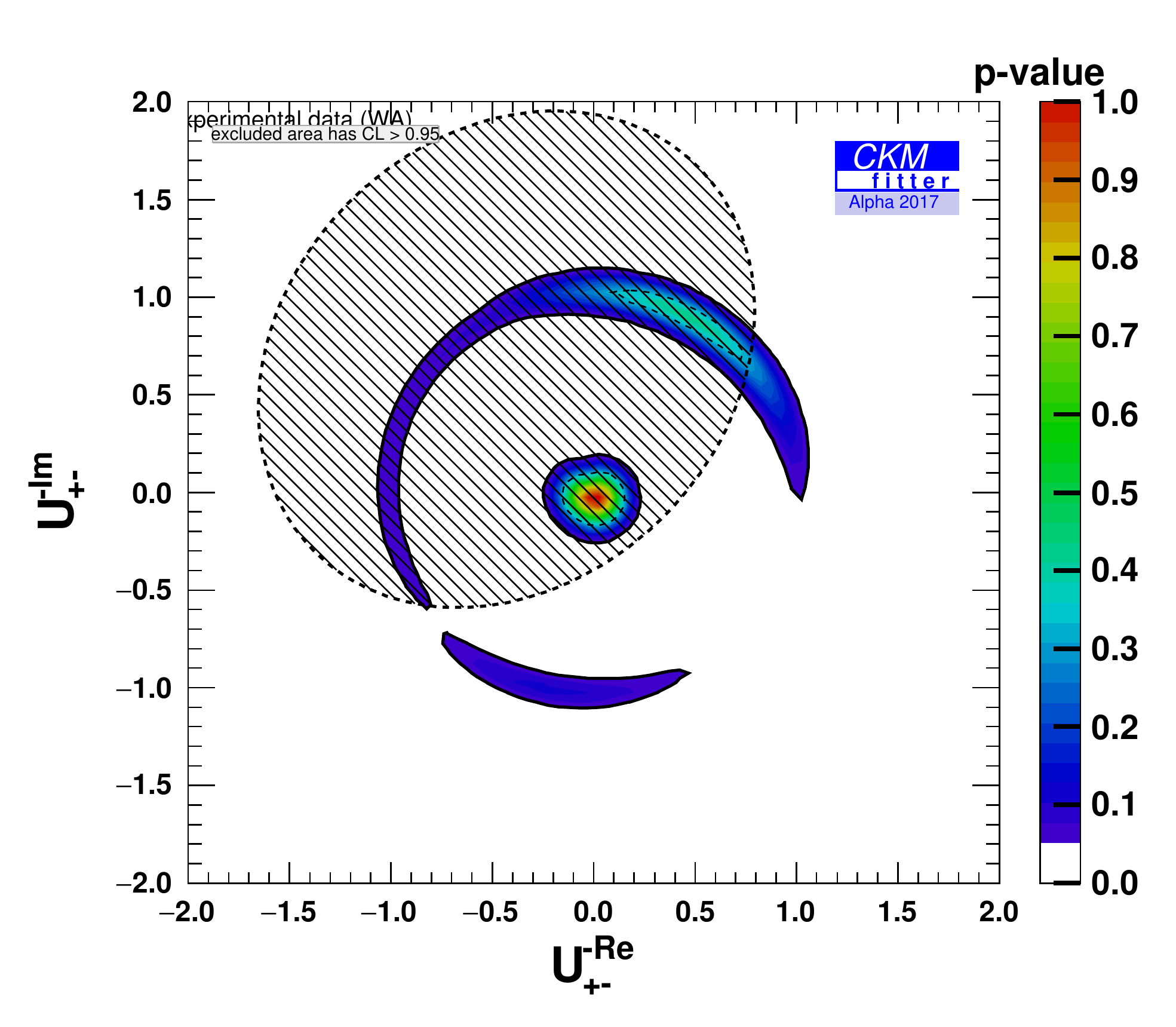}
    \includegraphics[width=12pc,height=12pc]{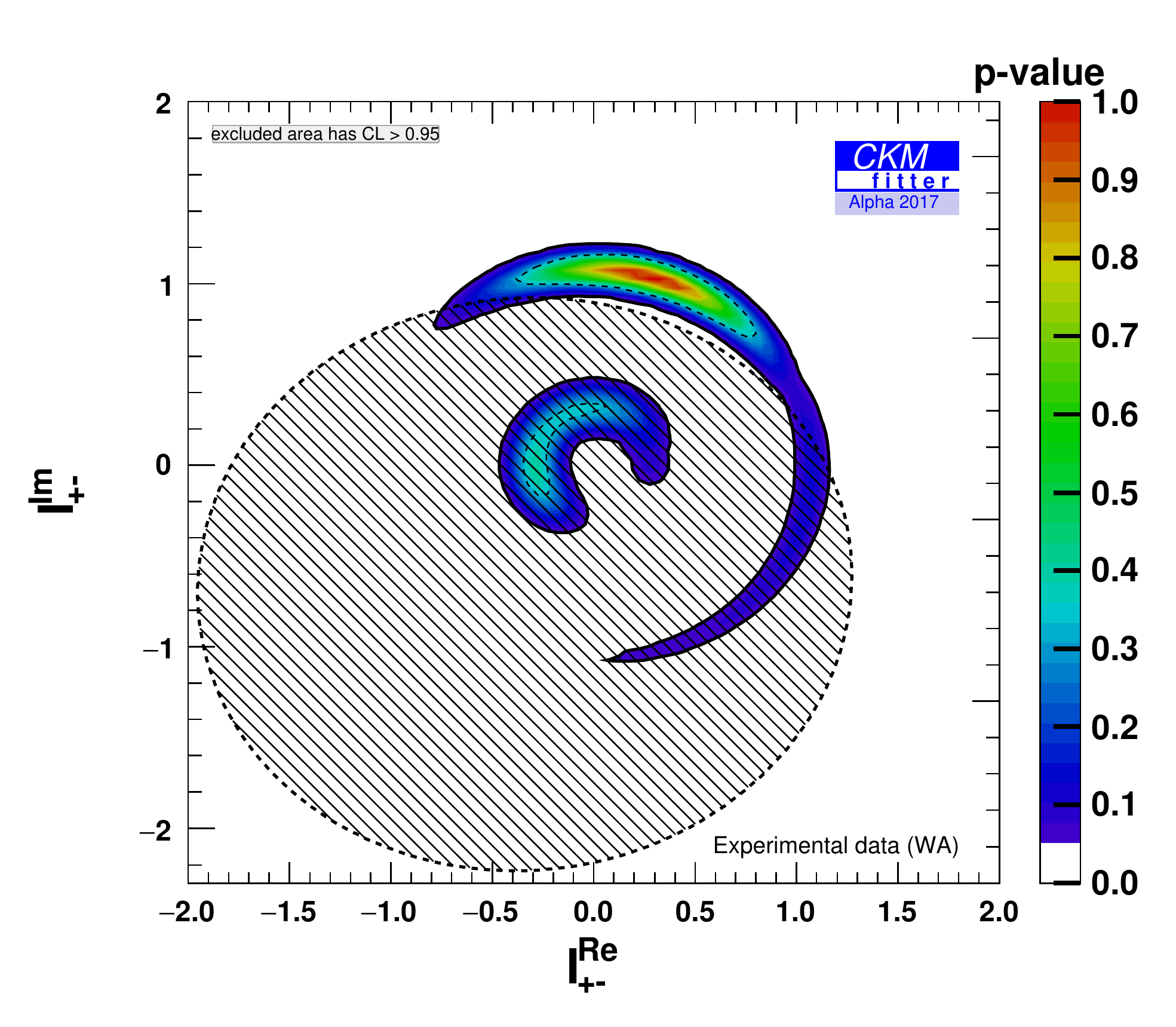}
    \includegraphics[width=12pc,height=12pc]{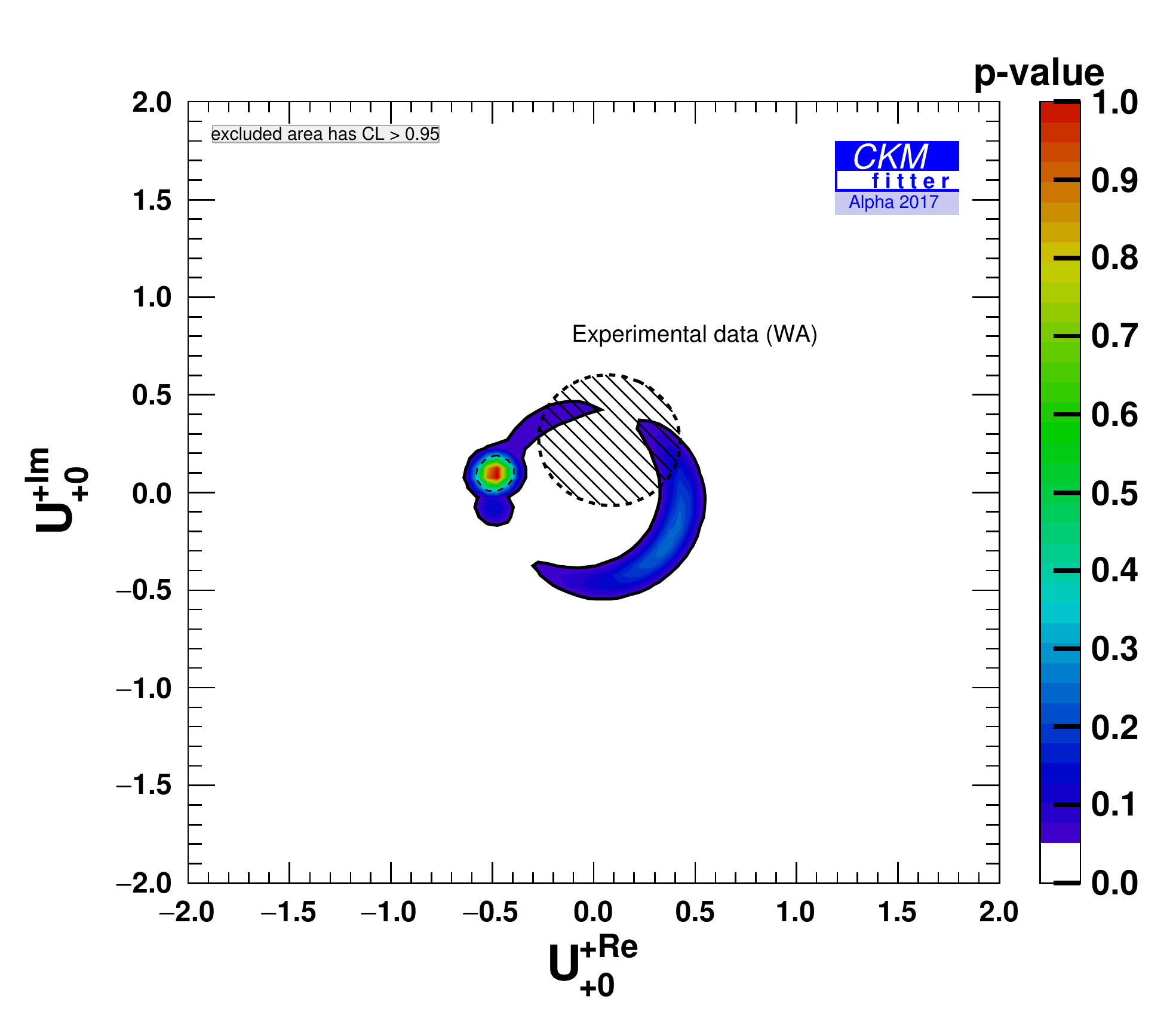}
    \includegraphics[width=12pc,height=12pc]{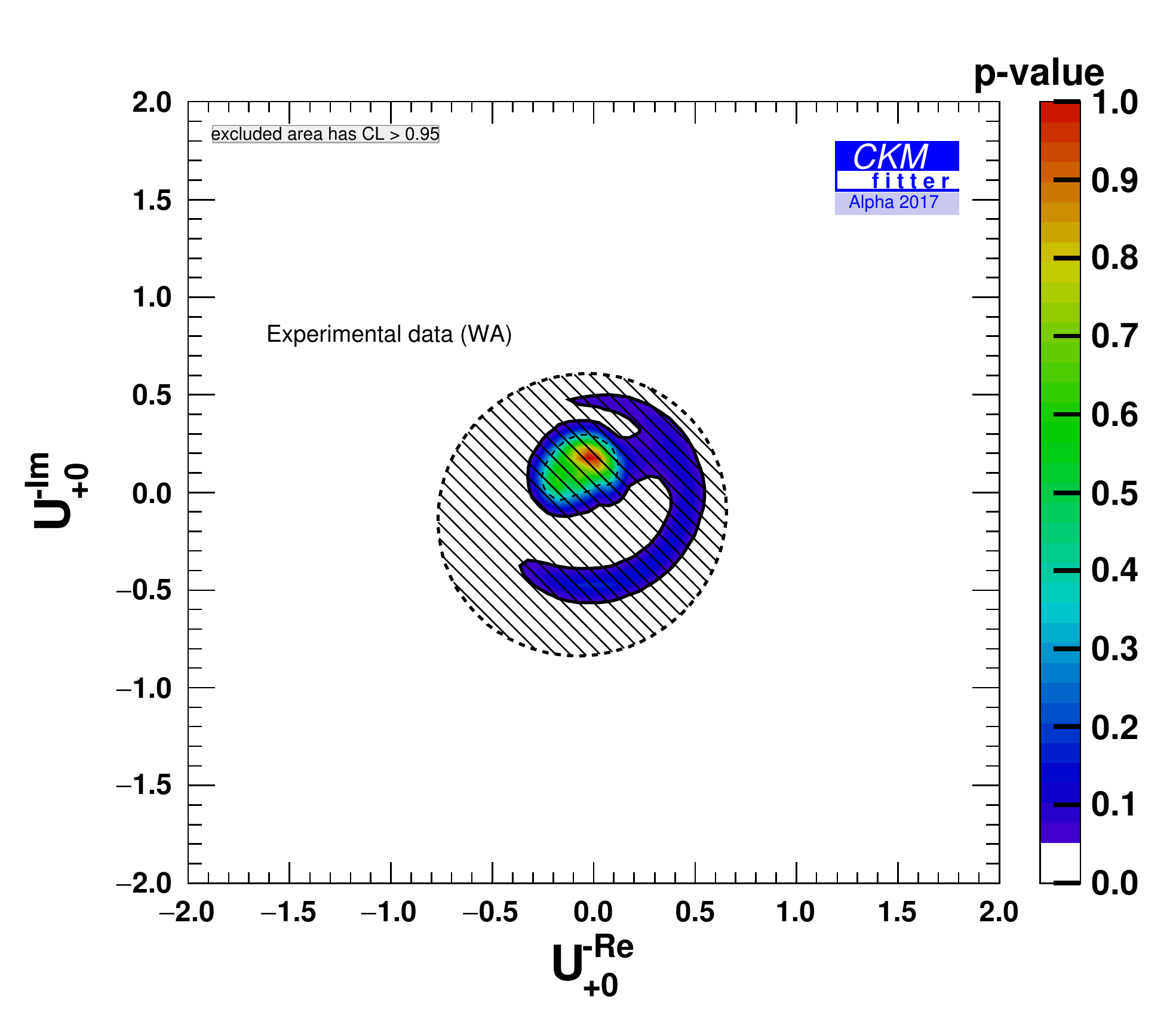}
    \includegraphics[width=12pc,height=12pc]{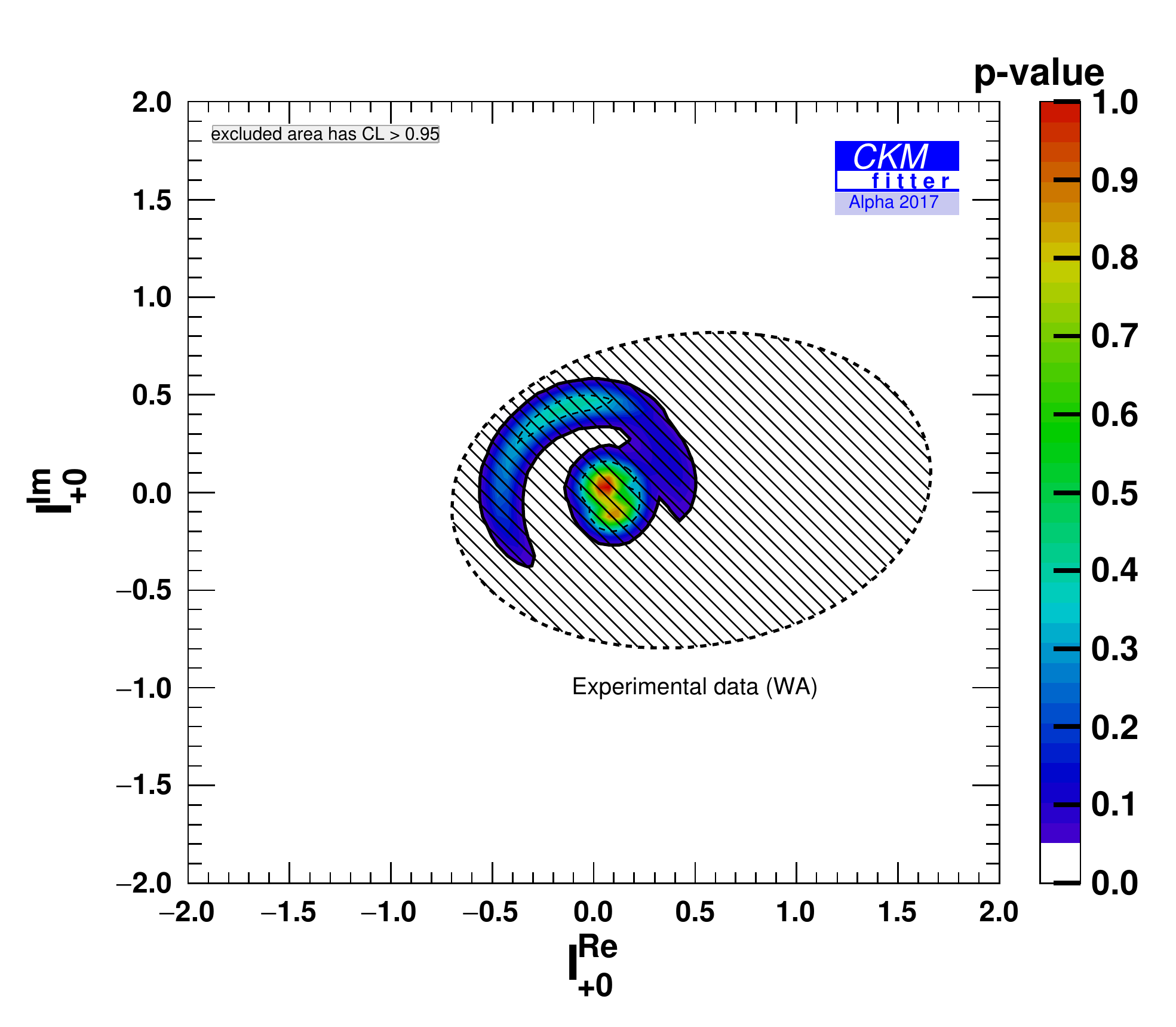}
    \includegraphics[width=12pc,height=12pc]{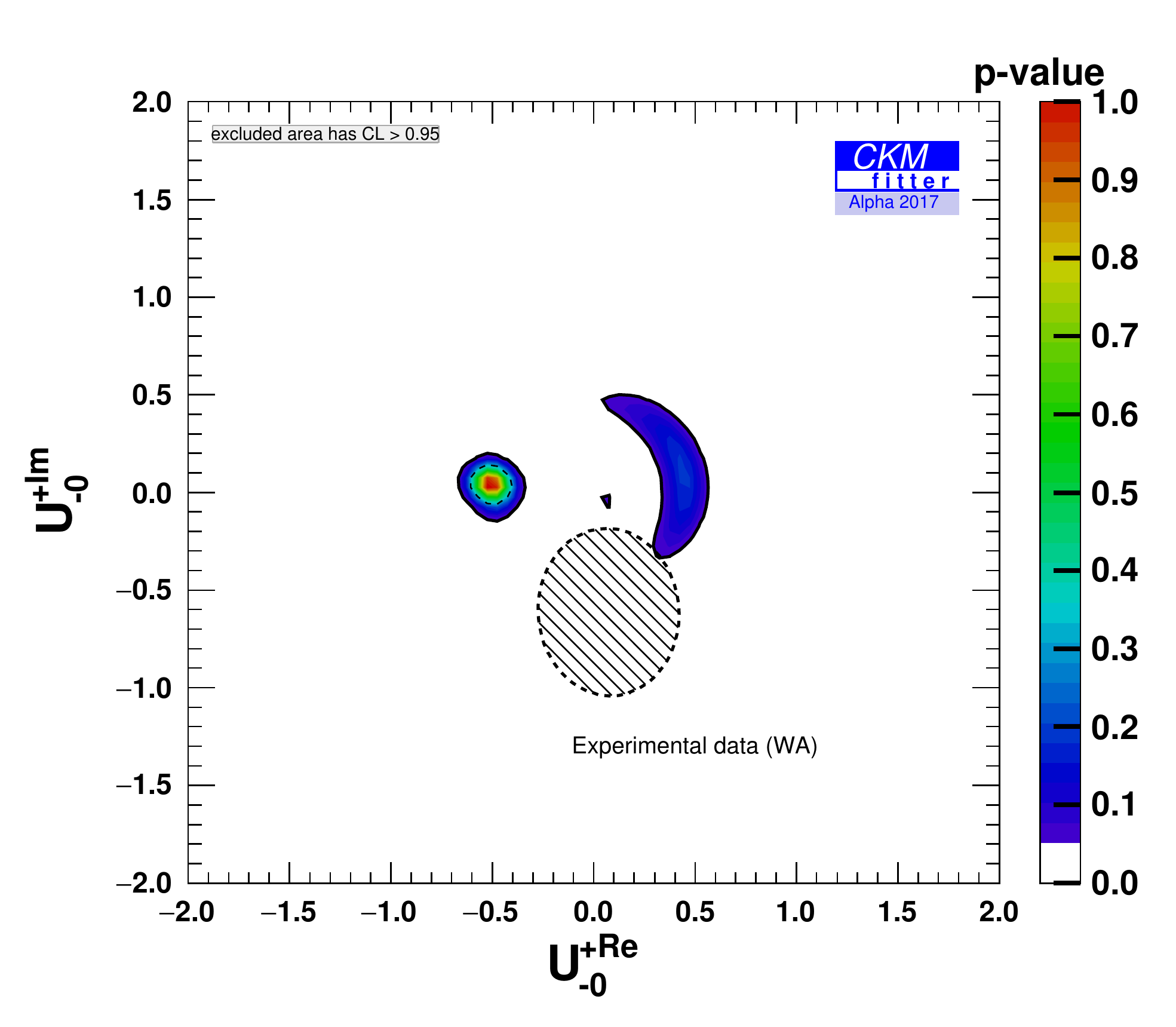}
    \includegraphics[width=12pc,height=12pc]{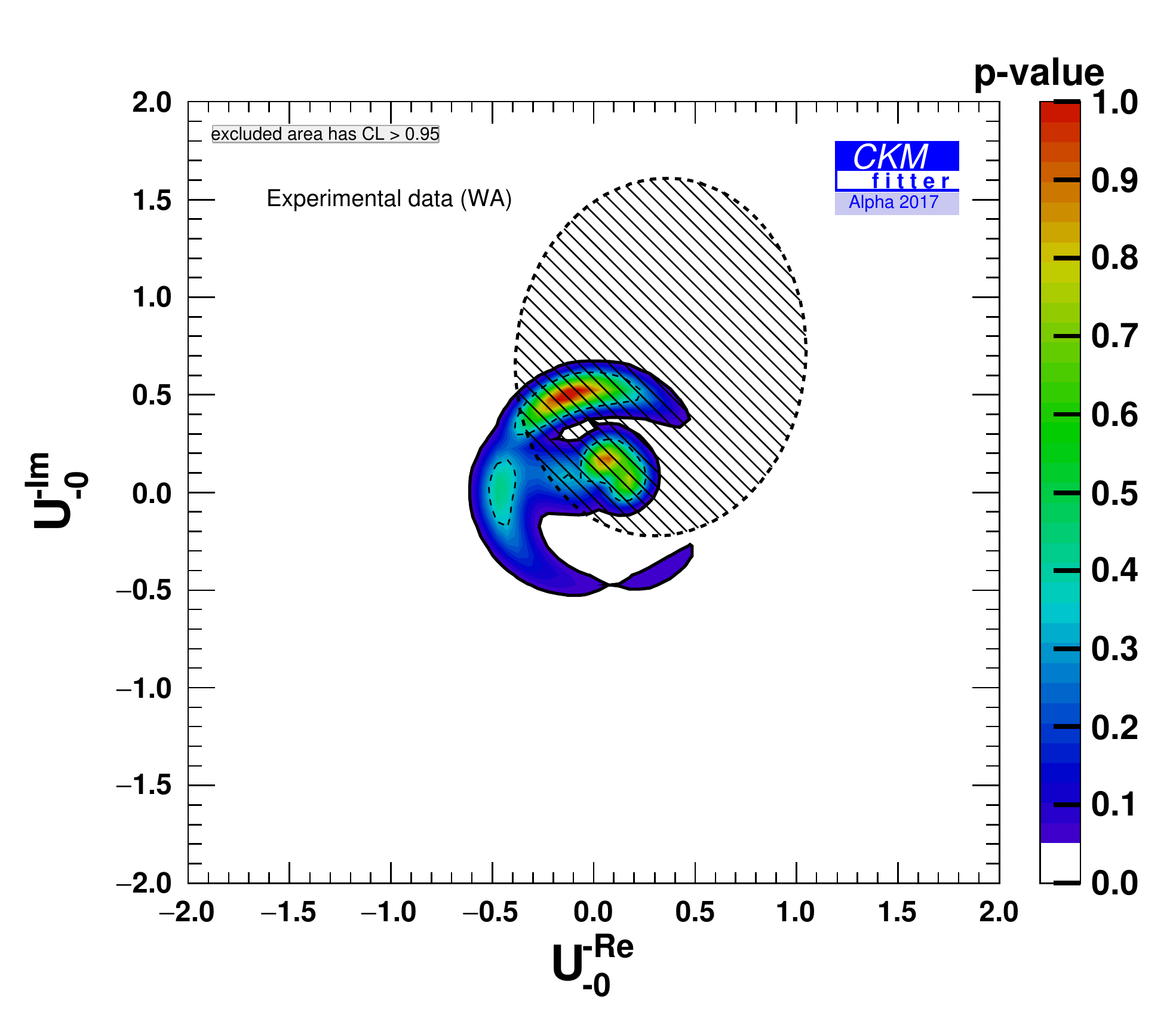}
    \includegraphics[width=12pc,height=12pc]{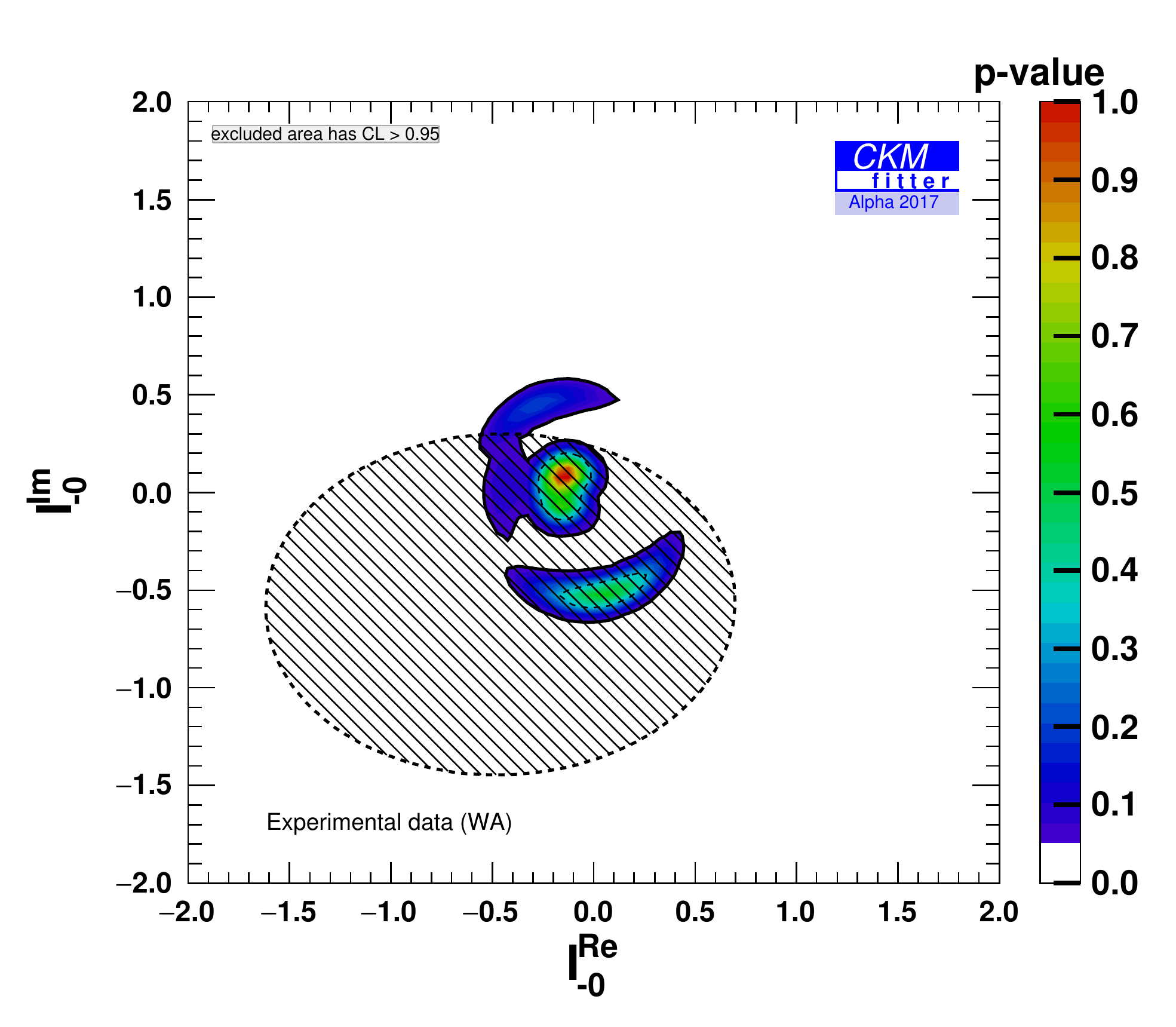}
\caption{\it\small  Two-dimensional \su{2} isospin constraint on the real and imaginary parts of interference-related coefficients \U and \I describing the \decay{B^0}{\pi^{+}\pi^{-}\pi^{0}} Dalitz decay. For each observable, the direct measurement, not included in the fit, is indicated by the shaded area. The largest tension is observed in the uppermost left figure related to the $B^0\to\rho^\pm\pi^\mp$ amplitude combination $(A^+A^{-^*}+ {\bar A}^+{\bar A}^{-^*})$.}\label{fig:UI_interf_RhoPi}
\end{center}       
\end{figure}

We can separate the coefficients related to the dynamics of the $\rho\pi$ intermediate state, $\vec{\cal O}_{\rho\pi}=(\U^+_i, \U^-_i, \I^i)_{(i=+,0,-)}$ (quasi-two-body terms or Q2B), from the parameters describing the interference pattern, $\vec{\cal O}_{\rm interf}=(\U^{\pm,\Re(\Im)}_i,\I^{\Re(\Im)}_{ij})$ (interference terms), as can be seen in Eq.~(\ref{eq:UandI}).
 The indirect determination of most of the Q2B coefficients, $\U^i_j$ and $\I_j$, related to the  \decay{B^0}{\rho^\pm\pi^\mp} intermediate states, deviates from the direct measurement by more than 2$\sigma$. This is also the case for the corresponding interference terms  $\U^{+\Re/\Im}_{ij}$, which represent the real and imaginary parts
of the combination of $B^0\to\rho^\pm\pi^\mp$ amplitudes, $(A^+A^{-^*}+ {\bar A}^+{\bar A}^{-^*})$.  The largest deviation is observed for the  $U^{+\Re}_{+-}$ coefficient which exceeds 3$\sigma$. Figs.~\ref{fig:UI_Q2B_RhoPi} and \ref{fig:UI_interf_RhoPi} display the indirect determinations of the form-factor bilinear coefficients for quasi-two-body  and interference coefficients, respectively.


In order to understand these discrepancies, we can rely on two parametrisations focusing on  the dynamics of the $\rho\pi$ intermediate state through Q2B observables. A first,  ``theoretical'', parametrisation was proposed in Ref.~\cite{LeDiberder:1991q2b}.
The time-dependent $CP$ asymmetry given by Eq.~(\ref{eq:acp}) is written independently for the three neutral \decay{B^0}{\rho^i\pi^j} intermediate states:
\begin{eqnarray}
a^{+}_{CP}(t)&=&\frac{\Gamma({\bar B}^0(t)\to\rho^+\pi^-) - \Gamma({B}^0(t)\to\rho^+\pi^-) }{\Gamma({\bar B}^0(t)\to\rho^+\pi^-) + \Gamma({B}^0(t)\to\rho^+\pi^-)}=\S^{+}\sin(\Delta m_d t)-\C^{+}\cos(\Delta m_d t)\,,\nonumber\\
a^{-}_{CP}(t)&=&\frac{\Gamma({\bar B}^0(t)\to\rho^-\pi^+) - \Gamma({B}^0(t)\to\rho^-\pi^+) }{\Gamma({\bar B}^0(t)\to\rho^-\pi^+) + \Gamma({B}^0(t)\to\rho^-\pi^+)}=\S^{-}\sin(\Delta m_d t)-\C^{-}\cos(\Delta m_d t)\,,\nonumber\\
a^{0}_{CP}(t)&=&\frac{\Gamma({\bar B}^0(t)\to\rho^0\pi^0) - \Gamma({B}^0(t)\to\rho^0\pi^0) }{\Gamma({\bar B}^0(t)\to\rho^0\pi^0) + \Gamma({B}^0(t)\to\rho^0\pi^0)}=\S^{0}\sin(\Delta m_d t)-\C^{0}\cos(\Delta m_d t)\,,
\end{eqnarray}
where $\C^{i}$  and $\S^{i}$ are direct and mixing-induced $CP$ asymmetries, respectively, and the subscript $i$ refers to the charge of the emitted $\rho^i$ meson.  

The $\rho^\pm\pi^\mp$ final state being a $CP$ admixture, the parameters $\C^{\pm}$ are not the only measurements of direct $CP$ violation.
Considering either the decay where the $\rho^\pm$ meson is emitted by the spectator interaction or by the $W$ exchange, we can use one of the following 
time-integrated flavour-independent asymmetries:
\begin{eqnarray}\label{eq:Apmtheo1}
A^{-}&=&\frac{|\A({\bar B}^0\to\rho^+\pi^-)|^2 - |\A({B}^0\to\rho^-\pi^+)|^2 }{|\A({\bar B}^0\to\rho^+\pi^-)|^2 + |\A({B}^0\to\rho^-\pi^+)|^2}\,,\nonumber\\
A^{+}&=&\frac{|\A({\bar B}^0\to\rho^-\pi^+)|^2 - |\A({B}^0\to\rho^+\pi^-)|^2 }{|\A({\bar B}^0\to\rho^-\pi^+)|^2 + |\A({B}^0\to\rho^+\pi^-)|^2}\,,
\label{eq:Apmtheo2}
\end{eqnarray}
to constrain the $B^0\to\rho^\pm\pi^\mp$ amplitude system completely.

\begin{figure}[hbtp]
\begin{center}
  \includegraphics[width=18pc]{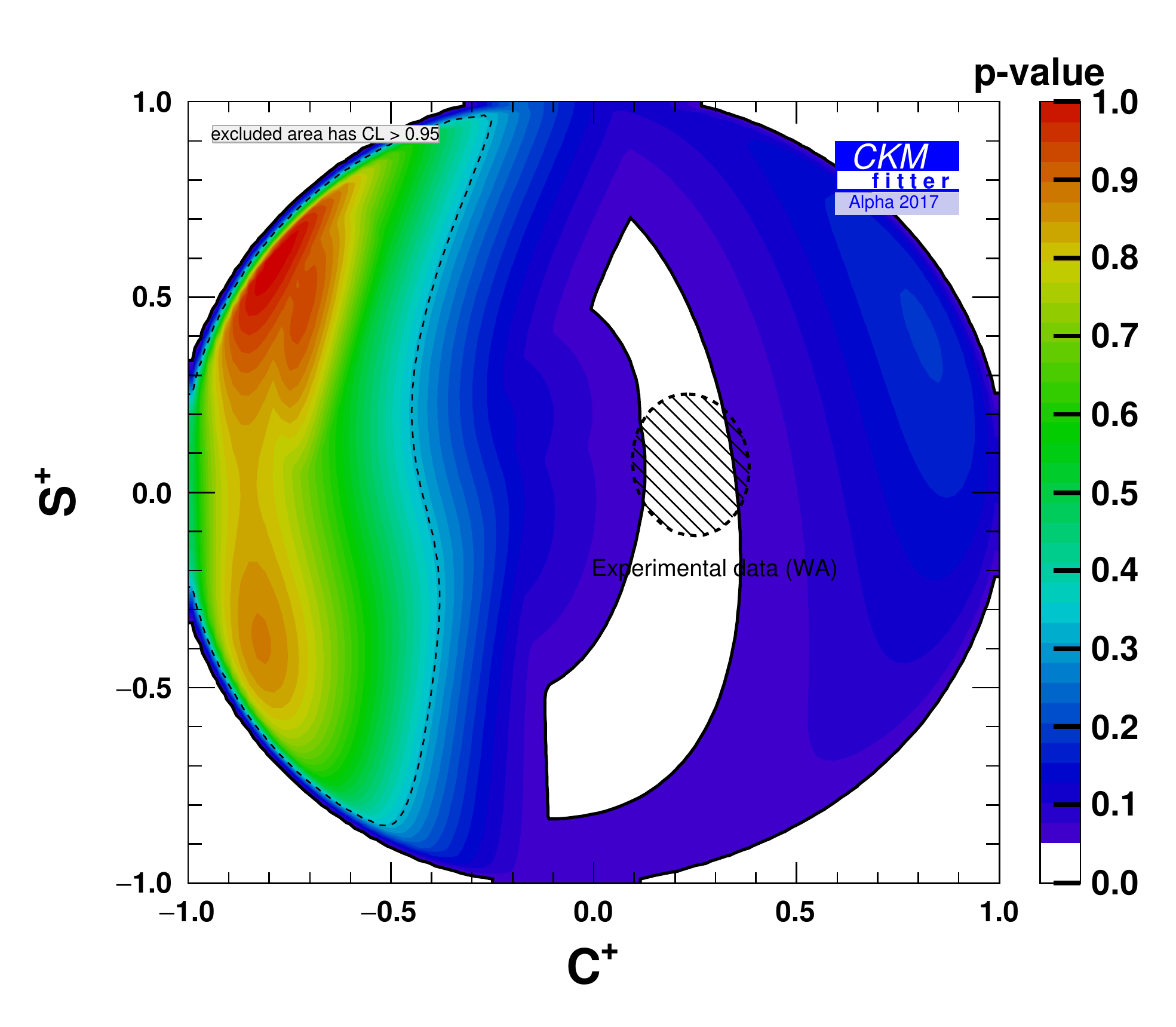}
  \includegraphics[width=18pc]{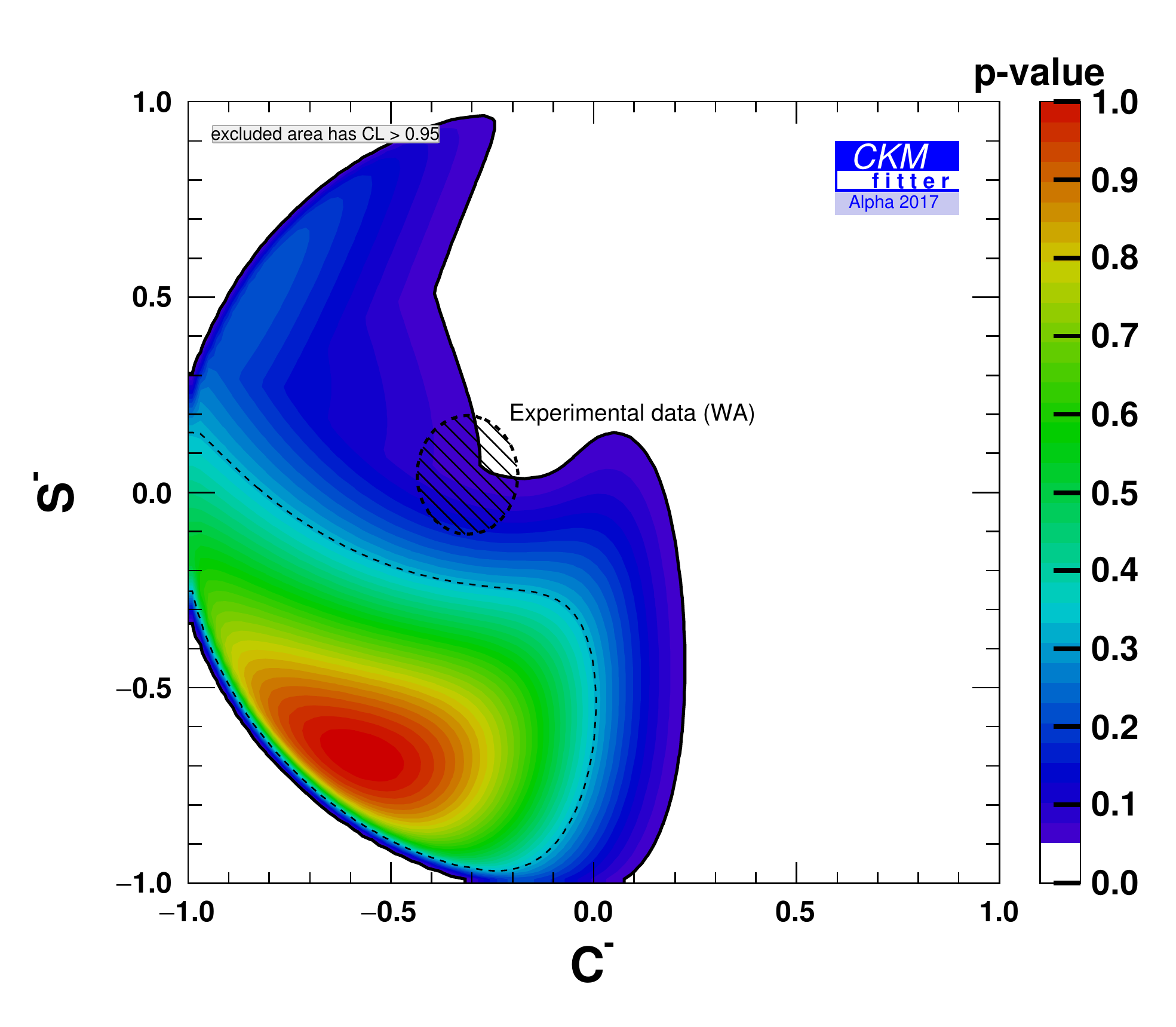}
  \includegraphics[width=18pc]{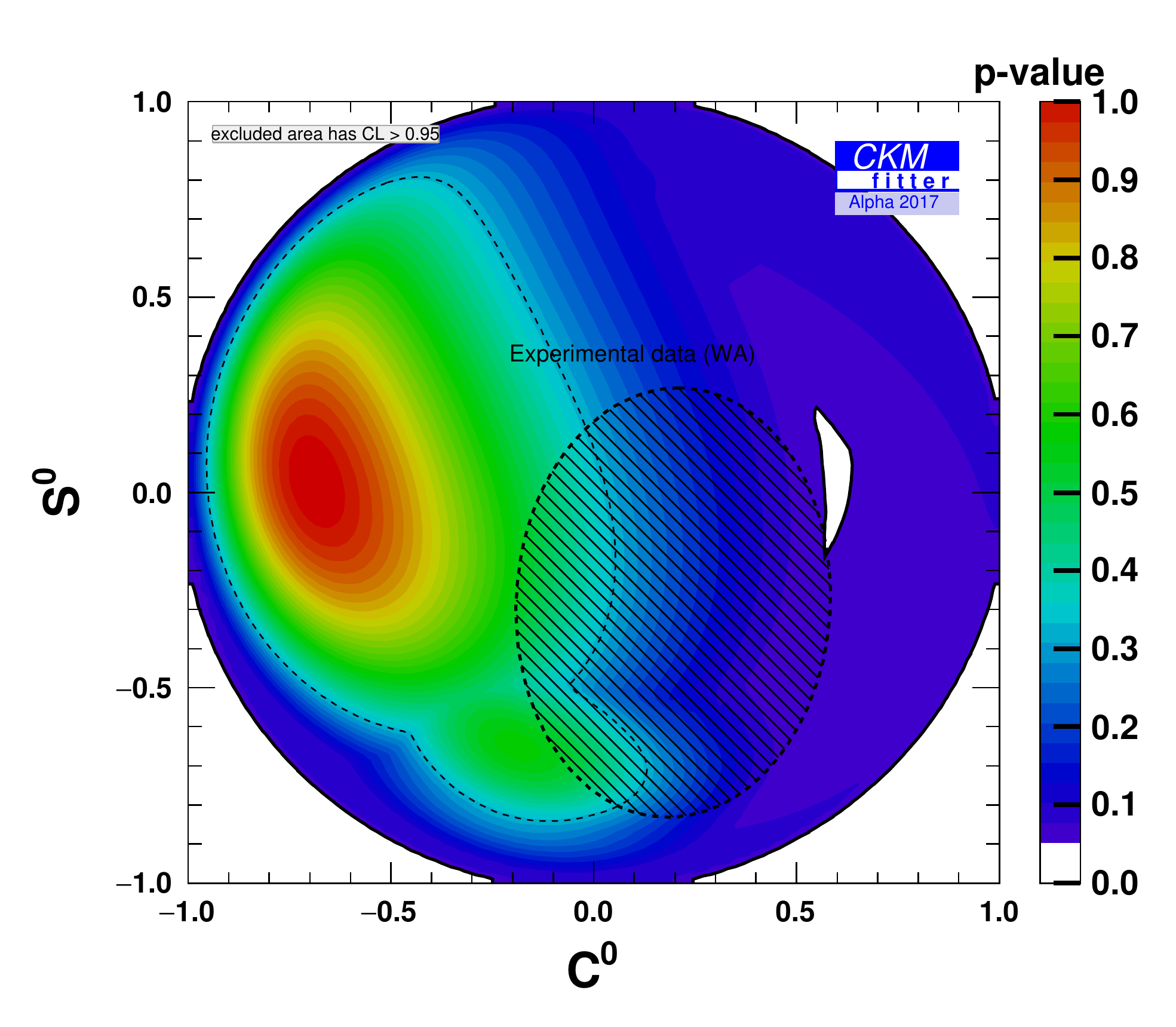}
  \includegraphics[width=18pc]{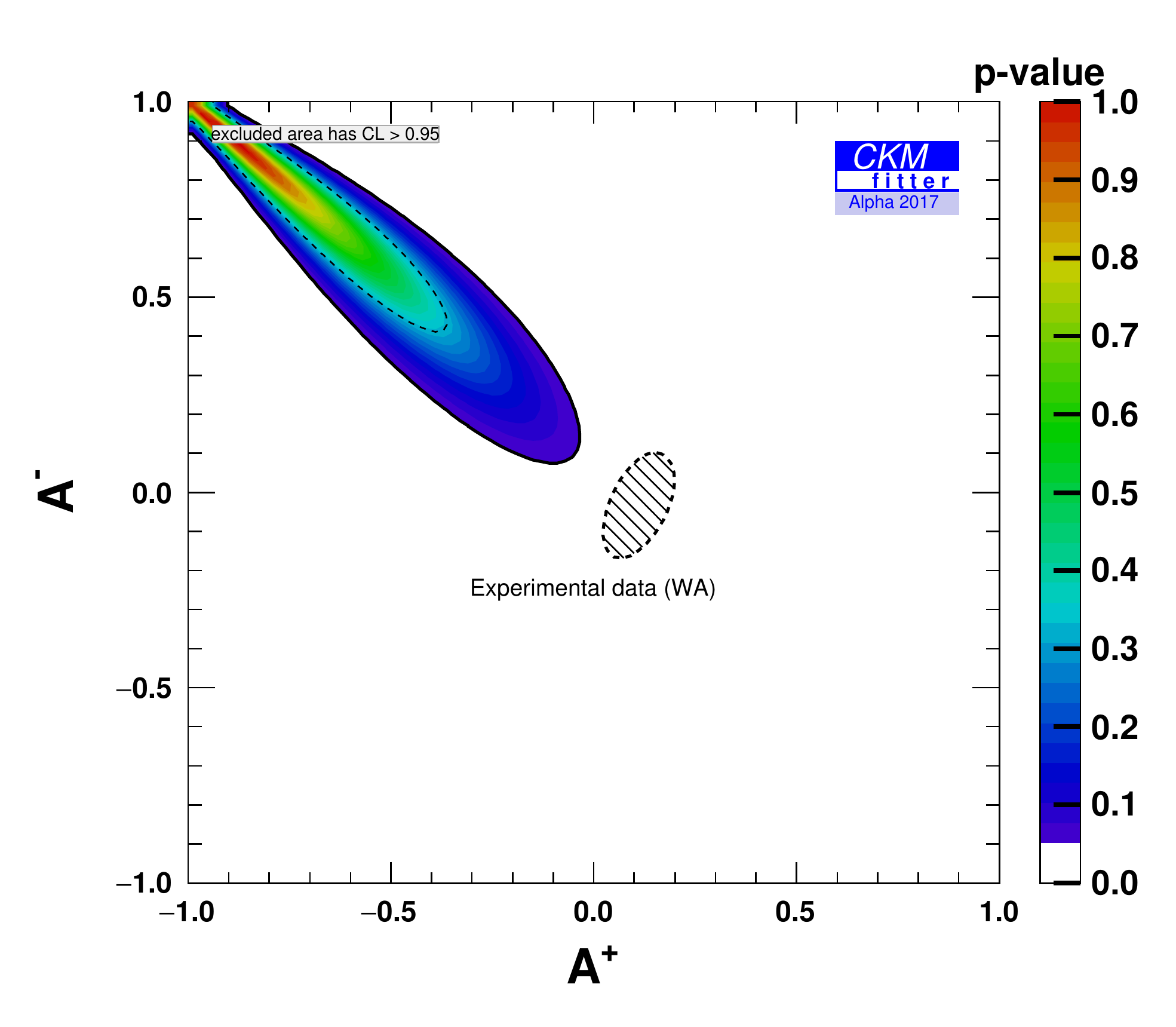}
\caption{\it\small  Two-dimensional constraints on the flavour-dependent, $(\C^i$,$\S^i)_{i=+,0,-}$ (top), and flavour-independent ($A^{+}$, $A^{-}$) (bottom) asymmetries. In each case, the   experimental value derived from the related \U or \I coefficients and not included in the fit is indicated as a shaded area.}
\label{fig:Q2B1_rhopi}
\end{center}       
\end{figure}

\begin{figure}[t]
\begin{center}
  \includegraphics[width=18pc]{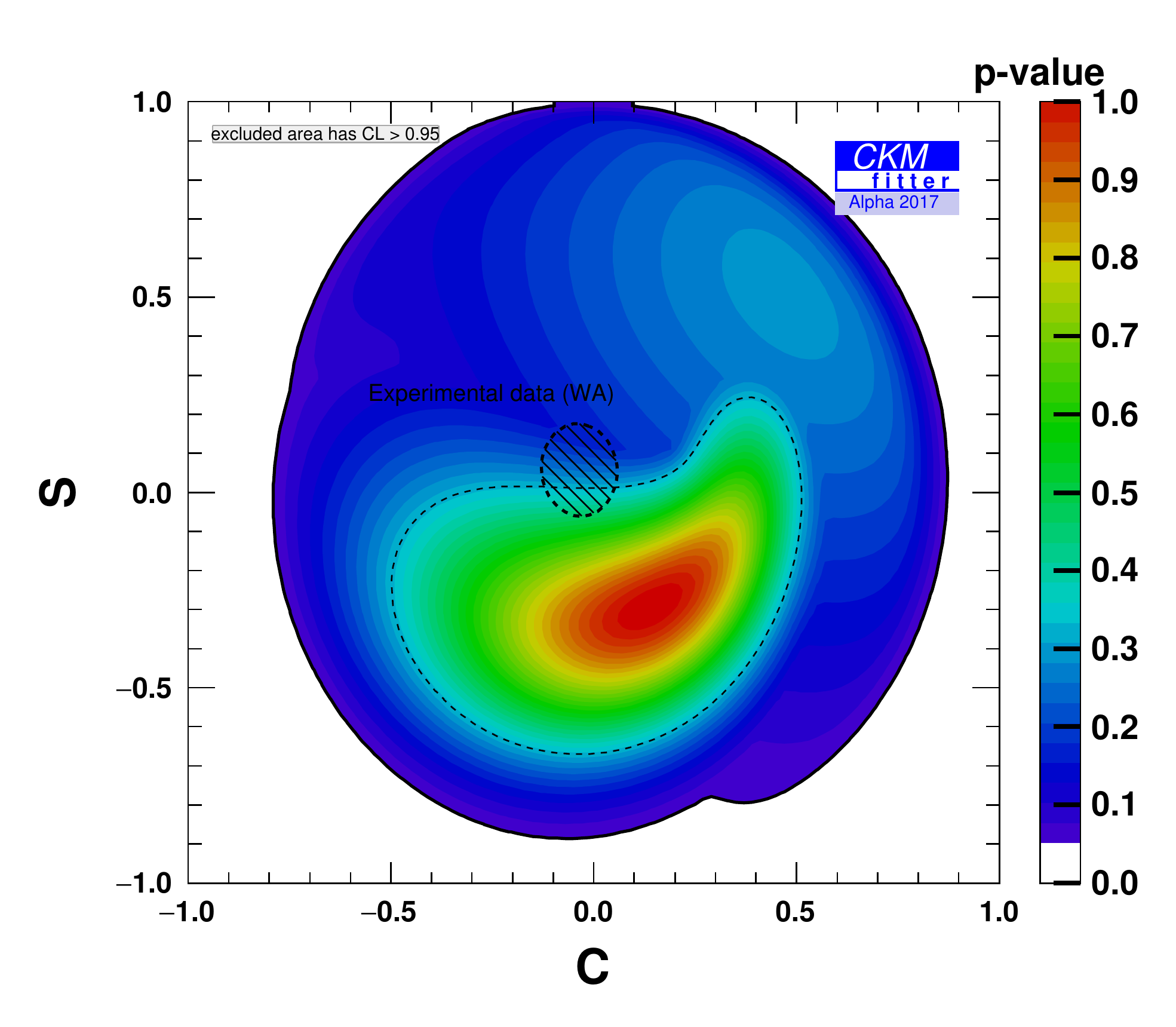}
  \includegraphics[width=18pc]{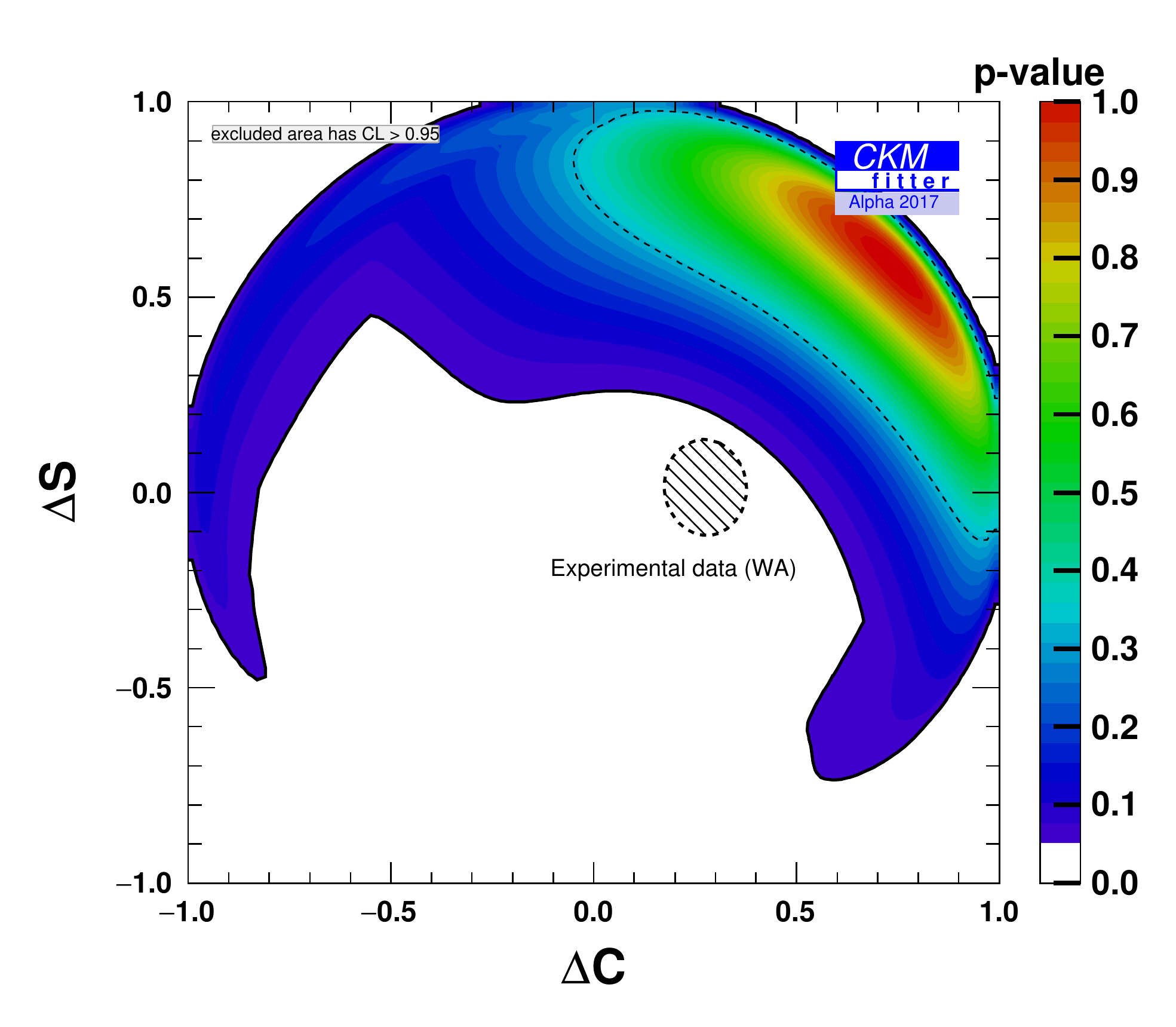}
  \includegraphics[width=18pc]{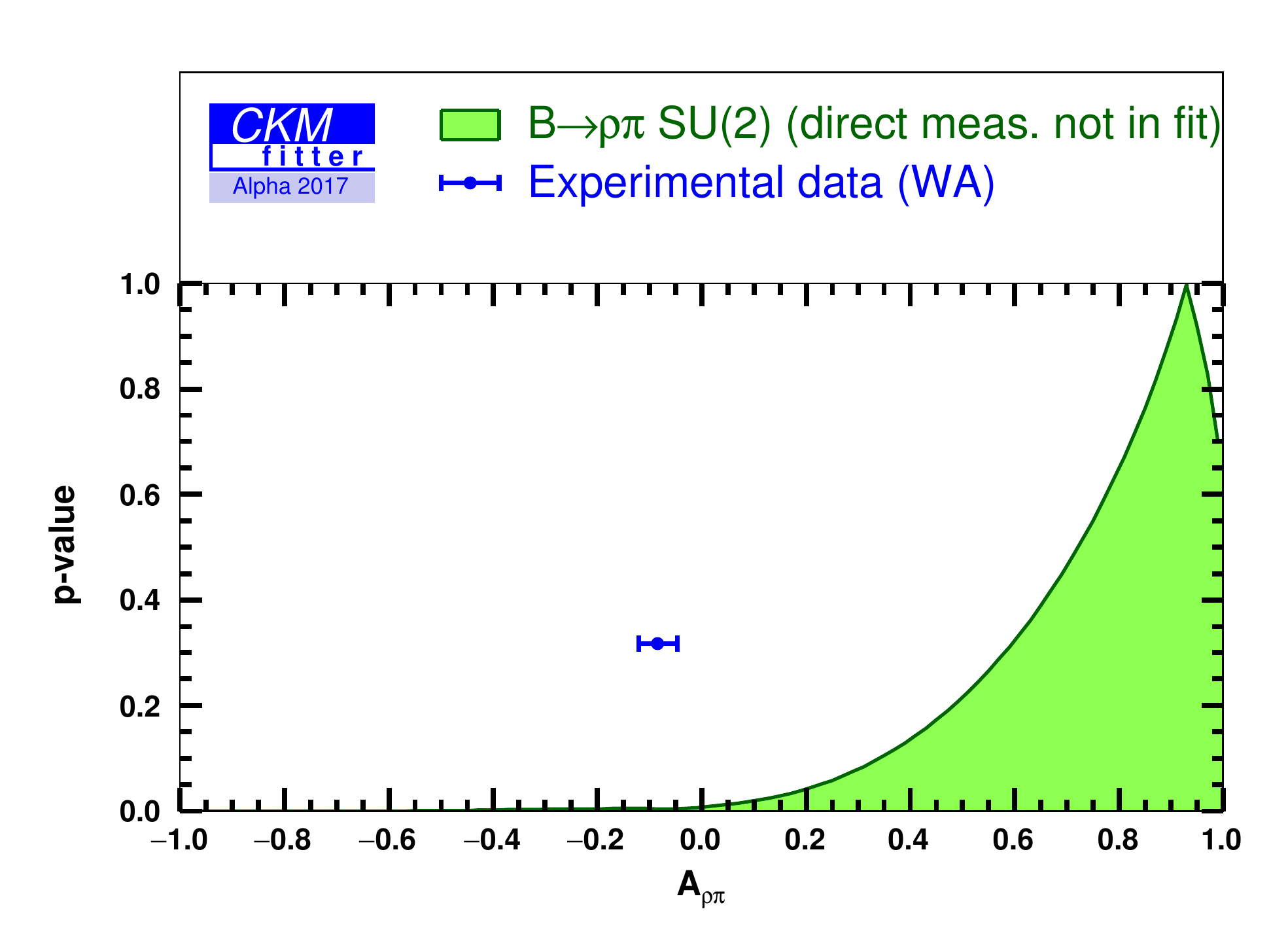}
  \includegraphics[width=18pc]{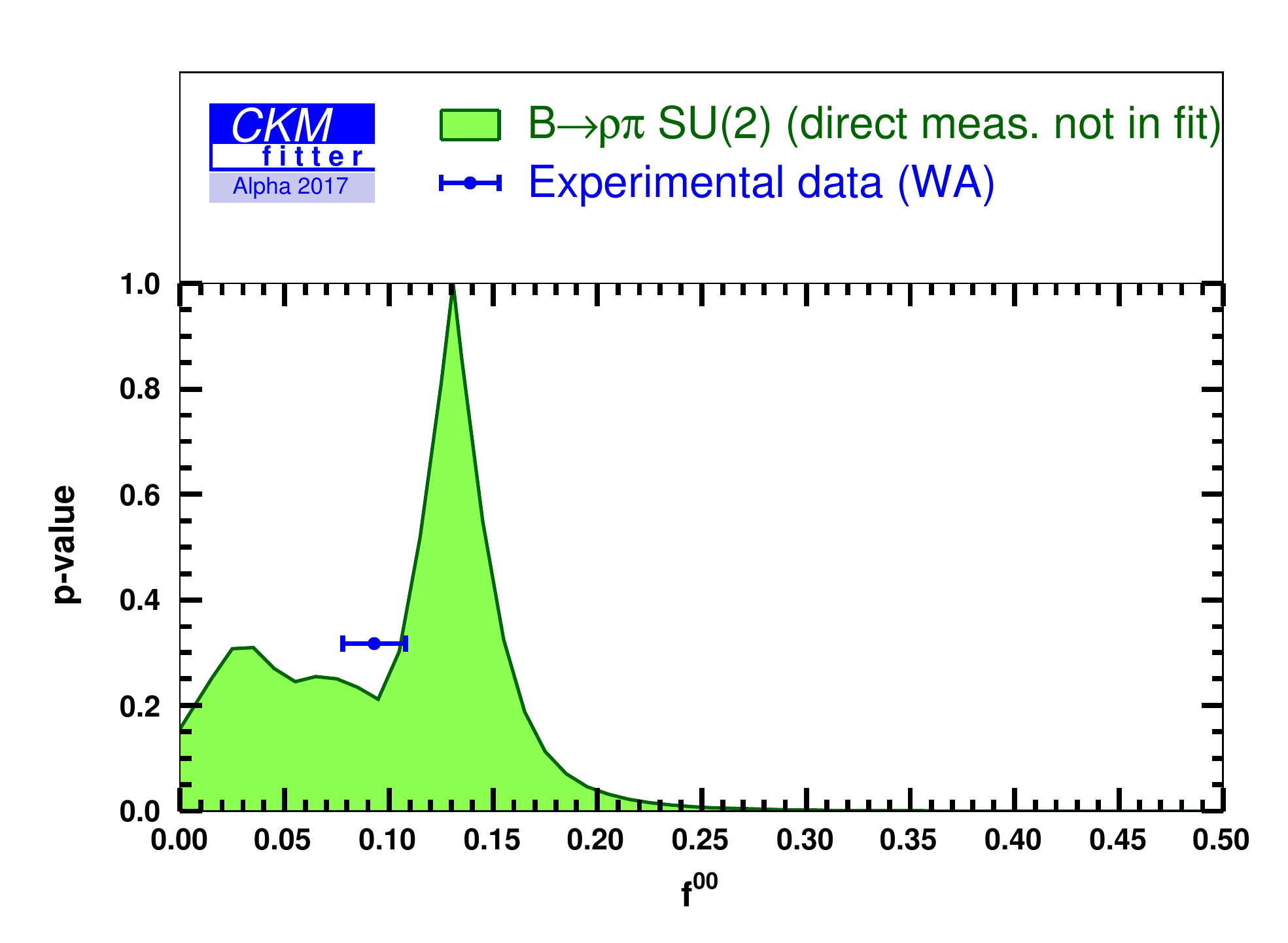}
\caption{\it\small  Two-dimensional constraints on the \decay{B^0}{\rho^\pm\pi^\mp} $CP$-violating parameters   ($\C_{\rho\pi}$,$\S_{\rho\pi}$) and $CP$-conserving parameters  ($\Delta\C_{\rho\pi}$,$\Delta\S_{\rho\pi}$)  (top) and one-dimensional constraints on the flavour-integrated charge asymmetry $\A_{\rho\pi}$ and the relative $B^0\to\rho^0\pi^0$ branching fraction $f^{00}$ (bottom). In each case, the corresponding experimental value derived from the related \cal U or \I coefficients, not included in each individual fit, is indicated either by a shaded area or an interval with a dot.}
\label{fig:Q2B2_rhopi}
\end{center}       
\end{figure}

 Due to the overlapping final states in the Dalitz analysis, the separation of the three intermediate states is experimentally challenging and prevents a clean direct measurement of the above Q2B asymmetries. The previous ``theoretical'' set of observables can be traded for a second,  ``experimental'', parametrisation in order to avoid this problem. Focusing only on the region of the charged $\rho$ meson  in the Dalitz plane and neglecting the interferences and the residual contribution from the colour-suppressed \decay{B^0}{\rho^0\pi^0} mode, 
a specific time-dependent measurement of the $CP$-mixed  \decay{B^0}{\rho^\pm\pi^\mp} intermediate state was first performed at $B$-factories in 2005 \cite{Babar_rp_pm,Belle_rp_q2b}. 
Introducing the flavour-integrated charge asymmetry:
\begin{equation}
A_{\rho\pi}=\frac{|\A({B}^0\to\rho^+\pi^-)|^2 + |\A({\bar B}^0\to\rho^+\pi^-)|^2 - |\A({B}^0\to\rho^-\pi^+)|^2 - |\A({\bar B}^0\to\rho^-\pi^+)|^2}{|\A({B}^0\to\rho^+\pi^-)|^2 + |\A({\bar B}^0\to\rho^+\pi^-)|^2 + |\A({B}^0\to\rho^-\pi^+)|^2 + |\A({\bar B}^0\to\rho^-\pi^+)|^2},
\end{equation}
the $B^0(t)\to\rho^\pm\pi^\mp$ decay rate is written as
\begin{eqnarray}
\Gamma(B^0(t)\to\rho^\pm\pi^\mp       )=(1+q_{\rho} A_{\rho\pi})e^{-\frac{t}{\tau_{B^0}}}(1&-&q_{B}(\S+q_{\rho}\Delta\S)\sin(\Delta m_d t)\nonumber\\
&+&q_{B}(\C+q_{\rho}\Delta\C)\cos(\Delta m_d t)),
\end{eqnarray}
where $q_B=+1 (-1)$ for $B^0({\bar B}^0)$ and $q_{\rho}$ is the electric charge of the emitted $\rho$-meson. The  $CP$-violating ($CP$-conserving) parameters $\C$ and $\S$ ($\Delta\C$ and $\Delta\S$) generate the flavour-dependent asymmetries:
\begin{equation}
 \C^{\pm}= \C \pm \Delta\C\,,\qquad
 \S^{\pm}= \S \pm \Delta\S\,,
 \end{equation}
while the flavour-independent $CP$ asymmetries defined in Eq.~(\ref{eq:Apmtheo2}) can be rewritten in terms of the experimental parameters as:
\begin{equation}\label{eq:Apmexp}
A^{+}=-\frac{A_{\rho\pi}+C+A_{\rho\pi}\Delta C}{1+\Delta C+A_{\rho\pi}C},\qquad
A^{-}=-\frac{A_{\rho\pi}-C-A_{\rho\pi}\Delta C}{A_{\rho\pi}-\Delta C-A_{\rho\pi}C}\,.
\end{equation}
In order to remove experimental biases due to the imperfect separation of the intermediate states, one can derive the Q2B parameters from the subset  $\vec{\cal O}_{\rho\pi}$ of the form-factor coefficients:
\begin{equation}
\C^i=\frac{\U^-_i}{\U^+_i}, \quad \S^i=\frac{2\I^i}{\U^+_i},\quad
A^\pm = \frac{ \pm(\U{-}{+}-\U{+}{+}) - (\U{-}{-}+\U{+}{-}) }{\pm(\U{+}{-}-\U{-}{-})+( \U{-}{+}+\U{+}{+}) },\quad
A_{\rho\pi} = \frac{ \U{+}{+} -\U{-}{+}}{\U{+}{+}+\U{-}{+}}\,.
\end{equation}

Finally, we can introduce the relative contribution of the colour-suppressed \decay{B^0}{\rho^0\pi^0} mode to the overall \decay{B^0}{(\rho\pi)^0} branching fraction:
\begin{equation}
f^{00}=\frac{\U^+_0}{\U^+_0 + \U^+_- + \U^+_+}.
\end{equation}

We can then reparametrise the $B^0\to\pi^+\pi^-\pi^0$ amplitude system in terms of a selected set of \U, \I and Q2B observables: on hand,  $\vec{\cal O}_{\rm interf}$, $\vec{\cal O}_{\rho^0\pi^0}=\{\C^0,S^0,f^{00}\}$ and on the other hand, either 
the theoretically motivated  set $\vec{\cal O}_{\rho^\pm\pi^\mp}=\{\C^\pm,S^\pm,A\}$ or the experimentally convenient choice $\vec{\cal O}_{\rho^\pm\pi^\mp}=\{C,S,\Delta C,\Delta S,A\}$ (where $A$ stands for $A^+$, $A^-$ or $A_{\rho\pi}$).
The world averages of the Q2B parameters derived from the measured \U and \I coefficients are  listed in Tab.~\ref{tab:Q2B_rhopi} in App.~\ref{sec:numerics}, together with their corresponding prediction in the \su{2} isospin framework. These results are also shown in Figs.~\ref{fig:Q2B1_rhopi} and \ref{fig:Q2B2_rhopi}.  

A reasonable agreement is observed for the parameters related to the $B^0\to\rho^0\pi^0$ mode.
The \su{2} isospin fit favours  large values for both direct and mixing-induced $CP$ asymmetry parameters in the $B^0\to\rho^\pm\pi^\mp$ decays, in contrast with  experimental data. The overall 3~$\sigma$ discrepancy is mostly reflected by the correlated flavour-independent $CP$ asymmetries, $A^\pm$, or equivalently the charge asymmetry, $A_{\rho\pi}$.
While the predicted $CP$-violating asymmetry averages, \C and \S, are in a reasonable agreement with the experimental data, the $CP$-conserving terms, $\Delta\C$ and $\Delta\S$, deviate by more than 2.5~$\sigma$.

\subsubsection{Role of the strong phases in the $B^0\to (\rho\pi)^0$ analysis\label{sec:rhopiphase}}

In order to understand this discrepancy in more detail, the effective weak angles associated with the $B^0\to\rho^+\pi^-$ and $B^0\to\rho^-\pi^+$ decays,
\begin{equation}
2\alpha^+={\rm Arg}\left[\frac{q}{p}\frac{\A({\bar B}^0\to\rho^-\pi^+)}{\A({B}^0\to\rho^+\pi^-)}\right]~~ {\rm and} ~~  2\alpha^-={\rm Arg}\left[\frac{q}{p}\frac{\A({\bar B}^0\to\rho^+\pi^-)}{\A({B}^0\to\rho^-\pi^+)}\right]\,,
\end{equation}
 constitute further  interesting quantities as they both reduce to  $\alpha$  in the limit of a vanishing penguin contribution. 
The time-independent Q2B observables in the  $B^0\to\rho^\pm\pi^\mp$  decays do not lead to a determination of each effective mixing angle: only the average $\alpha_{\rm eff}=(\alpha^+ + \alpha^-)/2$ can be measured  up to a fourfold ambiguity in the $[0,90]^\circ$ range \cite{Zupan:2004hv}. Indeed, the Q2B observables
can be combined to measure the two shifted phases:
\begin{equation}
2\alpha^\pm\pm2\delta={\rm Arg}\left[\frac{q}{p}\frac{\A({\bar B}^0\to\rho^\pm\pi^\mp)}{\A({B}^0\to\rho^\pm\pi^\mp)}\right]={\rm arcsin}\left[ \frac{\S\mp\Delta\S}{\sqrt{1-(\C\mp\Delta\C)^2}}\right]\,,
\end{equation}
where the phase shift:
\begin{equation}
2\delta = {\rm Arg}\left[\frac{\A({B}^0\to\rho^+\pi^-)}{\A({B}^0\to\rho^-\pi^+)}\right]
\end{equation}
 cannot be determined from the Q2B data alone but cancels in the average $\alpha_{\rm eff}$.
A bound on the deviation $\Delta\alpha$ between the effective angle $\alpha_{\rm eff}$ and the weak phase $\alpha$  can be derived \cite{Charles:2004jd,CharlesPhD,Quinn:2000by} from the penguin relation Eq.~(\ref{eq:Prhopi1}):
\begin{equation}
|\Delta\alpha|=|\alpha-\alpha_{\rm eff}|\leq \frac{1}{2}{\rm arccos}\left[ \frac{1}{\sqrt{1-A_{\rho\pi}^2}}\left(1-4\frac{\B^{00}}{\B^{\pm\mp}}\right)   \right]\,,
\end{equation}
where $\B^{00}$ and $\B^{\pm\mp}$ denote the branching fractions of the $B^0\to\rho^0\pi^0$ and $B^0\to\rho^\pm\pi^\mp$ decays, respectively.  This relation originates with the fact that the sum of the penguin amplitudes $({P}^{+-}+{ P}^{-+})=-{ P}^{00}$ is bounded by the branching fraction of the  colour-suppressed mode
while the difference $({ P}^{+-}-{ P}^{-+})$  vanishes in relation with the charge-integrated $CP$ asymmetry $A_{\rho\pi}$ (see also the discussion of the Q2B analysis of the charmless $B^0\to a_1^\pm\pi^\mp$ decay in Appendix~\ref{sec:a1pi}).

\begin{figure}[t]
\begin{center}
  \includegraphics[width=21pc]{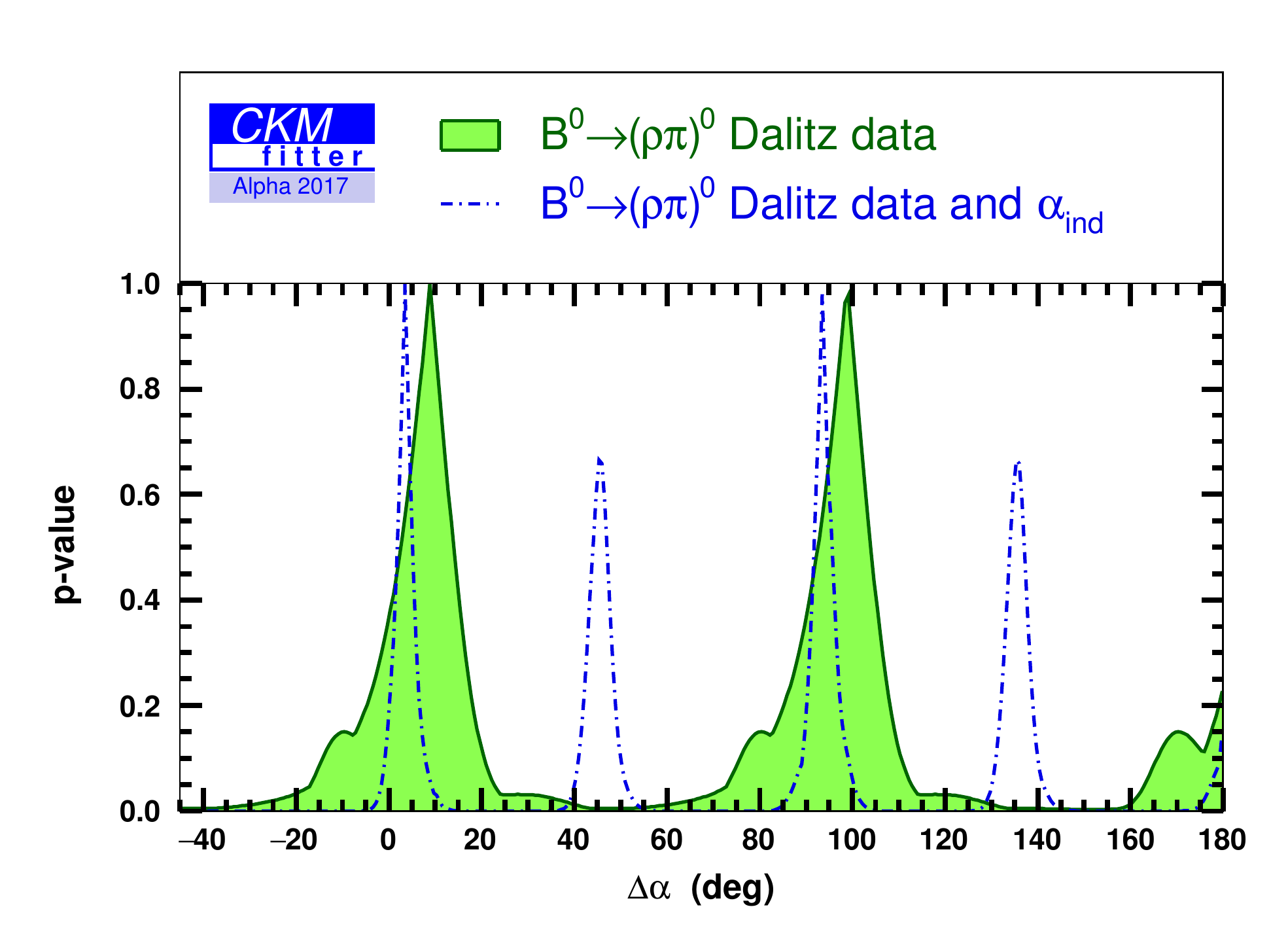}
\caption{\it\small  One-dimensional scan of  $\Delta\alpha=(\alpha-\alpha_{\rm eff})$ from the isospin analysis of  $B^0\to(\rho\pi)^0$ Dalitz data. The dashed curve 
includes  the indirect determination $\alpha_{\rm ind}$ as additional constraint. }
\label{fig:rhopi_deltaalpha}
\end{center}       
\end{figure}

\begin{figure}[t]
\begin{center}
  \includegraphics[width=18pc]{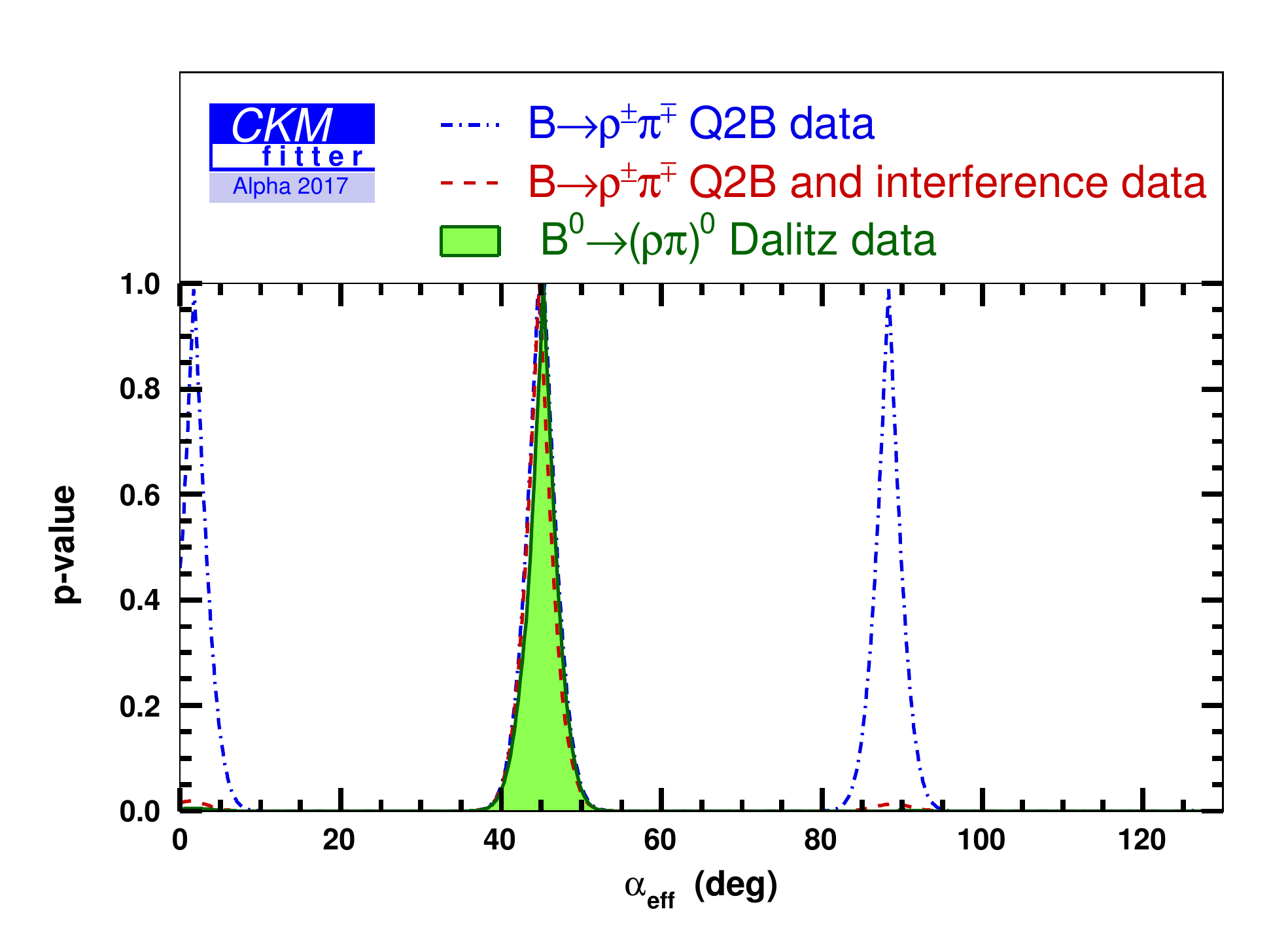}
  \includegraphics[width=18pc,height=14pc]{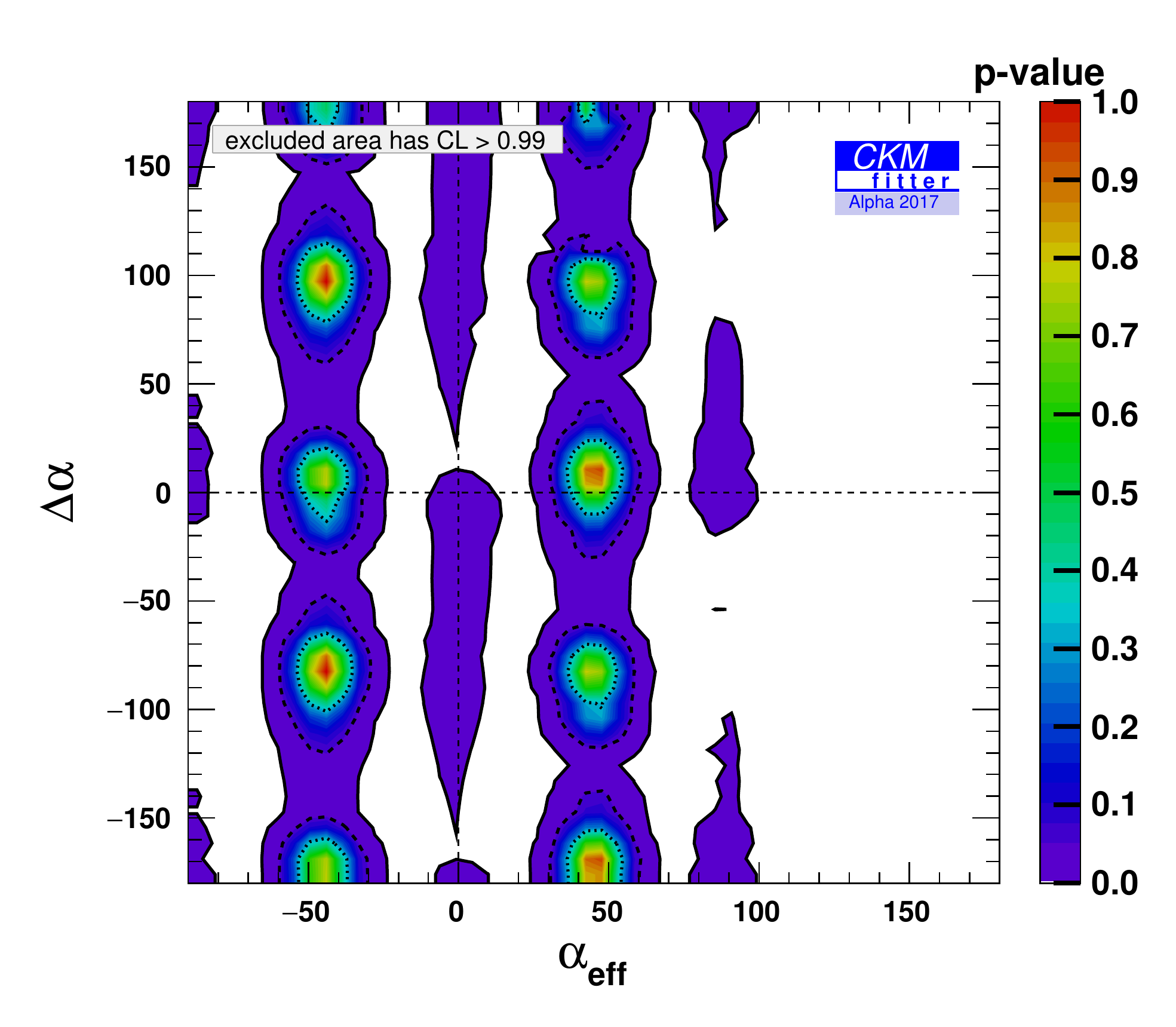}
\caption{\it\small  On the left: one-dimensional scan of the effective angle average $\alpha_{\rm eff}$ using  $B^0\to(\rho\pi)^0$ Dalitz data. The dot-dashed curve is obtained 
when using only the $B^0\to\rho^\pm\pi^\mp$ quasi-two-body observables, $\{\U^+_-,\U^-_-,\U^+_+,\U^-_+,\I_+,\I_-\}$. The dashed curve includes also the interference-related observables, $\{\U^{\pm\Re(\Im)}_{+-},\I_{+-}^{\Re(\Im)}\}$. On the right: two-dimensional constraint in the ($\alpha_{\rm eff}$,$\Delta\alpha$) plane obtained using the whole \U ~and \I observables. Contrary to other similar figures in this article (where the excluded region is above 2 $\sigma$ or 95\% CL), the excluded region corresponds here to a confidence level above 3$\sigma$ (99\% CL) in order to show the suppressed solution around $(90^\circ, 0^\circ)$.}
\label{fig:rhopi_alphaeff}
\end{center}       
\end{figure}

The constraint on $\Delta\alpha$ given in Fig.~\ref{fig:rhopi_deltaalpha} features 
a solution consistent with 0:
\begin{equation}\label{eq:DeltaAlpha}
|\Delta\alpha|_{\rm min}=\val{8.8}{+7.4}{-9.9}{$^\circ$}\,,
\end{equation}
and a mirror solution around $100^\circ$. Including the indirect determination $\alpha_{\rm ind}$, Eq.~(\ref{eq:alphaInd}), as an additional input, a tighter constraint is obtained, 
$|\Delta\alpha|_{\rm min}=$\val{3.5}{+2.4}{-2.6}{$^\circ$}, and a new mirror solution, disfavoured by the $B^0\to(\rho\pi)^0$ data, emerges around $45^\circ$.
As shown in Fig.~\ref{fig:rhopi_alphaeff},  when using only  the Q2B data set $\{\U^i_j,\I_j\}_{i,j=(+,-)}$, four solutions  are found for $\alpha_{\rm eff}$ around $0^\circ$, $45^\circ$ and $90^\circ$ (the peak near $45^\circ$ combines two quasi-degenerate solutions).  The solution  close to $90^\circ$ is consistent with the indirect determination of $\alpha$ Eq.~(\ref{eq:alphaInd}). However, only the solution for $\alpha_{\rm eff}$  near 45${}^\circ$ remains once the Q2B observables and the interference-related constraints for $B^0\to \rho^\pm\pi^\mp$ are considered altogether. The solutions close to $0^\circ$ and 90${}^\circ$ are disfavoured at  2.5$\sigma$. The rejection of these  solutions is slightly reinforced with the addition of  the   observables related to $B^0\to\rho^0\pi^0$.

\begin{figure}[t]
\begin{center}
  \includegraphics[width=24pc]{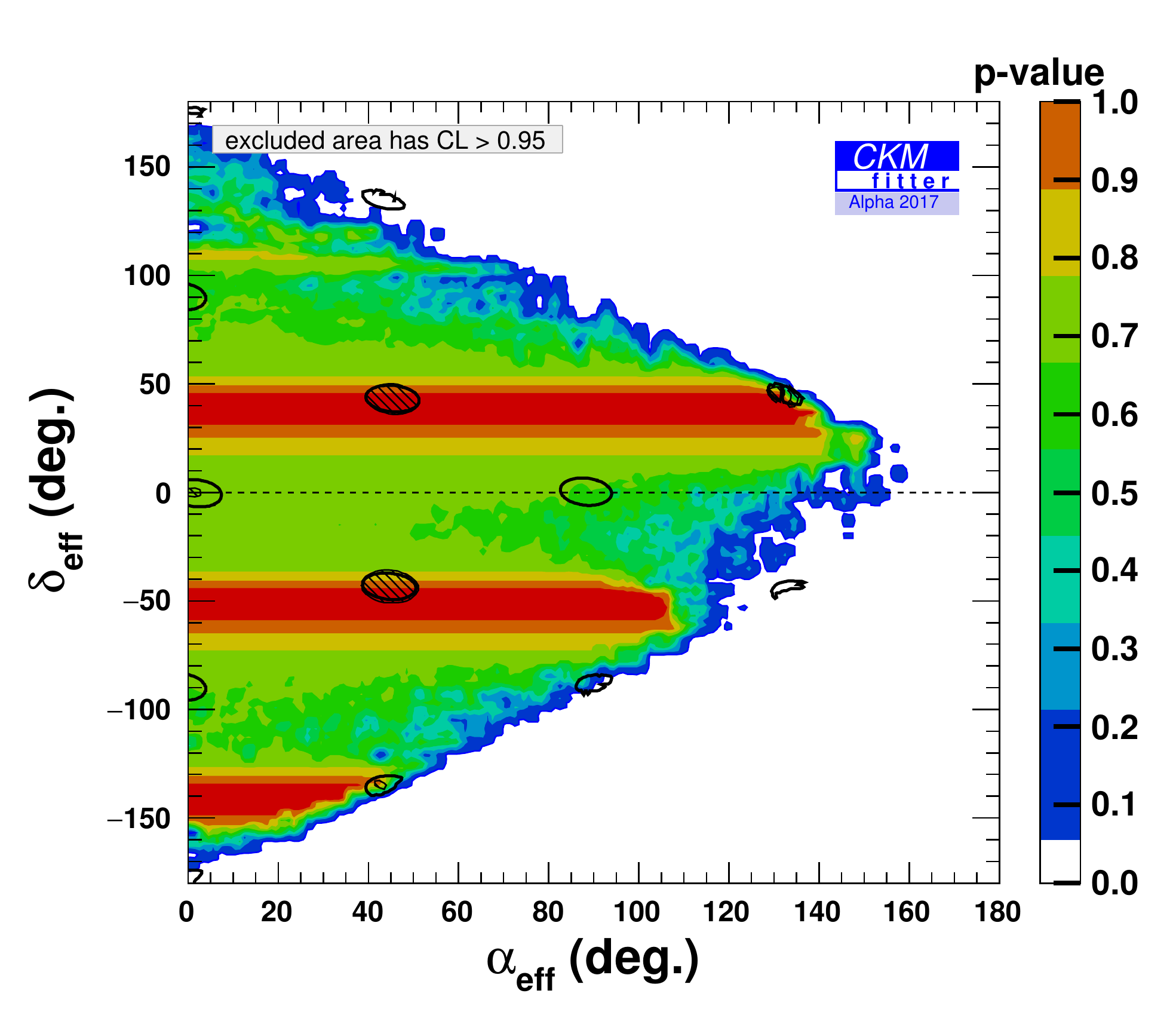}
\caption{\it\small  Two-dimensional constraint  in the ($\alpha_{\rm eff}$,$\delta_{\rm eff}$) plane. The coloured area is obtained from   the interference-related 
coefficients  $\U^{\pm\Im(\Re)}_{+-}$  only. The black ellipses come from the Q2B coefficients $\{\U^i_j,\I_j\}_{i,j=(+,-)}$. The shaded ellipses result from considering the whole \U and \I data set for $B^0\to (\rho\pi)^0$.}
\label{fig:rhopi_alphadeltaeff}
\end{center}       
\end{figure}

As can be expected, the interference observables provide further constraints  on the relative phases of the decay amplitudes. In particular,  one can easily show from Eq.~(\ref{eq:UandI}) that the phase shift $2\delta$, which vanishes in the absence of the penguin contribution, can be determined from the interference-related coefficients $\U^{\pm\Im(\Re)}_{+-}$:
\begin{equation}
\tan[2\delta] = \frac{ \U{+-}{+\Im} + \U{+-}{-\Im} }{\U{+-}{+\Re} + \U{+-}{-\Re}}\,.
\end{equation}
Simultaneously, the related phase between the conjugate modes:
\begin{equation}
2{\bar\delta} = 2\delta+(\alpha^+-\alpha^-) = {\rm Arg}\left[\frac{\A({\bar B}^0\to\rho^-\pi^+)}{\A({\bar B}^0\to\rho^+\pi^-)}\right]
\end{equation}
can be extracted from the same coefficients:
\begin{equation}
\tan[{2\bar\delta}] = -\frac{ \U{+-}{+\Im} - \U{+-}{-\Im} }{\U{+-}{+\Re} - \U{+-}{-\Re}}\,,
\end{equation}
meaning that the phase average $\delta_{\rm eff}=({\bar\delta}+\delta)/2$ can be extracted  independently using either the Q2B coefficients or the interference-related coefficients for $B^0\to\rho^\pm \pi^\mp$:
\begin{eqnarray}
4\delta_{\rm eff} &=& {\rm arctan}\left[\frac{ \U{+-}{+\Im} + \U{+-}{-\Im} }{\U{+-}{+\Re} + \U{+-}{-\Re}}\right]-{\rm arctan}\left[\frac{ \U{+-}{+\Im} - \U{+-}{-\Im} }{\U{+-}{+\Re} - \U{+-}{-\Re}}\right]\nonumber\\
&=& {\rm arcsin}\left[ \frac{2\I{-}}{\sqrt{\U{-}{+}{}^2-\U{-}{-}{}^2}}\right]-{\rm arcsin}\left[ \frac{2\I{+}}{\sqrt{\U{+}{+}{}^2-\U{+}{-}{}^2}}\right]\,.
\end{eqnarray}
The Q2B coefficients provide also a constraint on the average mixing angle $\alpha_{\rm eff}$:
\begin{equation}
4\alpha_{\rm eff}={\rm arcsin}\left[ \frac{2\I{-}}{\sqrt{\U{-}{+}{}^2-\U{-}{-}{}^2}}\right]+{\rm arcsin}\left[ \frac{2\I{+}}{\sqrt{\U{+}{+}{}^2-\U{+}{-}{}^2}}\right]\,.
\end{equation}
Fig.~\ref{fig:rhopi_alphadeltaeff} compares the two-dimensional constraints in the ($\alpha_{\rm eff}$,$\delta_{\rm eff}$) plane using either subset of observables.
Clearly, only the solutions for $\alpha_{\rm eff}$ near $45^\circ$ are favoured by both subsets.

\begin{figure}[t]
\begin{center}
  \includegraphics[width=18pc]{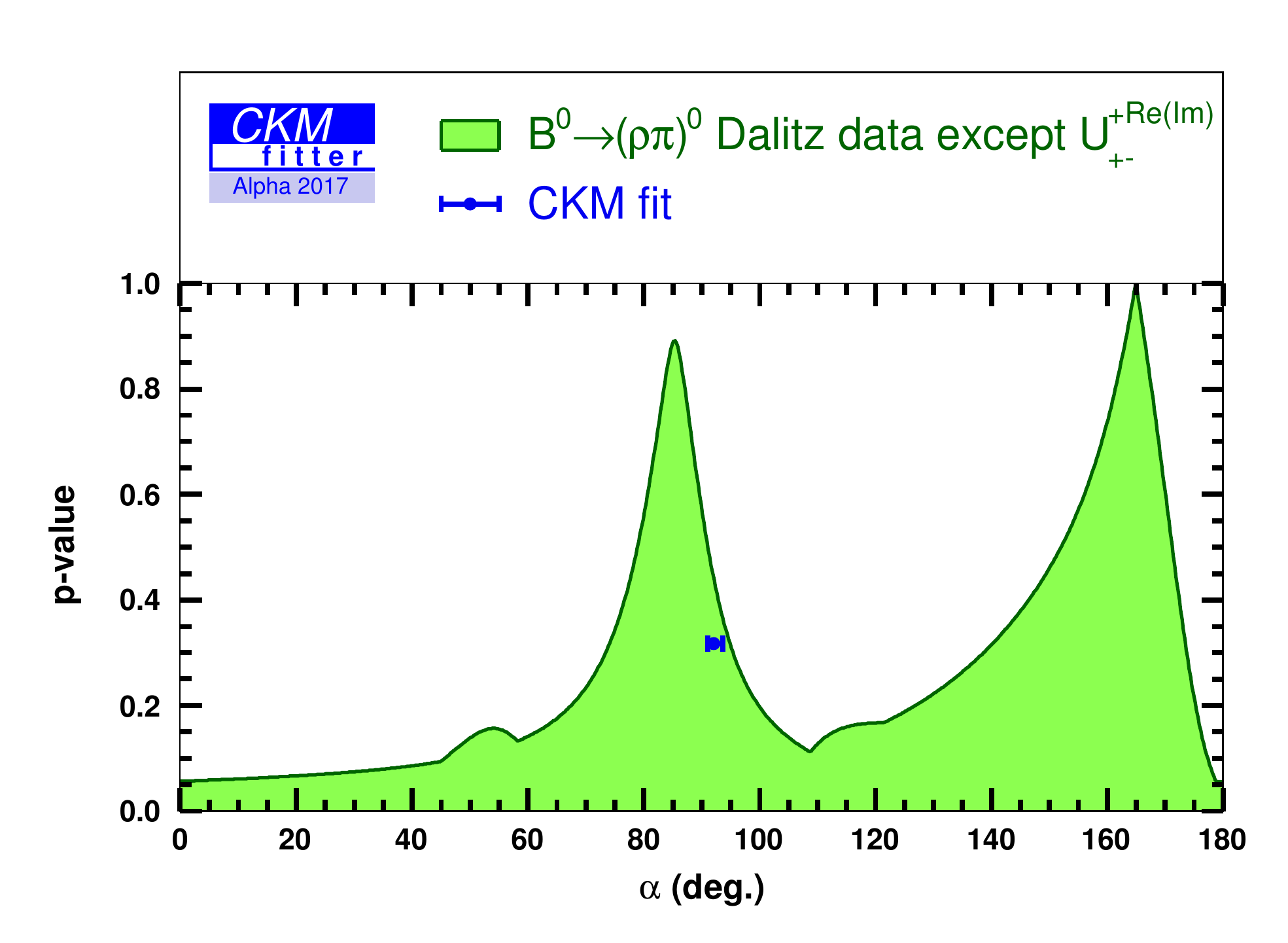}
  \includegraphics[width=18pc,height=14pc]{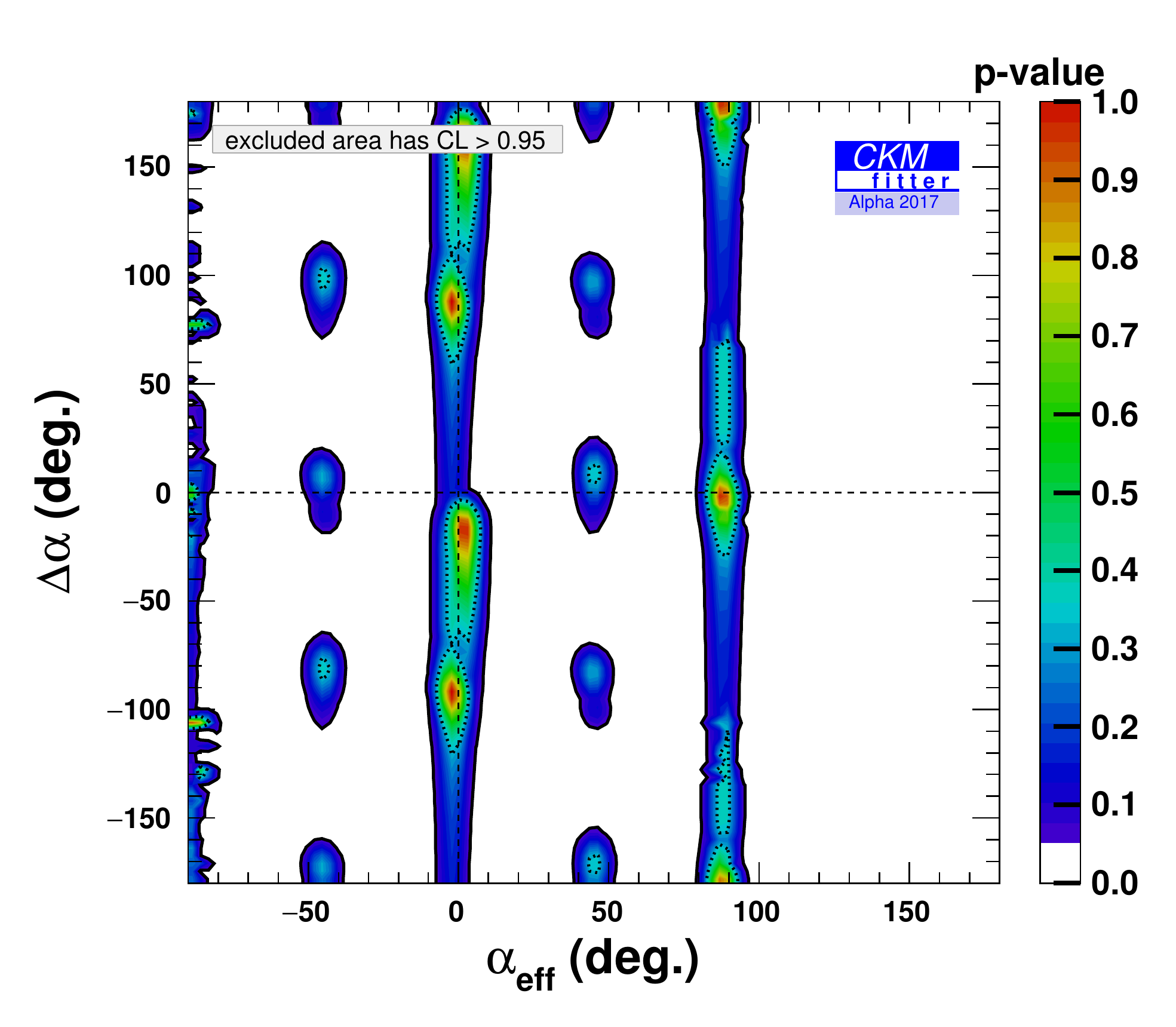}
\caption{\it\small  On the left: one-dimensional scan of the angle $\alpha$ using  the $B^0\to(\rho\pi)^0$ Dalitz data except the two coefficients  $\U^{+\Re}_{+-}$ and $\U^{+\Im}_{+-}$ that measure the real and imaginary parts of $(A^+A^{-^*}+ {\bar A}^+{\bar A}^{-^*})$ (left). The interval with a dot indicates the indirect  determination of $\alpha$ introduced in Eq.~(\ref{eq:alphaInd}). On the right: two-dimensional constraint in the ($\alpha_{\rm eff}$,$\Delta\alpha$) plane. As for other similar figures in this article, the excluded region corresponds to CLs above 2$\sigma$. The standard solution  around $(90^\circ, 0^\circ)$ is favoured with this set of observables.}
\label{fig:rhopi_noUPpm}
\end{center}       
\end{figure}

As discussed previously, the two  observables $\U^{+\Re(\Im)}_{+-}$ play an important role in the discrepancy between  the $B^0\to(\rho\pi)^0$ Dalitz data and the isospin hypothesis. It is interesting to show how removing these two observables impacts the analysis: the $\alpha_{\rm eff}$ solutions near $0^\circ$ and $90^\circ$ get favoured, and the direct determination of $\alpha$ agrees much better with the indirect value, as illustrated in Fig.~\ref{fig:rhopi_noUPpm}.
From the experimental side, the observables \U{+-}{$\pm$\Re(\Im)} and  \I{$\pm$}{\Re(\Im)}  are the  coefficients of the mixed form-factors bilinear ($f^+f^{-^*}$). Therefore, they are the only observables sensitive to a possible phase between the functionals describing the line-shape of  the $\rho$-mesons of opposite charge.  It would be very interesting to study the role played by the isobar approximation and to determine the possible re-interpretations of the experimental data, but this lies 
clearly  beyond the scope of the isospin analysis presented here, and we will refrain from drawing further conclusions on this discrepancy.

\subsubsection{\decay{B}{\rho\pi} pentagonal analysis\label{sec:pentagonalObservables}}

The absolute scale of the neutral $B^0\to\pi^-\pi^+\pi^0$ amplitudes can be determined by adding the measured branching fractions for the two charged modes 
$B^+\to\rho^0\pi^+$ and  $B^+\to\rho^+\pi^0$, related to the neutral modes via the pentagonal relation Eq.~(\ref{eqn:pentagonal}). Once the data listed in Tab.~\ref{input_rhopiC} is added to the fit, individual $B\to\rho\pi$ branching  fractions
and direct $CP$ asymmetries in the charged modes can be determined. The corresponding pulls are given in Tab.~\ref{tab:BR_RhoPi} in App.~\ref{sec:numerics}.

\begin{figure}[t]
\begin{center}
  \includegraphics[width=18pc]{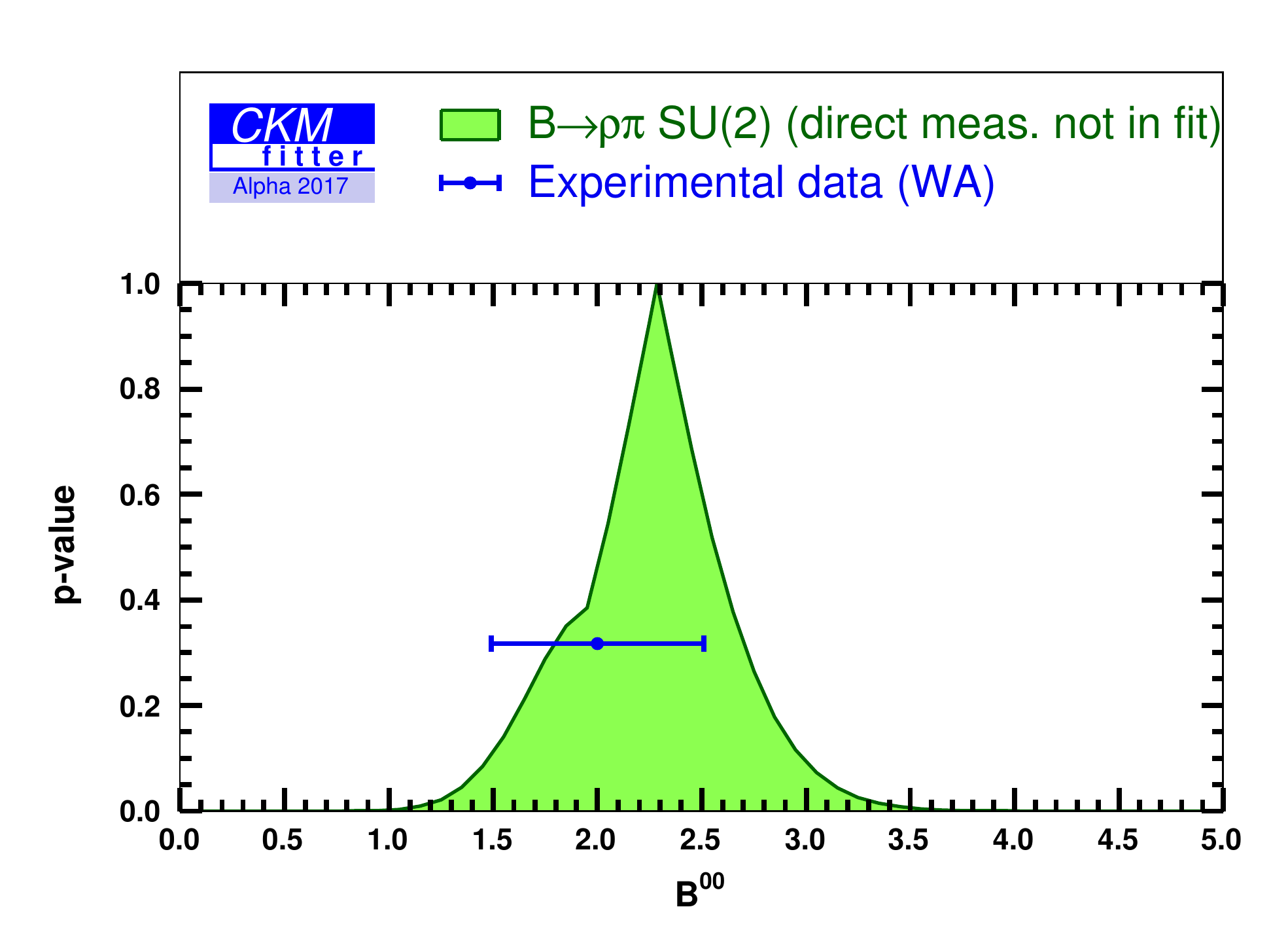}
  \includegraphics[width=18pc]{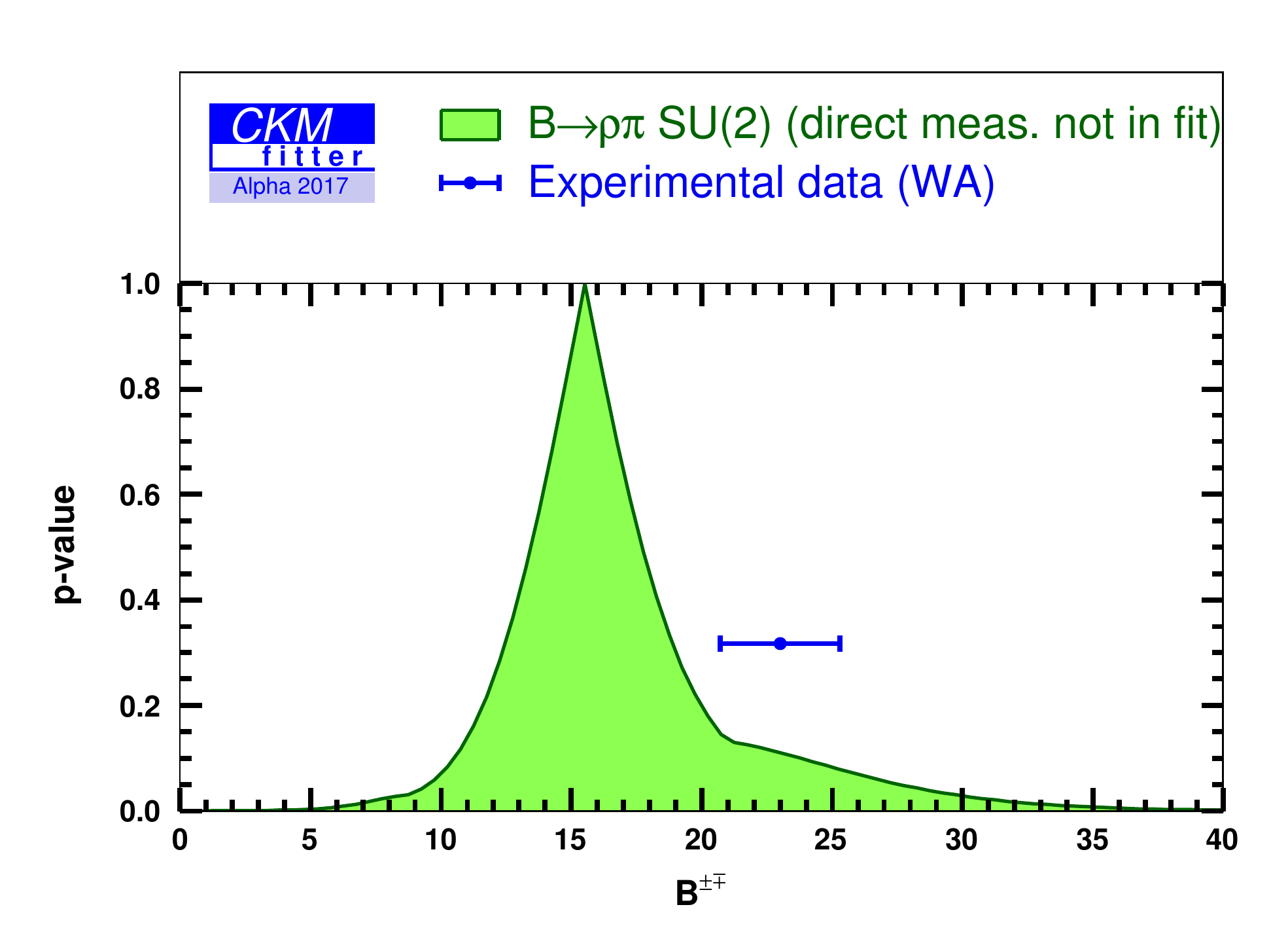}
  \includegraphics[width=18pc]{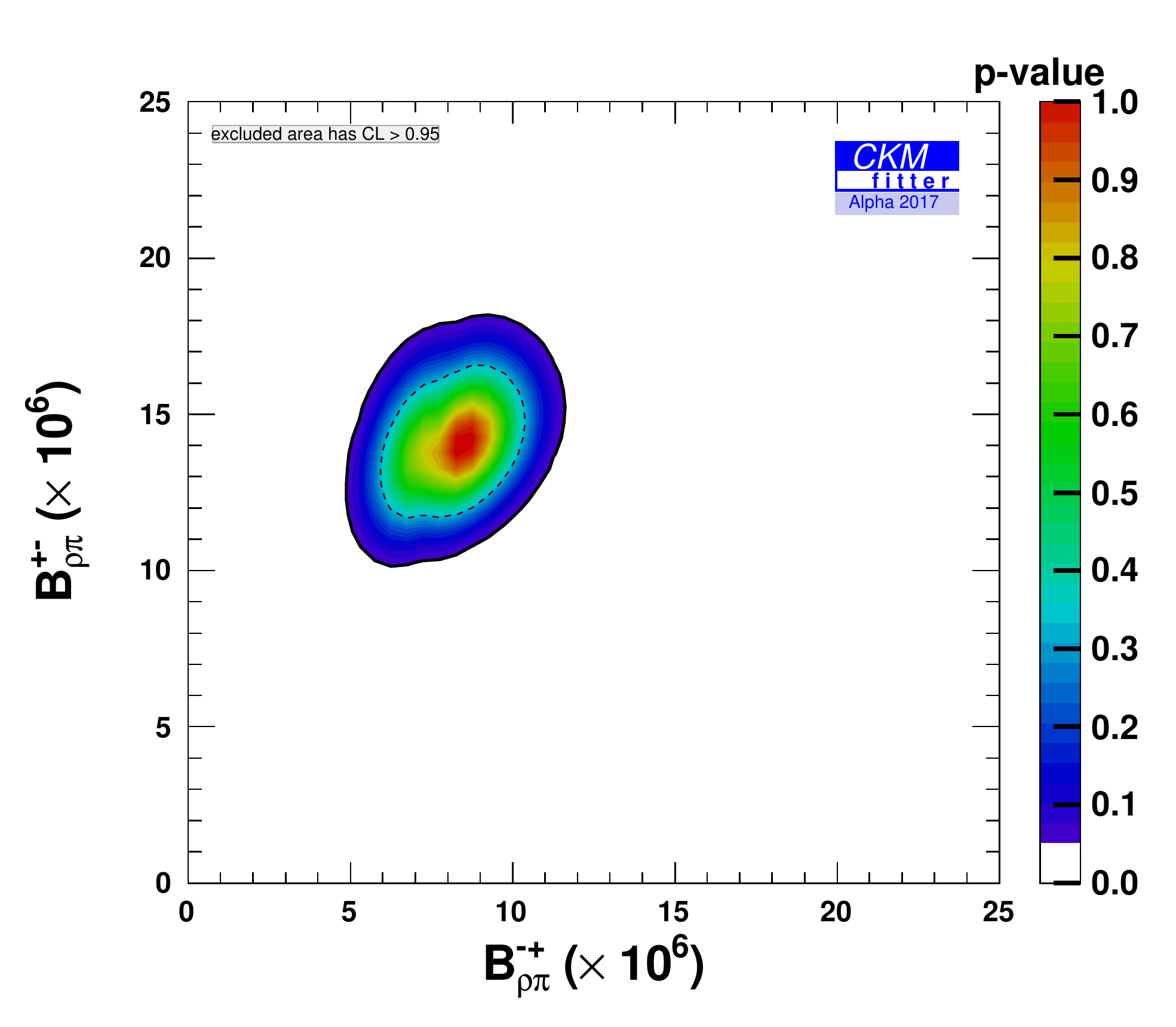}
\caption{\it\small  Constraint on the branching fraction of the neutral \decay{B^0}{\rho^0\pi^0} mode (top, left) and the mixed \decay{B^0}{\rho^\pm\pi^\mp} mode (top, right).
The direct measurement, not included in the fit, is indicated by an interval with a dot. The bottom figure displays the prediction for the branching fractions for   \decay{B^0}{\rho^\pm\pi^\mp}  in the (\decay{B^0}{\rho^+\pi^-}, \decay{B^0}{\rho^-\pi^+}) plane.}
\label{fig:BR_RhoPiN}
\end{center}       
\end{figure}

\begin{figure}[t]
\begin{center}
  \includegraphics[width=18pc]{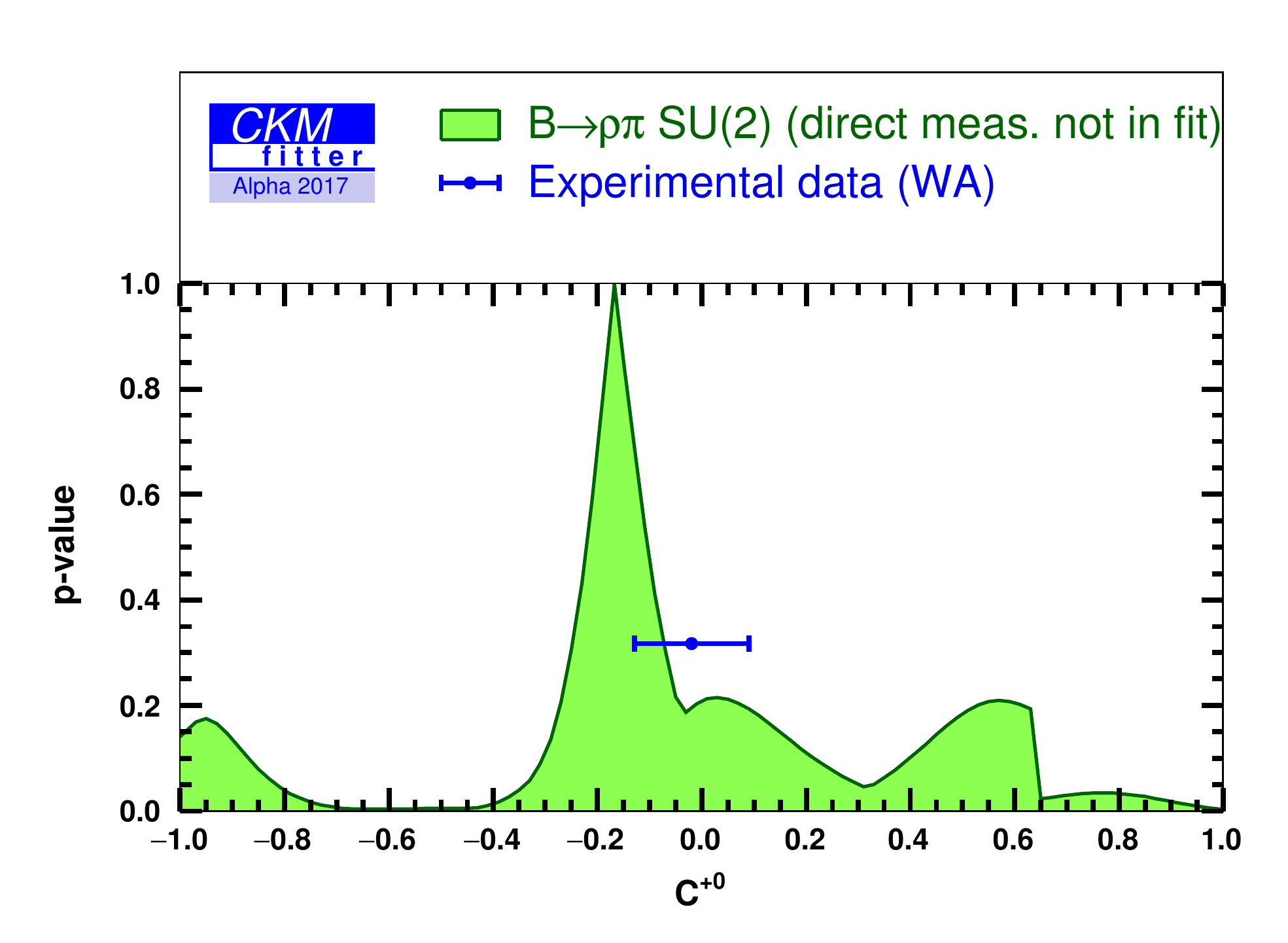}
  \includegraphics[width=18pc]{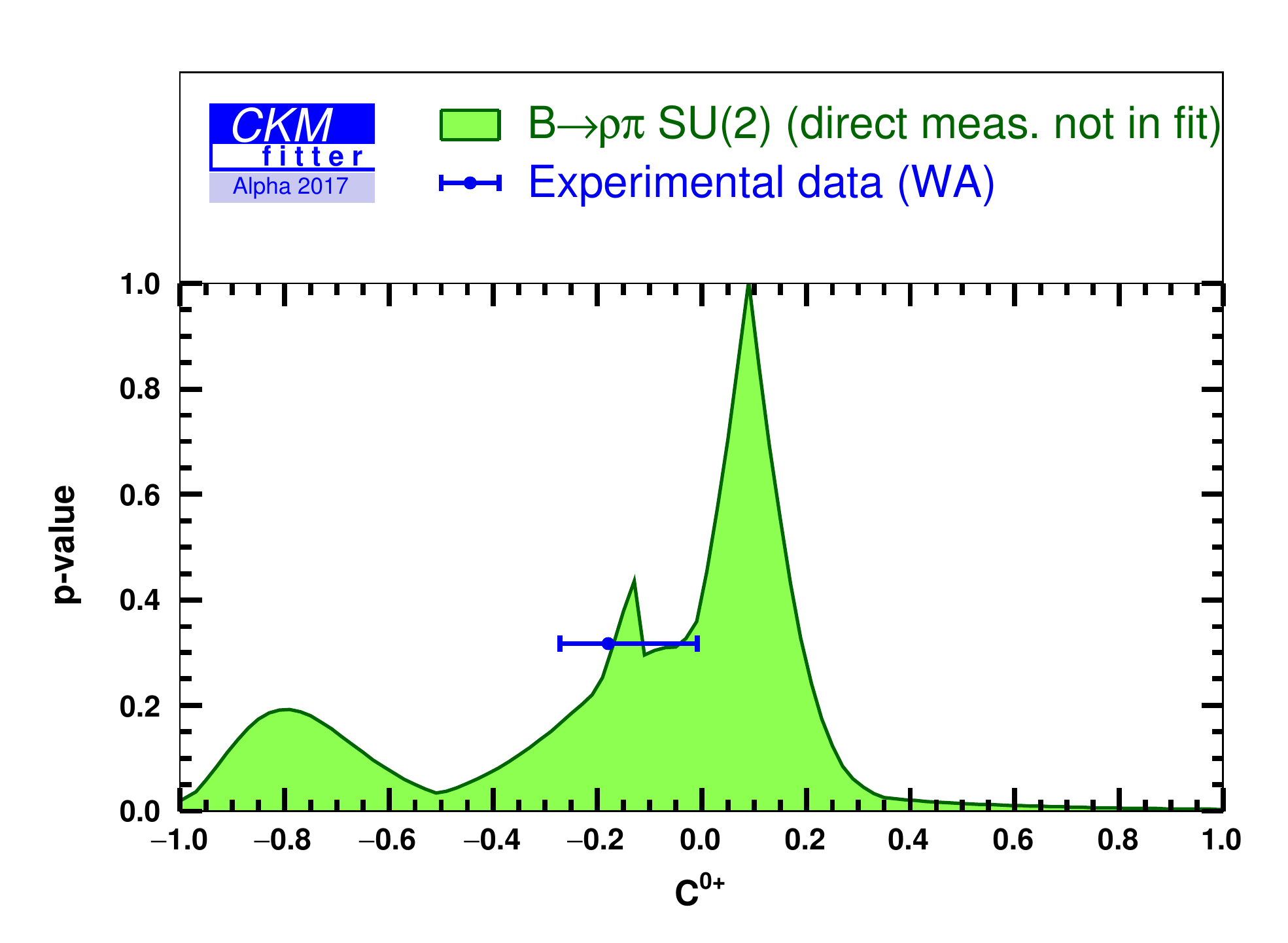}
  \includegraphics[width=18pc]{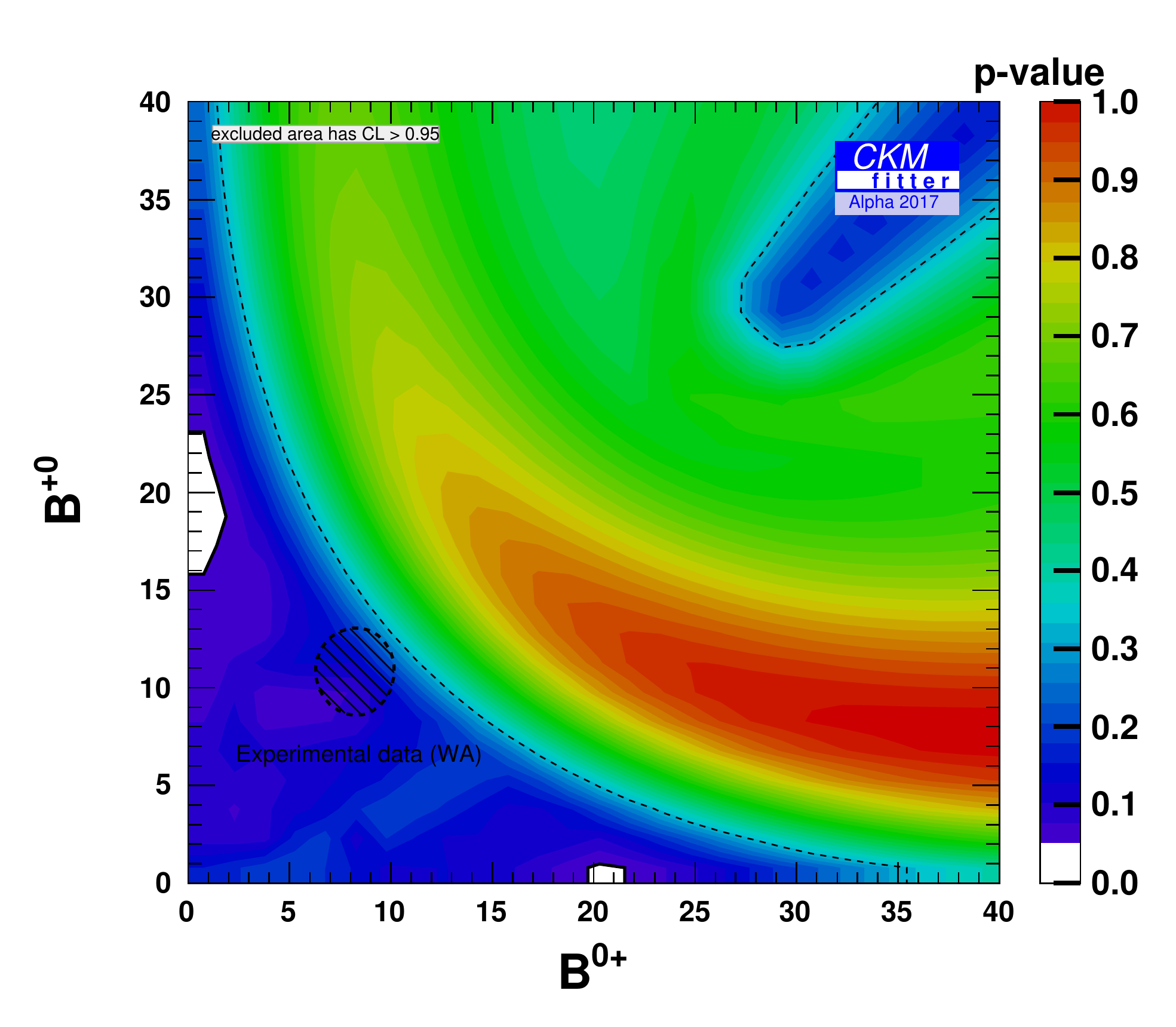}
\caption{\it\small  Constraint on the branching fraction and the direct $CP$ asymmetry of the charged \decay{B^+}{\rho^+\pi^0} mode (top, left) and \decay{B^+}{\rho^0\pi^+} mode (top, right). For each observable, the direct measurement, not included in the fit, is indicated by an interval with a dot.  The bottom figure displays the prediction for the branching fractions for \decay{B^+}{(\rho\pi)^+}  in the (\decay{B^+}{\rho^0\pi^+}, \decay{B^+}{\rho^+\pi^0}) plane compared to the experimental measurement indicated by the shaded area. }
\label{fig:BR_RhoPiC}
\end{center}       
\end{figure}

While the predicted branching fraction for the neutral $B^0\to\rho^0\pi^0$ mode is in very good agreement and competitive with the experimental measurement, 
a discrepancy is  observed in the balance of  branching fractions  for the charged modes and the mixed $B^0\to\rho^\pm\pi^\mp$ decays. The branching fraction of both charged modes are predicted to be larger than their measured value, which is compensated by a lower $B^0\to\rho^\pm\pi^\mp$ branching ratio.
Figs.~\ref{fig:BR_RhoPiN} and \ref{fig:BR_RhoPiC} display the \su{2} isospin constraints for the neutral and  charged modes, respectively.

\section{Prospective study} \label{sec:prospective}

In this section, we discuss how improved measurements of some $B\to\pi\pi$, $B\to\rho\rho$, $B\to\rho\pi$ observables can affect the accuracy on the CKM angle $\alpha$. So far, the experimental data is statistically limited for all the $B\to hh$ charmless modes considered here. Our prospective study aims at identifying the  specific decay channels worth measuring more accurately to improve the resolution on $\alpha$ significantly. 

For simplicity, we adopt a systematic approach rather than relying on the expected  performance of current or forthcoming flavour experiments such as LHCb or Belle II. 
We consider the subset of observables related to a specific $B\to h^ih^j$ decay, and we determine how much the accuracy on $\alpha$ is improved if we reduce the uncertainties for this subset of observables by the (arbitrary) factor $\sqrt{2}$, or if we take the radical limit of setting these uncertainties to zero. The central value of the current world-average measurements and correlation coefficients are kept unchanged. 
For the neutral modes $B^0\to h^+h^-$ and $B^0\to h^0h^0$, we distinguish the case of a counting analysis (C) that gives access to the branching ratio only (e.g., the LHCb measurement of $B^0\to\rho^0\rho^0$~\cite{LHCb_rr_b00}), and the case of flavour-tagged analyses: indeed, a time-integrated (TI)  analysis  extracts only the direct $CP$ asymmetry whereas a time-dependent (TD) analysis yields both the direct and  the mixing-induced $CP$-asymmetry parameters (e.g., the LHCb contribution to the study of $B^0\to\pi^+\pi^-$~\cite{LHCb_pp_cpm}). 

\begin{table}[t]
\begin{center}
\small
\begin{tabular}{|l|c|c|l||l|}
\hline
\small Decay          & Analysis     & Improved data  &   $\sigma_{\cal O}\to\sigma^{\scriptscriptstyle WA}_{\cal O}/\sqrt{2}$  &  $\sigma_{\cal O}\to 0$    \\
\hline 
\decay{B^0}{\rho^+\rho^-}   & C + TD   &   $\B^{+-},f_L^{+-},\C^{+-},\S^{+-}$    &  \val{91.8}{+4.4}{-4.0} \gain{-12} &  \val{92.2}{3.5}    \gain{-26}\\ 
\cline{2-5}
                               & C    &   $\B^{+-}, f_L^{+-}$                      &\val{91.8}{+4.9}{-4.5}      \gain{-1} &  \val{92.1}{+4.6}{-4.7} \gain{-2}\\ 
                               & TD   &   $\C^{+-}, S^{+-}$                        &\val{92.1}{+4.1}{-4.3}      \gain{-12}&  \val{92.1}{+3.5}{-3.7} \gain{-24}\\
\hline 
\decay{B^0}{\rho^0\rho^0}    & C + TD &   $\B^{00}, f_L^{00}, \C^{00}, \S^{00}$   &\val{91.8}{+4.1}{-4.0}   \gain{-15}&  \val{91.7}{3.1}     \gain{-35}\\ 
\cline{2-5}
                                & C  &    $\B^{00}, f_L^{00}$                      &\val{91.8}{+4.9}{-4.5}      \gain{-1} &\val{92.1}{+4.6}{-4.7}    \gain{-2}\\ 
                                & C + TD   &    $\C^{00}, \S^{00}$                 &\val{91.8}{+4.1}{-4.0}      \gain{-15}&\val{91.6}{3.1}           \gain{-35}\\ 
\hline 
\decay{B^+}{\rho^+\rho^0}     & C  &   $\B^{+0}$                                &\val{92.2}{+4.6}{-4.7}   \gain{-2}&\val{92.5}{+4.3}{-4.7}  \gain{-5}\\ 
\hline
\hline
\decay{B^{\pm,0}}{(\rho\rho)^{\pm,0}} & C + TI   &   $\B^{ij}, f_L^{ij}, \C^{ij}, \S^{ij}$  &  \val{92.0}{+3.3}{-3.4} \gain{-29} & -   \\
\hline
\end{tabular}
\caption{\small\it 68\% CL interval for $\alpha$ from the \su{2} isospin analysis of the $B\to\rho\rho$ data in the case that the uncertainty of some  observables is reduced by a factor $\sqrt{2}$ (third column) or  set to zero (fourth column). The relative gain in resolution  with respect to the current measurement is indicated within brackets. Only the preferred solution for $\alpha$ close to $90^\circ$ is reported.} \label{tab:prospect_RhoRho}
\end{center}
\end{table}

 The global combination of the  decay-specific determinations of $\alpha$ is so far dominated by the $B\to\rho\rho$ data that provides a constraint on $\alpha$ with a relative uncertainty at the level of 5\%, i.e., $\alpha_{\rho\rho}=$\val{92.1}{+4.6}{-4.9}{$^\circ$} for the solution near $90^\circ$ given in Eq.~(\ref{eq:alpharhorho}). 
 As shown in Tab.~\ref{tab:prospect_RhoRho}, if
 all other observables remain unchanged, improving the accuracy of the branching ratio of the charged mode $B^+\to\rho^+\rho^0$ would  improve the resolution on $\alpha$ only marginally, even in the case of a vanishing resolution (indicating that this observable has only a limited impact on the accuracy for $\alpha$). Improving the measurements for the neutral modes, in particular the colour-suppressed $B^0\to\rho^0\rho^0$ decay, has a larger impact, essentially driven by the $CP$-asymmetries parameters. Improving the time-dependent asymmetries in the $B^0\to\rho^0\rho^0$ is also worth investigating, e.g., in the second run of LHCb data taking. Reducing by $\sqrt{2}$ all the $B\to\rho\rho$  uncertainties would reduce the 68\% CL  interval for $\alpha$ by more than 1 degree.

The subleading contribution to the combined $\alpha$ determination is provided by the $B\to\pi\pi$ system, $\alpha_{\pi\pi}=(\val{93.0}{14.0})^\circ$, see Eq.~(\ref{eq:alphapipi}). As shown on Tab.~\ref{tab:prospect_PiPi}, any sizable improvement on $\alpha$ is  driven by the increased accuracy in the measurement of the direct $CP$ asymmetry in the colour-suppressed decay  $B^0\to\pi^0\pi^0$. A  reduction by $\sqrt{2}$ of the  uncertainty of this single observable would reduce the 68\% CL range for $\alpha$ by $1^\circ$. A similar improvement of all the measured $B\to\pi\pi$ observables would reduce by about $1.5^\circ$ the uncertainty on $\alpha$.

\begin{table}[t]
\begin{center}
\small
\begin{tabular}{|l|c|c|l||l|}
\hline
\small Decay          & Analysis     & Improved data  &   $\sigma_{\cal O}\to\sigma^{\scriptscriptstyle WA}_{\cal O}/\sqrt{2}$  &  $\sigma_{\cal O}\to 0$    \\
\hline 
\decay{B^0}{\pi^+\pi^-}            & C + TD      &$\B^{+-},\C^{+-},\S^{+-}$       &\val{93.0}{14.0} \gain{0} &\val{93.0}{14.0}  \gain{0}\\
\cline{2-5}
                                      & C           &$\B^{+-}$                      &\val{93.0}{14.0}   \gain{0}  &\val{93.0}{14.0}       \gain{0}\\
                                      & TD          &$\C^{+-}, S^{+-}$               &\val{93.0}{14.0}   \gain{0} &\val{93.0}{14.0}  \gain{0}\\
\hline 
\decay{B^0}{\pi^0\pi^0}            & C + TI       &$\B^{00},\C^{00}$               &\val{92.6}{13.0} \gain{-8}   &\val{84.1}{+4.5}{-3.5} \gain{-70}\\
\cline{2-5}
                                      & C           & $\B^{00}$                     &\val{93.0}{14.0}    \gain{0} &\val{93.5}{13.5}      \gain{-3.6} \\
                                      & TI          & $\C^{00}$                     &\val{92.0}{13.0}    \gain{-8} &\val{84.1}{+6.4}{-5.6}     \gain{-57}\\
\hline 
\decay{B^+}{\pi^+\pi^0}            & C            & $\B^{+0}$                     &\val{93.0}{14.0}  \gain{0}&\val{93.0}{14.0}   \gain{0}\\
\hline
\hline
\decay{B^{\pm,0}}{(\pi\pi)^{\pm,0}}  & C + TD or TI  &$\B^{ij},\C^{+-},\S^{+-},\C^{00}$&\val{92.5}{12.5}  \gain{-11}& -   \\
\hline
\end{tabular}
\caption{\small\it 68\% CL interval for $\alpha$ from the \su{2} isospin analysis of the $B\to\pi\pi$ data in case the uncertainty of some  observables is reduced by a factor $\sqrt{2}$ (third column) or set to zero (fourth column). The relative gain in resolution  with respect to the current measurement is indicated within brackets. Only the preferred  solution for $\alpha$ close to $90^\circ$ is reported.} \label{tab:prospect_PiPi}
\end{center}
\end{table}

\begin{figure}[t]
\begin{center}
  \includegraphics[width=25pc]{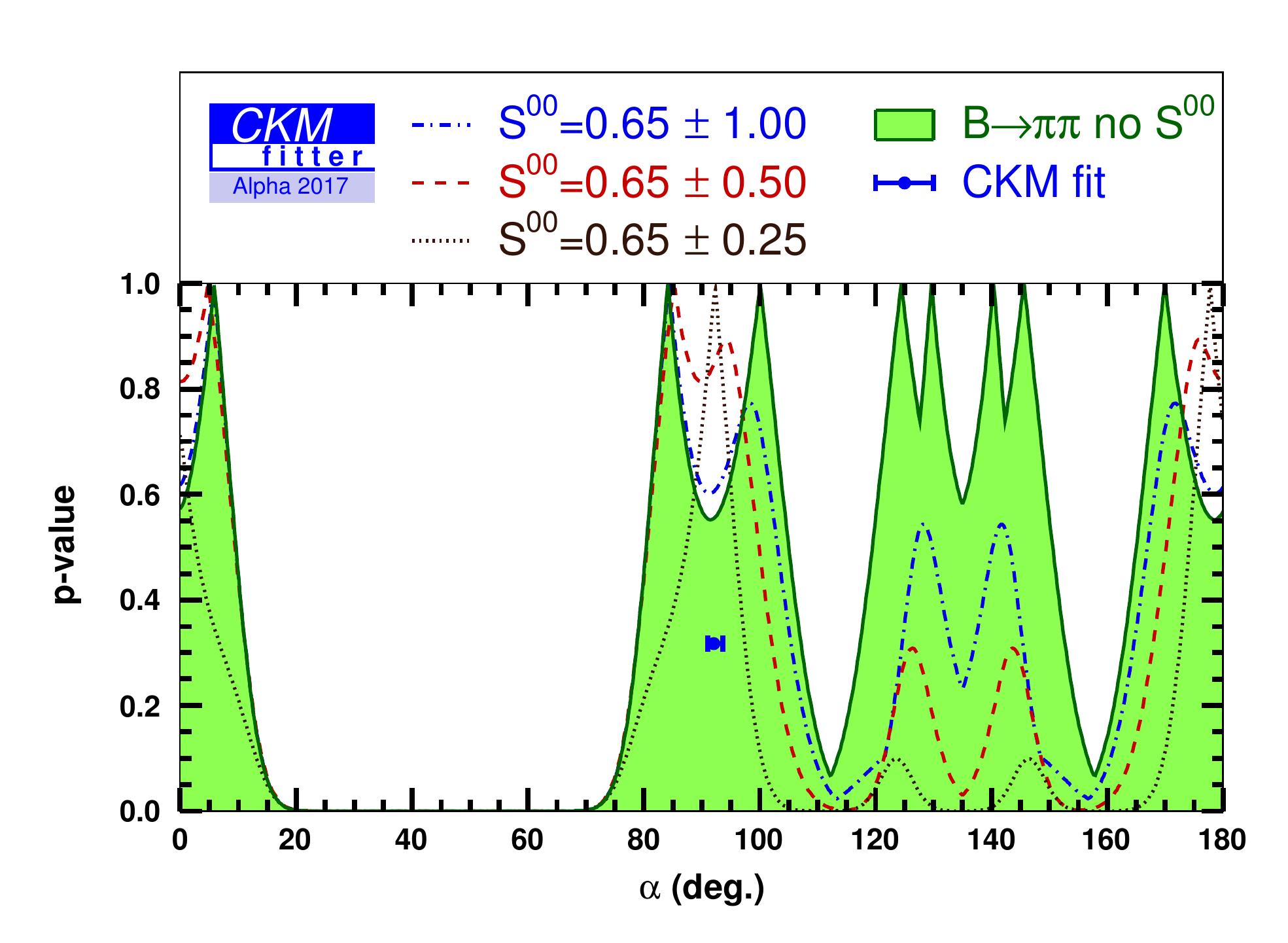}
\caption{\it\small  Expected impact of the measurement of the time-dependent asymmetry $\S^{00}$ on the $\alpha$ determination from the isospin analysis of the $∫B\to\pi\pi$ system. The central value for $\S^{00}$  corresponds to the best prediction from the \su{2} isospin analysis. }\label{fig:alphaPiPi_wS00}
\end{center}       
\end{figure}

No time-dependent analysis has become available so far for the colour-suppressed mode due to the dominant di-photon decay of the neutral pions that  hinders the measurement of the decay time, so that $\S^{00}_{\pi\pi}$ is not measured yet. However, the low branching fraction of the Dalitz decay $\pi^0\to\gamma e^+e^-$ or the photon conversion in the detector material may allow one to measure the $B^0$ decay time in the future high-statistics flavour experiments (LHCb upgrade, Belle II).  Fig.~\ref{fig:alphaPiPi_wS00} shows the $p$-value for $\alpha$ when adding the $S^{00}_{\pi\pi}$ observable under different hypotheses concerning the experimental resolution, and setting the central value as predicted in Eq.~(\ref{eq:s00_prediction}).  In addition to a significant improvement on $\alpha$, the measurement of  $\S^{00}_{\pi\pi}$ would reduce the number of mirror solutions in the $B\to\pi\pi$ isospin analysis (see the discussion around Eq.~(\ref{eq:S00pipi})).

The same prospective exercise is more delicate for the $B^0\to\pi^+\pi^-\pi^0$ system  due to the discrepancy between the direct measurement  $\alpha_{\rho\pi}$=\val{54.1}{+7.7}{-10.3}{$^\circ$}\and\val{141.8}{+4.8}{-5.4}{$^\circ$} (Eq.~(\ref{eq:alpharhopi})) and the indirect determination from the global CKM fit (Eq.~(\ref{eq:alphaInd})). New measurements of this decay would certainly aim first at a better understanding of this discrepancy, rather than improving the accuracy on $\alpha$ extracted from this channel. Assuming the central values given by the individual predictions for each \U and \I coefficient, as listed in Tab.~\ref{tab:UI_rhopi}, and keeping unchanged the current experimental resolution and correlations, we obtain a mild constraint on $\alpha$, consistent with $\alpha_{\rm ind}$, namely $\alpha_{\rho\pi}^{{\rm fit}~\U,\I}$=\val{82}{-33}{-48}{$^\circ$}. In this case, the reduction of the observable uncertainty by $\sqrt{2}$ results in a reduction by approximately $9^\circ$ for the 68\% CL interval for $\alpha$.
On the other hand, keeping the current world-average central values and reducing the uncertainty on all the \U and \I observables by a factor $\sqrt{2}$  leads to a more stringent 68\% CL interval $\alpha_{\rho\pi}^{\sigma\scriptscriptstyle/\sqrt{2}}$=\val{54.1}{+5.6}{-6.9}{$^\circ$}\and\val{141.0}{+1.9}{-1.9}{$^\circ$}, which remains difficult to interpret due to the discrepancies already discussed between the current data and our theoretical framework based on isospin symmetry. Additional measurements should hopefully provide a clearer and more consistent picture for the $B\to\rho\pi$ sector before discussing any improvement in the extraction of $\alpha$ from these modes.

\section{Conclusion\label{sec:conclusion}}

Quark flavour transitions provide particularly stringent tests of the Standard Model, both through rare decays and \CP-violating processes. An accurate knowledge of the Cabibbo--Kobayashi--Maskawa matrix is essential for these studies and it requires the combination of many precise constraints. We have focussed on the determination of the $\alpha$ angle, which can be extracted with a high accuracy from two-body charmless $B$-meson decays extensively studied  at $B$-factories and LHCb.

We recalled that this extraction can be done from $B\to \pi\pi$, $B\to\rho\pi$ and $B\to\rho\rho$ decays but it
is affected by the presence of penguin contributions. We explained how \su{2} isospin symmetry can be used to constrain the structure of hadronic penguin and tree amplitudes, enabling the extraction of $\alpha$ from branching ratios and \CP asymmetries. We gave details on the analyses of $B\to \pi\pi$, $B\to\rho\pi$ and $B\to\rho\rho$ systems separately, before combining these results to reach an accuracy around 4$^\circ$ on the direct determination of $\alpha$.  The $B\to \pi\pi$  and $B\to\rho\rho$ systems dominate the combination and they favour solutions in good agreement with the indirect determination of $\alpha$ from a global CKM fit analysis,
Eq.~(\ref{eq:alphaInd}),  it is not the case for the $B\to\rho\pi$ system which favours different ranges of values for $\alpha$ with a discrepancy at the level of  3~$\sigma$ compared to the indirect determination. The combination of the three channels is dominated by $B\to \rho\rho$, and to a lesser extent $B\to \pi\pi$, resulting in the 68\% CL confidence intervals given in Eq.~(\ref{eq:alphadir}).

We have then studied several uncertainties that may affect this extraction. We have
tested the hypotheses underlying the \su{2} isospin analysis: setting the $\Delta I=\ofrac{3}{2}$ electroweak penguins to zero, neglecting
the difference of light-quark masses generating $\pi^0-\eta-\eta'$ mixing, setting the $\rho$ width to zero to cancel $\Delta I=1$ contributions thanks to Bose--Einstein symmetry. These effects may shift the central value of $\alpha_{\rm dir}$ by around $2^\circ$, while keeping the uncertainty around $4^\circ$ to $5^\circ$, thus remaining within the statistical uncertainty quoted in Eq.~(\ref{eq:alphadir}). In addition, we have discussed a few aspects concerning the statistical treatment used in order to extract the $p$-value, comparing two statistical approaches to test the impact of hadronic nuisance parameters on coverage. The approach based on Wilks' theorem and mainly used here proves to be more conservative than the bootstrap method for $B\to\pi\pi$ and $B\to\rho\rho$, but less conservative in the case of $B\to\rho\pi$. The comparison of the coverage properties of the two approaches leads to a further uncertainty of around $1^\circ$. We stress that this uncertainty is only attached to the combination of the three direct determinations of $\alpha$ with the current data: it is likely to be reduced if one combines the direct determination of $\alpha$ with other observables, leading to a more accurate determination of $\alpha$ (which is in particular the typical case of global CKM fits). 

Assuming the validity of the Standard Model and taking as an input the indirect determination of $\alpha$ from the global CKM fit, we used
the observables in $B\to \pi\pi$, $B\to\rho\pi$ and $B\to\rho\rho$ decays to extract information on ratios of hadronic amplitudes (penguin-to-tree and colour-suppressed), finding results in broad agreement with expectations from QCD factorisation, apart from the ratio of the colour-suppressed to colour-allowed tree $\pi\pi$ contributions: indeed both the phase and the modulus of this ratio do not agree well with theoretical expectations. It would be interesting to widen this discussion and see how various theoretical approaches to non-leptonic two-body decays can reproduce the patterns of hadronic amplitudes that we have extracted from the data. 

Under the same hypotheses, we have also performed the indirect determinations of observables of interest (using all the other measurements available), comparing the pulls for those already observed and predicting the values of the remaining ones. The compatibility between direct measurements and indirect determinations is very good for the observables in the $B\to\pi\pi$ and $B\to\rho\rho$ systems, whereas we could identify a subset of $B\to \rho\pi$ observables likely to be responsible for the discrepancies observed with respect to the Standard Model expectations in these modes.  Among many other quantities, we have predicted the yet-to-be-measured mixing-induced \CP asymmetry in the \decay{B^0}{\pi^0\pi^0} decay; see Eq.~(\ref{eq:S00pipi}).

Finally, we have performed a prospective study to analyse how improved measurements for some subsets of observables
can improve the uncertainty of $\alpha$. In particular, we have noticed that an improved accuracy for the time-dependent asymmetries in $B^0\to \rho^0\rho^0$ and the measurement of $\S^{00}_{\pi\pi}$ would reduce the uncertainty on $\alpha$ in a noticeable way.

We have seen that the extraction of $\alpha$ is now possible to a high accuracy, using many different channels and experimental sources. It would be very interesting to measure the remaining observables that we can accurately predict using the data already available. The current accuracy reached by $\alpha$ makes it a particularly useful constraint both for the Standard Model and for searches of New Physics, although the  theoretical computations for these transitions remain challenging. Some of these channels will be improved by the LHCb experiment. In addition, the advent of the Belle-II experiment will certainly lead to new and improved measurements for a large set of branching ratios and \CP asymmetries, providing an opportunity to understand better the results obtained for the $B\to \rho\pi$ system and allowing us to extract hadronic parameters with a higher accuracy. Both avenues should be highly beneficial for the upcoming studies of flavour physics in the quark sector.

\section*{Acknowledgements}
We would like to thank all our collaborators from the CKMfitter group for many useful discussions on the statistical issues covered in this article, and in particular K. Trabelsi and L. Vale Silva for a careful reading of the manuscript. SDG acknowledges partial support from Contract FPA2014-61478-EXP. This project has received funding from the European Union's Horizon 2020 research and innovation programme under Grant agreements nos 690575, 674896 and 692194.

\cleardoublepage
\appendix
\cleardoublepage
\section{Results from the \babar and Belle experiments} \label{sec:experiments}

Both \babar and Belle experiments have measured the relevant observables for an extraction of the weak phase $\alpha$ in each of  the three charmless decay systems. We have provided an analysis of their average in Sec.~\ref{sec:alpha}, and we briefly discuss the separate results from each $B$-factory in this appendix.

\subsection{$B\to\pi\pi$ and $B\to\rho\rho$ analyses}
\begin{figure}[htpb]
  \begin{center}
    \includegraphics[width=18pc]{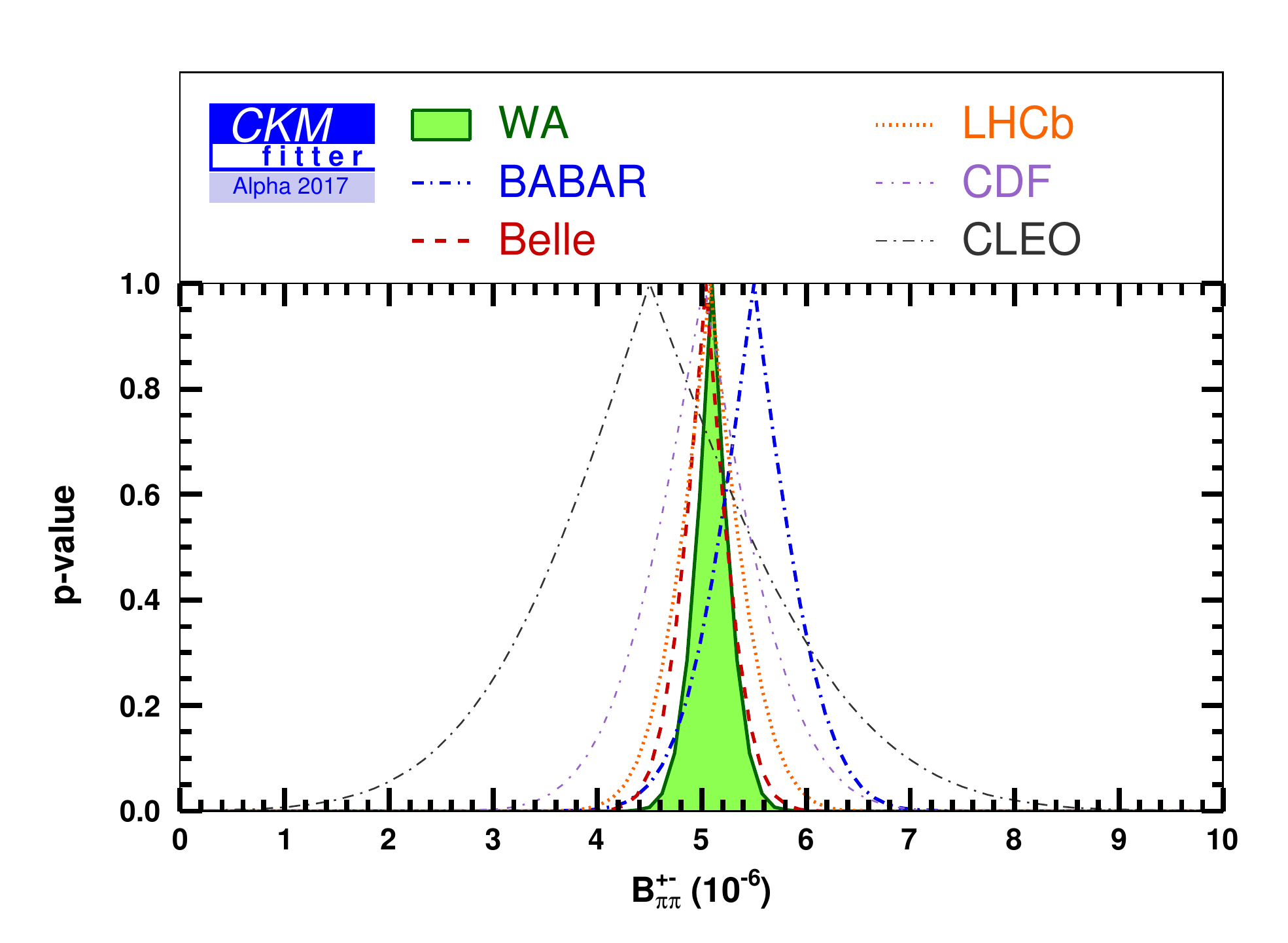}
    \includegraphics[width=18pc]{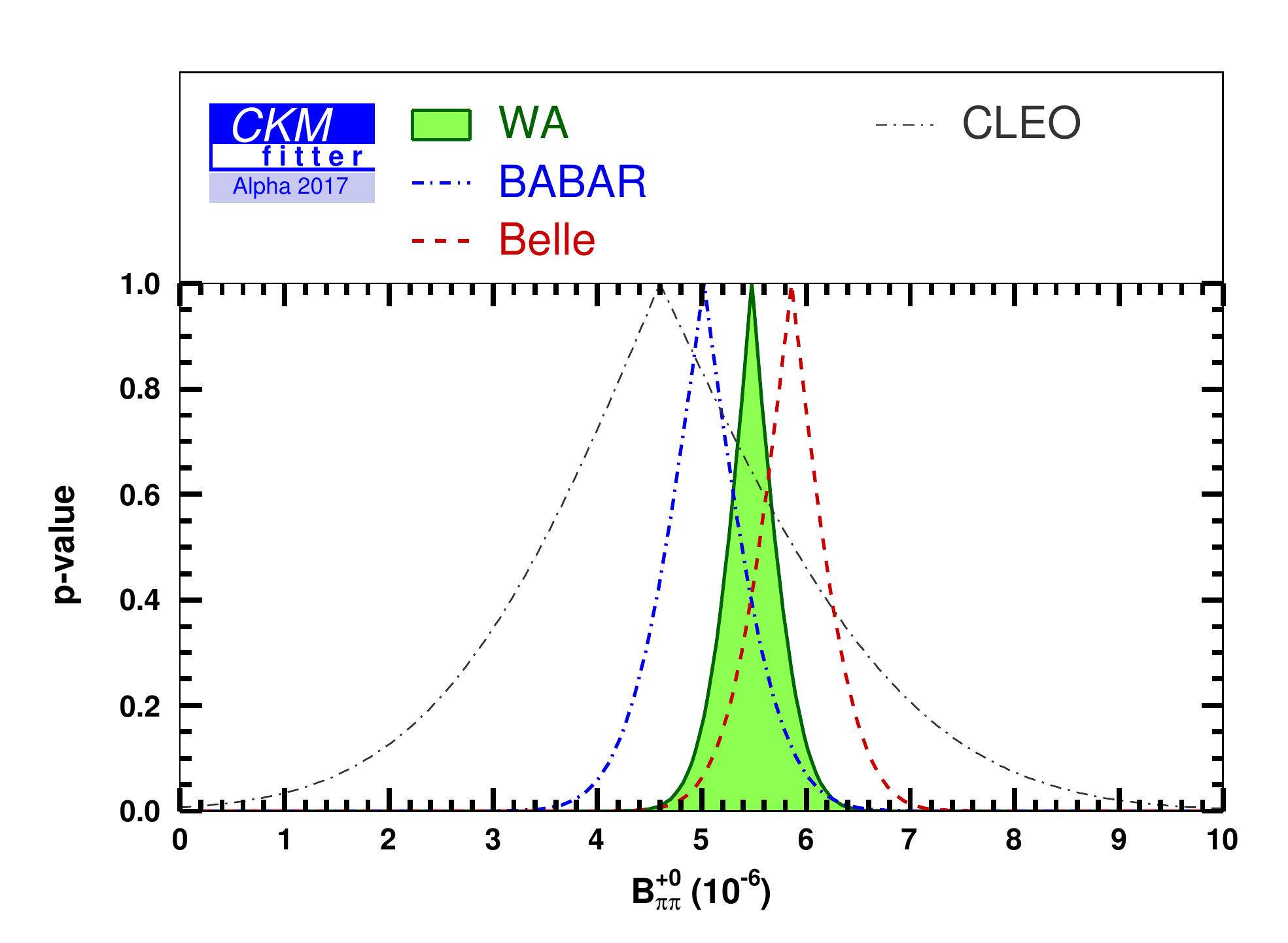}
    \includegraphics[width=18pc]{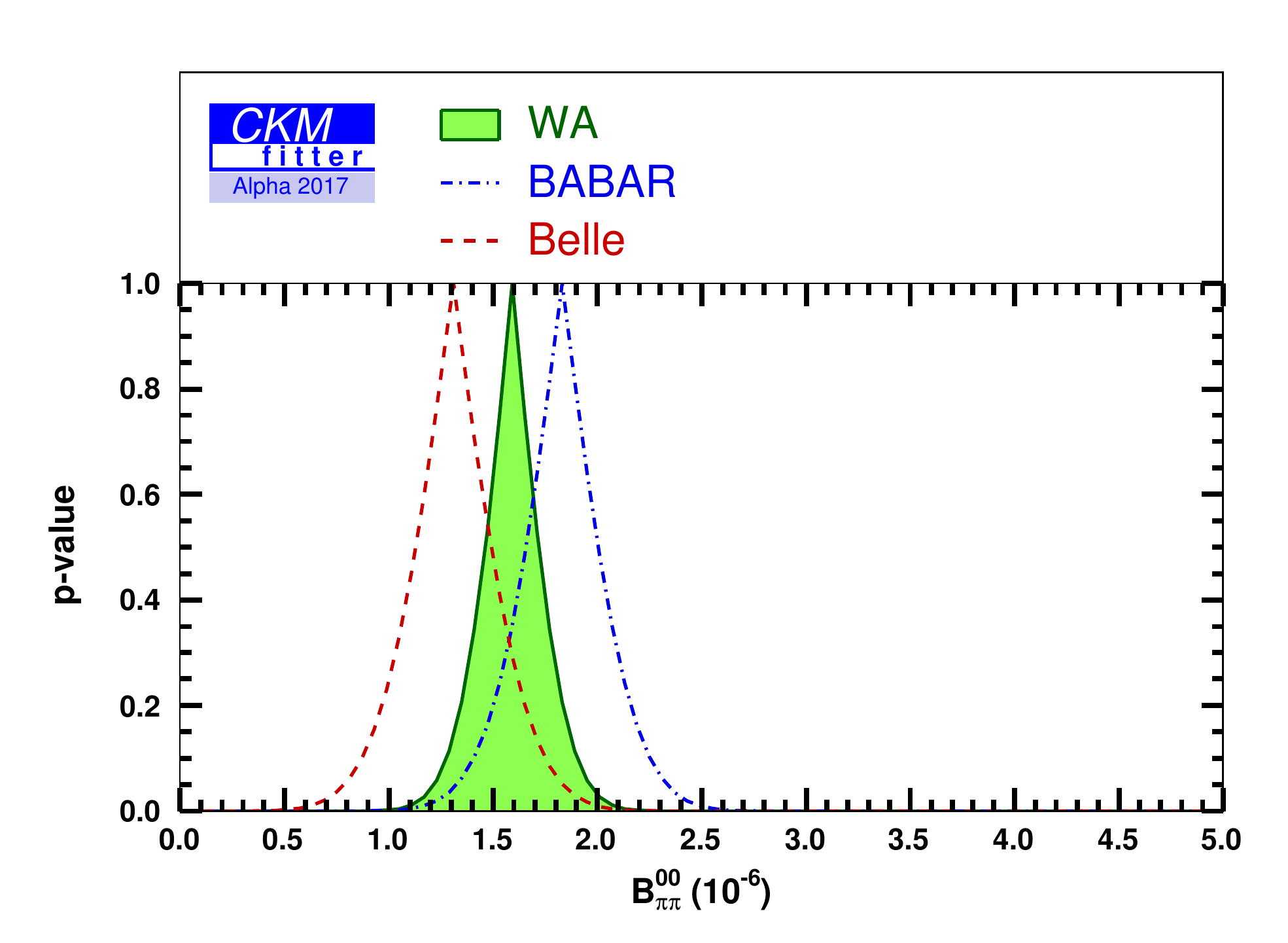}
    \caption{\it\small  Measurements of the $B\to\pi\pi$ branching fractions from \babar (blue curve) and Belle (red curve).
When available, other experimental contributions are shown: LHCb (orange), CDF (purple) and CLEO (black).  The green shaded area represents the world average. }\label{fig:pipi_exp1}
\end{center}       
\end{figure}

\begin{figure}[t]
  \begin{center}    
    \includegraphics[width=18pc]{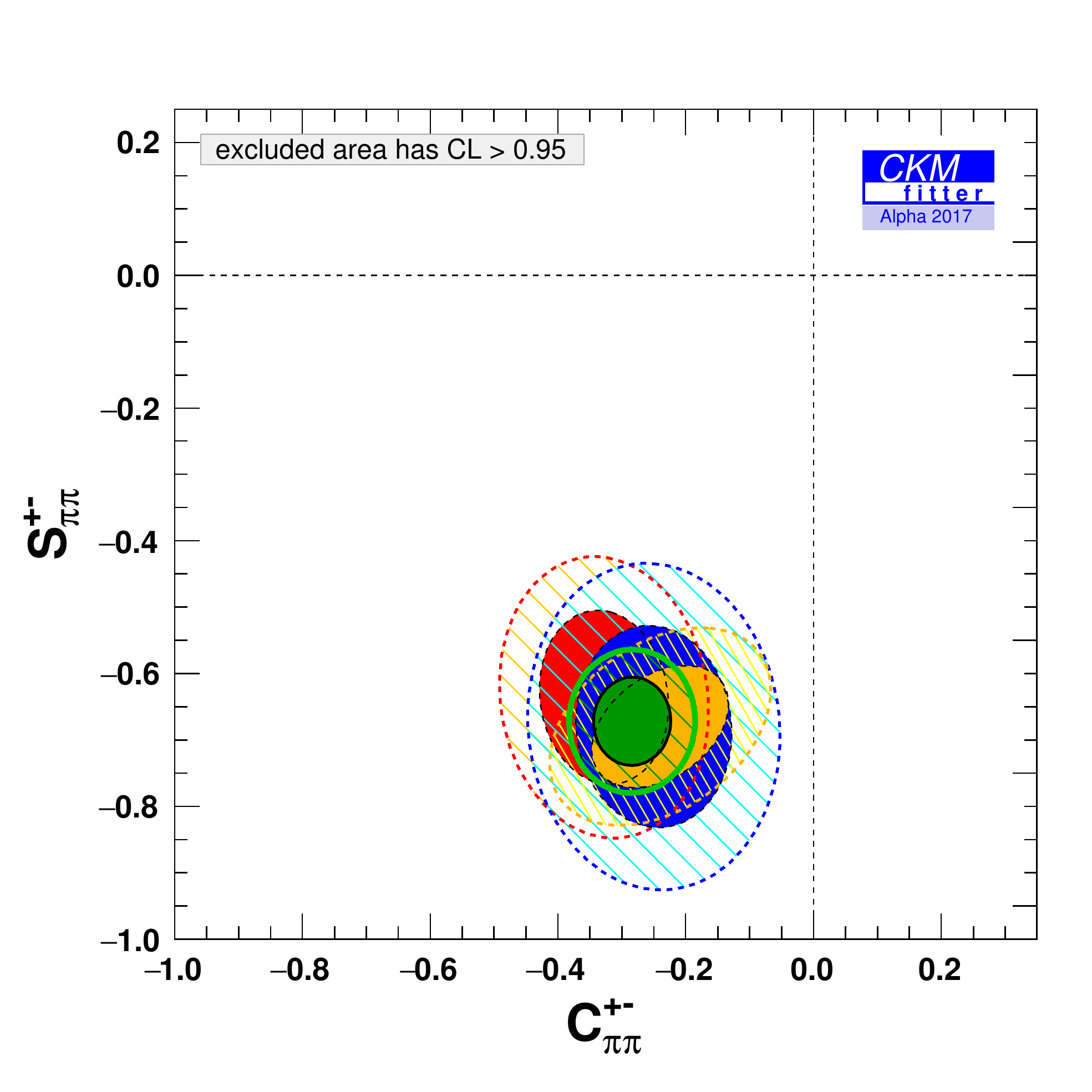}
    \includegraphics[width=18pc]{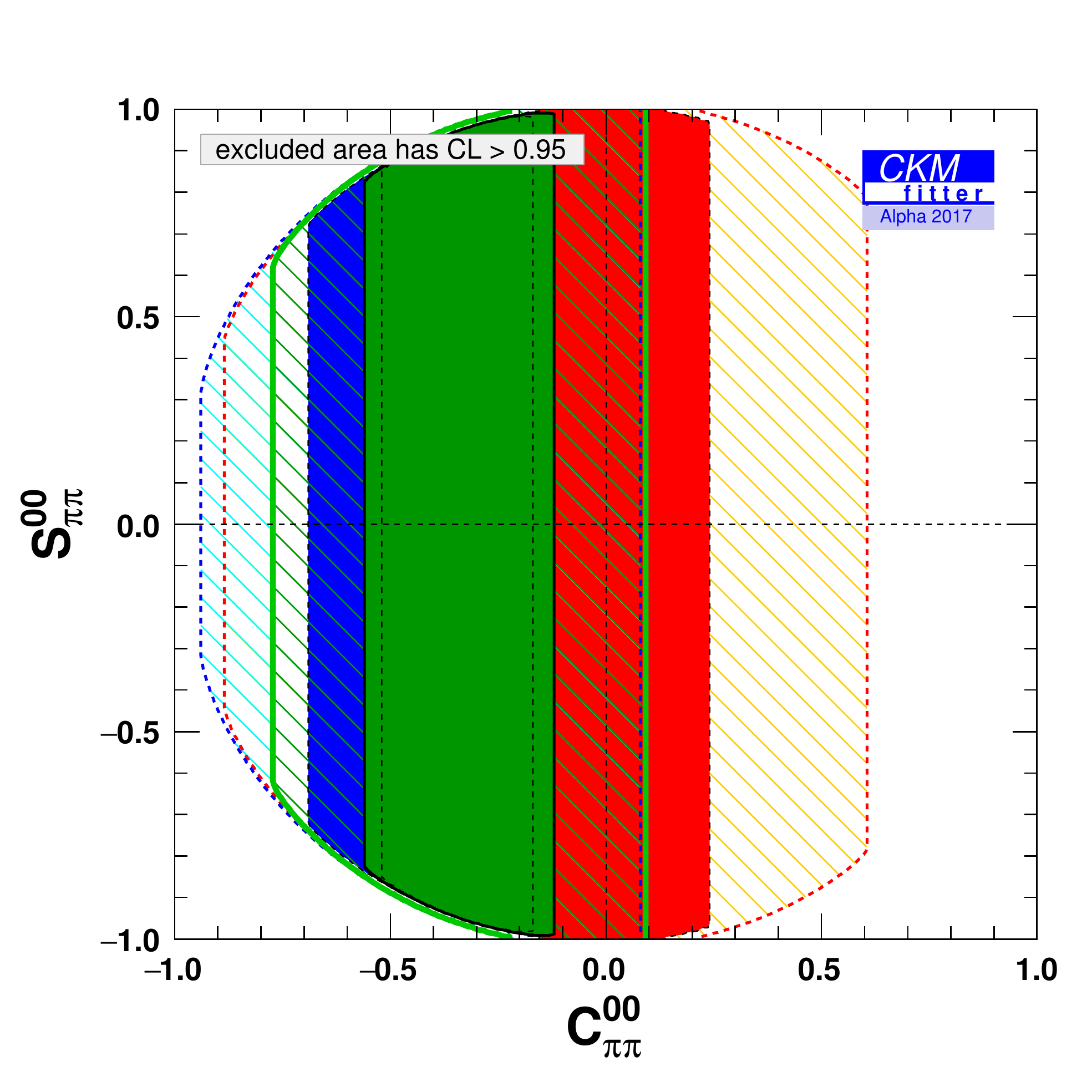}
\caption{\it\small  Measurements of the $B\to\pi\pi$ $CP$ asymmetries from \babar (blue curve) and Belle (red curve).
When available, other experimental contributions are shown: LHCb (orange), CDF (purple) and CLEO (black).  The green shaded area represents the world average. }\label{fig:pipi_exp2}
\end{center}       
\end{figure}

\begin{figure}[t]
  \begin{center}
    \includegraphics[width=18pc]{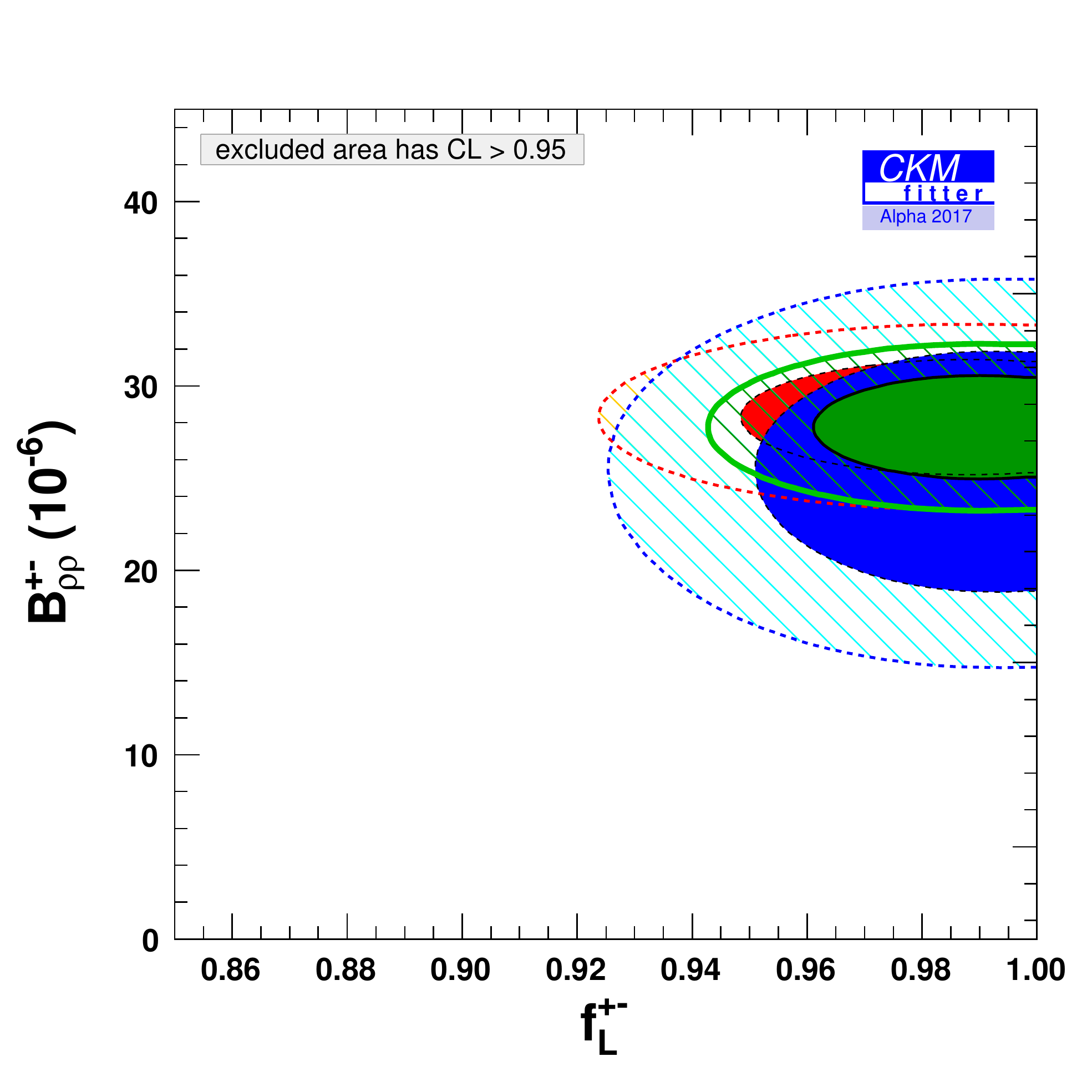}
    \includegraphics[width=18pc]{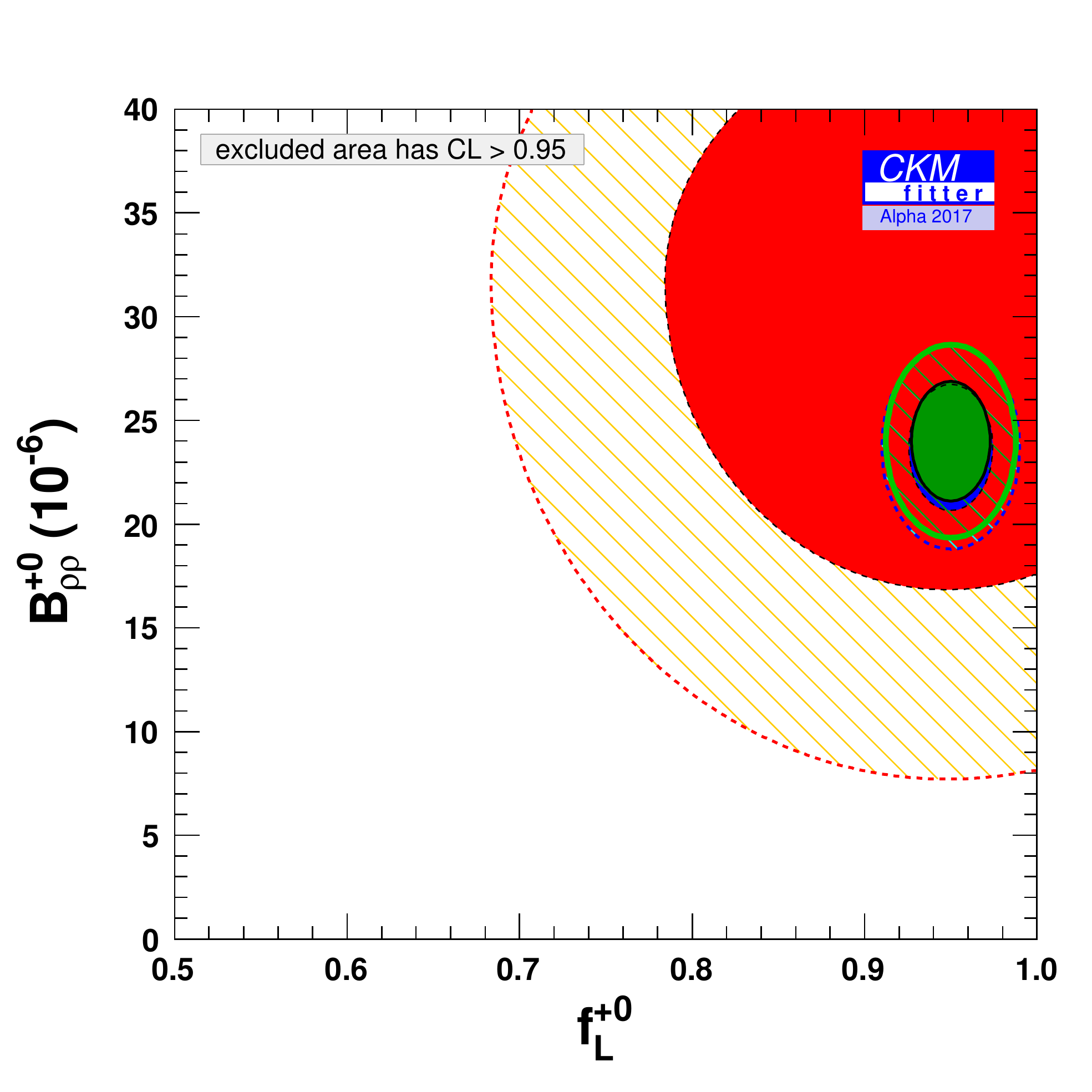}
    \includegraphics[width=18pc]{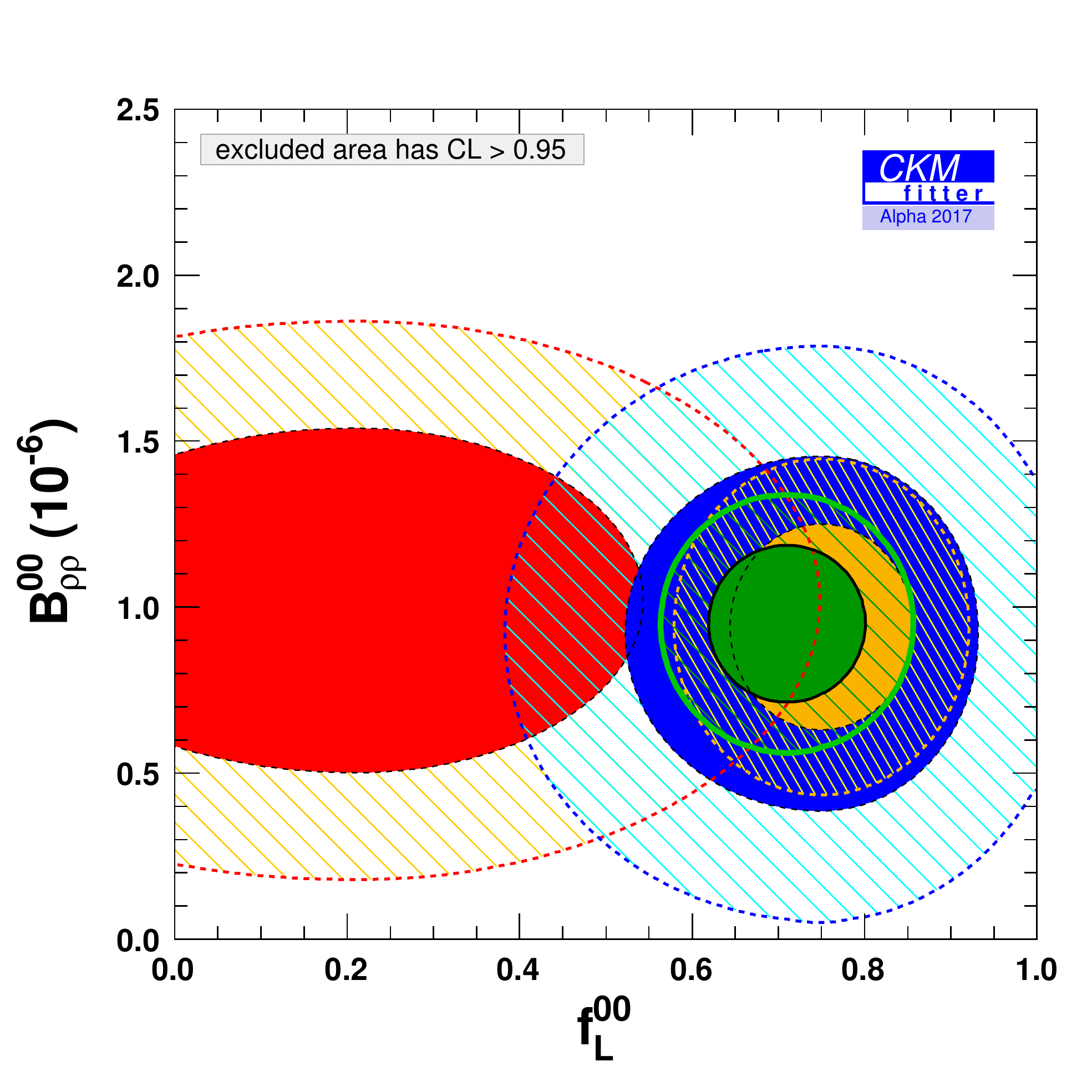}
    \caption{\it\small  Measurements of the $B\to\rho\rho$ branching fractions from \babar (blue area) and   
Belle (red area). The  LHCb contribution to the $B\to\rho^0\rho^0$ branching fraction  is indicated by the orange area.  The green shaded area represents the world average.}
\label{fig:rhorho_exp1}
\end{center}       
\end{figure}

\begin{figure}[t]
  \begin{center}
    \includegraphics[width=18pc]{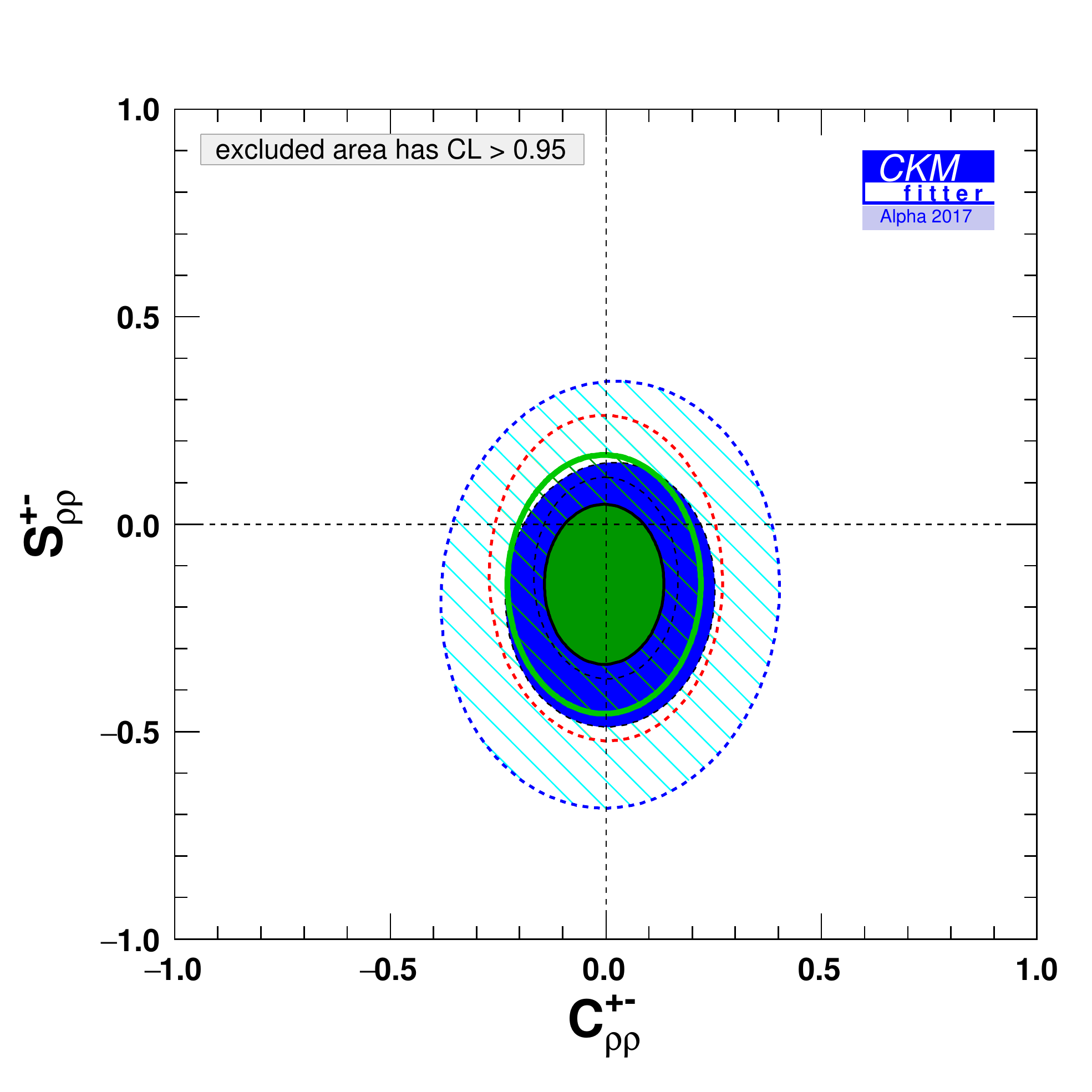}
    \includegraphics[width=18pc]{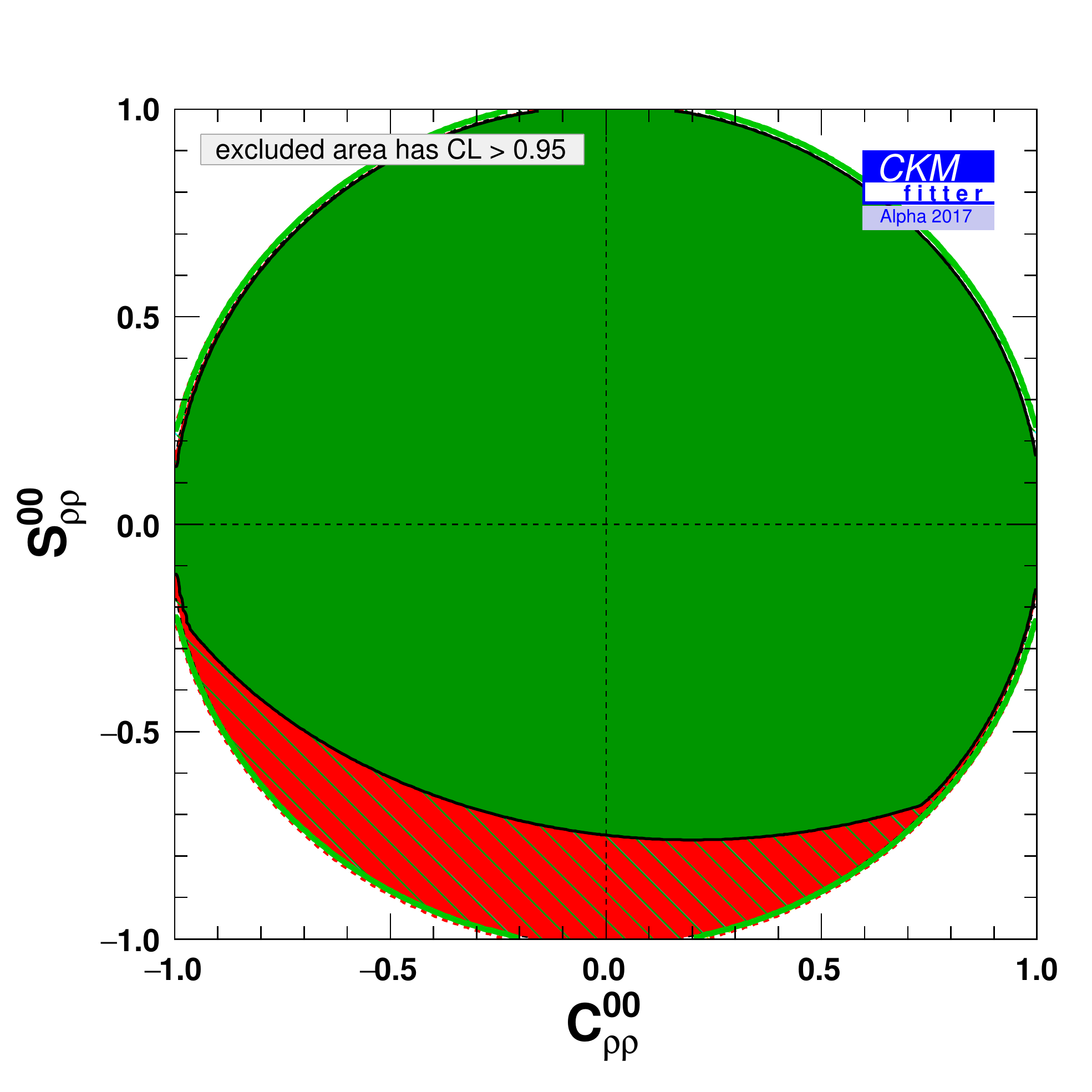}
\caption{\it\small  Measurements of the $B\to\rho\rho$ $CP$ asymmetries from \babar (blue area) and   
Belle (red area). The  LHCb contribution to the $B\to\rho^0\rho^0$ branching fraction  is indicated by the orange area.  The green shaded area represents the world average.}
\label{fig:rhorho_exp2}
\end{center}       
\end{figure}

The individual  measurements of the $B\to\pi\pi$ and  $B\to\rho\rho$ observables are shown in Figs.~\ref{fig:pipi_exp1} and \ref{fig:pipi_exp2}, and Figs.~\ref{fig:rhorho_exp1} and \ref{fig:rhorho_exp2}, respectively. 
For each observable, a good agreement is observed between the different sources of experimental data (\babar, Belle, and when available, other experiments).

\begin{figure}[t]
\begin{center}
  \includegraphics[width=18pc]{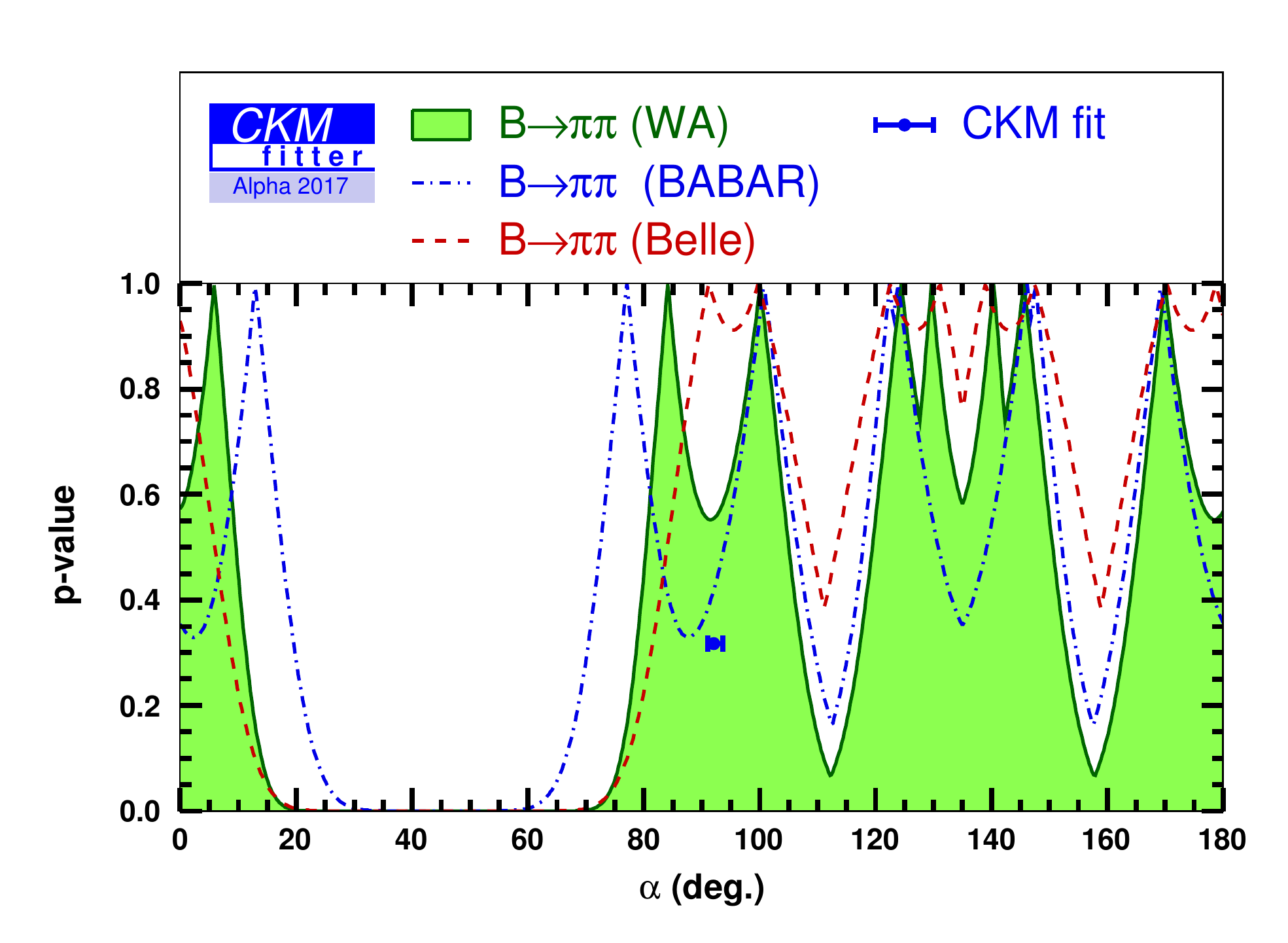}
  \includegraphics[width=18pc]{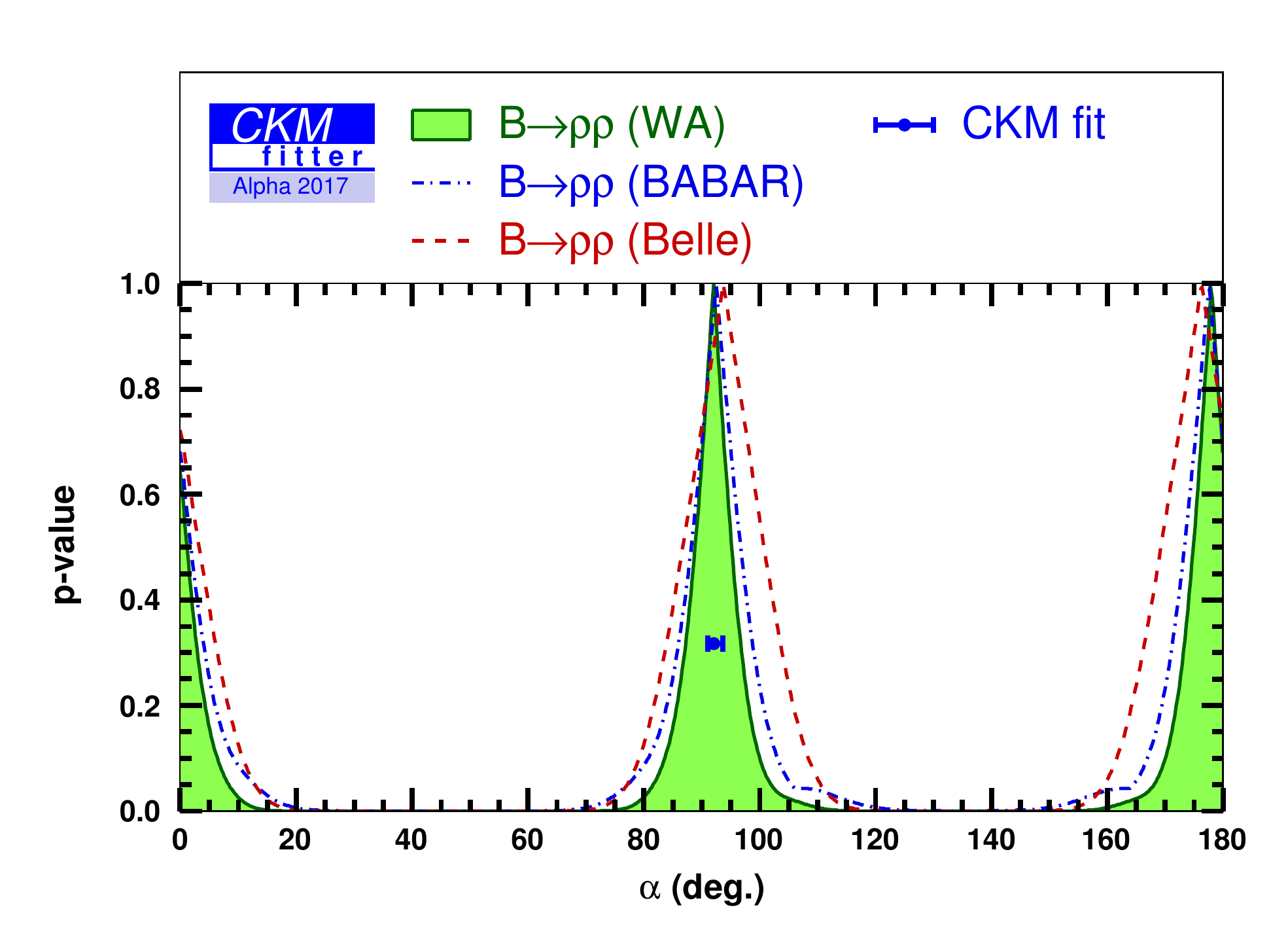}
\caption{\it\small  Constraint on $\alpha$ from the isospin analysis of the  $B\to\pi\pi$  (left) and  $B\to\rho\rho$ system (right) using \babar data (blue curve) and Belle data (red curve). The green shaded area represents the determination based on the world average for these observables. The interval with a dot indicates the indirect determination introduced in Eq.~(\ref{eq:alphaInd}).}\label{fig:exp_alphaHH}
\end{center}       
\end{figure}

The  individual determinations of $\alpha$ based on the \babar and Belle data separately are shown in Fig.~\ref{fig:exp_alphaHH}.
The corresponding 68\%  CL intervals are:
\begin{eqnarray}
\alpha_{\pi\pi}  &{\rm\scriptstyle(\babar)}&:~\val{77.1}{+32.1}{-6.6}{$^\circ$} ~~\and~~ (\val{135.0}{19.0})^\circ ~~\and~~\val{169.3}{+40.3}{-8.4}{$^\circ$}\,,\\
                &{\rm\scriptstyle(Belle)}& :~ (\val{134.8}{53.8})^\circ  \,,\\
\alpha_{\rho\rho}&{\rm\scriptstyle(\babar)}&  :~\val{92.5}{+6.3}{-6.5}{$^\circ$} ~~\and~~ \val{177.6}{+6.6}{-6.3}{$^\circ$}\,,\\
                &{\rm\scriptstyle(Belle)}& :~ \val{93.7}{+9.9}{-9.8}{$^\circ$} ~~\and~~ \val{176.3}{+9.8}{-9.9}{$^\circ$} \,.
\end{eqnarray}

\subsection{$B\to\rho\pi$ analysis}

\begin{figure}[t]
\begin{center}
    \includegraphics[width=12pc,height=11pc]{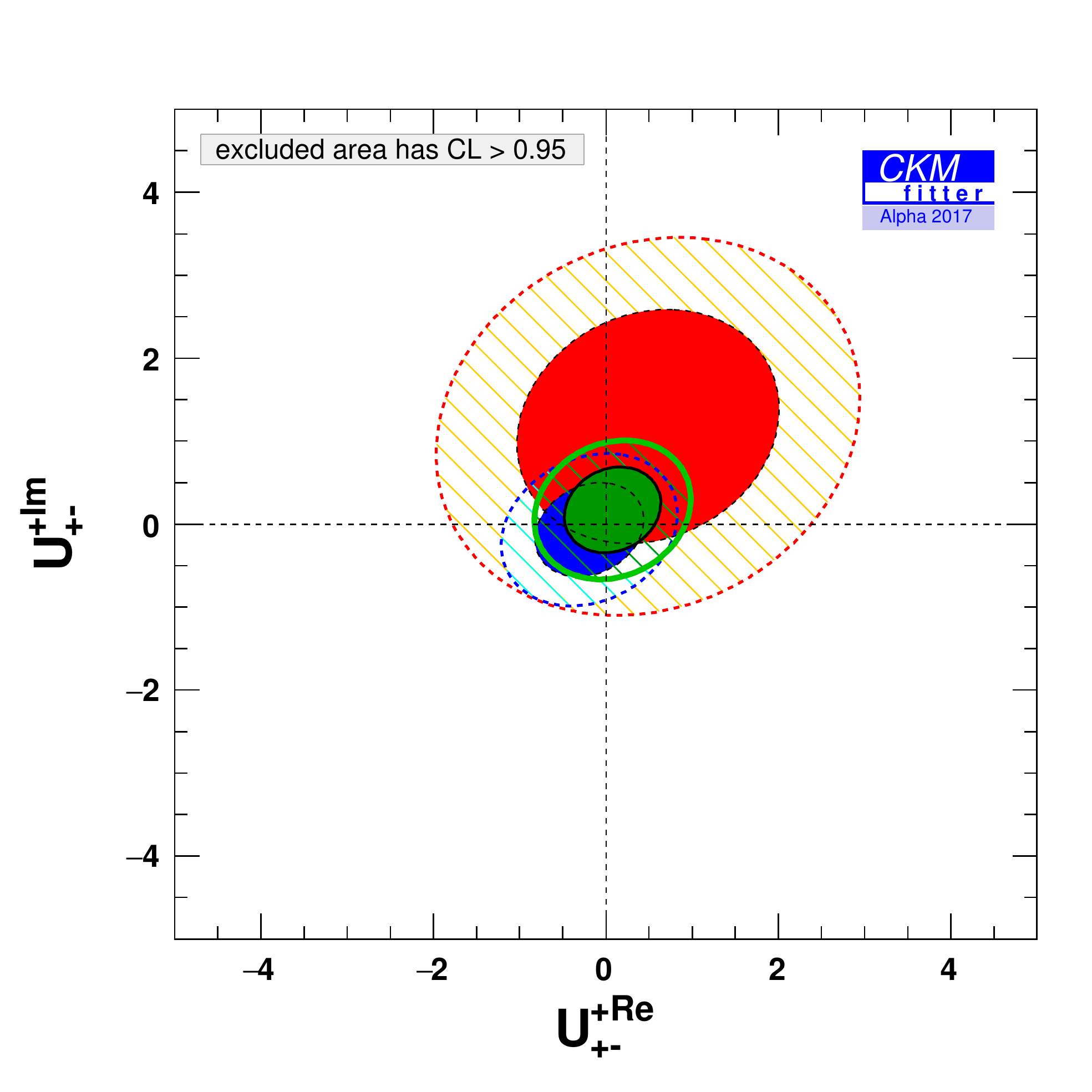}
    \includegraphics[width=12pc,height=11pc]{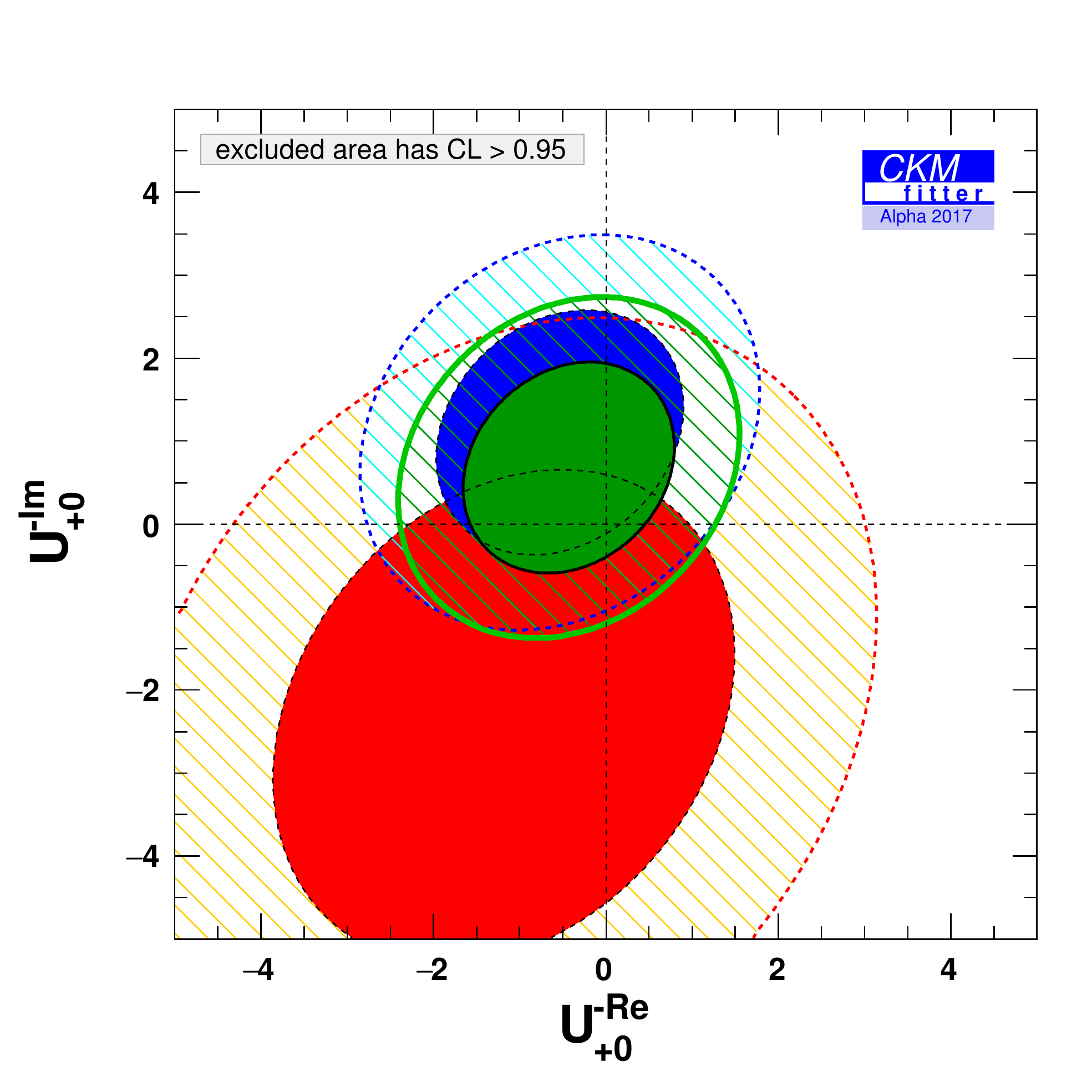}
    \includegraphics[width=12pc,height=11pc]{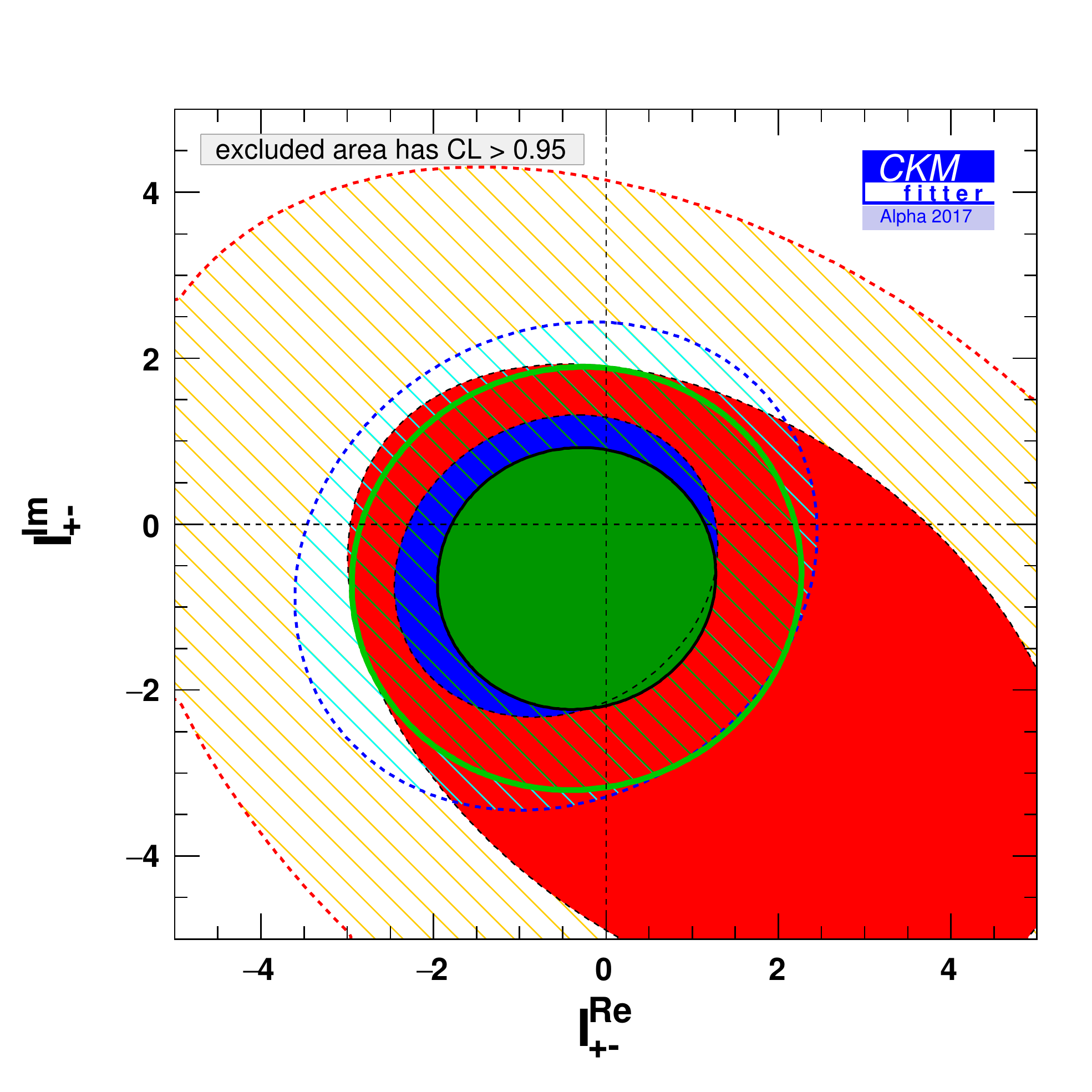}
    \includegraphics[width=12pc,height=11pc]{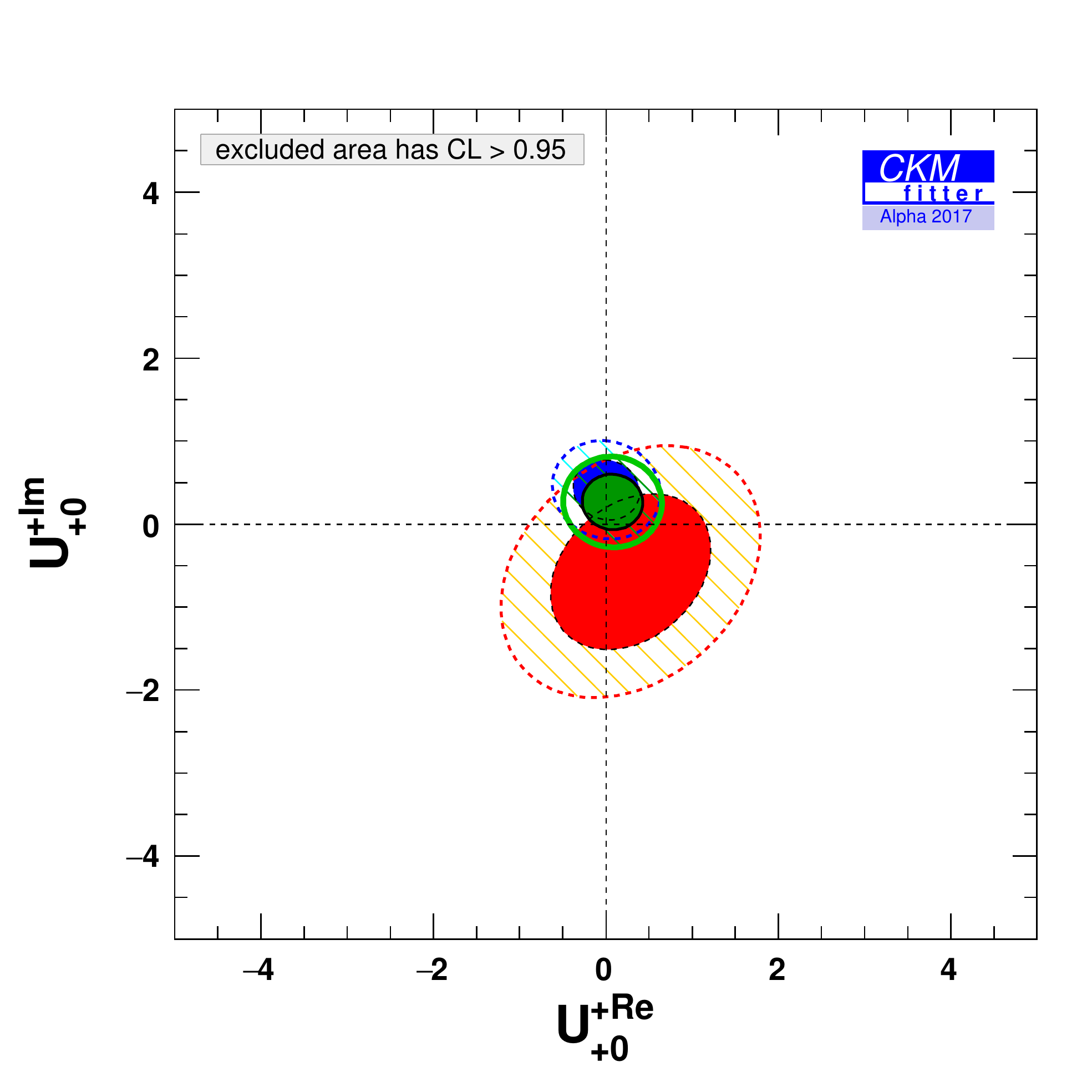}
    \includegraphics[width=12pc,height=11pc]{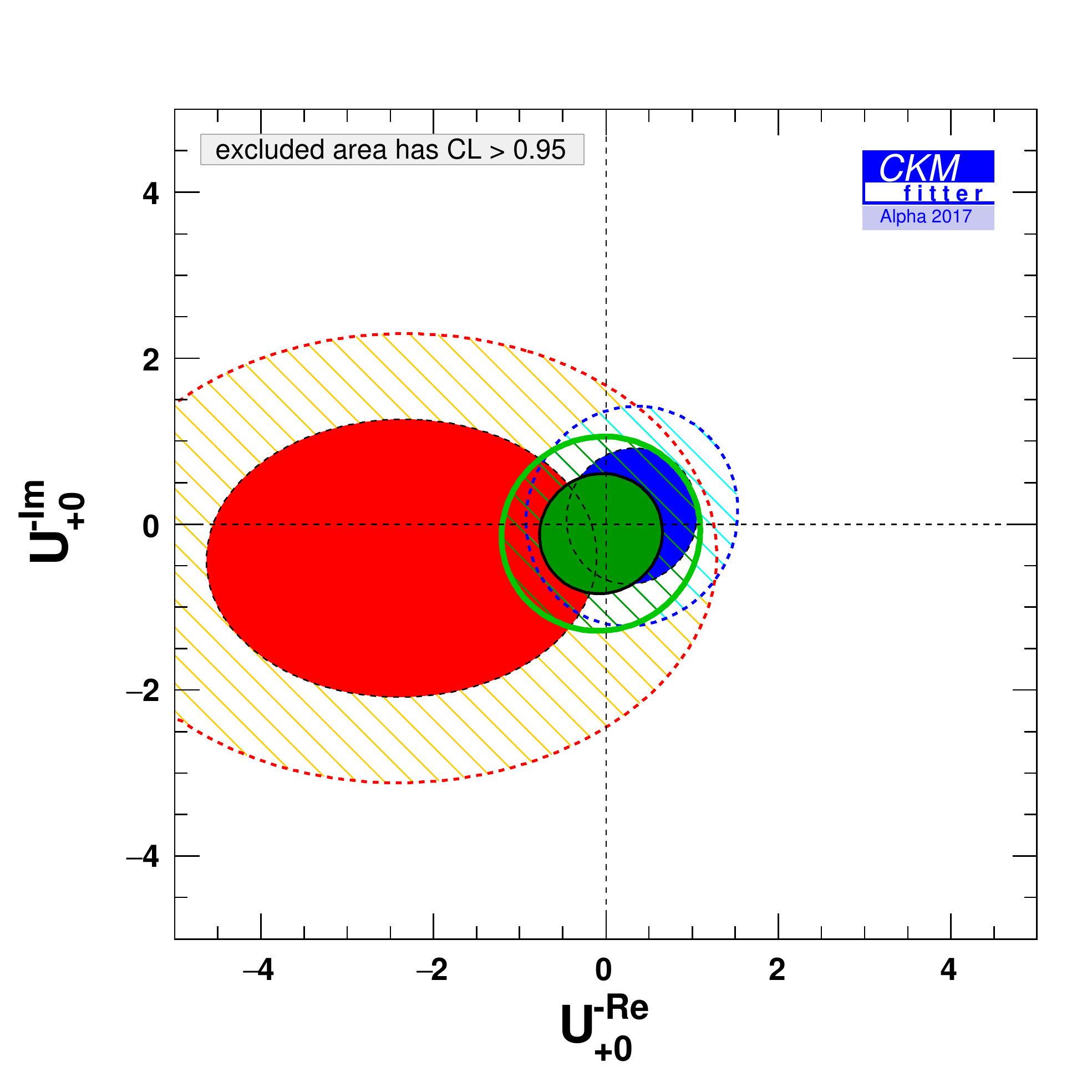}
    \includegraphics[width=12pc,height=11pc]{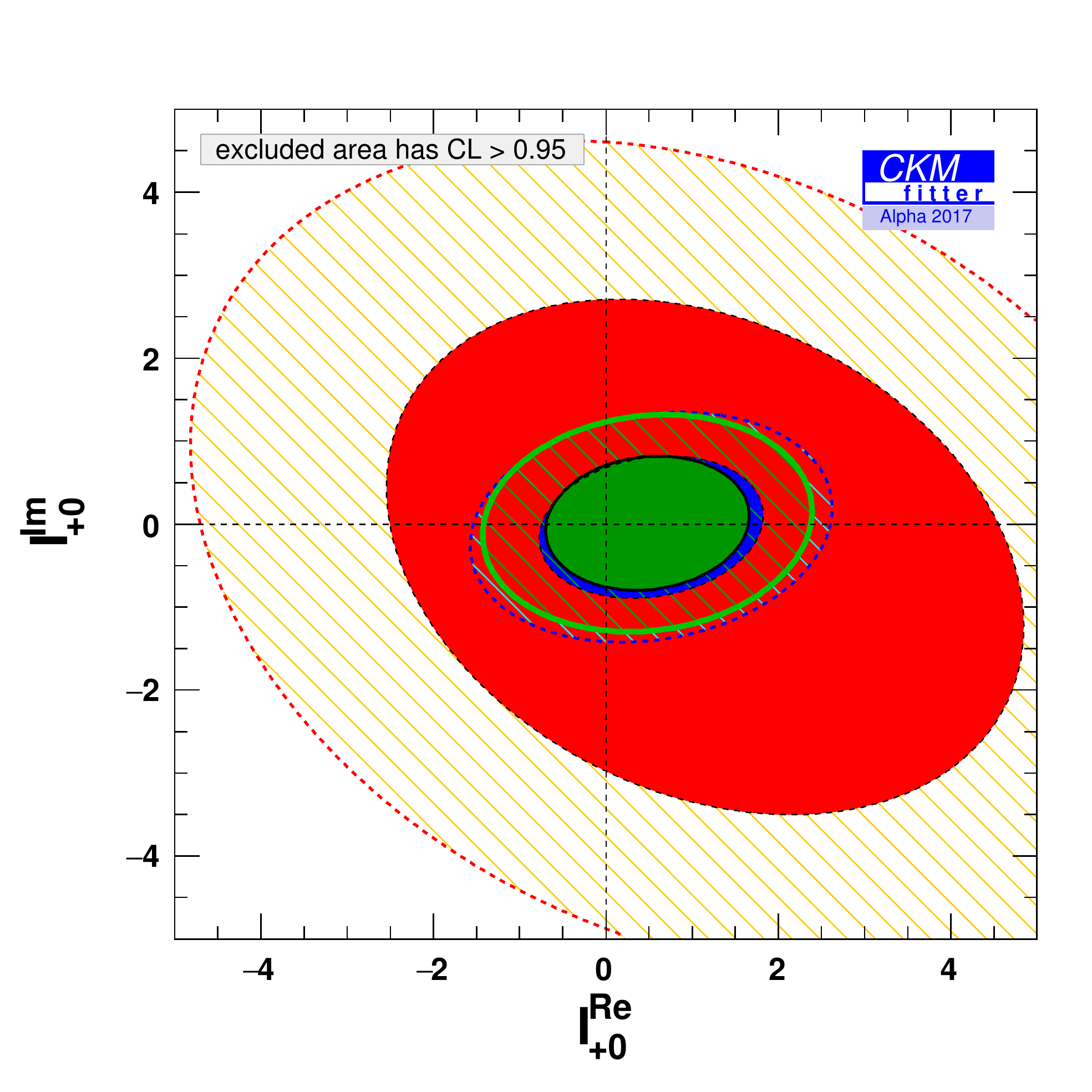}
    \includegraphics[width=12pc,height=11pc]{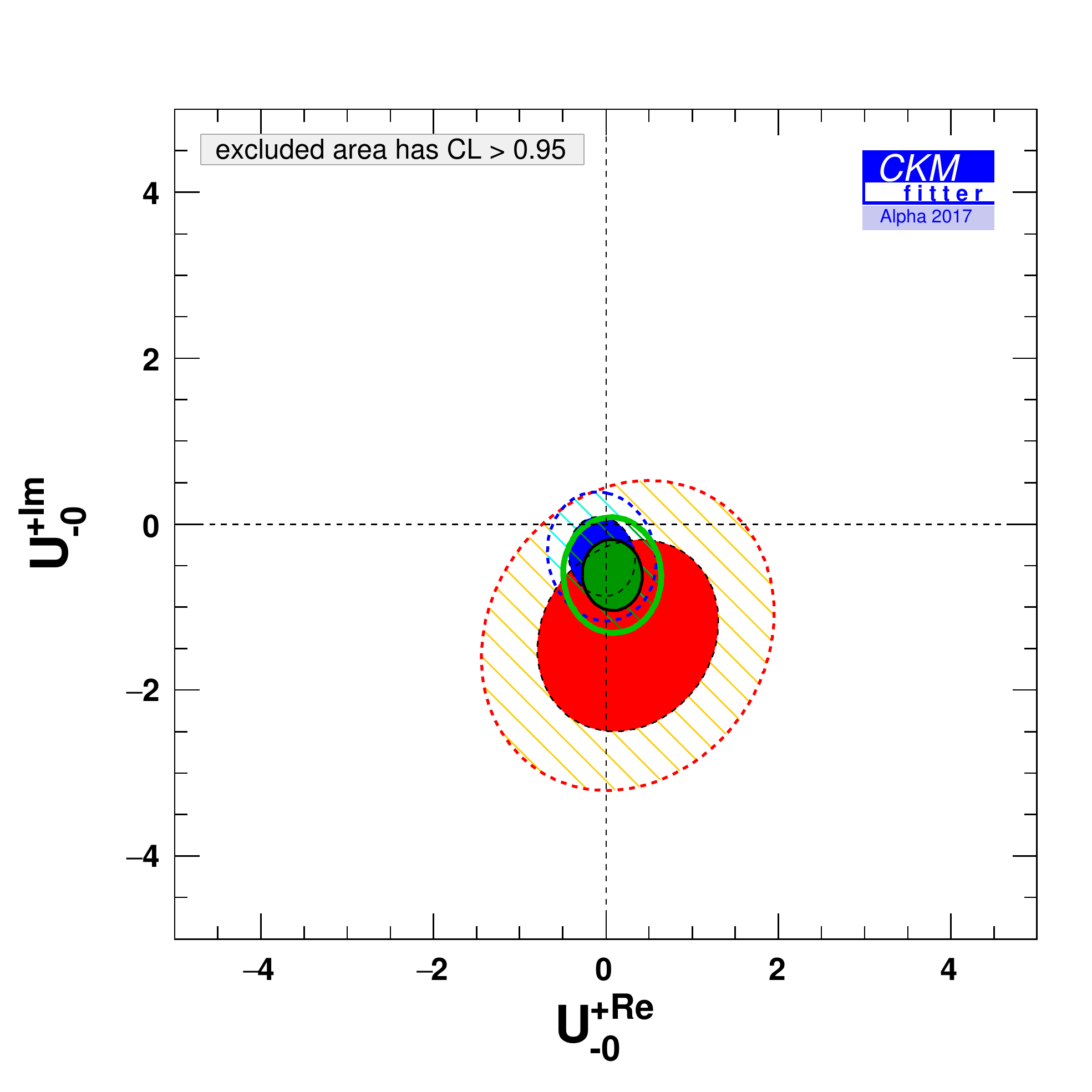}
    \includegraphics[width=12pc,height=11pc]{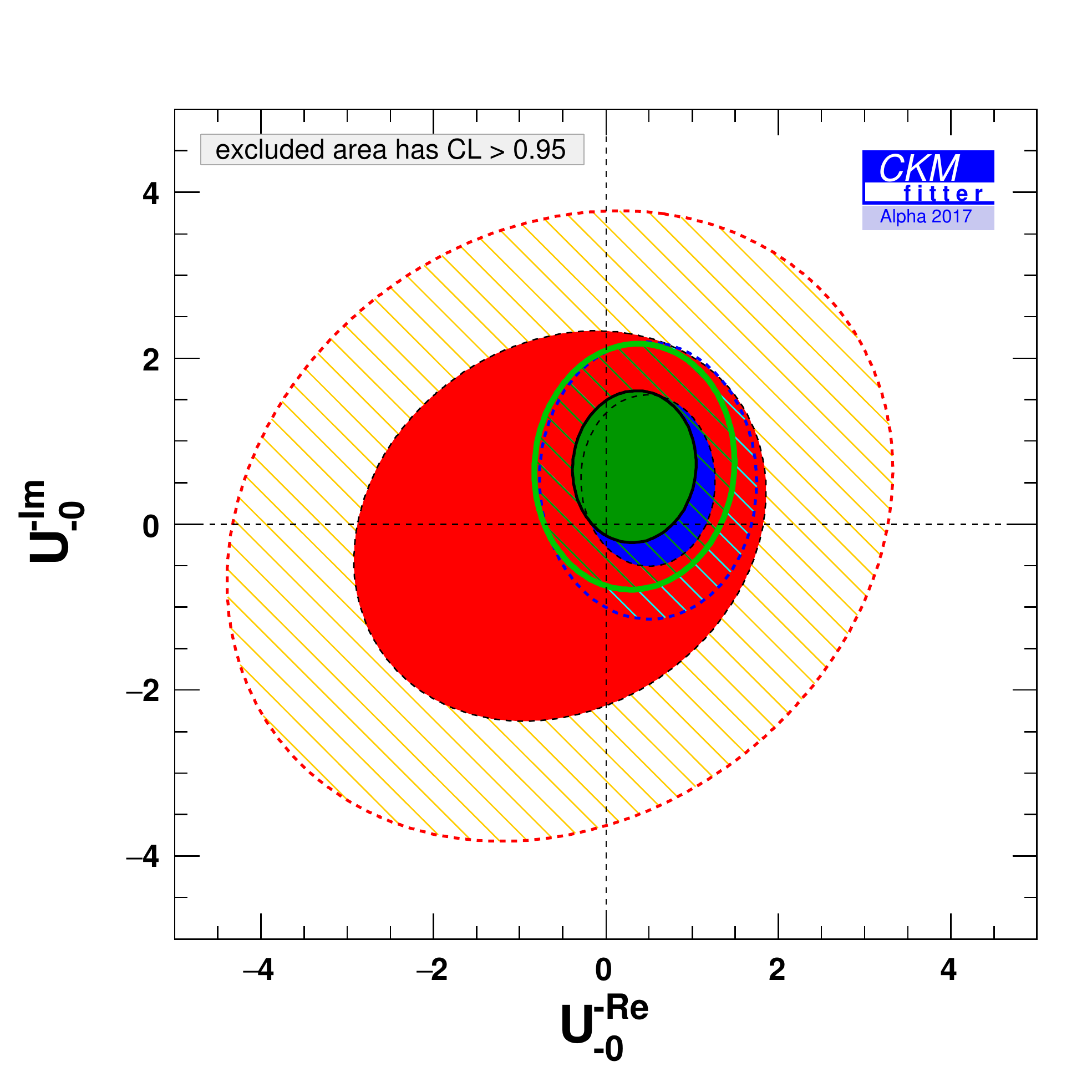}
    \includegraphics[width=12pc,height=11pc]{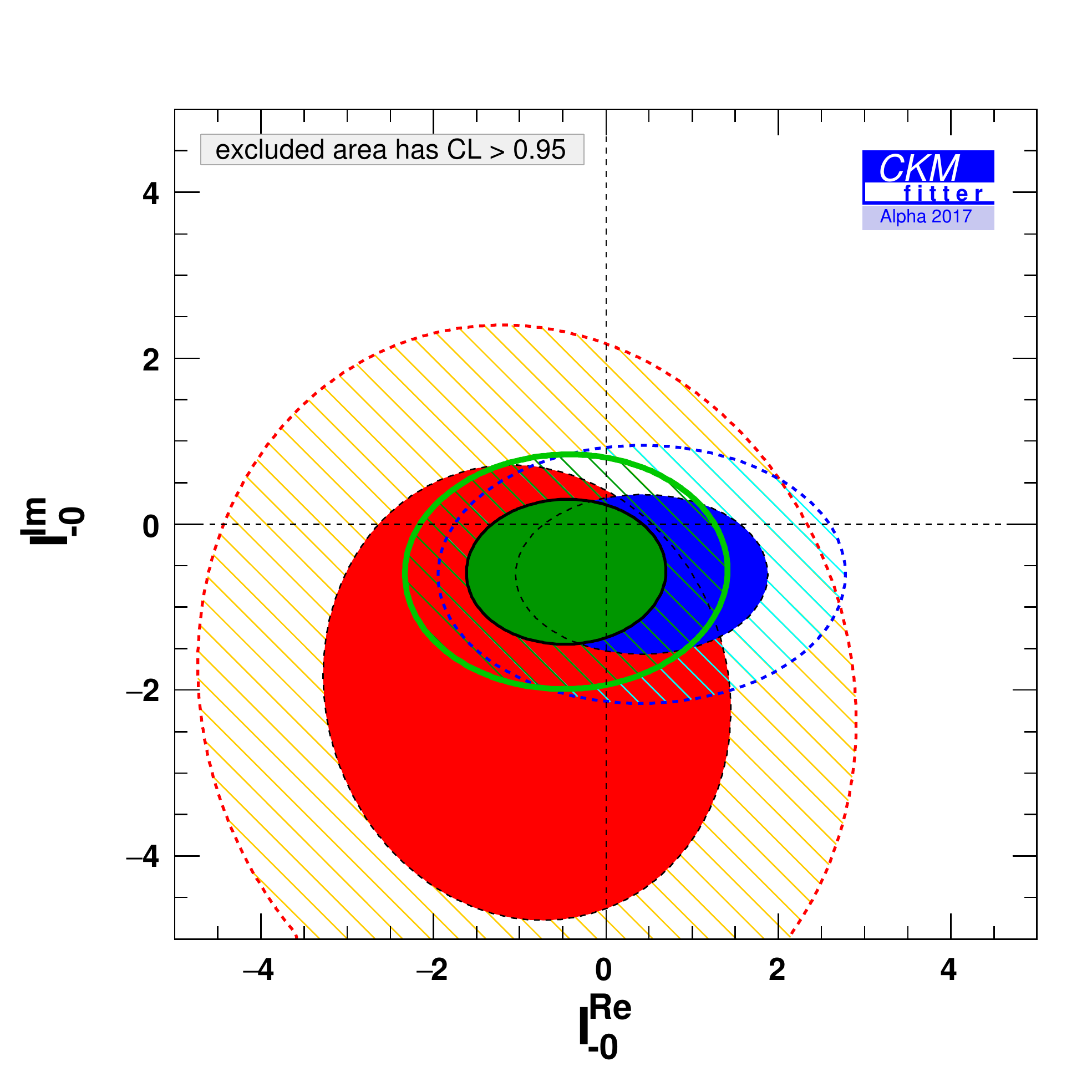}
\caption{\it\small  Experimental measurements  of the interference-related  \U and \I parameters from \babar (blue area) and Belle (red area). The green shaded area represents the world average.}\label{fig:exp_UI_interf}
\end{center}       
\end{figure}

\begin{figure}[t]
  \begin{center}
    \includegraphics[width=18pc,height=12pc]{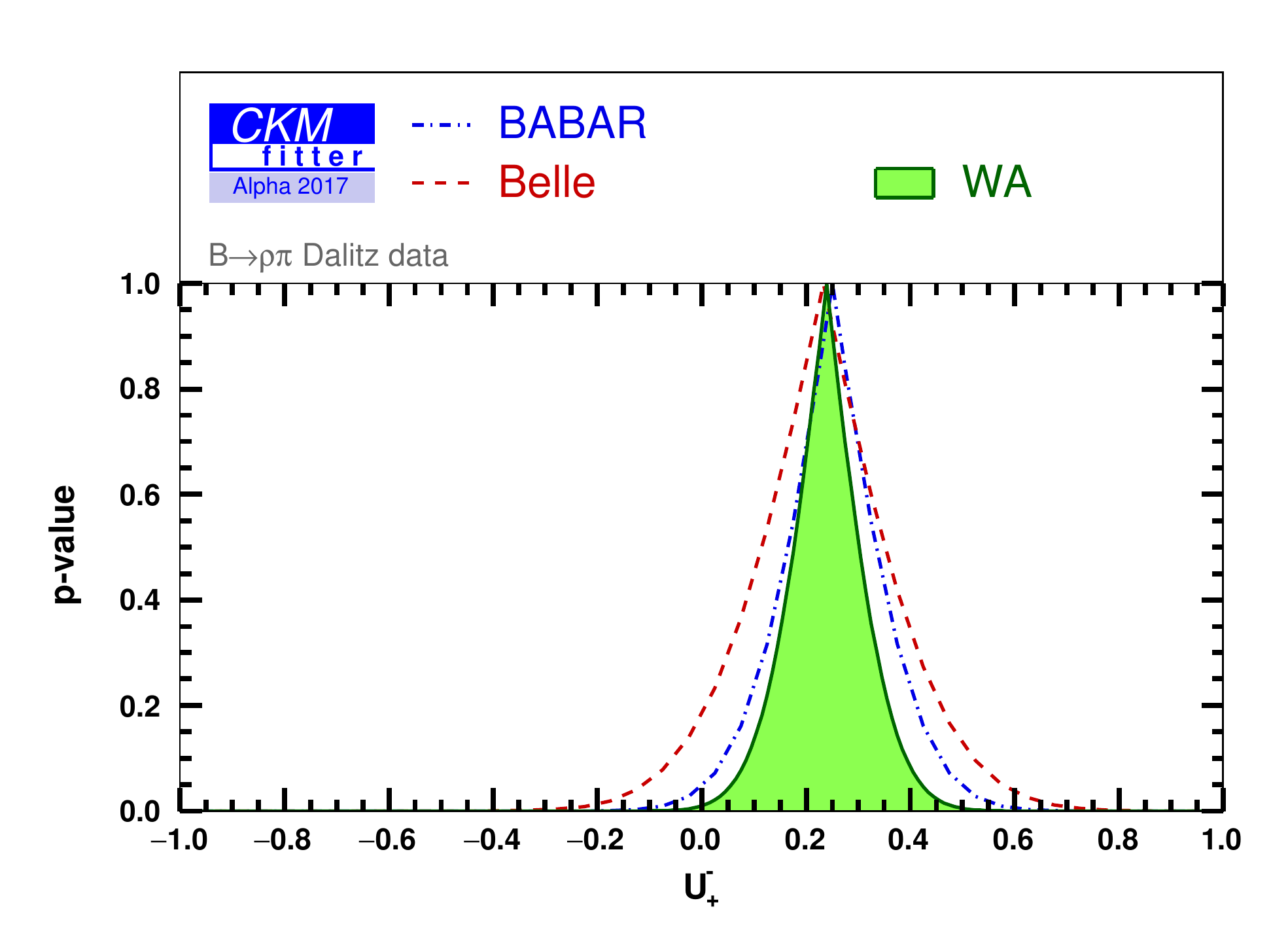}
    \includegraphics[width=18pc,height=12pc]{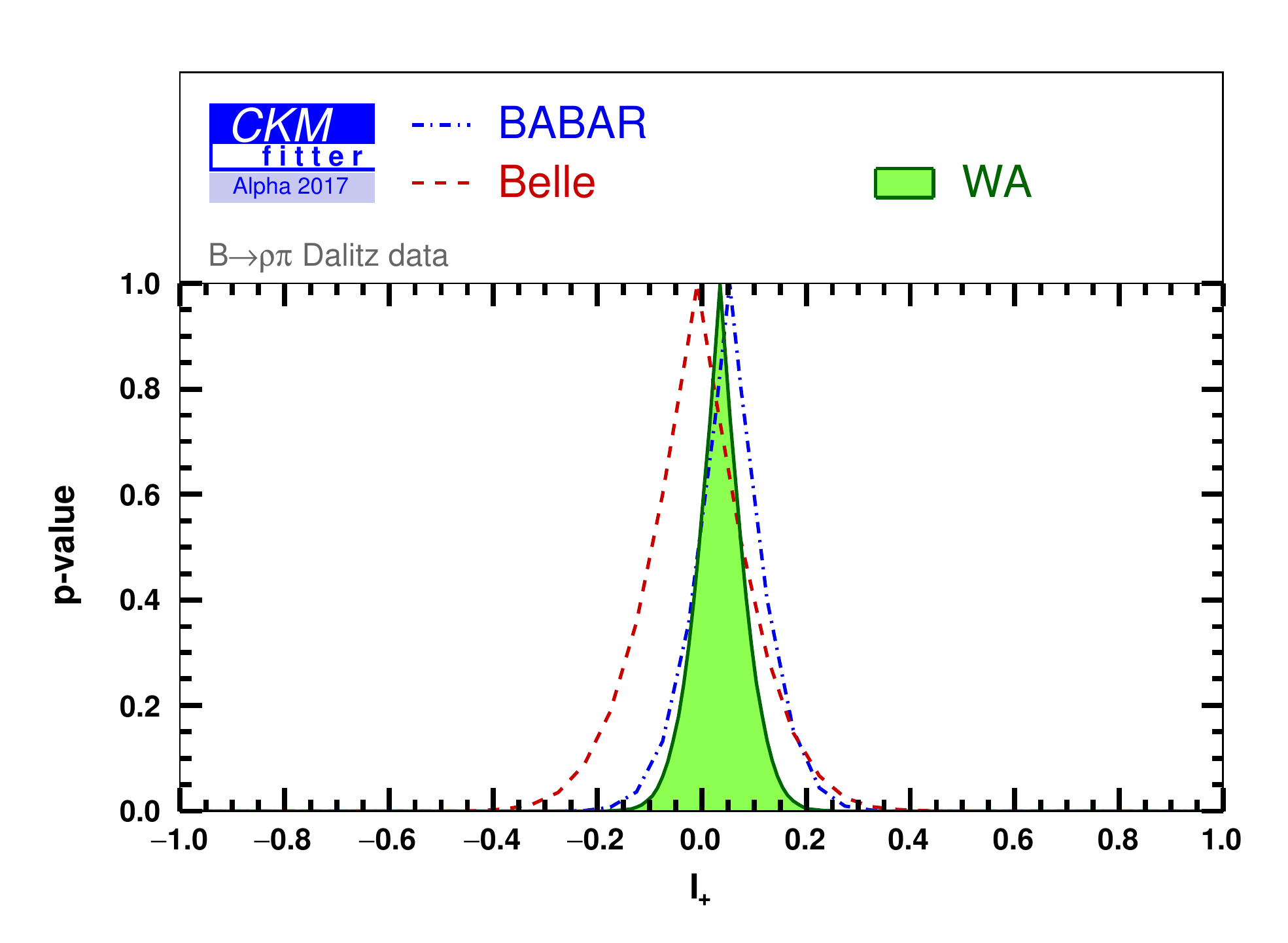}
    \includegraphics[width=18pc,height=12pc]{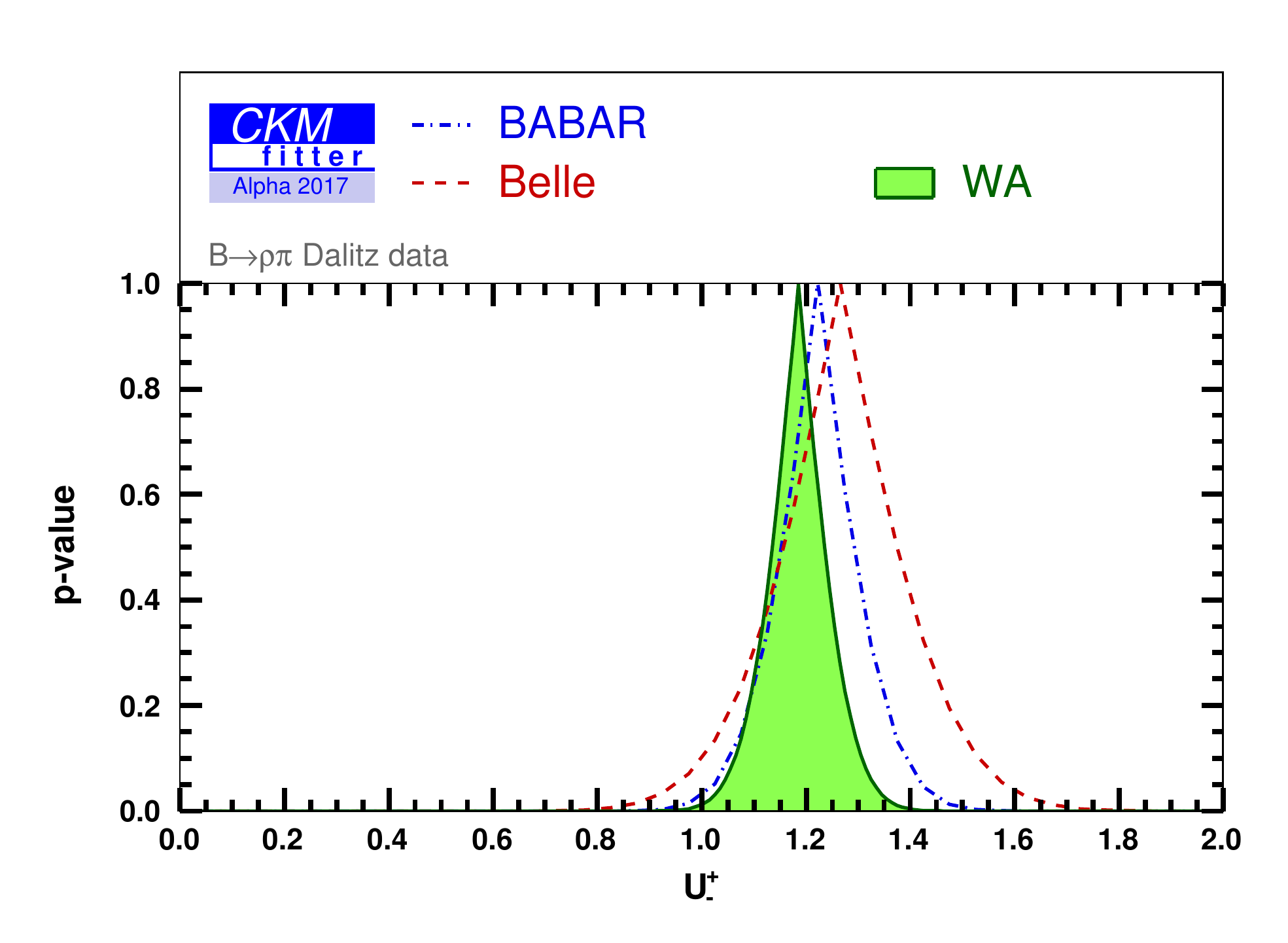}
    \includegraphics[width=18pc,height=12pc]{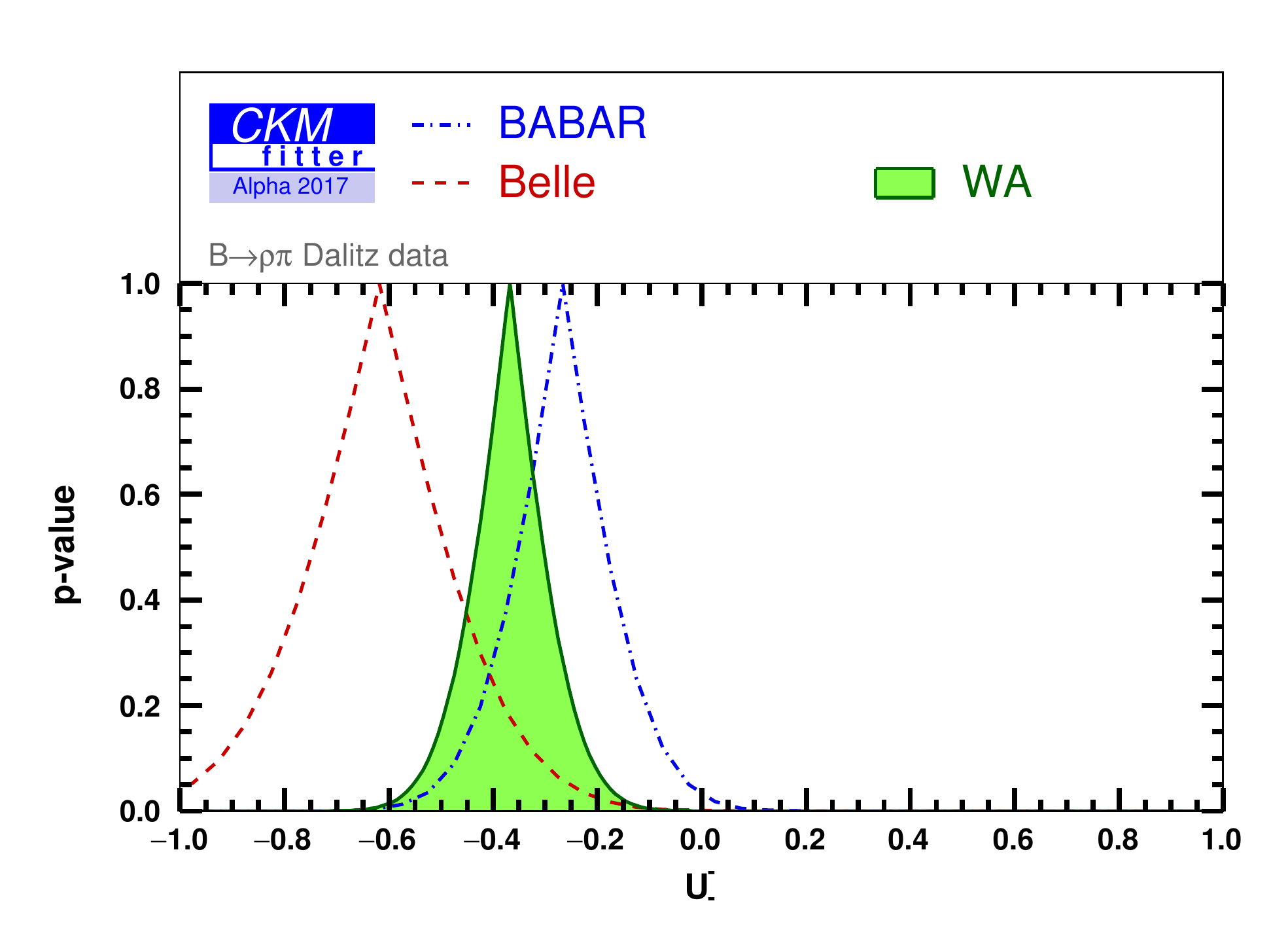}
    \includegraphics[width=18pc,height=12pc]{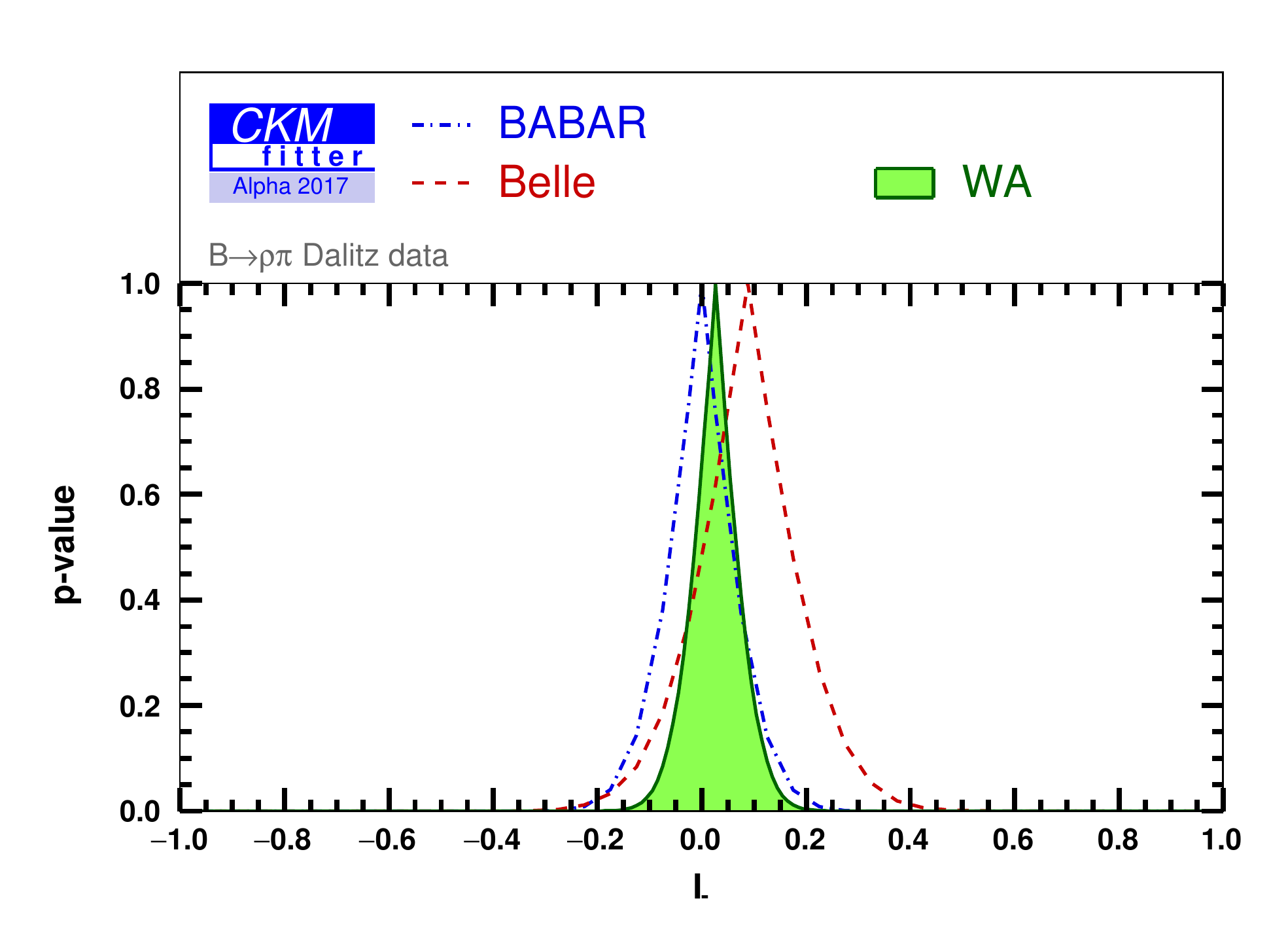}
    \includegraphics[width=18pc,height=12pc]{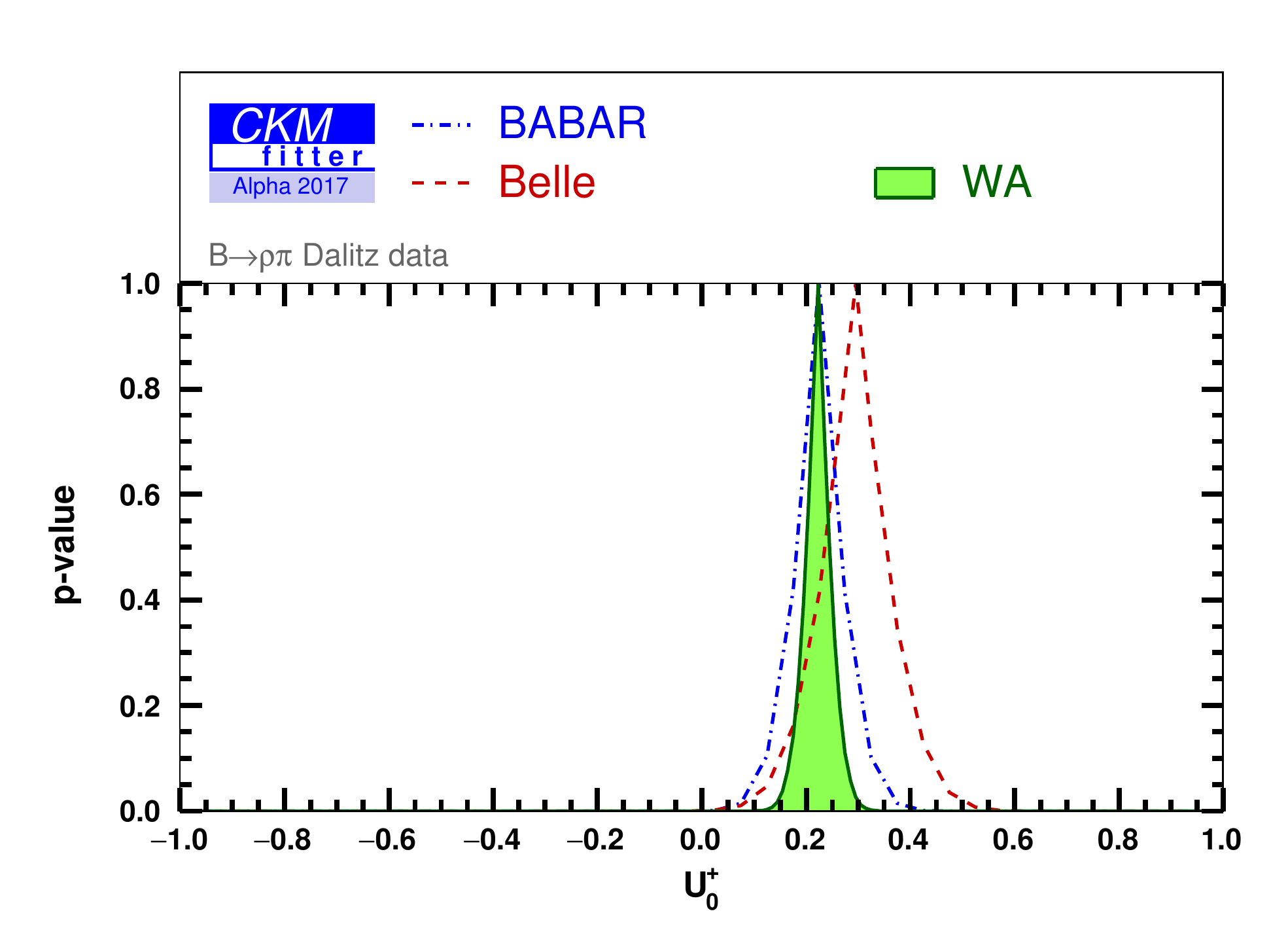}    
    \includegraphics[width=18pc,height=12pc]{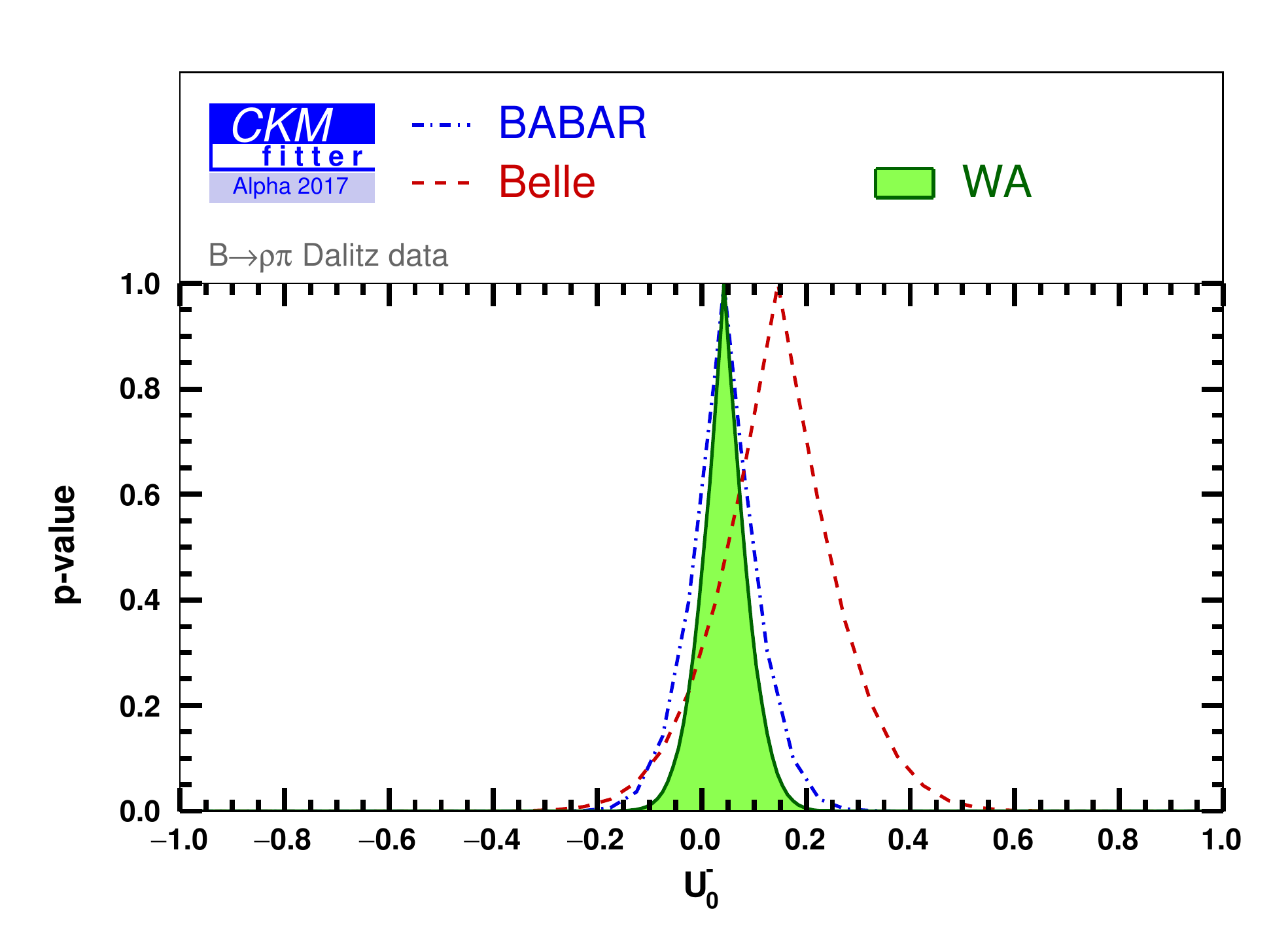}
    \includegraphics[width=18pc,height=12pc]{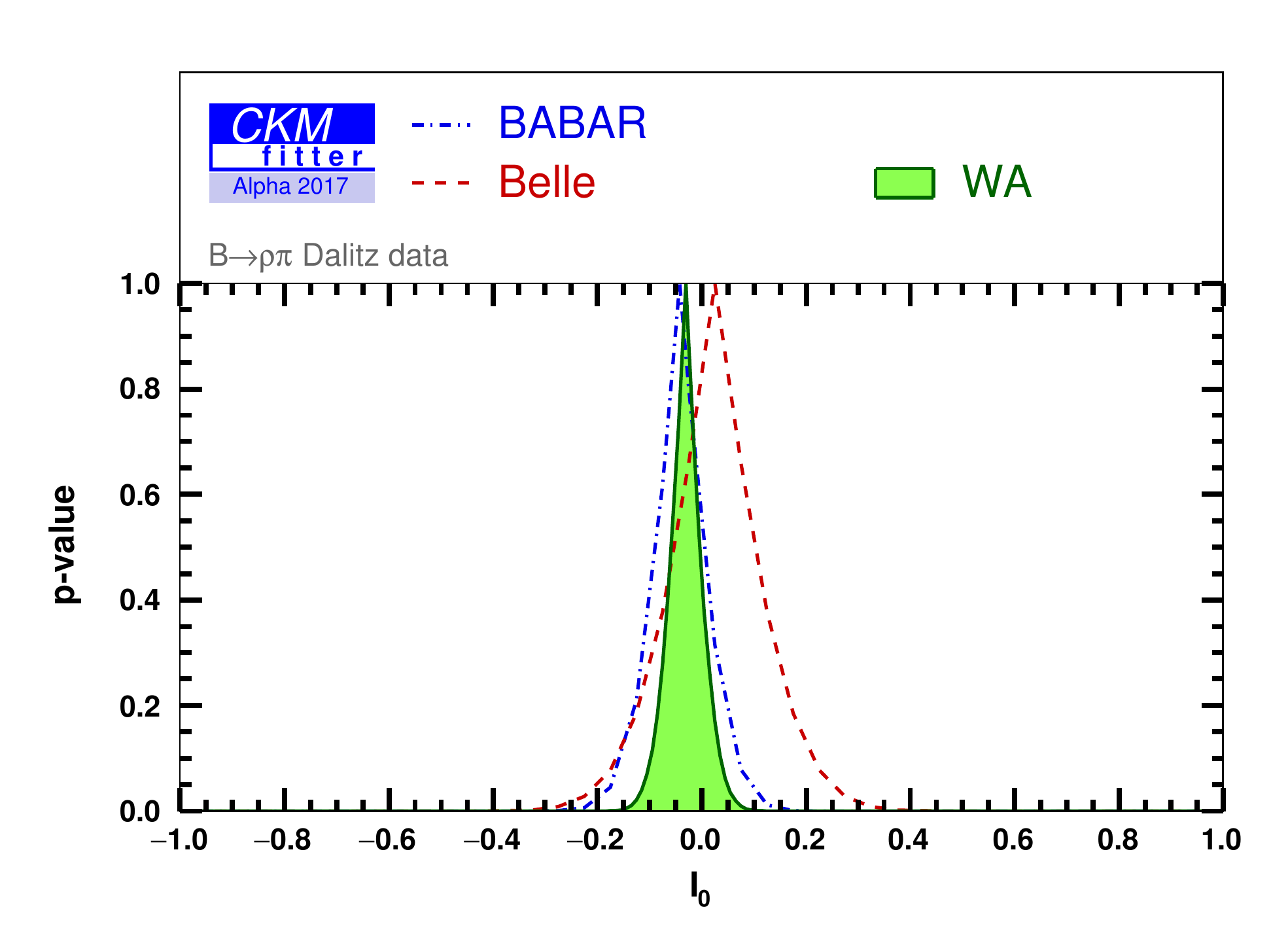}
\caption{\it\small Experimental measurements  of the Q2B-related  \U and \I parameters from \babar (blue curve) and Belle (red curve). The green shaded area represents the world average.}\label{fig:exp_UI_Q2B1}
\end{center}       
\end{figure}

Both B-factories have performed a full Dalitz analysis of the $B^0\to\pi^+\pi^-\pi^0$ decay using the same \U and \I observables.
The  measurements of the Q2B- and interference-related coefficients are summarised in Fig.~\ref{fig:exp_UI_Q2B1}  and 
Fig.~\ref{fig:exp_UI_interf}, respectively. Good agreement between the two experiments is observed. 
The correlated average gives $\chi^2/n_{\rm dof}=18.1/26$.

\begin{figure}[t]
\begin{center}
  \includegraphics[width=18pc]{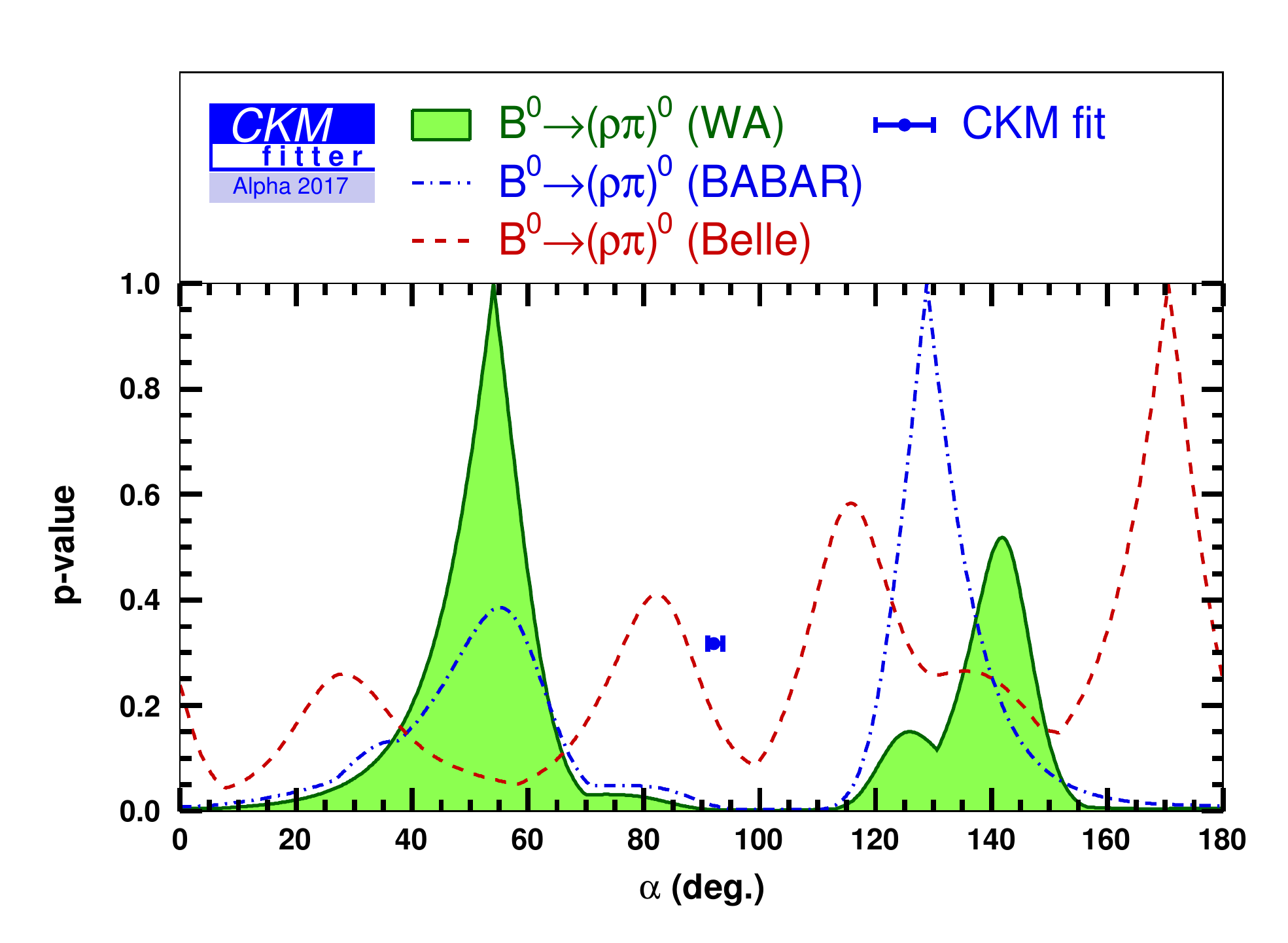}
  \includegraphics[width=18pc]{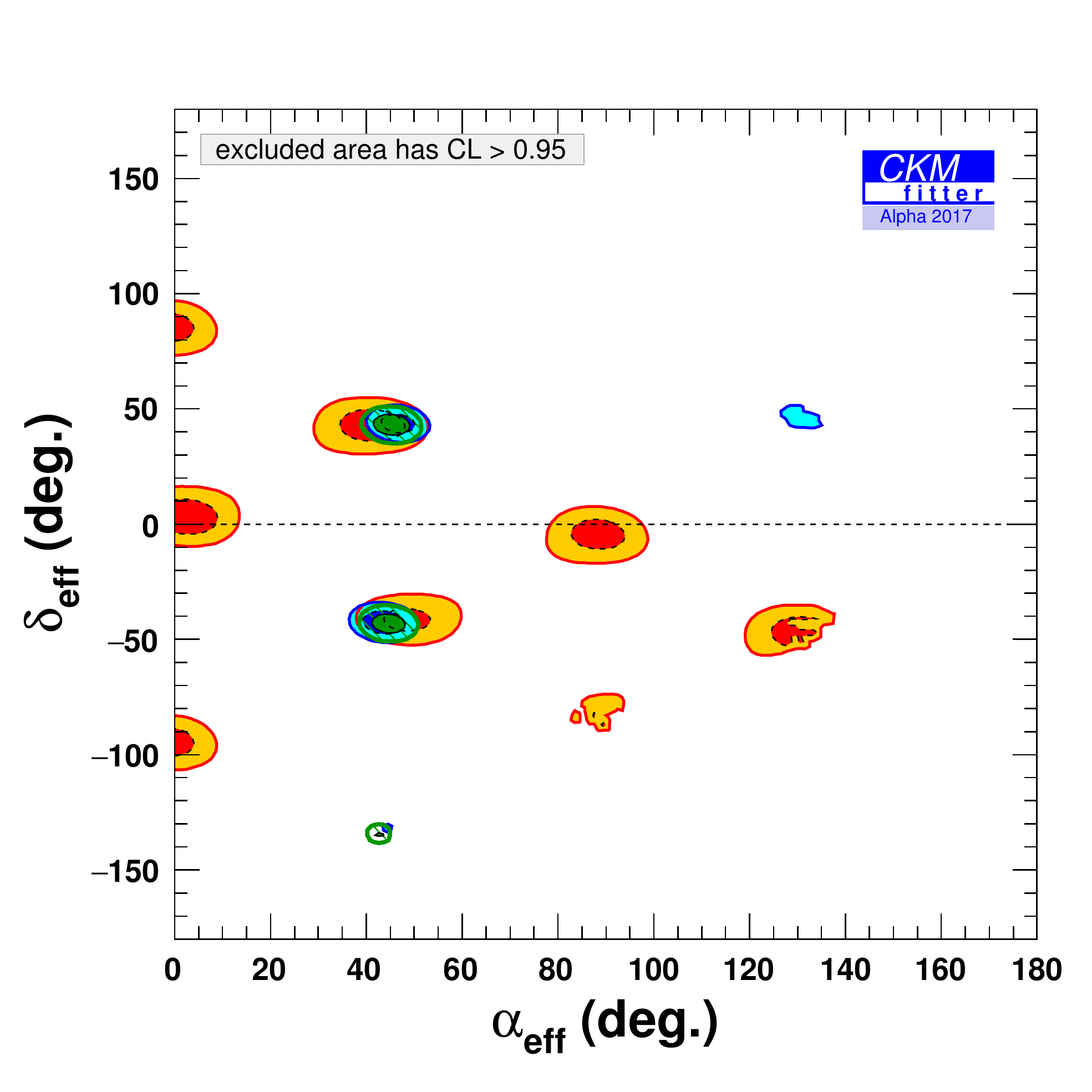}
\caption{\it\small  Constraint on $\alpha$ from the \su{2} isospin analysis of the  $B^0\to(\rho\pi)^0$ system (left) using \babar data (blue curve) and Belle data (red curve). The green shaded  area represents the determination based on the world average.  The interval with a dot indicates the indirect $\alpha$ determination introduced in Eq.~(\ref{eq:alphaInd}). The right panel represents the two-dimensional constraint in the ($\alpha_{\rm eff}$,$\delta_{\rm eff}$) plane  with the same color code.}\label{fig:exp_alpha_rhopi}
\end{center}       
\end{figure}

The  individual determinations of $\alpha_{\rho\pi}$ based on  \babar and Belle data separately are shown in Fig.~\ref{fig:exp_alpha_rhopi} (left panel).
The corresponding 68\% intervals are
\begin{eqnarray}
\alpha_{\rho\pi}  &{\rm\scriptstyle(\babar)}&:\val{55.3}{+4.7}{-5.8}{$^\circ$} ~~\and~~ \val{128.9}{+9.5}{-7.1}{$^\circ$} \,,\\ 
                &{\rm\scriptstyle(Belle)}& :\val{82.3}{+5.2}{-5.7}{$^\circ$} ~~\and~~ \val{115.7}{+9.8}{-8.0}{$^\circ$}~~\and~~\val{170.5}{+8.3}{-11.2}{$^\circ$} \,.
\end{eqnarray}
The $B^0\to\pi^+\pi^-\pi^0$  data from Belle and \babar are consistent with  the indirect determination $\alpha_{\rm ind}$ at the level of 1.3 and 2.6 standard deviations, respectively. Being the projection of the combination of constraints in a multi-dimensional parameter space, the constraint on $\alpha$ that results from the averaged  data cannot be  interpreted as the direct combination of  the one-dimensional constraints.  A more explicit picture is obtained when representing the constraint  in the  ($\alpha_{\rm eff}$,$\delta_{\rm eff}$) plane, as shown in  Fig.~\ref{fig:exp_alpha_rhopi} (right panel). The  preferred $\alpha$ solution  near $50^\circ$  clearly appears as the solution favoured by both Belle and \babar experiments.

\subsection{Combined analysis}
\begin{figure}[t]
\begin{center}
  \includegraphics[width=22pc]{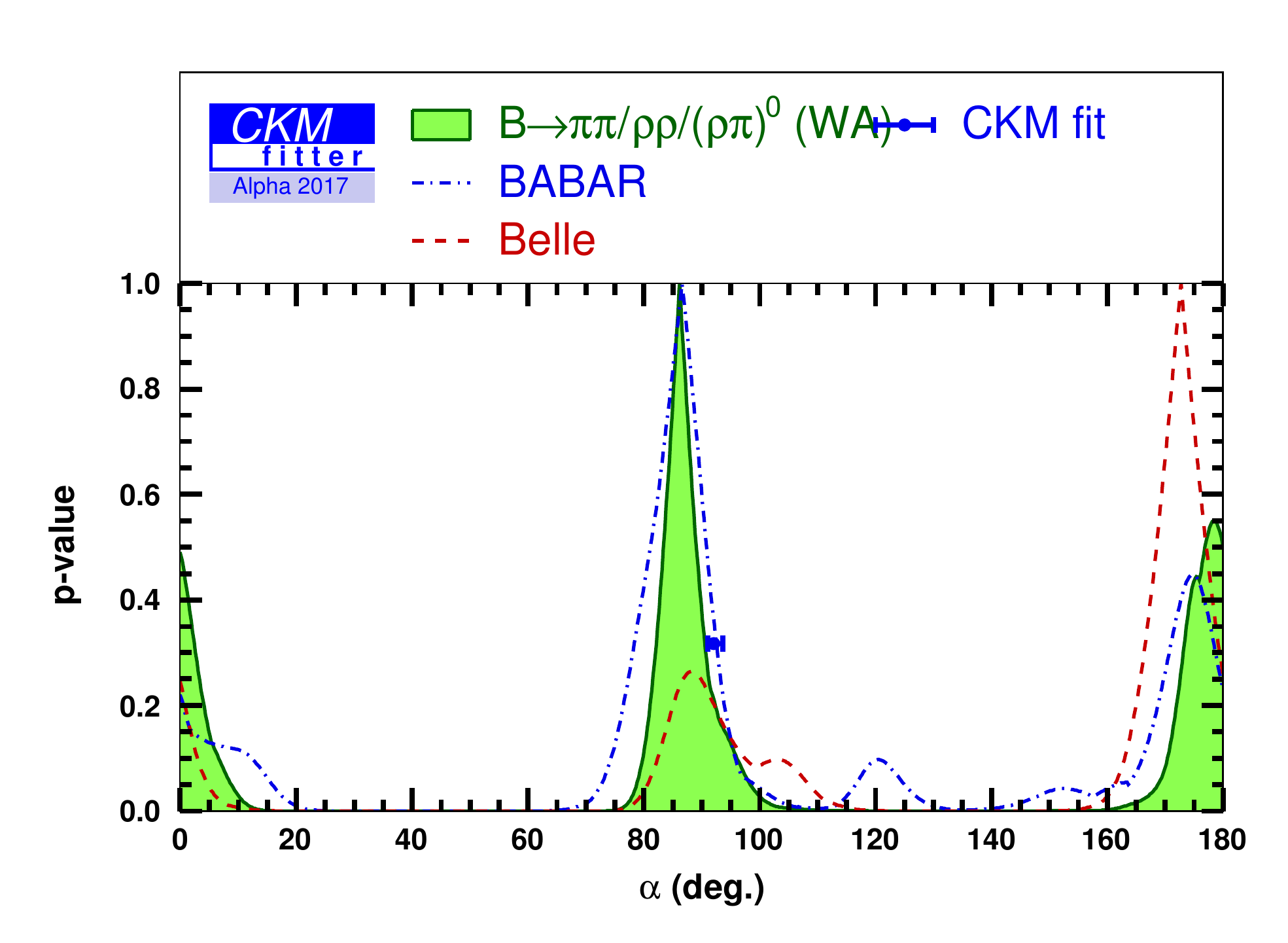}
\caption{\it\small
 Combined constraint on $\alpha$ from the \su{2} isospin analysis of $B\to\pi\pi$,  $B\to\rho\rho$ and  $B^0\to(\rho\pi)^0$  systems using \babar data (blue curve) and Belle data (red curve). The green shaded area represents the determination based on the world average for these observables. The interval with a dot indicates the indirect determination introduced in Eq.~(\ref{eq:alphaInd}).}\label{fig:exp_UU}
\end{center}       
\end{figure}

The combined constraints on $\alpha$ using \babar and Belle data separately are shown in Fig.~\ref{fig:exp_UU}. The corresponding 68\% CL intervals are
\begin{eqnarray}
\alpha_{\rm dir}  &{\rm\scriptstyle(\babar)}&:~\val{86.6}{+5.9}{-8.3}{$^\circ$} ~~\and~~ \val{174.8}{+3.6}{-3.8}{$^\circ$}  \,,\\ 
                &{\rm\scriptstyle(Belle)}& :~\val{172.7}{+6.5}{-6.1}{$^\circ$}\,.
\end{eqnarray}
The agreement with the indirect $\alpha_{\rm ind}$ determination is 0.9\SIG and 1.5\SIG for  \babar and Belle data, respectively. One can also notice that  \babar and Belle data do not favour the same peak around $0^\circ$ or $90^\circ$, even though both intervals in $\alpha$ are acceptable at 95\% CL.

\cleardoublepage
\section{Numerical tables for the prediction of observables} \label{sec:numerics}

In the following tables, we present the results for the indirect determination of various observables discussed in Sec.~\ref{sec:observables}. For each quantity, we present the world average of the available measurements, the indirect determination from the \su{2} isospin analysis without including this observable,  and the compatibility pull between the two determinations, obtained by comparing the minimum value of the \chisq function with and without the experimental measurement of this quantity, see Eq.~(\ref{eq:pull}) (Tabs.~\ref{tab:BR_PiPi}, \ref{tab:BR_RhoRho}, \ref{tab:UI_rhopi}, \ref{tab:Q2B_rhopi}, \ref{tab:BR_RhoPi}).

\begin{table}[h]
\begin{center}
\small
\begin{tabular}{|l|c||c|c||c|}
\hline
\small Observable & \small Experimental value & \multicolumn{2}{c|}{ \su{2} constraint (dir. measurement not in fit)} & \Pull \\
\cline{3-4}
           & \small (world average)     &  68\% CL  & 95\% CL                                 &\\
\hline
 \Obs{\B}{+-}{\pi\pi} $(\times 10^6)$  & \val{5.10}{0.19}     & \val{5.9}{+2.2}{-1.3}\and \val{15.4}{+1.7}{-2.2}  & \val{15.4}{+3.8}{-12.0}      & 0. \\
 \Obs{\C}{+-}{\pi\pi}                  & \val{-0.304}{0.047}  & \val{-0.2}{+0.3}{-0.2}\and  \val{0.5}{0.1}        & \val{-0.2}{+0.8}{-0.3}       & 0. \\
 \Obs{\S}{+-}{\pi\pi}                  & \val{-0.662}{0.062}  & \val{-0.6}{+0.3}{-0.2}\and  \val{0.5}{+0.2}{-0.3} & \val{-0.6}{+0.5}{-0.3}\and \val{0.5}{+0.3}{-0.5} & 0. \\
\hline
 \Obs{\B}{+0}{\pi\pi} $(\times 10^6)$  & \val{5.48}{0.34}     & \val{5.0}{+0.7}{-0.9}\and  \val{0.4}{+0.3}{-0.1}  & \val{5.0}{+1.3}{-4.9}         & 0. \\
\hline
 \Obs{\B}{00}{\pi\pi} $(\times 10^6)$  & \val{1.59}{0.18}    & \val{1.3}{+1.0}{-0.3} \and \val{9.9}{+4.7}{-6.8}   & $>0.7$                        & 0. \\
 \Obs{\C}{00}{\pi\pi}                  & \val{-0.34}{0.22}   & \val{-0.49}{+0.12}{-0.08} \and \val{0.82}{+0.05}{-0.06} & \val{-0.49}{+0.34}{-0.16} \and \val{0.82}{+0.10}{-0.18} & 0. \\
\hline
\end{tabular}
\caption{\small\it Indirect \su{2} isospin determination of \decay{B}{\pi\pi} observables compared to their experimental measurement (world average).} \label{tab:BR_PiPi}
\end{center}
\end{table}

\begin{table}[h]
\begin{center}
\small
\begin{tabular}{|l|c||c|c||c|}
\hline
\small Observable & \small Experimental value & \multicolumn{2}{c|}{ \su{2} constraint (dir. measurement not in fit)} & \Pull \\
\cline{3-4}
           & \small (world average)     &  68\% CL  & 95\% CL                                 &\\
\hline
 \Obs{\B}{+-}{\rho\rho} $(\times 10^6)$   & \val{27.76}{1.84}      & \val{29.7}{+31.7}{-4.1} &\val{29.7}{+37.9}{-7.2}   & 0.39\\
 \Obs{f}{+-}{L}                          & \val{0.990}{0.020}      & $>0.91$                 &$>0.79$                   & 0.22\\
 \Obs{\C}{+-}{\rho_L\rho_L}              & \val{-0.00}{0.09}       &\val{-0.03}{+0.19}{-0.17} &\val{-0.03}{+0.35}{-0.31} & 0.14\\
 \Obs{\S}{+-}{\rho_L\rho_L}              & \val{-0.15}{0.13}       &\val{-0.14}{0.15}         &\val{-0.14}{+0.30}{-0.33} & 0.00 \\
\hline
 \Obs{\B}{+0}{\rho\rho} $(\times 10^6)$   & \val{24.9}{1.9}       &\val{21.4}{+2.0}{-13.0}   &\val{21.4}{+3.6}{-14.2}  & 0.39\\
 \Obs{f}{+0}{L}                           & \val{0.950}{0.016}    &$>0.34$                   &$>0.29$                  & 0.39\\
\hline
 \Obs{\B}{00}{\rho\rho}  $(\times 10^6)$  & \val{0.93}{0.14}      &$>0.6$                    &$>0.2$                   & 0.37\\
 \Obs{f}{00}{L}                            & \val{0.71}{0.06}     &$>0.48$                   &$>0.20$                  & 0.37\\
 \Obs{\C}{00}{\rho_L\rho_L}                & \val{0.2}{0.9}       &\val{0.02}{0.46}          &\val{0.02}{+0.79}{-0.81} & 0.18\\
 \Obs{\S}{00}{\rho_L\rho_L}                & \val{0.3}{0.7}       &\val{0.21}{+0.48}{-0.55}  &\val{0.21}{+0.76}{-1.10} & 0.10\\
\hline
\end{tabular}
\caption{\small\it Indirect \su{2} isospin determination of the \decay{B}{\rho\rho} observables compared to their experimental measurement (world average).} \label{tab:BR_RhoRho}
\end{center}
\end{table}

\begin{table}[h]
\begin{center}
\begin{tabular}{|l|c||c|c||c|}
\hline
\small Obs. & \small Experimental value & \multicolumn{2}{c|}{ \su{2} constraint (dir. measurement not in fit)} & \Pull \\
\cline{3-4}
           & \small (world average)     &  68\% CL  & 95\% CL                                 &\\
\hline
\U{+}{-}     & \val{0.24}{0.09} &\val{ 0.98}{+0.04}{-0.15}   &\val{ 0.98}{+0.06}{-0.60}                                         & 2.1\\   
\I{+}        & \val{0.04}{0.06} &  \val{ 0.48}{+0.03}{-0.10}  & \val{ 0.48}{+0.06}{-1.00}                              & 1.6\\
\hline
\U{-}{-}     & \val{-0.37}{0.09} &\val{-1.04}{+0.32}{-0.16}   &\val{-1.04}{+2.35}{-0.24}                                         & 1.8\\
\U{-}{+}     & \val{1.19}{0.07} &\val{ 0.38}{+0.24}{-0.14}   &\val{ 0.38}{+0.62}{-0.25}                                         & 2.2\\
\I{-}        & \val{0.03}{0.06} &  \val{-0.38}{+0.18}{-0.16}  & \val{-0.38}{+0.34}{-0.24}                              & 2.3\\
\hline
\U{0}{+}     & \val{0.22}{0.03} &\val{ 0.09}{+0.09}{-0.07}\and\val{0.31}{0.04}{-0.05}   &\val{ 0.09}{+0.32}{-0.11}              & 1.3\\
\U{0}{-}     & \val{0.04}{0.06} &\val{-0.13}{+0.11}{-0.08}   &\val{-0.13}{+0.24}{-0.14}\and\val{0.21}{0.01}{0.02}      &1.4\\
\I{0}        & \val{-0.03}{0.04} &  \val{-0.07}{0.03}          & \val{-0.07}{+0.22}{-0.07}                              & 1.0\\
\hline
\U{+0}{+\Re} & \val{0.08}{0.23} &\val{-0.49}{0.04}           &\val{-0.49}{0.08}\and\val{ 0.43}{+0.02}{-0.04}                    & 2.4\\
\U{+0}{+\Im} & \val{0.27}{0.21} &\val{-0.40}{+0.19}{-0.08}   &\val{-0.40}{+0.88}{-0.13}                                         & 1.7\\
\U{+0}{-\Re} & \val{-0.06}{0.47} &\val{-0.18}{0.25}\and\val{0.01}{+0.06}{-0.02} &\val{-0.18}{0.10}\and\val{ 0.01}{+0.22}{-0.05}  & 0.3\\
\U{+0}{-\Im} & \val{-0.12}{0.47} &\val{ 0.18}{+0.06}{-0.12}   &\val{ 0.18}{+0.12}{-0.22}                                         & 1.0\\ 
\I{+0}{\Re}  & \val{0.48}{0.78} &  \val{ 0.06}{+0.10}{-0.06}  & \val{ 0.06}{+0.19}{-0.55}                              & 0.8\\
\I{+0}{\Im } & \val{ 0.01}{0.53} &  \val{-0.10}{+0.19}{-0.04}  & \val{-0.10}{+0.63}{-0.10}                              & 0.6\\
\hline
\U{-0}{+\Re} & \val{0.07}{0.22} &\val{-0.50}{0.05}           &\val{-0.50}{0.09}\and\val{ 0.36}{+0.05}{-0.03}                    & 2.3\\
\U{-0}{+\Im} & \val{-0.61}{0.28} &\val{ 0.36}{+0.12}{-0.34}   &\val{ 0.36}{+0.82}{-0.80}\and\val{-0.46}{0.14}                    & 1.9\\
\U{-0}{-\Re} & \val{0.33}{0.47} &\val{-0.13}{+0.23}{-0.18}   &\val{-0.13}{+0.48}{-0.42}                                         & 0.8\\
\U{-0}{-\Im} & \val{0.69}{0.60} &\val{ 0.17}{0.06}\and\val{0.51}{0.05} &\val{ 0.17}{+0.11}{-0.17}\and\val{ 0.51}{+0.09}{-0.21}  & 0.8\\
\I{-0}{\Re}  & \val{-0.46}{0.76} &  \val{-0.15}{+0.40}{-0.07}  & \val{-0.15}{+0.55}{-0.13}                              & 0.8\\
\I{-0}{\Im}  & \val{-0.57}{0.57} &  \val{ 0.09}{+0.06}{-0.07}  & \val{ 0.09}{+0.11}{-0.23}\and\val{0.43}{0.08}          & 1.3 \\
\hline
\U{+-}{+\Re} & \val{0.08}{0.37} &\val{-1.03}{+0.10}{-0.05}   &\val{-1.03}{+0.32}{-0.10}                                         & 3.3\\ 
\U{+-}{+\Im} & \val{0.17}{0.34} &\val{ 1.04}{+0.04}{-0.08}   &\val{ 1.04}{+0.09}{-0.29}\and\val{-1.00}{+0.14}{-0.06}            & 2.9\\
\U{+-}{-\Re} & \val{-0.43}{0.81} &\val{ 0.44}{+0.35}{-0.54}   &\val{ 0.44}{+0.57}{-1.47}                                         & 0.8\\
\U{+-}{-\Im} & \val{0.68}{0.84} &\val{-0.03}{0.07}           &\val{-0.03}{0.14}\and\val{ 0.97}{+0.10}{-0.35}                    & 1.5\\
\I{+-}{\Re}  & \val{-0.34}{1.06} &  \val{-0.28}{+0.11}{-0.08}\and\val{0.33}{0.35}{-0.42}  & \val{-0.28}{+1.42}{-0.36}   & 0.4\\
\I{+-}{\Im}  & \val{-0.66}{1.05} &  \val{ 1.04}{+0.07}{-0.17}  & \val{ 1.04}{+0.11}{-0.46}\and\val{-0.02}{+0.44}{-0.25} & 1.8\\
\hline
\end{tabular}
\caption{\small\it Indirect \su{2} isospin determination of the \U and \I coefficients compared to their experimental measurement (world average). \label{tab:UI_rhopi}}
\end{center}
\end{table}

\begin{table}[h]
\begin{center}
\begin{tabular}{|l|c||c|c||c|}
\hline
\small Observable & \small Experimental value & \multicolumn{2}{c|}{ \su{2} constraint (dir. measurement not in fit)} & \Pull \\
\cline{3-4}
           & \small (world average)     &  68\% CL  & 95\% CL                         &\\
\hline
$\C^+$    & \val{ 0.24}{0.10}  & \val{-0.86}{+0.28}{-0.13} &  \val{-0.86}{+0.70}{-0.14}\and \val{0.87}{+0.12}{-0.24} & 2.4\\
$\S^+$    & \val{ 0.07}{0.12}  & \val{ 0.95}{+0.04}{-0.29} &$[-1.,+1]$                & 1.4\\
$\C^-$    & \val{-0.31}{0.08}  & \val{-0.99}{+0.25}{-0.01} &  \val{-0.99}{+1.10}{-0.01} & 1.8\\
$\S^-$    & \val{ 0.04}{0.10}  & \val{-0.68}{+0.32}{-0.25} &\val{-0.85}{+0.77}{-0.15} & 2.2\\
$\C^0$    & \val{ 0.19}{0.25}  & \val{-0.62}{+0.44}{-0.25} &  \val{-0.62}{+0.98}{-0.36} & 1.4\\
$\S^0$    & \val{-0.28}{0.36}  & \val{-0.70}{+0.96}{-0.16} &$[-1.,+1]$                & 0.4\\
$f^{00}$   & \val{ 0.09}{0.03}  & \val{ 0.13}{+0.03}{-0.03} &\val{ 0.13}{+0.07}{-0.13} & 0.7\\
$A^{+}$   & \val{ 0.11}{0.06}  & \val{-0.93}{+0.35}{-0.07} &  \val{-0.93}{+0.74}{-0.07} &   2.8 \\
$A^{-}$   & \val{-0.04}{0.09}  & \val{ 0.94}{+0.06}{-0.33} &  \val{ 0.94}{+0.06}{-0.70} &   2.8 \\
\hline
$\C$          & \val{-0.04}{0.06}  & \val{ 0.36}{+0.13}{-0.16} &  \val{ 0.36}{+0.49}{-1.12} & 1.2\\
$\Delta\C$    & \val{ 0.28}{0.07}  & \val{ 0.96}{+0.04}{-0.15} &  \val{ 0.96}{+0.04}{-0.42}\and \val{-0.96}{+0.10}{-0.04} & 2.6\\
$\S$          & \val{ 0.06}{0.08}  & \val{ 0.34}{+0.23}{-0.23} &  \val{-0.34}{+1.30}{-0.46}  & 1.6\\
$\Delta\S$    & \val{ 0.01}{0.08}  & \val{ 0.82}{+0.13}{-0.23} &  \val{ 0.82}{+0.18}{-0.52}   & 2.7\\
$A_{\rho\pi}$  & \val{-0.09}{0.04}  & \val{ 0.93}{+0.07}{-0.33} &  \val{-0.93}{+0.07}{-0.72} & 2.8\\
\hline
\end{tabular}
\caption{\small\it Indirect \su{2} isospin determination of the $B^0\to\pi^+\pi^-\pi^0$ parameters derived from the combined \U and \I  coefficients, compared to their experimental measurement (world average).\label{tab:Q2B_rhopi}}
\end{center}
\end{table}

\begin{table}[h]
\begin{center}
\small
\begin{tabular}{|l|c||c|c||c|}
\hline
\small Observable & \small Experimental value & \multicolumn{2}{c|}{ \su{2} constraint (dir. measurement not in fit)} & \Pull \\
\cline{3-4}
           & \small (world average)     &  68\% CL  & 95\% CL                                 &\\

\hline
\Obs{\B}{$\pm\mp$}{\rho\pi} $(\times 10^6)$  & \val{23.0}{2.3}    & \val{15.5}{+3.4}{-3.1}   &\val{15.5}{+12.5}{-6.1} & 1.6\\
\Obs{\B}{00}{\rho\pi} $(\times 10^6)$  & \val{2.0}{0.51}        & \val{2.3}{+0.4}{-0.5}    &\val{2.3}{0.9}          & 0.5\\
\hline
\Obs{\B}{0+}{\rho\pi} $(\times 10^6)$  & \val{8.3}{1.2}{-1.3}   & \val{22.3}{+67.6}{-8.5}   & $<99.0$               & 1.9\\
\Obs{\B}{+0}{\rho\pi} $(\times 10^6)$  & \val{10.9}{1.3}{-1.5}  & \val{55.0}{+31.0}{-36.0}  &\val{55}{+47}{-44}     & 2.0\\
\Obs{\C}{0+}{\rho\pi}                  & \val{-0.18}{0.17}{0.09}& \val{0.09}{+0.10}{-0.13}\and\val{-0.13}{+0.02}{-0.04}  &\val{0.09}{+0.22}{-0.56}\and\val{0.80}{+0.26}{-0.16}&0.8\\
\Obs{\C}{+0}{\rho\pi}                  & \val{-0.02}{0.11}      & \val{-0.17}{+0.10}{-0.08} &\val{-0.17}{+0.08}{-0.18}\and\val{-0.95}{+0.14}{-0.05}&1.0\\
\hline
\end{tabular}
\caption{\small\it Indirect \su{2} isospin determination of the branching fractions of the charged and neutral \decay{B}{\rho\pi} decays and direct $CP$ asymmetries for the charged modes compared to their experimental measurement (world average).} \label{tab:BR_RhoPi}
\end{center}
\end{table}

\cleardoublepage
\section{Quasi-two-body analysis of $B^0\to a_1^\pm\pi^\mp$ \label{sec:a1pi}}

\begin{figure}[t]
\begin{center}
  \includegraphics[width=18pc]{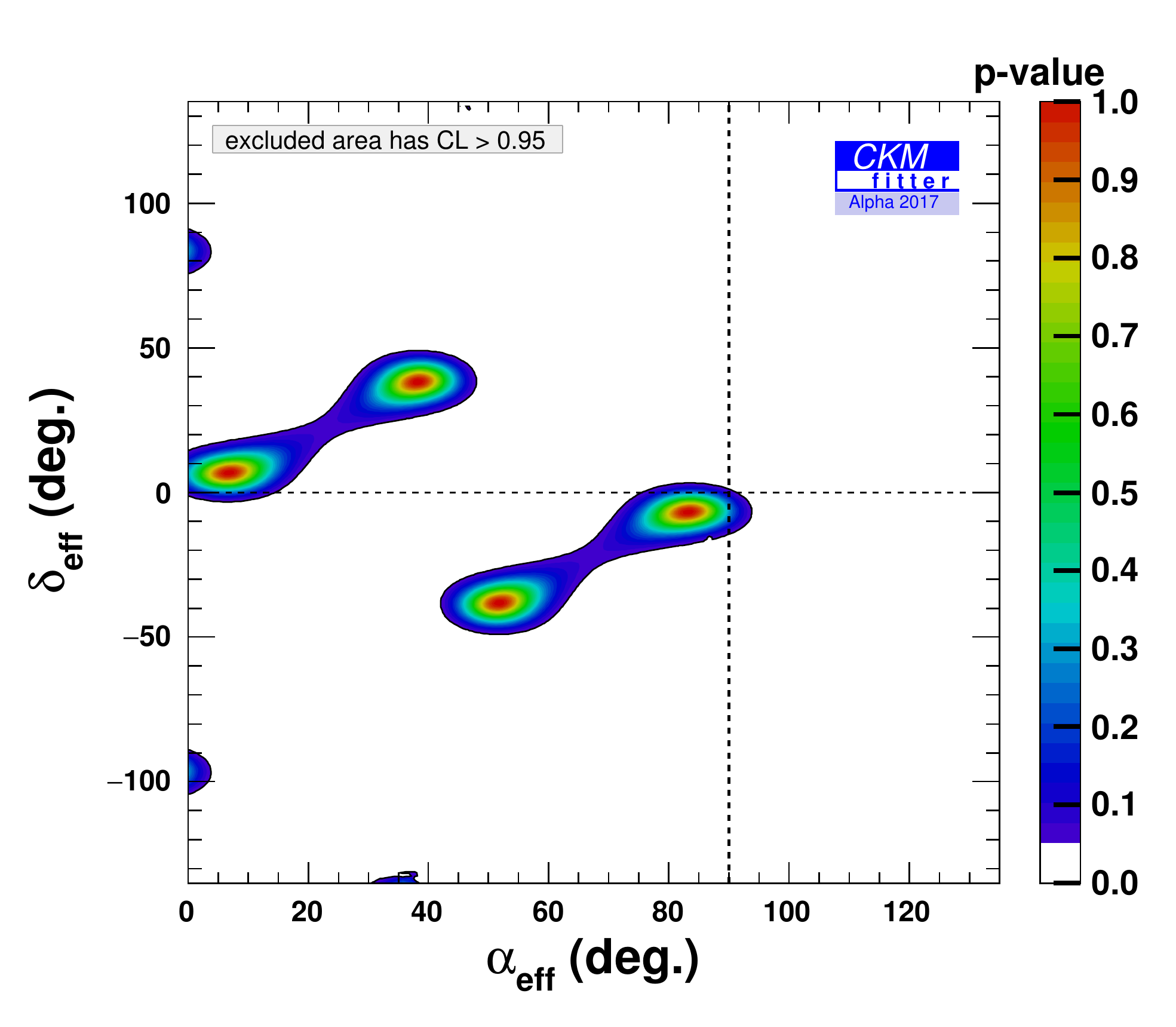}
  \includegraphics[width=18pc]{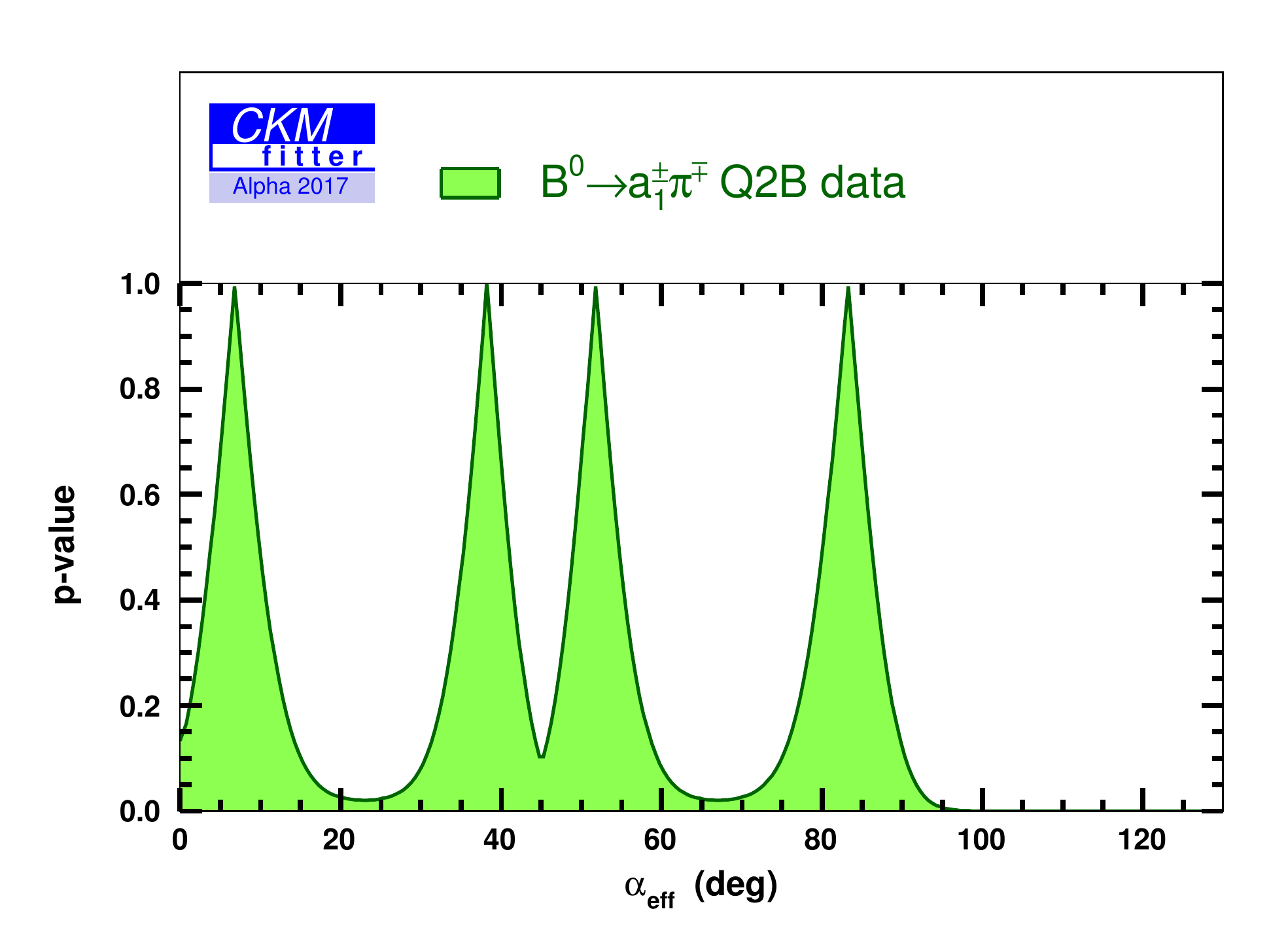}
\caption{\it\small Two-dimensional constraint in the ($\alpha_{\rm eff}$,$\delta_{\rm eff}$) plane from the Q2B analysis of the $B^0\to a_1^\pm\pi^\mp$ decay (left) and one-dimensional projection for $\alpha_{\rm eff}$ (right).}
\label{fig:a1pi}
\end{center}       
\end{figure}

As discussed in Sec.~\ref{sec:rhopiphase}, the quasi-two-body (Q2B) analysis of the neutral $B^0\to\rho^\pm\pi^\mp$ decay provides enough information to extract both the average mixing angle $\alpha_{\rm eff}$, which coincides with the CKM angle $\alpha$ in the limit of a vanishing penguin contribution, and the phase shift $\delta_{\rm eff}$, corresponding to the phase between $\rho^+\pi^-$ and $\rho^-\pi^+$ decay amplitudes. A similar analysis can be performed in the case of the charmless $B^0\to a_1^\pm\pi^\mp$ decay. Fig.~\ref{fig:a1pi} displays the two-dimensional constraint in the ($\alpha_{\rm eff}$,$\delta_{\rm eff}$) plane (left panel) and the corresponding one-dimensional projection on $\alpha_{\rm eff}$ (right panel) when using the world-average $B^0\to a_1^\pm\pi^\mp$ Q2B parameters collected in Ref.~\cite{Amhis:2016xyh}. The 68\% CL intervals on $\alpha_{\rm eff}$ are:
\begin{equation}
\alpha_{\rm eff}(B^0\to a_1^\pm\pi^\mp)=\val{6.8}{+4.7}{-4.3}{$^\circ$}~~\and~~\val{38.2}{+4.0}{-4.4}{$^\circ$}~~\and~~\val{51.8}{+4.4}{-4.0}{$^\circ$}~~\and~~\val{83.2}{+4.3}{-4.7}{$^\circ$}\,.
\end{equation}
The two mirror solutions close to $0^\circ$ and $90^\circ$ are consistent with a vanishing average phase shift $\delta_{\rm eff}$.

\begin{figure}[t]
\begin{center}
  \includegraphics[width=18pc,height=16pc]{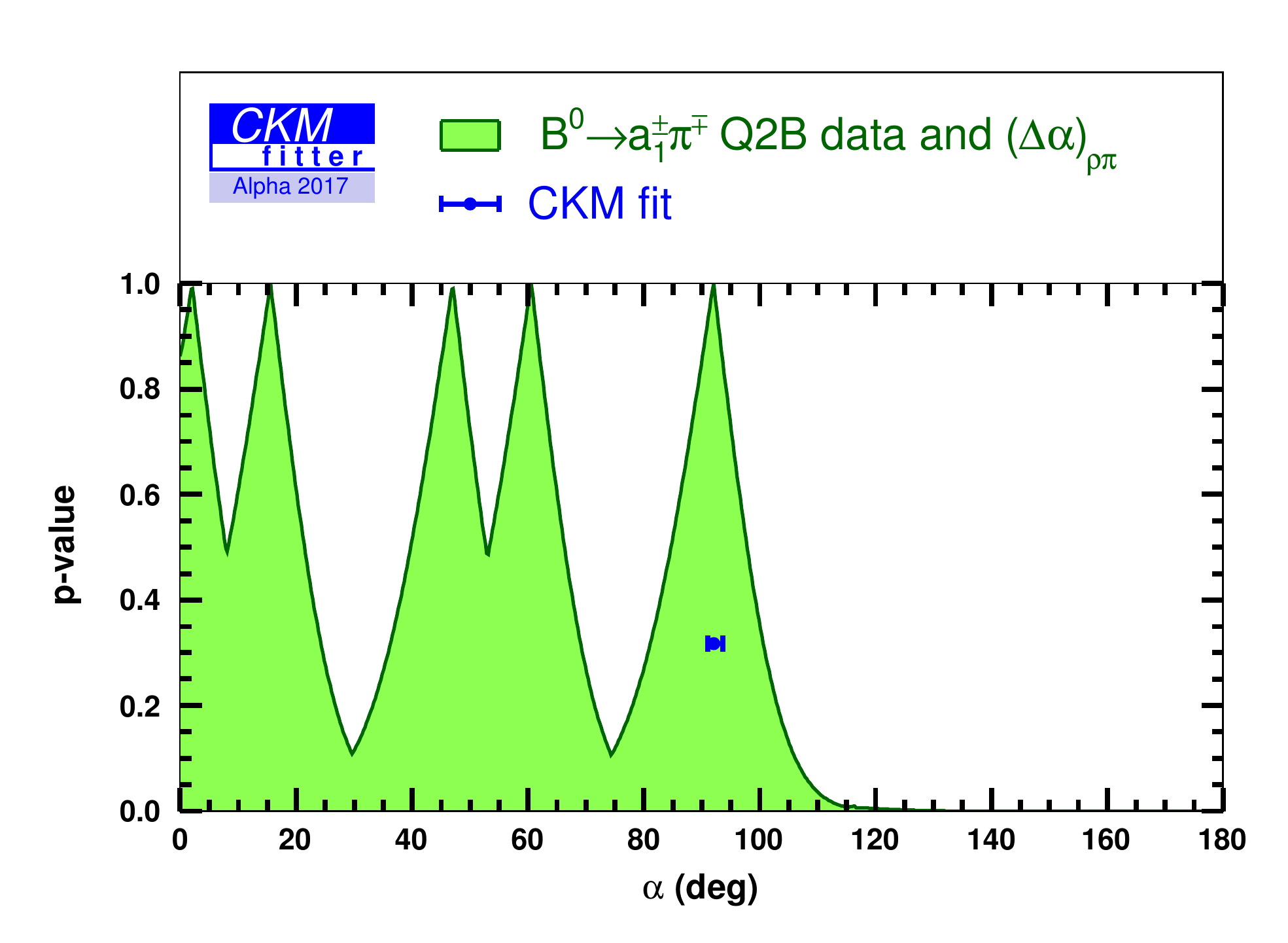}
\caption{\it\small One-dimensional constraint on $\alpha$ derived from the Q2B $B^0\to a_1^\pm\pi^\mp$ analysis, assuming that $\Delta\alpha=(\alpha-\alpha_{\rm eff})$ has the same value as in the case of the $B^0\to (\rho\pi)^0$ Dalitz analysis. The interval with a dot indicates the indirect  determination introduced in Eq.~(\ref{eq:alphaInd}).}\label{fig:a1pi_alpha}
\end{center}       
\end{figure}

The current experimental limit \cite{Amhis:2016xyh} on the branching fraction of the colour-suppressed mode, $\B(B^0\to a_1^0\pi^0) < 1.1\times10^{-3} $ at 90\% CL,  does not allow us to derive a \su{2} bound on the difference $\Delta\alpha=\alpha-\alpha_{\rm eff}$. Considering the flavour-related modes $B\to a_1 K$, a bound based on the  \su{3} symmetry can, however, be defined \cite{Gronau:2005kw}. Such an analysis, performed by the \babar collaboration \cite{Aubert:2009ab}, leads to the constraint $|\Delta\alpha|=|\alpha-\alpha_{\rm eff}({a_1\pi})|<13^\circ$ at 90\% CL, consistent with the value from Eq.~(\ref{eq:DeltaAlpha})  obtained for the $B^0\to(\rho\pi)^0$ decay. Assuming for simplicity that the latter deviation $(\Delta\alpha)_{\rho\pi}$ has the same value in the case of the $B\to a_1\pi$ decay, the constraint on $\alpha$ illustrated by Fig.~\ref{fig:a1pi_alpha} gives~:
\begin{equation}
\alpha_{a_1\pi}:\val{9.0}{+16.0}{-18.0}{$^\circ$}~~\and~~\val{60.6}{+8.5}{-24.3}{$^\circ$}~~\and~~\val{92.0}{+8.7}{-10.8}{$^\circ$} 
\qquad (68\%\ {\rm CL})
\end{equation}
The solution near $90^\circ$ is consistent with the indirect $\alpha$ determination given by Eq.~(\ref{eq:alphaInd}). However, this constraint is based on a hypothesis with a poor theoretical motivation, and it will not be included in our combination of direct $\alpha$ determinations (whereas we include information from the $\pi\pi$, $\rho\rho$ and $\rho\pi$ modes).

\cleardoublepage

\end{document}